\documentclass[12pt,beltcrest]{ociamthesis} 

\usepackage{graphicx,amsmath,amssymb,amstext}
\usepackage{amssymb,amsbsy,amsfonts,amsthm}
\usepackage{url}
\usepackage{graphicx}
\usepackage{dcolumn}
\usepackage{bm}
\usepackage[usenames, dvipsnames]{color}
\usepackage[normalem]{ulem}
\usepackage{subcaption}
\usepackage[]{algorithm2e}
\usepackage{sidecap}
\usepackage[colorlinks,bookmarks]{hyperref}
\definecolor{linkblue}{rgb}{0,0,0.8}
\definecolor{linkgreen}{rgb}{0,0.5,0}

\hypersetup{pdfpagemode=UseNone, pdfstartview=FitH, linkcolor=linkblue, %
            citecolor=linkgreen, urlcolor=linkblue}
\usepackage{amssymb}

\title{Origin \& evolution of the Universe}   

\author{Darsh Kodwani}             
\college{St Hugh's College}  

\degree{Doctor of Philosophy}     
\degreedate{Michaelmas 2019}         

\begin{document}


\setcounter{secnumdepth}{3}
\setcounter{tocdepth}{3}

\maketitle                  
\begin{dedication}
\emph{The best teacher in life is experience.
Experience comes with time.
Time is in the future; which is where I will spend the rest of my life...}
\end{dedication}        
\begin{acknowledgements}

Firstly I would like to thank my main supervisors at both Oxford and Toronto, Pedro Ferreira, David Alonso and Ue-Li Pen. 
In particular, David is almost single handedly responsible for developing my computational skills, and having the patience to help me with many of the silly mistakes I made along the way, for which I will always be indebted to him. 
It has been a real pleasure to work with them throughout my DPhil and it wouldn't have been possible without their constant support and encouragement to pursue my ideas. 
I have also had the pleasure of collaborating with many other amazing scientists throughout my studies, in particular:  Daan Meerburg and Xin Wang have both taken the time to help me in various projects and immensely augmented my research abilities. 
Research in physics is driven by constantly using creative ideas to solve challenging problems. 
I have had the pleasure of working with many of the most creative people in physics; none more than I-Sheng Yang whom I have worked with on several projects.
I-Sheng's ability to solve problems from many different areas of physics and constantly challenging the basic assumptions is something that has always amazed me and this is the approach I have tried to take during my research as well. 
His initial support and guidance in helping me formulating new problems have had the most direct impact on me having a successful research career so far and I will always be grateful for that. 
In Oxford, Harry Desmond has played a similar role. 
His ability to understand the fundamentals of a problem and suggest new approaches to solve them has always been something I have admired. 
Moreover, he has also been a close friend and mentor for me, not to mentioning the constant schooling of new vocabulary that he has given me!
In addition to working on research projects, I have always maintained a friendship with all my collaborators which has made my DPhil an extremely enjoyable endeavour. Thank you all!

Maintaining a good work-life balance and a calm mental state is crucial to being creative and solving challenging problems that inevitably come up during research. 
A lot of credit for me being able to do this has to go the friends I have made along the way. 
Firstly, my office mates in Toronto; Derek and Dana with whom I shared many coffee breaks and stories about our struggles with research, were always a source of comfort and support for me. Also Daniel, my academic twin in Toronto and also coauthor, Leonardo, Tharshi (for making me most welcome in Toronto at first), Vincent, Sonia (for putting up with my lame physics jokes), Gunjan, Simon, Alex (for teaching me smash bros!), Phil (with whom I am also writing a never ending paper) have all been a source of fun and support throughout my time in Toronto and I am sure will continue to be so in the future. 
In Oxford, my incoming class of fellow students have always been a source of comfort and support and being around them is something that I have always enjoyed. In particular, Dina (my academic twin), Christiane, Alvaro, Theresa (and our many ``therapy" sessions in uni parks), Sam, Eva, Shahab, Andrea, Sergio, Emilio, Max (also soon to be co-author), Ryo, Richard (another soon to be co-author), Zahra, Stefan have made my time in Oxford extremely enjoyable and I am sure we will have many enjoyable times in the future as well.

Finally, without the unwavering confidence, support and faith of my parents none of this would be possible. 
They have always encouraged me to pursue my interests and not worry about anything else. 
It is this freedom that has lead to my pursuit of understanding the origin and evolution of the universe.
I am also indebted to my family in India for their affection and support throughout. 
In particular my nephews and niece's, Maitri, Pari, Parth, Dhriti, who have always brightened up even the darkest days. 
Figure \ref{me_fam} shows the people most responsible for keeping me going through the journey of understanding the origin and evolution of the universe. 
\begin{figure}[h!]
\begin{centering}
\includegraphics[scale=0.25]{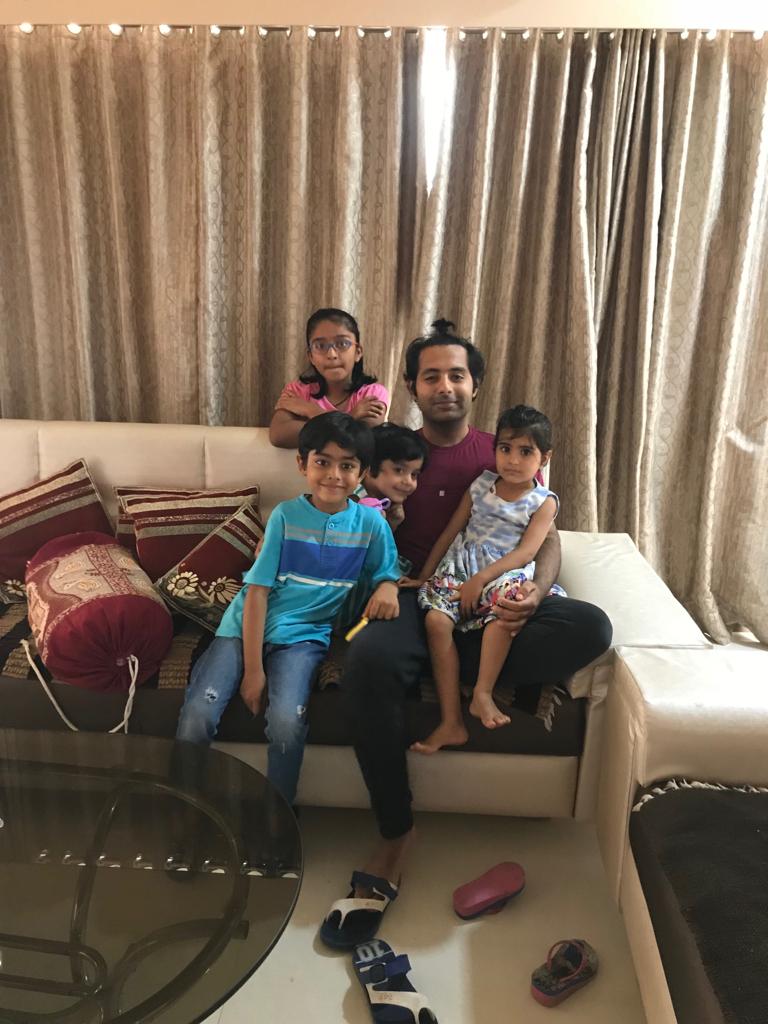}
\caption{Dhriti, Parth, Maitri, Pari and Me}
\label{me_fam}
\end{centering}
\end{figure}

\end{acknowledgements}   
\begin{abstract}

The aim of this thesis is to question some of the basic assumptions that go into building the $\Lambda$CDM model of our universe. The assumptions we focus on are the initial conditions of the universe, the fundamental forces in the universe on large scales and the approximations made in analysing cosmological data. 
For each of the assumptions we outline the theoretical understanding behind them, the current methods used to study them and how they can be improved and finally we also perform numerical analysis to quantify the novel solutions/methods we propose to extend the previous assumptions. 

The work on the initial conditions of the universe focuses on understanding what the most general, gauge invariant, perturbations are present in the beginning of the universe and how they impact observables such as the CMB anisotropies. 
We show that the most general set of initial conditions allows for a decaying adiabatic solution which can have a non-zero contribution to the perturbations in the early universe. 
The decaying mode sourced during an inflationary phase would be highly suppressed and should have no observational effect, thus, if these modes are detected they could potentially rule out most models of inflation and would require a new framework to understand the early universe such as a bouncing/cyclic universe. 

After studying the initial conditions of the universe, we focus on understanding the nature of gravity on the largest scales. 
It is assumed that gravity is the only force that acts on large scales in the universe and we propose a novel test of this by cross-correlating two different types of galaxies that should be sensitive to fifth-force's in the universe. 
By focusing on a general class of scalar-tensor theories that have a property of screening, where the effect of the fifth force depends on the local energy density, we show that future surveys will have the power to constrain screened  fifth-forces using the method we propose. 

Finally, to test theoretical models with observations a complete understanding of the statistical methods used to compare data with theory is required. 
The goal of a statistical analysis in cosmology is usually to infer cosmological parameters that describe our theoretical model from observational data. 
We focus on one particular aspect of cosmological parameter estimation which is the covariance matrix used during an inference procedure. 
The usual assumption in modelling the covariance matrix is that it can be computed at a fiducial point in parameter space, however, this is not self-consistent. 
We check this claim explicitly by calculating the effect of including the parameter dependence in the covariance matrix on the constraining power of cosmological parameters. 

\end{abstract}          

\begin{romanpages}          
\tableofcontents            
\listoffigures              
\end{romanpages}            

\chapter{Introduction}

\section{A brief history of the Universe}

\begin{figure}
\begin{centering}
\includegraphics[scale=0.4]{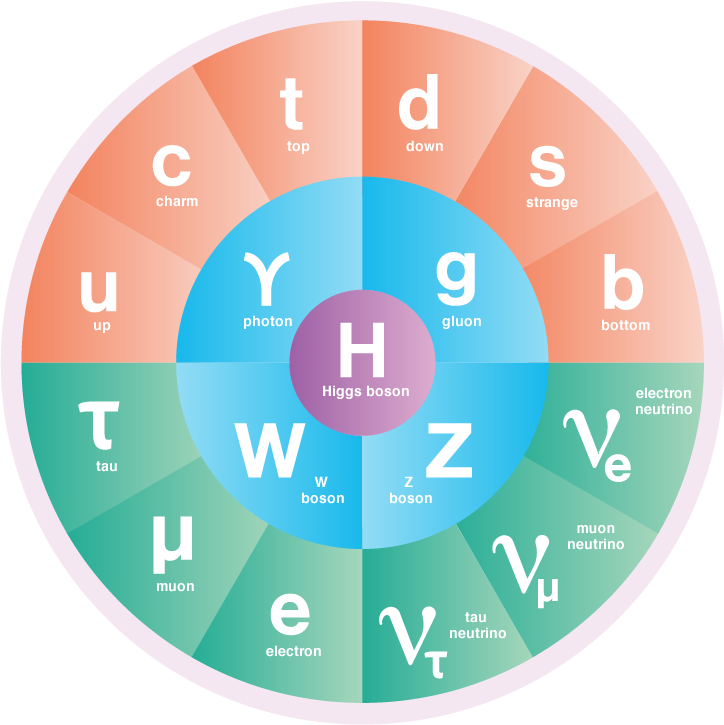}
\caption{Schematic diagram of the Standard Model of particle physics \cite{SM_schem}}
\label{Schematic_SM}
\end{centering}
\end{figure}

This thesis is about the origin and evolution of the Universe.
The dynamics of all the particles traversing through spacetime under the forces of nature is what we call the Universe.
In the Standard Model of particle physics we have a well established, quantum mechanical, framework to understand the fundamental particles and the forces that lead to interactions between them. 
In particle physics the convention is to classify the fundamental particles into quarks and leptons and the forces are characterised by force carrying particles called bosons. 
There are 12 fundamental particles, and 12 conjugate anti-particles, (6 quarks and 6 leptons) and 5 force carrying gauge bosons (4 spin 1 bosons and 1 spin 0 boson) corresponding to three fundamental forces: electromagnetism, weak nuclear force and strong nuclear force. 
A schematic diagram showing the Standard Model is shown in figure \ref{Schematic_SM}.
This picture of the Universe has been tested and confirmed in experiments spanning over 5 decades \cite{review_pp}. 
A summary of the properties of the Standard Model particles is given in table \ref{SM_summary}.

\begin{table}
\begin{centering}
\begin{tabular}{ p{3cm}  p{3cm} p{3cm}}
\hline
\hline
Particle & Mass & Spin \\
\hline
\hline
t & 173 GeV & 1/2 \\
b & 4 GeV & 1/2 \\
c & 2 GeV & 1/2 \\
s & 100 MeV & 1/2 \\
d & 5 MeV & 1/2 \\
u & 2 MeV & 1/2 \\
$\tau$ & 1777 MeV & 1/2 \\
$\mu$ & 106 MeV & 1/2 \\
e & 511 keV  & 1/2 \\
$\nu_\tau$ & $<$ 0.6 eV & 1/2 \\
$\nu_\mu$ & $<$ 0.6 eV & 1/2 \\
$\nu_e$ & $<$ 0.6 eV & 1/2 \\
$W^{\pm}$ & 80 GeV & 1 \\
$Z$ & 91 GeV & 1 \\
$\gamma$ & 0 & 1 \\
$g$ & 0 & 1 \\
$H$ & 125 GeV & 0 \\
\hline \hline
\end{tabular}
\caption{Summary of the mass and spin of the Standard Model particles}\label{SM_summary}
\end{centering}
\end{table}

There is one basic limitation of the Standard Model of particle physics, however, which is that it doesn't contain gravity. 
In particular, there is no self-consistent quantum field theory of a spin 2 gauge boson that can describe gravity.  
The current best description of gravity is given by general relativity (GR). 
In GR gravity isn't described as a force. 
Rather the fundamental description of gravitation relies on the existence of spacetime: a four dimensional hyperbolic geometric manifold that accounts for our three dimensions of space and one dimension of time. 
The two pillars upon which GR describes gravity are the following:
\begin{enumerate}

\item Particles traverse through spacetime on geodesics.

\item Particles in spacetime will change the geometry of spacetime. 

\end{enumerate}
The motion of particles through spacetime can be further divided into three categories. 
Particles that travel on timelike, null or spacelike geodesics.
All the particles which have a positive mass follow timelike geodesics. 
In the Standard Model of particle physics all particles have a positive mass (anti-particles have a positive mass just the opposite electric charge to the conjugate particle) except two of the gauge bosons, the photon and gluon, which are massless (although gluons are never observed to be massless as Quantum-Chromo-Dynamics (QCD) is asymptotically free) and they follow null geodesics.
Particles that follow spacelike geodesics must have an imaginary mass and are termed tachyons however these have never been observed in nature.
The theory of GR has stood the test of all experiments conducted till today spanning over a century \cite{Will2014}. 
More impressively GR has been tested over a large range of energy scales, from the lab \cite{Gundlach_2005} to cosmological scales \cite{Ishak:2018his}. 
Most recently the detection of gravitational waves have provided further spectacular experimental confirmation of GR \cite{TheLIGOScientific:2016src}. 

Cosmology is the study of all of the particles and forces in the Standard Model of particle physics on the largest scales. 
On the largest scales gravity is the most important force that needs to be accounted for, even though it is the weakest of all the forces, in order to determine the dynamics of objects. 
There are a few exceptions to this where the conditions can be more extreme when the other forces play a role, for instance in the very early Universe prior to the electroweak and QCD phase transitions and also in neutron stars and other compact objects. 
We will mention these exceptions when necessary, however, for the most part we will focus on the dynamics under gravity. 
Cosmology has come from being an unexplored territory, typically left to philosophers, to being one of the most precisely tested areas of experimental science. 
While GR and its implications have been known for around a century, its only in the last few decades that cosmology has become a field open to experimental investigation. 
Initially the experimental tests of cosmology were driven by measurements of the cosmic microwave background radiation (CMB) \cite{Penzias:1965wn, wmap, Aghanim:2018eyx}. 
More recently measurements of the galaxy clusters \cite{galaxy_clustering} and the lensing of light from galaxies \cite{Mandelbaum:2017jpr} have augmented the cosmological information obtained by the CMB by orders of magnitude making cosmology not only an experimental science but a science that is in the era of precision measurements. 
While these are the probes that will be the focus of this thesis, there are several other probes that also provide us with crucial cosmological information. 
In particular the measurement of the primordial abundance of elements and the measurements of distances using supernovae are crucial to understanding several cosmological parameters. 
As cosmology has entered the realm of precision science, the use of advanced statistical methods has become mandatory in order to extract cosmological information from increasingly large and complex datasets.
We describe the details of the cosmological probes and statistical methods in upcoming sections, however, before doing that we describe the understanding of the Universe these cosmological observations have provided us with.

According to our current understanding the Universe starts roughly 13.7 billion years ago where the spacetime is expanding. 
All the particles in the Standard Model are relativistic in the very early Universe and we describe that phase to be the radiation dominated era. 
In this phase all the particles have perturbations in their energy densities as well as perturbations in the spacetime itself. 
The current standard model of cosmology, called the $\Lambda$CDM model, is the story of how the Universe starts from this initial radiation dominated era and forms the structure we see in it today. 

\section{$\Lambda$CDM model}

\subsection{Background}

As GR is at the heart of the $\Lambda$CDM model we start by writing down its action.
\begin{equation}
	S_{\text{GR}} = \frac{1}{16 \pi G} \int d^4x \ \sqrt{-g} R
\end{equation}
we are using the standard notation where $g_{\mu \nu}$ is the metric, $g$ is the determinant of the metric, $R_{\mu \nu}$ is the Riemann tensor, $R = g^{\mu \nu} R_{\mu \nu}$ is the Ricci scalar. 
For generic matter content described by fields $\Phi_i$, the action in the presence of gravity can be written as 
\begin{equation}
	S_{\text{matter}} = \int d^4x \ \sqrt{-g} \ \mathcal{L}_{\text{matter}}(g, \Phi_i). 
\end{equation}
If the Standard Model of particle physics is the only matter content in the Universe, which is what has been proven so far, then the action that describes the full content of particles and spacetime in our Universe is given by the sum of the GR action and the Standard Model action $S_{\text{SM}}$\footnote{The full form of the SM action can be found here \cite{SM_action} - given its complexity and length we do not write it out in full here.}
\begin{equation}
	S_{\text{Universe}} = S_{\text{GR}} + S_{\text{SM}}.
\end{equation}
The equation of motion for the metric or the particles can be obtained by varying the equation w.r.t to the appropriate field. 
The Einstein Field Equations (EFE) that describe the dynamics of the metric are given by varying the $S_{\text{GR}}$ w.r.t $g_{\mu \nu}$ 
\begin{equation}
	G_{\mu \nu} = 8 \pi G T_{\mu \nu}
\end{equation}
where $G_{\mu \nu} \equiv R_{\mu \nu} - \frac{1}{2} Rg_{\mu \nu}$ with the stress-energy tensor 
\begin{equation}
	T_{\mu \nu} = \frac{2}{\sqrt{-g}} \frac{\delta \left[ \sqrt{-g} \mathcal{L}_{\text{matter}} \right] }{\delta g^{\mu \nu}}.
\end{equation}
One of the main assumptions in the $\Lambda$CDM model is that the Universe is isotropic (preserves SO(3) symmetry) and is homogenous on the largest scales. 
This is known as the \emph{cosmological principle}. 
The solution of the EFE under these symmetries is given by the Friedmann-Robertson-Walker (FRW) metric, which in spherical coordinates, can be written as 
\begin{equation}
	ds^2 = -dt^2 + a(t)^2 \left( \frac{dr^2}{1 - \kappa r^2} + r^2 d \Omega_2^2 \right). \label{frw_cosmic}
\end{equation}
Here we have defined $a(t)$ as the scale factor,  which describes the expansion of the three spatial dimensions. $r$ is the unit-less comoving radius normalised by the current radius of the Universe, $\kappa$ is the curvature of spatial surfaces and $d \Omega_2^2 = d \theta^2 + \sin^2 \theta d \phi^2$ is the solid angle element.
The curvature $\kappa$ can take on three possible values: 
\begin{itemize}
\item greater than zero for \emph{positive} curvature (spherical),
\item less than zero for \emph{negative} curvature (hyperbolic),
\item zero for \emph{zero} curvature. 
\end{itemize}
While all three of these values are consistent with the cosmological principle, observationally we find the Universe is consistent with being flat.
The dynamics of the Universe can therefore be determined by scale factor $a(t)$. 
To solve for the scale factor we need information about the stress-energy tensor. 

The most important stress-energy tensor in cosmology is that of a perfect fluid given by 
\begin{equation}
	T^{\mu \nu} = (\rho + P) u^\mu u^\nu + P g^{\mu \nu},
\end{equation}
$\rho$ is the energy density of the fluid, $p$ is the pressure of the fluid and $u^\mu$ is the 4-velocity of the fluid.
The time-time component of the EFE gives the Friedmann equation
\begin{equation}
	H^2 = \frac{8 \pi G}{3} \rho - \frac{\kappa}{a^2} + \frac{\Lambda}{3} \label{friedman1}
\end{equation}
where $H \equiv \frac{\dot{a}}{a}$.
In this equation we have also included $\Lambda$ which represents the cosmological constant. 
This can be thought of as a fluid with a given equation of state, as we describe below, or as an additive constant to the Einstein-Hilbert action. 
It is often convenient to define the critical density of the Universe, at which there is no spatial curvature and no cosmological constant, as 
\begin{equation}
	\rho_c \equiv \frac{3 H^2_0}{8 \pi G}.
\end{equation}
Using this definition the density of different particle species is commonly defined as 
\begin{equation}
	\Omega_i = \frac{\rho_i}{\rho_c}. \label{density_omega}
\end{equation}
In a cosmological context the particle species are defined in a different way than in the Standard Model of particle physics. 
For the purposes of cosmological evolution the particle species are defined just by their equation of state (as we describe below). 
The particle content in our cosmological model are radiation, matter, curvature and dark energy (cosmological constant). 
There may be further species depending on what context the Friedmann equation is being solved, however we will introduce these new species as and when needed. 
It is also worth mentioning the distinction between different types of matter. 
In particular we know there is matter that interacts with light, which we call baryonic matter and matter that does not interact with light called dark matter. 
Our current understanding leads us to believe that dark matter does not carry any significant kinetic energy and thus termed \emph{cold dark matter} (CDM). 

The particle species in cosmology are usually defined in terms of their stress-energy tensor.
The equation of motion of a fluid can be obtained from the conservation of the stress-energy tensor
\begin{equation}
	\nabla_\mu T^\mu_\nu = 0 \Rightarrow \dot{\rho} + 3 H ( \rho + P) = 0. \label{continuity}
\end{equation}
This is called the continuity equation.
The space-space part of the EFE gives the only other equation (as there are no time-space terms), as called the acceleration equation, 
\begin{equation}
	\frac{\ddot{a}}{a} = - 4 \pi G (\rho + 3P) + \frac{\Lambda}{3}.
\end{equation}
This can also be obtained by the combination of the Friedman equation in Eq \ref{friedman1} and the continuity equation in Eq \ref{continuity} and therefore is not an independent equation. 
In order to fully solve for the dynamics of the scale factor in the FRW metric one also needs to specify a relation between the energy density and pressure. 
This is typically written as 
\begin{equation}
	{P = w \rho}
\end{equation}
where $w$ is called the equation of state parameter.
The most common equations of state are for radiation where $w = \frac{1}{3}$, dust $w = 0$ and the cosmological constant $w = -1$. 
For any of these constant values for $w$ we can solve the continuity equation to get 
\begin{equation}
	\rho = \rho_0 a^{-3(1 + w)}.
\end{equation}
In the case of radiation we see that $\rho \propto a^{-4}$ and $\rho \propto a^{-3}$ for dust, or pressure-less matter, as we expect.
Solving the Friedmann and continuity equations can uniquely determine the evolution of any cosmology with a given matter content. 
Using the definition of the densities given in Eq \ref{density_omega} we can rewrite the Friedmann equation for the different particle species described above
\begin{equation}
	H^2 = H_0^2 \left[ \Omega_{r} \left(\frac{a_0}{a} \right)^4  + \Omega_m \left(\frac{a_0}{a} \right)^3 + \Omega_\kappa \left(\frac{a_0}{a} \right)^2 + \Omega_{\Lambda} \right]. \label{friedmann_hubble}
\end{equation}
Here we have defined the curvature as a density as well $\Omega_\kappa \equiv - \frac{\kappa}{(a_0 H_0)^2}$. All quantities with a zero subscript are evaluated at $z=0$, i.e today. 
Using Eq \ref{friedmann_hubble} we can solve for the scale factor as a function of time. 
Before looking at some of the solutions to this equation it is worth defining a new time coordinate called conformal time. $\tau$. 
It is related to the cosmic time variable $t$ defined in the metric in Eq \ref{frw_cosmic} by $d\tau \equiv \frac{dt}{a(t)}$. 
In these coordinates the FRW metric becomes
\begin{equation}
	ds^2 = a(\tau)^2 \left( -d \tau^2 + \left( \frac{dr^2}{1 - \kappa r^2} + r^2 d \Omega_2^2 \right) \right).
\end{equation}
By solving the Friedmann equation we can solve the dynamics of the scale factor, and thus the Universe on large scales, for three different regimes which correspond to our observed Universe. 
\begin{itemize}
\item Radiation domination: This is when the Universe consists of a fluid of relativistic particles. The early Universe is extremely hot and dense and thus a single relativistic fluid is a good approximate description of the very early Universe. 
The equation of state is $\omega = \frac{1}{3}$, the energy density evolution is $\rho \propto a^{-4}$ and the evolution w.r.t cosmic and conformal time is $a(t) \propto t^\frac{1}{2}$ and $a(\tau) \propto \tau$ respectively. 

\item Matter domination: This is when the Universe consists only of a pressure-less fluid. As the Universe expands and cools the phase after the radiation dominated phase is the matter dominated phase. 
The equation of state is $\omega = 0$, the energy density evolution is $\rho \propto a^{-3}$ and the evolution w.r.t cosmic and conformal time is $a(t) \propto t^\frac{2}{3}$ and $a(\tau) \propto \tau^2$ respectively. 

\item Vacuum energy ($\Lambda$) domination: We observe that the current evolution of the Universe is described by a vacuum energy dominated phase.
The equation of state is $\omega = -1$, the energy density evolution is $\rho = $ constant and the evolution w.r.t cosmic and conformal time is $a(t) \propto e^{Ht} $ and $a(\tau) \propto -\frac{1}{\tau}$ respectively. 

\end{itemize}
Now that we know the evolution of the Universe we can analyse the interactions of the particles.
The interactions between particles can be calculated in quantum field theory and depends on the coupling strength between the particles. 
The interaction rate is typically written as $\Gamma$. 
In the early Universe, when the temperature and density is high, the Universe is a thermal soup of all radiation with all the particles being constantly created an annihilated. 
When the rate of the interaction between a specific species of particles drops below the Hubble expansion $H$, then the particle species will decouple from the primordial plasma.
Different particle species have different interaction rates and thus decouple from the primordial plasma at different times.
This is a crucial concept that allows us to calculate when important events in the history of the Universe happen and thus test our model of the Universe.

We can briefly describe the important events in the history of the Universe chronologically. 
The origin of the Universe is described by a yet unknown mechanism, but the most widely accepted theory for this is \emph{inflation}. Inflation encompasses a set of theories that predict there was a period prior to radiation domination in our Universe where the Universe expands at an accelerated rate (mostly assumed to be an exponential expansion) \cite{Guth:1980zm}. 
There are large number of theories that can be written in the context of inflation, however no single model has as yet been established as the true theory describing the beginning of the Universe \cite{Martin:2013tda}.
Typically inflation is assumed to happen below the Planck scale (the Planck time is $\approx 5 \times 10^{-44}$ s). 
There are, however, other models that can be used to describe the beginning of the Universe such as bounce/cyclic models of the Universe \cite{Khoury:2001wf, Lehners:2008vx}. 
Following some process that sets up the initial conditions for the evolution of the Universe, we expect a process called \emph{Baryogenesis} to take place. 
This process that describes the origin of matter in our Universe. 
If there is equal amount of matter and antimatter in the Universe initially, then the Universe should just be filled with photons as all the particles annihilate. 
However, that is not what we see. 
It appears there is one additional matter particle to every $10^9$ anti-matter particle. 
This is usually quantified as the ratio of baryons (which is most of the observed matter in the Universe) to photons that we observer in the Universe, 
\begin{equation}
	\eta \equiv \frac{\#_{baryons}}{\#_\gamma} \approx 10^{-9}.
\end{equation}
The origin of the asymmetry in matter and anti-matter is still a mystery with many possible solutions being speculated \cite{Riotto:1999yt}.
We know that when the temperature of the primordial plasma reaches 100 GeV the {particles} will get their mass through the Higgs mechanism and this is known as the \emph{Electroweak} (EW) phase transition. 
At 150 MeV the quarks and gluons become coupled and form composite structures: baryons and mesons. 
One of the mysteries of the $\Lambda$CDM model is the origin and nature of dark matter. 
It is typically assumed that dark matter is a thermal relic, i.e is a particle that was present at the origin of the Universe and decouples from the primordial plasma at early times, after which is it traverses our Universe un-impeded. 
The most popular theory of dark matter is that it consists of weakly interacting massive particles (WIMPs). 
These types of particles only interact through the weak force and typically decouple from the plasma at $\sim$ 1 MeV.\footnote{The precise values depend on the model being used to describe the WIMP.} 
Neutrinos also interact only through the weak force (however are not massive enough to account from the required dark matter) and given their mass they decouple from the primordial plasma at 0.8 MeV. 
Following this the electrons and positrons are no longer in thermal equilibrium after the temperature reaches $\sim 500$ keV. 
At this stage the energy in the electrons and positrons gets transferred into the photons they produce, however the neutrinos do not receive any of the energy as they have already decoupled (this is why the temperature of the photons is higher than the temperature of the neutrinos today). 
At a temperature of around 100 keV the Universe has cooled enough such that the first nuclei start forming. 
This process is called \emph{Big Bang Nucleosynthesis} (BBN). 
From the Friedmann equation in Eq \ref{friedmann_hubble} we can see that there will come at time at which the radiation will start to become subdominant to the matter in the Universe and thus the behaviour of the scale factor will change. 
This happens at $z \sim 3400$ and is called the \emph{matter-radiation equality}. 
Below this redshift the first structures can start to form under the collapse of matter. 
At a temperature of $\sim$ 0.3 eV the photons and electrons decouple and the mean free path of light suddenly increases to $\sim \infty$. 
At this time, $z \approx 1100$, the Universe first becomes transparent and this is called \emph{recombination}. 
This is followed by a phase known as the ``dark ages". In this phase there are no stars and thus no light is being produced.
Neutral hydrogen that is present in the Universe after recombination (produced during BBN) is ionised again by high energy photons in a process called \emph{reionisation}. 
The source of these high energy photons is not well understood, however, popular candidates for ionising photons are dwarf galaxies, Quasars and population III {stars}. The redshift at which this happens is also unclear however the most popular ideas suggest it happens between $\sim$ 5-20 \cite{Loeb:2000fc}.

Finally, again from the Friedmann equation we see there must also come a phase when dark energy dominates the energy content of the Universe and thus starts accelerating the expansion of the Universe.
This happens at a redshift of $\sim$ 0.4. 

\begin{figure}
\begin{centering}
\includegraphics[scale=0.4]{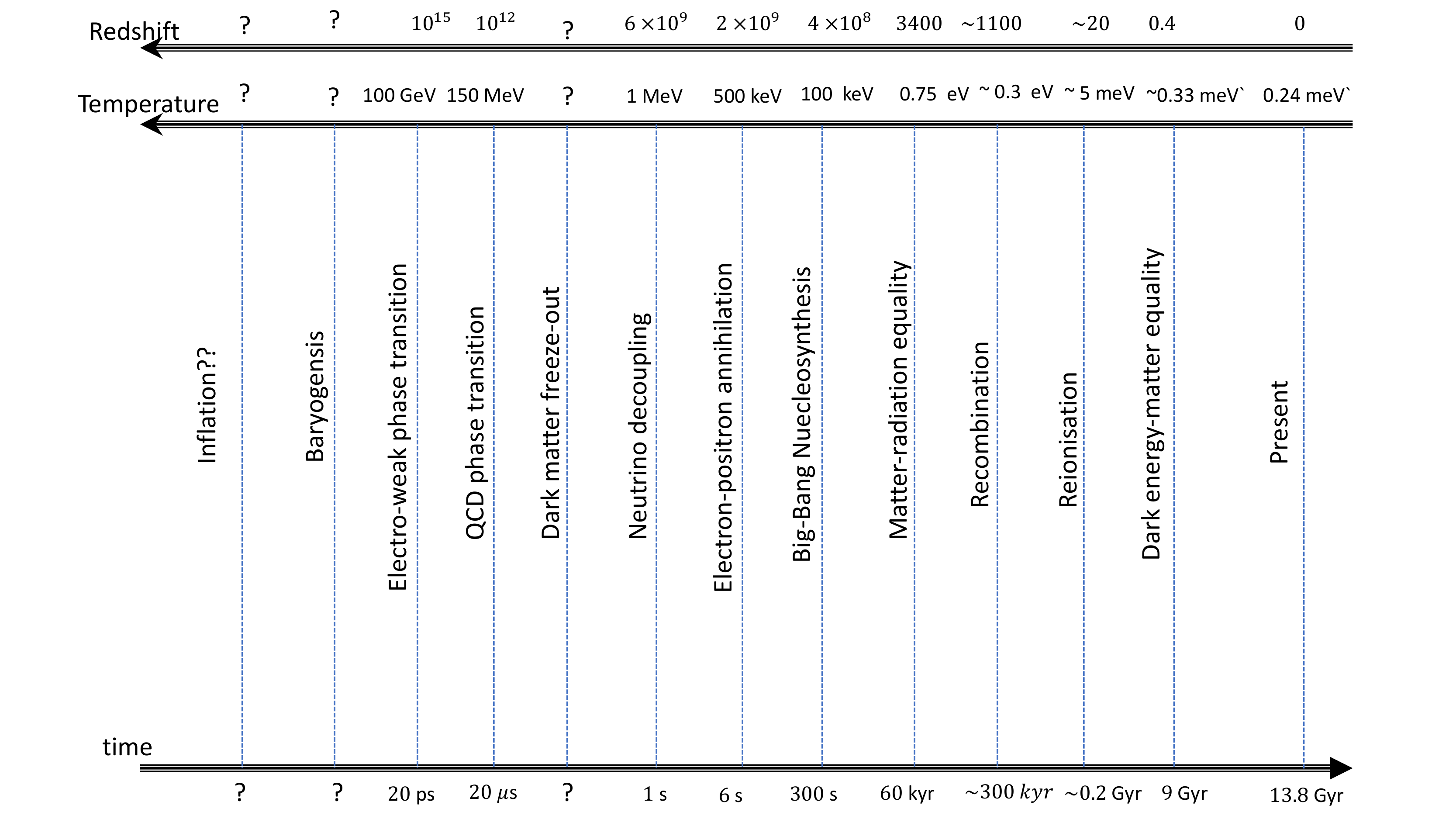}
\caption{Schematic diagram showing the key events in the history of the Universe according to the $\Lambda$CDM model.}
\label{cosmo_hist}
\end{centering}
\end{figure}

The precise values of the time/temperature at which these events happens can be used to test our cosmological model and infer the parameters that best fit the model given observational data. 
A summary of each of these events is given in figure \ref{cosmo_hist}. 
In the next section we briefly outline how we go about calculating predictions based on $\Lambda$CDM by describing perturbation theory in the context of $\Lambda$CDM.

\subsection{Perturbations}

In order to make quantitative predictions from the $\Lambda$CDM model, we need to write down the perturbations to Einstein's equations, 
\begin{equation}
	\delta G_{\mu \nu} = 8 \pi G \delta T_{\mu \nu}.
\end{equation}
We start by analysing the metric first.
We already know the metric is well approximated on large scales by the FRW metric. 
In general the metric has 10 degrees of freedom, in the absence of curvature, (as it is a 4x4 symmetric matrix) and the most general set of {linear} perturbations to the FRW metric can be written as \cite{cosmo_lecs}
\begin{equation}
	ds^2 = a^2(\tau) \left[ - (1 + 2A) d \tau^2 + 2 B_i dx^i d \tau - (\delta_{ij} + h_{ij} ) dx^i dx^j \right].
\end{equation}
It is typical to perform a scalar-vector-tensor (SVT) decomposition of the perturbations in order to identify the scalar, vector and tensor perturbations individually.
A generic rank-2 symmetric matrix, $h_{ij}$, can be decomposed into a scalar, vector and tensor variable as follows
\begin{equation}
	h_{ij} = 2C\delta_{ij} + 2\left( \partial_i \partial_j - \frac{1}{3} \delta_{ij} \nabla^2 \right) E + \left(\partial_i \hat{E}_j + \partial_j \hat{E}_i \right) + 2 \hat{E}_{ij},
\end{equation}
with $\partial^i \hat{E}_i \equiv 0, \partial^i \hat{E}_{ij} \equiv 0, \hat{E}^i_i = 0$.
A generic vector can be written in terms of a scalar and a divergence-less vector
\begin{equation}
	B_i = \partial_i B + \hat{B}_i
\end{equation}
where $\partial^i \hat{B}_i \equiv 0$. 
Using these transformations we see that there are indeed 10 degrees of freedom in the metric where 4 of them are scalars, $A, B, C, E$, 4 are in the vectors $ \hat{B}_i, \hat{E}_i$ and 2 are in the tensor $\hat{E}_{ij}$. 
However, in GR one needs to make sure the terms we write down are the same in any coordinate system as any physical quantity should not change by the redefinition of coordinates (also called gauge). 
It can be shown that in the case of the FRW metric perturbations defined above, the following variables are invariant under any gauge transformation
\begin{eqnarray}
	\Psi & \equiv & A + \mathcal{H}(B - E')  + (B-E')', \nonumber \\
	\Phi & \equiv & -C - \mathcal{H}(B - E') + \frac{1}{3} \nabla^2 E, \nonumber \\
	\hat{\Phi}_i & \equiv & \hat{E}'_i - \hat{B}_i, \nonumber \\
	\hat{E}_{ij}. & & 
\end{eqnarray}
These are typically called the \emph{Bardeen variables}. 

One of the most common gauge choices is the \emph{Newtonian} gauge which is useful for analysing scalar perturbations and is defined by setting $B = E = 0, A = \Psi$ and $C = - \Phi$
\begin{equation}
	ds^2 = a^2(\tau) \left( - (1 + 2 \Psi) d \tau^2 + (1 + 2 \Phi) \delta_{ij}  dx^i dx^j \right). \label{metric_newtonian}
\end{equation}
We will use this when we come to describe the initial conditions of the Universe. 
There is one further simplification that is usually used: we assume there is no anisotropic stress, which implies $\Phi = \Psi$. This however is not true, for example in the early Universe after neutrino decoupling as neutrinos then free stream and generate anisotropic stress.
Along with this, a popular gauge for computational purposes is the synchronous gauge. 

Similarly we can write down the general first order perturbations to the stress energy tensor of a perfect fluid as, $\delta T^\mu_\nu$
\begin{equation}
	\delta T^\mu_\nu = (\delta \rho + \delta P) \bar{U}^\mu \bar{U}_\nu + (\bar{\rho} + \bar{P})(\delta U^\mu \bar{U}_\nu + \bar{U}^\mu \delta U_\nu) - \delta P\delta^\mu_\nu - \Pi^\mu_\nu.
\end{equation}
The quantities with a bar on top correspond to the background values. 
By applying the conservation of stress energy we arrive at the relativistic Euler and continuity equations respectively
\begin{eqnarray}
	& & \vec{v}' + 3 \mathcal{H} \left( \frac{1}{3} - \frac{\bar{P}'}{\bar{\rho}'} \right) \vec{v} = - \frac{\vec{\nabla} \delta P}{\bar{\rho} + \bar{P}} - \vec{\nabla} \Phi \nonumber \\
	& & \delta' + 3 \mathcal{H} \left( \frac{\delta P}{\delta \rho} - \frac{\bar{P}}{\bar{\rho}} \right) \delta = - \left( 1 + \frac{\bar{P}}{\bar{\rho}} \right) (\vec{\nabla} \cdot \vec{v} - 3 \Phi')
\end{eqnarray}
where $'$ is the derivative w.r.t conformal time and $\mathcal{H}$ is the conformal Hubble parameter.
The perturbed Einstein equations for the time-time, time-space and the trace part are respectively given by
\begin{eqnarray}
	& & \nabla^2 \Phi = 4 \pi G a^2 \bar{\rho} \delta + 3 \mathcal{H} ( \Phi' + \mathcal{H} \Phi), \\
	& & \Phi' + \mathcal{H} \Phi = - 4 \pi G a^2(\bar{\rho} + \bar{P}) v, \\
	& & \Phi'' + 3 \mathcal{H} \Phi' + (2 \mathcal{H}' + \mathcal{H}^2) \Phi = 4 \pi G a^2 \delta P.
\end{eqnarray}
These equations hold for any perfect fluid without anisotropic stress. 
We will use these equations to build an intuition for the decaying adiabatic perturbation that we will describe in the chapter discussing the initial conditions of the Universe. 

\section{Probes of cosmology}

The aim of this section is to outline the cosmological probes that can be used to test the $\Lambda$CDM model.
We will discuss three probes in more detail that are used in the research presented in this thesis. 
There are several other probes that we only mention in passing, however, they also provide extremely valuable information for testing the $\Lambda$CDM model.

\subsection{Cosmic microwave background}

The Universe becomes transparent after recombination at a redshift of around 1100. 
This light was first detected in 1965 \cite{Penzias:1965wn}. 
Followup experiments such as COBE \cite{Jaffe:2000tx} measured the statistical properties of the temperature of this light and found it to be the same in all parts of the sky to {$\sim$1 part in $10^4$}. 
{This is often interpreted as confirmation of the cosmological principle, however this is not the case as the to test this requires measurements of the CMB from at least two different points in spacetime. 
A more thorough discussion of this can be found in \cite{Clarkson:2010uz} and references therein - for the purposes of this thesis we assume the cosmological principle holds and thus the FRW metric is a fair representation of spacetime on large scales. }
The small differences in temperature, on the other hand, can further test our models as we can use perturbation theory to calculate the statistical properties of these temperature fluctuations for a given cosmological model. 
In recent years the temperature anisotropies have been mapped in exquisite detail by the WMAP \cite{wmap} and Planck satellites \cite{Aghanim:2018eyx}.
In addition to temperature anisotropies, they have also mapped polarisation anisotropies from the CMB.

\begin{figure}
\begin{centering}
\includegraphics[scale=0.6]{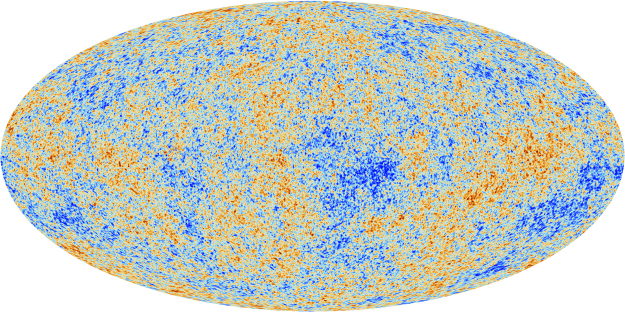}
\caption{All sky map of the CMB {taken from} the Planck satellite \cite{planck_allskymap}.}
\label{planck_allskymap}
\end{centering}
\end{figure}

\subsubsection{Temperature anisotropies}

We start by analysing how the temperature anisotropies are produced in the CMB. 
As photons traverse through spacetime they will feel the effects of gravity and it is these effects that we can calculate and test against observations. 
Fundamentally, the temperature of a CMB photon measured by an observer on Earth is given by the dot product of the observer's four vector and the photons {four momentum}\footnote{In units where $\hbar = k_B = 1.$}. 
The observer is usually situated on the Earth thus it will be in the reference frame of the Earth and its motion. 
To calculate the four vector of the photon observed today we need to account for the motion of the photon as it traverses along a geodesic towards the Earth. 
The photons follow null geodesics and the energy of a photon is simply given by the $0$ component four momenta. 
Using the perturbed Newtonian metric in Eq \ref{metric_newtonian} and the geodesic equation
\begin{equation}
	\frac{\partial P^\mu}{\partial \tau} = - \Gamma^\mu_{\alpha \beta} P^\alpha P^\beta
\end{equation}
where $P^\mu$ is the four momentum, we can calculate the relative perturbation of the photon energy of CMB photons observed today
\begin{equation}
	\left(\frac{\delta \omega}{\omega} \right)_{0} = \left(\frac{\delta \omega}{\omega} \right)_{R} + \Psi_{R} - \Psi_{0} + \int^{\tau_0}_{\tau_*} (\Psi' - \Phi') d \tau + \hat{n} \cdot\vec{V}. \label{photon_temp}
\end{equation}
$\omega$ is the energy of the photon, the $R$ index stands for quantities that are evaluated recombination and $0$ stands for quantities evaluated today. 
Here $\hat{n}$ is the direction of the photon and $\vec{V}$ is the relative velocity of the observer compared to the CMB rest frame. 
All of these terms have an intuitive physical meaning. 
First we see that there is a term that accounts for the energy of the photon when it is emitted, which is the first term in the equation above.
The second term accounts for how the energy of the photon changes due to the gravitational redshift caused by the gravitational potentials at recombination and the third term is similar except it accounts for the integrated effect over the variation of the gravitational potentials (as they will also evolve with time as structures form in the Universe). 
These terms are called the \emph{Sachs-Wolfe} (SW) and \emph{Integrated Sachs-Wolfe} (ISW) effect respectively. 
The final term is a \emph{Doppler} effect coming from the relative velocity of the observer. 
While the temperature of a photon coming from any point on the sky can be calculated by Eq \ref{photon_temp}, this is not an immediately useful quantity.
This is because we cannot predict the properties of a photon coming from a specific location on the sky. 
Rather, only the statistical properties of the distribution of photons. 
This is a fundamental point as each photon will be a realisation of the underlying distribution describing all the photons and measuring individual photons cannot be used to test our cosmological model.
While Eq \ref{photon_temp} allows us to calculate the temperature of the photons given the gravitational potentials, we still need to calculate the gravitational potentials at recombination and how they evolve with time. 
This can be complicated as the particle species will couple to gravity and thus the evolution of all that needs to be accounted for as well. 
This is typically done by solving the thermodynamic Boltzmann equations in the presence of gravity.
This is now a textbook subject and therefore we only briefly review them here following \cite{dodelson}. 
Schematically, the Boltzmann equation can be written in an innocent looking form
\begin{equation}
	\frac{df}{dt} = C[f].
\end{equation}
$f$ is the distribution function of the particle species that we are interested in, $C[f]$ is the collision term which accounts for the interactions of the particle species under consideration with all the other particle species. 
When the collision term is zero the equation above is simply a manifestation of Liouville's theorem in statistical mechanics which says that the phase space of particles in a closed system is conserved. 
The closed system here consists of gravitational interactions as well and to account for these we usually expand the total time derivative as follows
\begin{equation}
	\frac{df}{dt} = \frac{\partial f}{\partial t} + \frac{\partial f}{\partial x^i} \frac{dx^i}{dt} + \frac{\partial f}{\partial p} \frac{dp}{dt} + \frac{\partial f}{\partial \hat{p}^i} \frac{d \hat{p}^i}{dt}
\end{equation}
where $x^i$ are the spatial positions, $p$ is the magnitude of the momentum and $\hat{p}^i$ is the momentum unit vector.
By looking at first order perturbations to the photon distribution of the form
\begin{equation}
	f(\vec{x}, p, \hat{p}, t) = \frac{1}{\exp{ \left( \frac{p}{T(t)(1 + \Theta(\vec{x}, \hat{p}, t))} \right)} - 1 }
\end{equation}
and using the perturbed FRW metric in the Newtonian gauge shown in Eq \ref{metric_newtonian} gives the leading order correction to the Boltzmann equation
\begin{equation}
	\frac{df}{dt}|_{(1)} = - p \frac{\partial f^{(0)}}{\partial p} \left[ \frac{\partial \Theta}{\partial t} + \frac{\hat{p}^i}{a} \frac{\partial \Theta}{\partial x^i} + \frac{\partial \Phi}{\partial t} + \frac{\hat{p}^i}{a} \frac{\partial \Psi}{\partial x^i} \right].
\end{equation}
A few comments are in order here. 
The terms containing the gravitational potentials show how the distribution of photons is affected by gravity. 
The first two terms account for free streaming of the photons after they are emitted. 
These {affect} the anisotropies of photons on small angular scales. 
The physical distance in the FRW metric is given by $a\vec{x}$ and thus all $x^i$ factors come with a factor of the scale factor. 
The collision term will depend on the species under consideration. 
For instance, in the case of photons, Compton scattering of electrons is a key scattering process in the formation of the CMB
\begin{equation}
	e^-(\vec{q}) + \gamma(\vec{p}) \leftrightarrow e^-(\vec{q}') + \gamma(\vec{p}').
\end{equation}

\begin{figure}[h]
\begin{centering}
\includegraphics[scale=0.6]{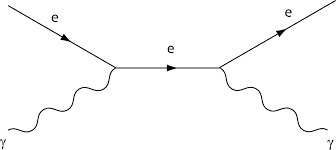}
\caption{Feynmann diagram of Compton scattering.}
\label{Compton_scattering}
\end{centering}
\end{figure}
The collision term of Compton scattering of photons is given by 
\begin{equation}
	C[f(\vec{q})] = - p \frac{\partial f^{(0)}}{\partial p} n_e \sigma_T \left( \Theta_0 - \Theta(\vec{p}) + \hat{p} \cdot \vec{v}_b \right)
\end{equation}
$v_b$ is the bulk velocity of the electrons, $n_e$ is the number density of electrons, $\sigma_T$ is the Thompson cross section computed from QED and we have defined the monopole of the perturbation as 
\begin{equation}
	\Theta_0(\vec{x}, t) \equiv \frac{1}{4 \pi} \int \ d \Omega' \Theta( \hat{p}', \vec{x}, t).
\end{equation}
This is typically solved in Fourier space and conformal time, so we define
\begin{equation}
	\Theta(\vec{x}) \equiv \int \frac{d^3k}{(2 \pi)^3} e^{i \vec{k} \cdot \vec{x}} \tilde{\Theta}(\vec{k}).
\end{equation}
It is also convenient to define the cosine of the angle between the photon direction and the wavenumber $\vec{k}$
\begin{equation}
	\mu \equiv \frac{\vec{k} \cdot \hat{p}}{k},
\end{equation}
and the optical depth 
\begin{equation}
	\tau(\eta) \equiv \int^{\eta_0}_{\eta} d \eta' \ n_e \sigma_T a
\end{equation}
to finally get the full Boltzmann equation for photons
\begin{equation}
	\dot{\tilde{\Theta}} + i k \mu \tilde{\Theta} + \dot{\tilde{\Phi}} + i k \mu \tilde{\Psi} = - \dot{\tau} \left[ \tilde{\Theta}_0 - \tilde{\Theta} + \mu \tilde{v}_b \right]. 
\end{equation}
We can similarly calculate Boltzmann equations for all the particle species\footnote{The full derivation can be found in \cite{dodelson}.}
\begin{eqnarray}
	\dot{\theta} + i k \mu \Theta & = & - \dot{\Phi} - ik \mu \Psi - \dot{\tau} \left[ \Theta_0 - \Theta + \mu v_b - \frac{1}{2} \mathcal{L}_2(\mu) \Pi \right] \\
	\Pi & = & \Theta_2 + \Theta_{P2} + \Theta_{P0} \\
	\dot{\Theta}_P + ik\mu \Theta_P & = & - \dot{\tau} \left[ - \Theta_P + \frac{1}{2} (1 - \mathcal{L}_2(\mu)) \Pi \right] \\
	\dot{\delta} + ik v & = & -3 \dot{\Phi} \\
	\dot{v} + \frac{\dot{a}}{a} v & = & -ik \Psi \\
	\dot{\delta}_b + i k v_b & = & - 3 \dot{\Phi} \\
	\dot{v}_b + \frac{\dot{a}}{a} v_b & = & -ik \Psi + \frac{\dot{\tau}}{R} \left[v_b + 3i \Theta_1 \right], \hspace{5mm} R \equiv \frac{3}{4} \frac{\rho_b^{(0)}}{\rho_\gamma^{(0)}}  \\
	\dot{\mathcal{N}} + i k \mu \mathcal{N} & = & - \dot{\Phi} - ik\mu \Psi
\end{eqnarray}
where we have defined the general multipolar expansion of the temperature field
\begin{equation}
	\Theta_\ell \equiv \frac{1}{(-i)^\ell} \int^1_{-1} \frac{d \mu}{2} \mathcal{L}_\ell (\mu) \Theta(\mu)
\end{equation}
the $\mathcal{L}_\ell$ is the Legendre polynomial of order $\ell$, $\mathcal{N}$ stands for the neutrino perturbations. 
We have also dropped the tilde's over the Fourier transformed variables as we will only be looking at Boltzmann equations in Fourier space from now on. 
Here we have assumed the {neutrinos} have zero mass, however the generalisation of that can be found in many resources \cite{dodelson}.
There are numerical codes such as CAMB\footnote{https://camb.info} and CLASS\footnote{http://class-code.net} that solve these Boltzmann equations in order to calculate various cosmological observables such as the angular power spectrum of the CMB anisotropies. 
The solutions to the Boltzmann equations can be divided into two distinct classes of solutions.
\begin{itemize}
\item Adiabatic solutions: When all particle species have the same {fractional density perturbations}. 
\item Isocurvature solutions: When different particle species can have different relative fractional densities. Typically the difference is defined w.r.t the energy density of photons. 
\end{itemize}
We will describe these more in later chapters, however it is important to know that these two categories of solutions exist and lead to different anisotropies in the CMB. 

The usual approach to analysing the CMB anisotropies is to decompose the temperature field into spherical harmonics 
\begin{equation}
	\frac{\Delta T(\vec{x}, \hat{p}, \eta)}{\bar{T}} = \Theta(\hat{n}, \vec{p}, \eta) = \sum_{\ell m} a_{\ell m}(\vec{x}, \eta) Y_{\ell m}(\hat{p}) \label{temp_1}
\end{equation}
As the temperature field can only take on real numbers it must satisfy $T^*_{lm} = (-1)^mT_{l, -m}$.
The $a_{\ell m}$'s are the amplitudes of the temperature fluctuation at a given location on the sky.
We can rearrange Eq \ref{temp_1} to get 
\begin{equation}
	a_{\ell m}(\vec{x}, \eta) = \int \frac{d^3k }{(2 \pi)^3} e^{i \vec{k} \cdot \vec{x}} \int d \Omega \ Y^*_{\ell m} (\hat{p}) \Theta(\vec{k}, \hat{p}, \eta).
\end{equation}
where $\vec{x}$ is the position of the observer.
As we can only make statistical statements about the measurements, we can look at different moments of the distribution of $a_{\ell m}$'s. 
The first moment of the distribution is the mean and we expect this to be zero. 
The second moment is the variance and is typically denoted
\begin{equation}
	\langle a_{\ell m} a^*_{\ell' m'} \rangle = \delta_{\ell \ell'} \delta_{m m'} C_\ell.
\end{equation}
In addition we assume that distribution of the $a_{\ell m}$'s is Gaussianly distributed, {although current constraints of non-guassianity can be found in \cite{Akrami:2019izv}}. 
Thus the variance and the mean account for all the statistical information in the map of the CMB. 
Interestingly, we also notice another fact which is that for each $\ell$ there will be $2 \ell + 1$ $m$'s and thus for low $\ell$'s there are fewer samples of the distribution. 
This makes sense intuitively as a lower value for $\ell$ corresponds to a higher angular size of the sky and we have fewer large angular samples of the sky then smaller ones. 
Thus there is a fundamental limit to how much information is accessible on the largest scales which is know as \emph{cosmic variance}
\begin{equation}
	\left[ \frac{\Delta C_\ell}{C_\ell} \right]_{\text{cosmic variance}} = \left( \frac{2}{2\ell+1} \right)^\frac{1}{2}.
\end{equation}
Solving for the $C_\ell$'s requires knowledge of the cosmological parameters which describe the $\Lambda CDM$ model. 
In the context of Boltzmann codes that compute these spectra, the anisotropies are typically computed as follows
\begin{equation}
	C_\ell^{XY} = 4 \pi \int \frac{dk}{k} \ P_\Phi(k)^{XY} \ | \Delta^X_\ell(k) \Delta^Y_\ell(k)|^2. \label{clform}
\end{equation}
The $X,Y$ stand for the different initial conditions, i.e adiabatic or isocurvature, as they will have different transfer functions and are also parameterised by different $P_\Phi(k)$ parameters.
$P_\Phi(k)$ is called the primordial power spectrum\footnote{We have added a $\Phi$ subscript here to distinguish the primordial power spectrum from the matter power spectrum that is defined later in this chapter. In subsequent chapters on the initial conditions we will use $P(k)$ for the primordial power spectrum as the matter power spectrum is not mentioned in those chapters.} (PPS) and is defined as the two point correlation function of the primordial gravitational fluctuations
\begin{equation}
	P_\Phi(k) \equiv \frac{2 \pi^2}{k^3} \langle \Phi(k)^2\rangle.
\end{equation}
The PPS is usually parametrised as a power law of the form
\begin{equation}
	P_\Phi(k) = A_s \left( \frac{k}{k_*} \right)^{n_s -1}.
\end{equation}
$A_s$ is the amplitude of scalar fluctuations and $n_s$ is the spectral index. 
The transfer function, $\Delta_\ell(k)$ contains the solutions of the Boltzmann equations and projection factors to compute the angular power spectrum. 
It is usually separated into a source function, $S(k, \tau)$ and a projection factor, $\mathcal{P}_\ell(k, \tau)$, of the form 
\begin{equation}
	\Delta^X_\ell(k) = \int d \tau  \ S^X(k, \tau) \mathcal{P}_\ell(k, \tau).
\end{equation}
The source function contains the other cosmological parameters such as the energy densities of the different particle species, the Hubble constant etc.

\begin{figure}
\minipage{0.33\textwidth}%
  \includegraphics[width=\linewidth]{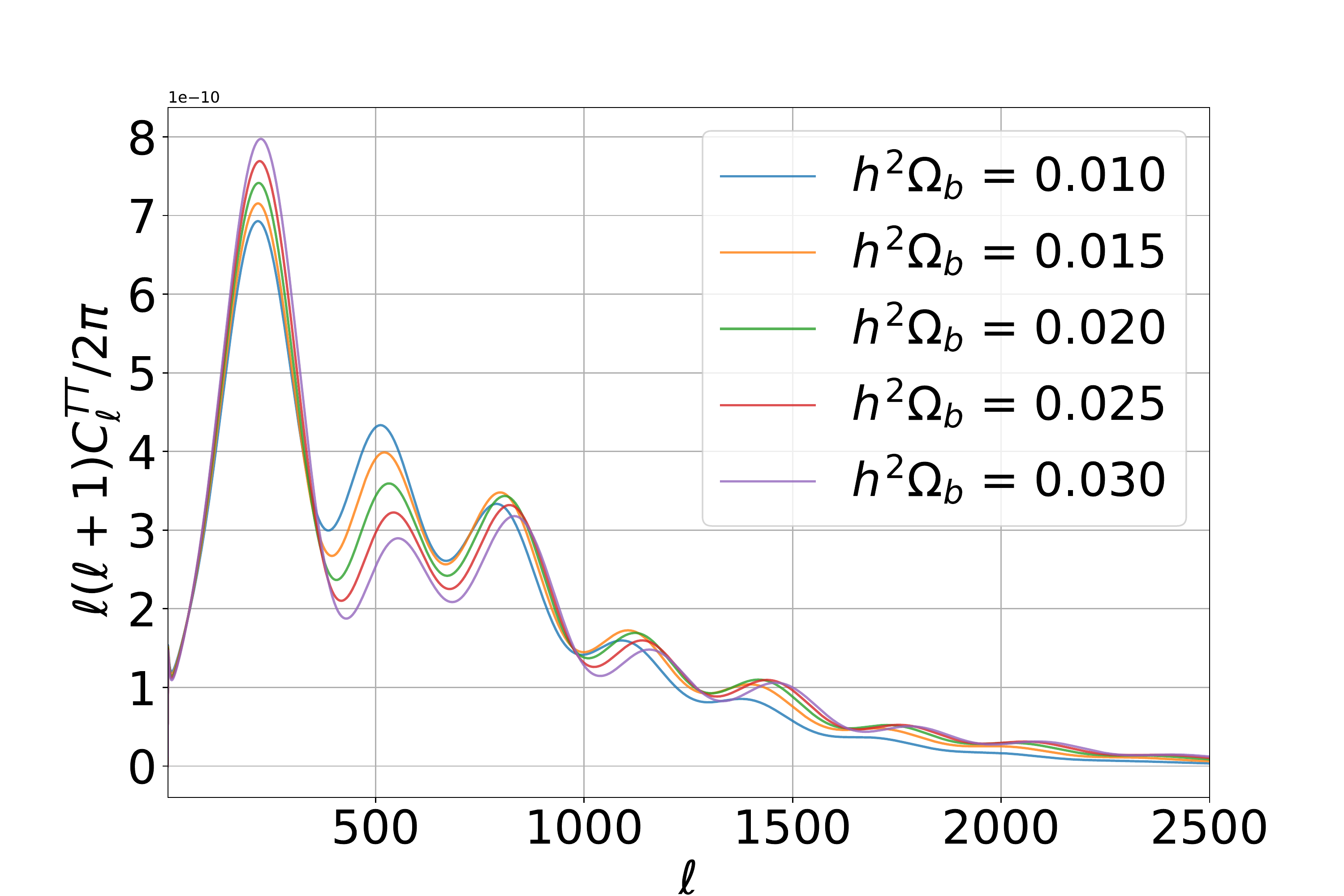}
  \caption{}\label{fig:cl_omegab}
\endminipage
\minipage{0.33\textwidth}
  \includegraphics[width=\linewidth]{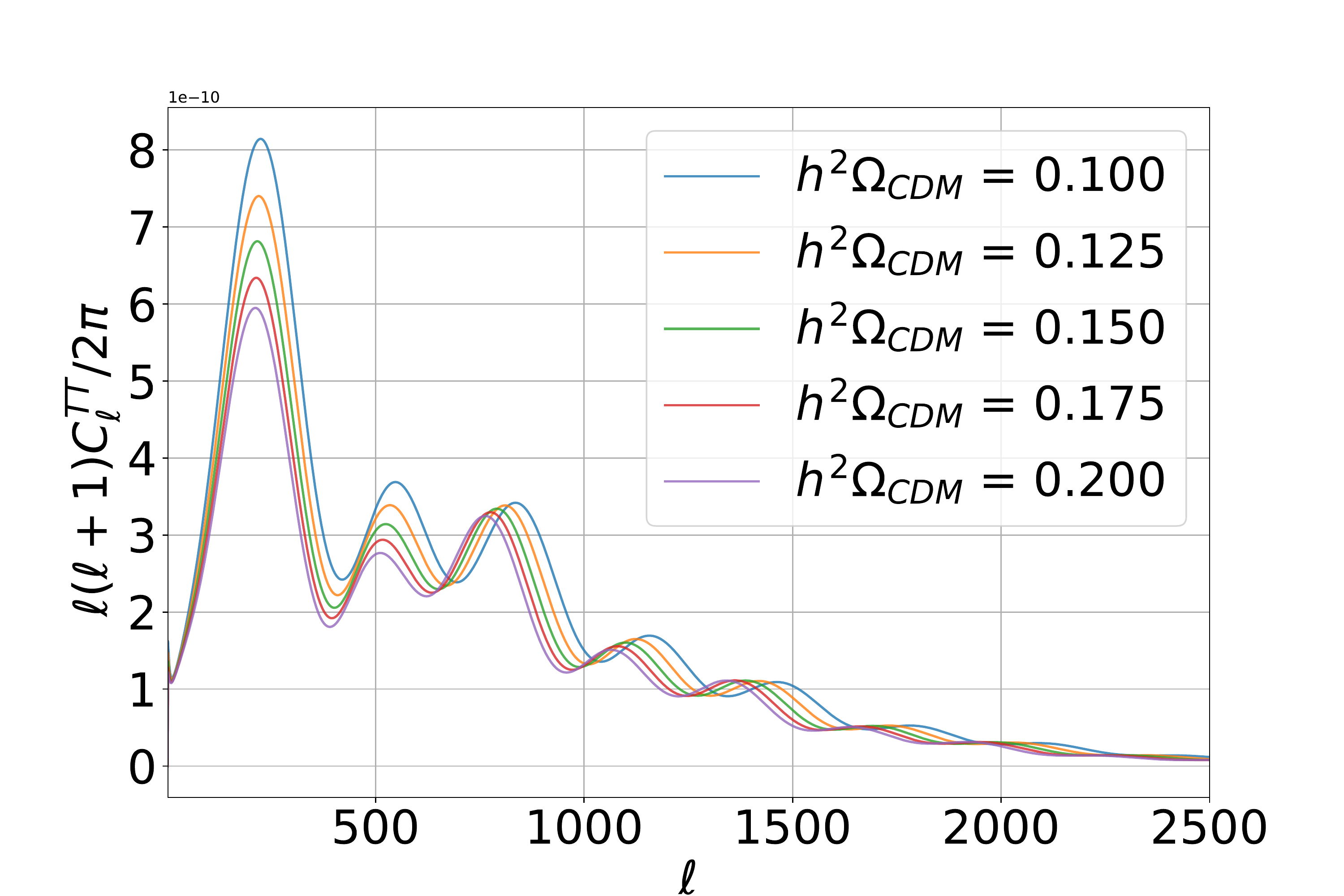}
  \caption{}\label{fig:cl_omegacdm}
\endminipage\hfill
\minipage{0.33\textwidth}
  \includegraphics[width=\linewidth]{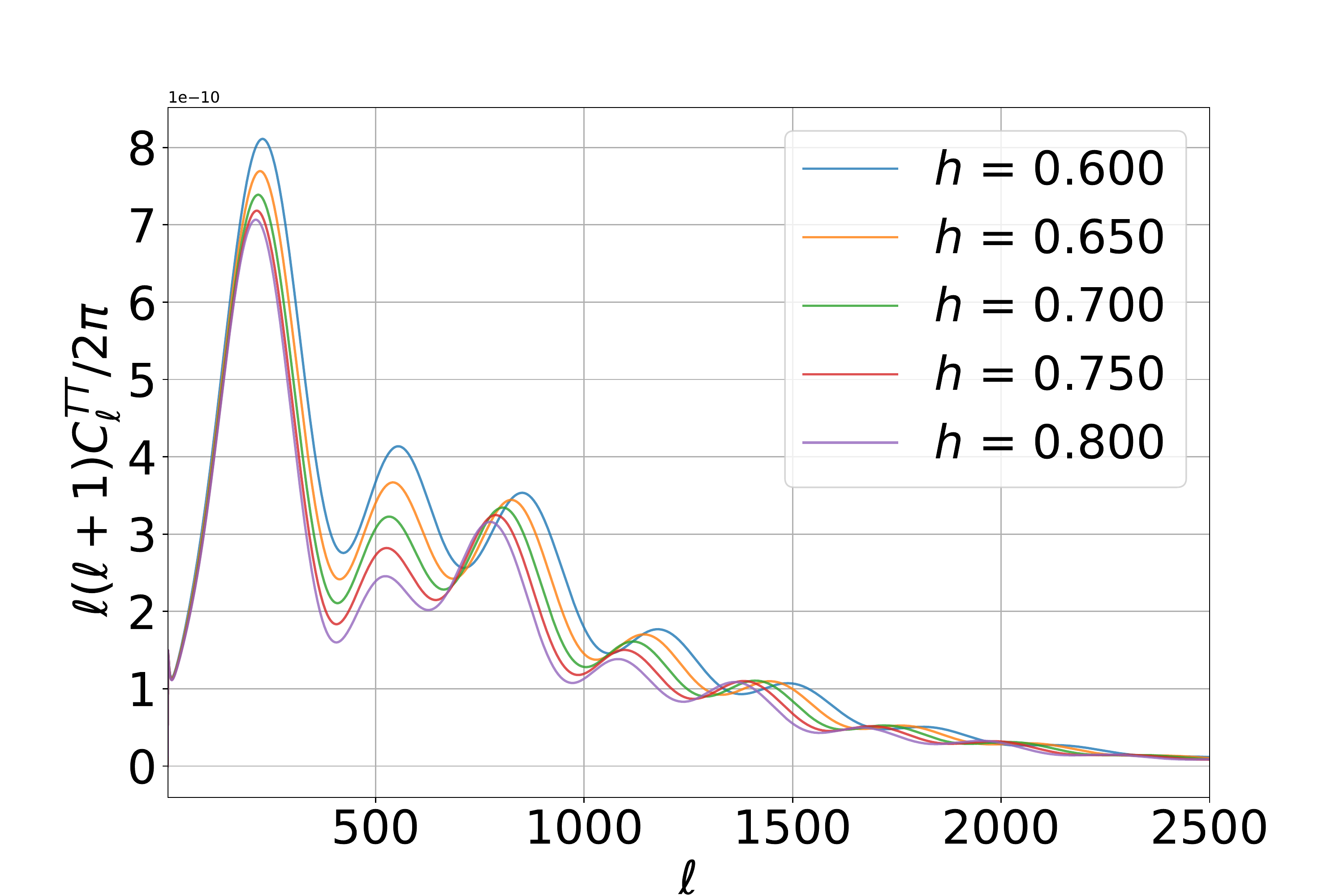}
  \caption{}\label{fig:cl_h}
\endminipage\hfill
\caption{{The left figure shows how Baryons effect the CMB peaks. They predominantly effect the gravitational potential in which the acoustic oscillations happen. The middle figure shows how dark matter effect the CMB peaks - predominantly by changing the time of matter-radiation equality. The figure on the right shows how the Hubble factor changes the CMB peaks.}}
\end{figure}

We can briefly explore the effect some of the cosmological parameters have on the CMB temperature anisotropies. 
\begin{itemize}
\item $\Omega_b h^2$ : The baryonic density affects the amplitudes of the CMB peaks. Increasing it will increase the amplitude of the first peak while it lowers the second peak. 
Intuitively this is easy to understand as the first peak happens when the first sound waves reaches its maximum compression and increasing the $\Omega_b$ will compress the sound wave more. 
However the second peak is due to {rarefaction} (outward motion) of the sound wave and that is suppressed as the $\Omega_b$ will add gravity and act against the pressure of the radiation fluid. We see this in figure \ref{fig:cl_omegab}.

\item $\Omega_c h^2$: Increasing the amount of dark matter means that overall there is more matter in the Universe and thus matter-radiation equality happens earlier. This means there is less time for the radiation fluid to oscillate and thus the peaks are generally shifted down in their amplitude as is shown in figure \ref{fig:cl_omegacdm}. 

\item $H_0$: The Hubble parameter can be inferred by the angular scale of the first acoustic peak (as it affects the inferred distance to it). Thus, it isn't directly observed by the CMB, rather is inferred from angular size of the peak and the measurement of $\Omega_m$ (which defined the sound horizon). 
Thus changing $H_0$ changes the location of the peaks as seen in figure \ref{fig:cl_h}.
Interestingly this is also the effect changing initial conditions has on the peaks of the CMB. For instance, CDM isocurvature initial conditions lead to acoustic peaks that are shifted.

\end{itemize}

\subsubsection{Polarisation anisotropies}

Understanding the details of how polarisation is generated in the CMB is a complex topic and thus we only describe it briefly here. 
More detailed derivations and calculations can be found in \cite{cmb_durrer}.
The polarisation anisotropies are generated by the quadrupole moment due to photon diffusion before last scattering. 
This is typically much smaller in amplitude when compared to the temperature anisotropies as the photon distribution is mostly isotropic. 
There are two times at which there is a local quadrupole anisotropy in the photon distribution that can generate polarisation: once at recombination and once at reionisation \cite{Pritchard:2004qp, Zaldarriaga:1996xe}. 

The formalism to describe polarisation is given by decomposing the polarisation tensor of the electric field, {$\vec{E}$}, in terms of the {Stokes} parameters $I, Q$ and $U$ (defined in Cartesian coordinates below), 
\begin{eqnarray}
	P_{ij} & \propto & \langle |\vec{E}|^2 \rangle \propto \begin{pmatrix} I + Q & U \\ -U & I - Q.\end{pmatrix} \nonumber \\
	I & \equiv & \langle |E_x|^2 \rangle + \langle |E_y|^2 \rangle \nonumber \\
	Q & \equiv & \langle |E_x|^2 \rangle - \langle |E_y|^2 \rangle \nonumber \\
	U & \equiv & \langle 2 \text{Re}(E_x E_y^*) \rangle.
\end{eqnarray}
We have implicitly assumed there is no circular polarisation here. 
Polarisation is a spin 2 field and thus it can be decomposed in spin weighted spherical harmonic functions, $_{\pm 2}Y_{\ell m}$ as follows
\begin{equation}
	Q(\hat{n}) \pm i U(\hat{n}) = \sum_{\ell, m} (E_{\ell m} \pm i B_{\ell m}) _{\pm 2} Y_{\ell m} (\hat{n}).
\end{equation}
Scalar perturbations can only generate temperature and E-mode perturbations. 
Furthermore, since E-modes have even parity, they can be correlated with the temperature anisotropies to give the cross correlation between the temperature and E-mode polarisation photons across the sky. 
B-mode polarisation can only be generated by primordial vector or tensor perturbations.
We won't discuss vector perturbations here however a description of them can be found in \cite{dodelson, cmb_durrer}. 
We discuss tensor perturbations in more detail in chapter \ref{ICOU_tensors} when we discuss decaying tensor modes, however the reason they are usually considered interesting is that they can be generated in a pre-radiation phase of inflation and thus provide a signature of very high energy physics that could have described our Universe at such an early stage of its evolution. 
Interestingly, the B-modes have an odd parity and therefore at leading order they do not have a cross correlation term with the E-modes or temperature anisotropies. 
As B-modes can only be generated by tensor fluctuations (or in principle also vectors, however as mentioned before, we will not consider them here), detecting them could be sign of primordial gravitational waves. 
They can also be produced by lensing of CMB photons and thus to disentangle the primordial B-mode signal we need to accurately model the lensing signal as well. 
The anisotropies for polarisation can also be written in the form of Eq \ref{clform}. 

\subsection{Galaxy clustering}

\begin{figure}[h!]
\begin{centering}
\includegraphics[scale=0.4]{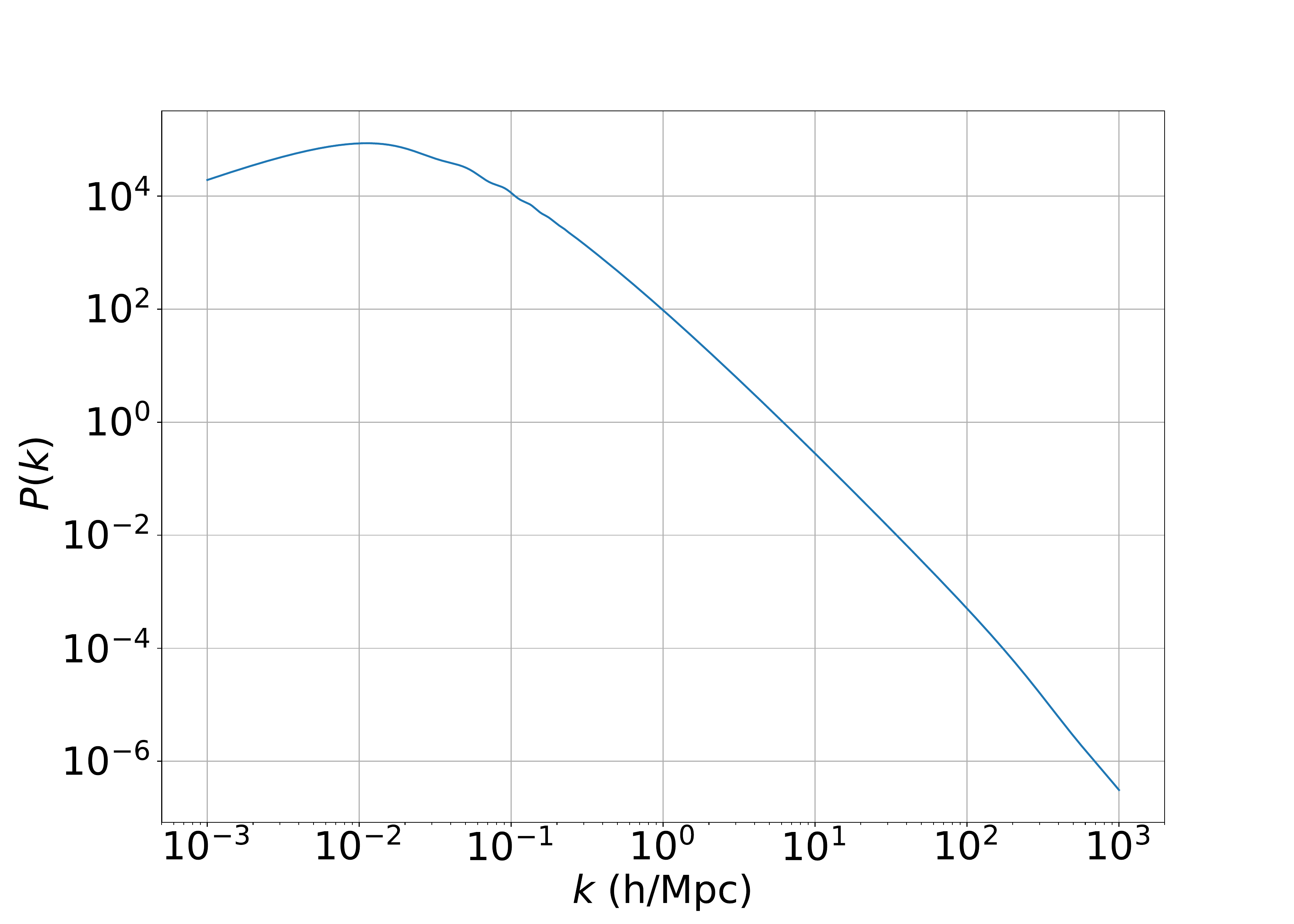}
\caption{Matter power spectrum in $\Lambda$CDM model at redshift zero.}
\label{Pk}
\end{centering}
\end{figure}

Once the Universe is in matter domination, the primordial density perturbations seeded in the early Universe start to collapse under gravity and form compact structures such as galaxies. 
At a fundamental level we can describe this by tracking the evolution of the gravitational potentials from early times to late times and then relating them to the density perturbations using Poisson's equation. 
The evolution of the primordial gravitational potentials in the Newtonian gauge, $\Phi_P$, to the late time gravitational potential is given by 
\begin{equation}
	\Phi(\vec{k}, a) = \frac{9}{10} \Phi_P (\vec{k}) T(k) \frac{D(a)}{a}.
\end{equation}
where $T(k)$ is the transfer function defining the scale independent transition from radiation to matter dominated eras of different $k$ modes. 
The transfer function has to be calculated numerically and Boltzmann codes are used for that.
There are also numerical fitting functions for the transfer function such as the commonly used BBKS transfer function \cite{Bardeen:1985tr}:
\begin{equation}
	T(x = k/k_{eq}) = \frac{ \ln [1 + 0.171x]}{0.171x} \left[ 1 + 0.284x + (1.18x)^2 + (0.399x)^3 + (0.490x)^4 \right]^{-0.25}
\end{equation}
The growth factor $D(a)$ accounts for the overall growth as a function of time of fluctuations. 
Using Poisson's equation in Fourier space we can relate the gravitational potential to the density perturbations $\delta$
\begin{equation}
	\Phi(\vec{k}, a) = \frac{4 \pi \rho_m a^2 \delta}{k^2}
\end{equation}
using $\rho_m = \frac{\Omega_m \rho_{cr}}{a^3}$ and $4 \pi G \rho_{cr} = \frac{3}{2} H_0^2$ we can write 
\begin{equation}
	\delta( \vec{k}, a). =\frac{3}{5} \frac{k^2}{\Omega_m H_0^2} \Phi_p(\vec{k}) T(k) D(a). \label{density_1}
\end{equation}
To fully solve for the dynamics of the density matter perturbations one would need to solve the full set of coupled Boltzmann equations.
As we discussed in the context of CMB anisotropies, we cannot make predictions about individual density perturbations as we only know about the probability distribution they come from, which in the case of density perturbations is assumed to be Gaussian. 
Thus the canonical approach is to calculate the two point correlation function of the density perturbations. 
In Fourier space this is called the power spectrum, $P(k,a)$, which can be calculated from Eq \ref{density_1}
\begin{equation}
	P(k,a) = 2 \pi^2 A_s\frac{k^n}{H_0^{n+3}} T^2(k) \left( \frac{D_(a)}{D(a = 1)} \right)^2.
\end{equation}
Equivalently, one can define the dimensionless power spectrum 
\begin{equation}
	\Delta^2(k, a) \equiv \frac{k^3 P(k,a)}{2 \pi^2}.
\end{equation}
Now we have shown how the matter power spectrum of the matter density perturbations can be calculated. Figure \ref{Pk} shows the matter power spectrum for the standard $\Lambda$CDM model at redshift zero. 
However, there remains an obvious question which is how can we measure the density perturbations. Of course, they cannot be observed directly, instead the best we can do is to look for things that correlate or trace the underlying density perturbations. 
An obvious candidate, in the late Universe once structures have started to form, are galaxies. Thus we need to find a way to relate the matter density power spectrum to an observable related to the observed galaxies. 

In a galaxy survey, one can observe galaxies at different locations in the sky across various depths (or redshifts). 
Thus we can correlated the angular positions of galaxies in the sky and relate that to the underlying density perturbations. 

\begin{figure}[h!]
\begin{centering}
\includegraphics[scale=0.4]{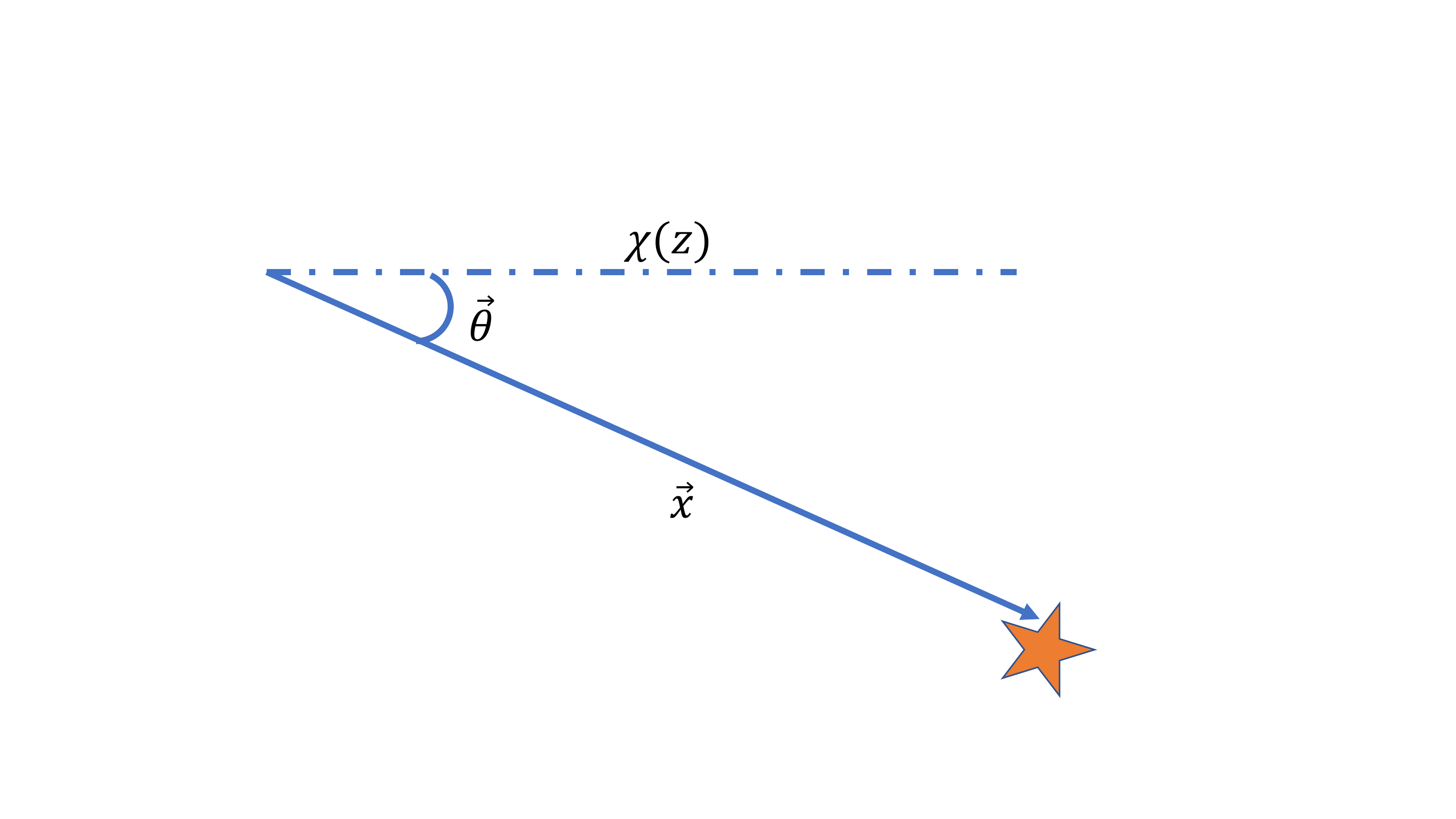}
\caption{Angular position of galaxies observed in a galaxy survey.}
\label{angular_gc}
\end{centering}
\end{figure}

A galaxy at comoving distance $\chi(z)$ can be described by a three dimensional vector, as shown schematically in figure \ref{angular_gc} and can be written as 
\begin{equation}
	\vec{x}(\chi(z), \vec{\theta}) = \chi(z) (\theta_1, \theta_2, 1).
\end{equation}
Here we have assumed a flat sky and defined the angular vector $\vec{\theta} \equiv (\theta_1, \theta_2)$.\footnote{This approximation breaks down when galaxies are not close to the $z$ axis.} 
We can define the two dimensional overdensity, $\delta_2$, at location $\vec{\theta}$ as the integral over the line of sight of the three dimensional density
\begin{equation}
	\delta_2(\vec{\theta}) \equiv \int^{\chi_\infty}_0 d \chi W(\chi) \delta(\vec{x}(\chi, \vec{\theta}) ). \label{delta2}
\end{equation}
Here we have defined a selection function which accounts for the redshift distribution of the sample. For instance, in a galaxy survey there will be some galaxies that are faint and will not be detected. This function is normalised as 
\begin{equation}
	\int^{\chi_{\infty}}_0 \ d\chi \ W(\chi) \equiv 1.
\end{equation}
We can perform a two dimensional Fourier transform of the two dimensional density field as follows
\begin{equation}
	\tilde{\delta}_2(\vec{l}) = \int d^2 \theta e^{-i \vec{l} \cdot \vec{\theta}} \ \delta_2(\vec{\theta}),
\end{equation}
where $\vec{l}$ is the conjugate variable to $\vec{\theta}$. The two-dimensional power spectrum is defined analogously to the three dimensional power spectrum
\begin{equation}
	P_2(l) \equiv \int d^2 l' \frac{\langle \tilde{\delta}_2(\vec{l}) \tilde{\delta}^*_2(\vec{l}') \rangle}{2 \pi^2}. \label{p2l}
\end{equation}
We have assumed SO(2) symmetry in writing this equation and also assumed all modes are independent.
Using Eq \ref{delta2} and performing a series of integrals, Eq \ref{p2l} can be written in terms of the three dimensional power spectrum and the selection function. 
\begin{equation}
	P_2(l) = \frac{1}{l} \int^\infty_0 dk \ P(k) \ W^2(l/k).
\end{equation}
Equivalently in real space we have 
\begin{eqnarray}
	w(\theta) \equiv \int^\infty_0 dk \ k \ P(k) \ F(k, \theta), \nonumber \\
	F(k, \theta) \equiv \frac{1}{k} \int^\infty_0 \frac{dl}{2\pi} \ J_0(l \theta) W^2(l/k)
\end{eqnarray}
where $F(k, \theta)$ is the kernel of the angular correlation function where $J_0(l \theta)$ is the Bessel function of order zero (that comes from the inverse Fourier transform). 

Thus far we have focused on calculations based on the underlying density perturbations. 
In a galaxy survey we can only measure the number of galaxies at a given redshift and position $\hat{n}$ in the sky. 
This can be related to the underlying density field as follows \cite{Bonvin:2011bg}
\begin{eqnarray}
	\delta_z(\hat{n}, z) & \equiv & \frac{\rho(\hat{n}, x) - \langle \rho \rangle (z)}{\langle \rho \rangle (z)}, \nonumber \\
	& = & \frac{N(\hat{n}, z) - \langle N \rangle(z)}{\langle N \rangle (z)} - \frac{\delta V(\hat{n}, z)}{V(z)}.
\end{eqnarray}
Where we have defined the volume mapped by a survey at a given redshift as $V(\hat{n}, z)$. 
The perturbations in the metric will change the volume observed at a given location (for instance the lensing of photons will distort the past light cone as seen by an observer on Earth). 
This change be parameterised as 
\begin{equation}
	V(\hat{n}, z) \equiv V(z) + \delta V(\hat{n}, z).
\end{equation}	
Thus we can define
\begin{equation}
	\Delta(\hat{n}, z) \equiv \frac{N(\hat{n}, z) - \langle N \rangle (z)}{\langle N \rangle(z)} = \delta_z(\hat{n}, z) + \frac{\delta V(\hat{n}, z)}{V(z)}.
\end{equation}
The $\Delta(\hat{n},z)$ can be directly observed and the theoretical computation of it follows an analogous calculation to the one we did to obtain Eq \ref{photon_temp}. 
We do not reproduce the calculation for the $\Delta(\hat{n},z)$ here due to its length and complexity, it can be found in \cite{Classgal, Bonvin:2011bg}. The final result, to leading order in perturbations, is 
\begin{eqnarray}
	\Delta(\hat{n}, z) & = & D_g + \Phi + \Psi + \frac{1}{\mathcal{H}} \left( \dot{\Phi} + \partial_r(\vec{V} \cdot \hat{n}) \right) \nonumber \\
	& + & \left( \frac{\dot{\mathcal{H}}}{\mathcal{H}^2} + \frac{2}{r_S\mathcal{H}} \right) \left( \Psi + \vec{V} \cdot \hat{n} + \int^{r_S}_0 d \lambda \ (\dot{\Phi} + \dot{\Psi}) \right)  \nonumber \\
	& + & \frac{1}{r_S} \int^{r_S}_0 d \lambda \ \left(2 - \frac{r_s - r}{r} \Delta_\Omega \right) (\Phi + \Psi) \label{galaxy_number_density}
\end{eqnarray}
where the $S$ index stands for the source, $D_g$ is the density fluctuation on a uniform curvature hypersurface and 
\begin{equation}
	\Delta_\Omega \equiv \left( \text{cot} \theta \ \partial_\theta + \partial^2_\theta + \frac{1}{\sin^2{\theta}} \partial^2_\phi \right).
\end{equation}
Using Eq \ref{galaxy_number_density} we can directly compare theoretical predictions to observed values from a galaxy survey.
As we are interested in the statistical properties, it is common to decompose the observed number density field in spherical harmonics and compute the angular power spectra of the galaxy number density, which takes the form of Eq \ref{clform}. 
In the case of galaxy number counts, the transfer function is given by Eq \ref{galaxy_number_density} convolved with the appropriate projection factors. 
These can also be found, with full derivations, in \cite{Classgal}.


\subsection{Galaxy lensing}

\begin{figure}[h!]
\begin{centering}
\includegraphics[scale=0.4]{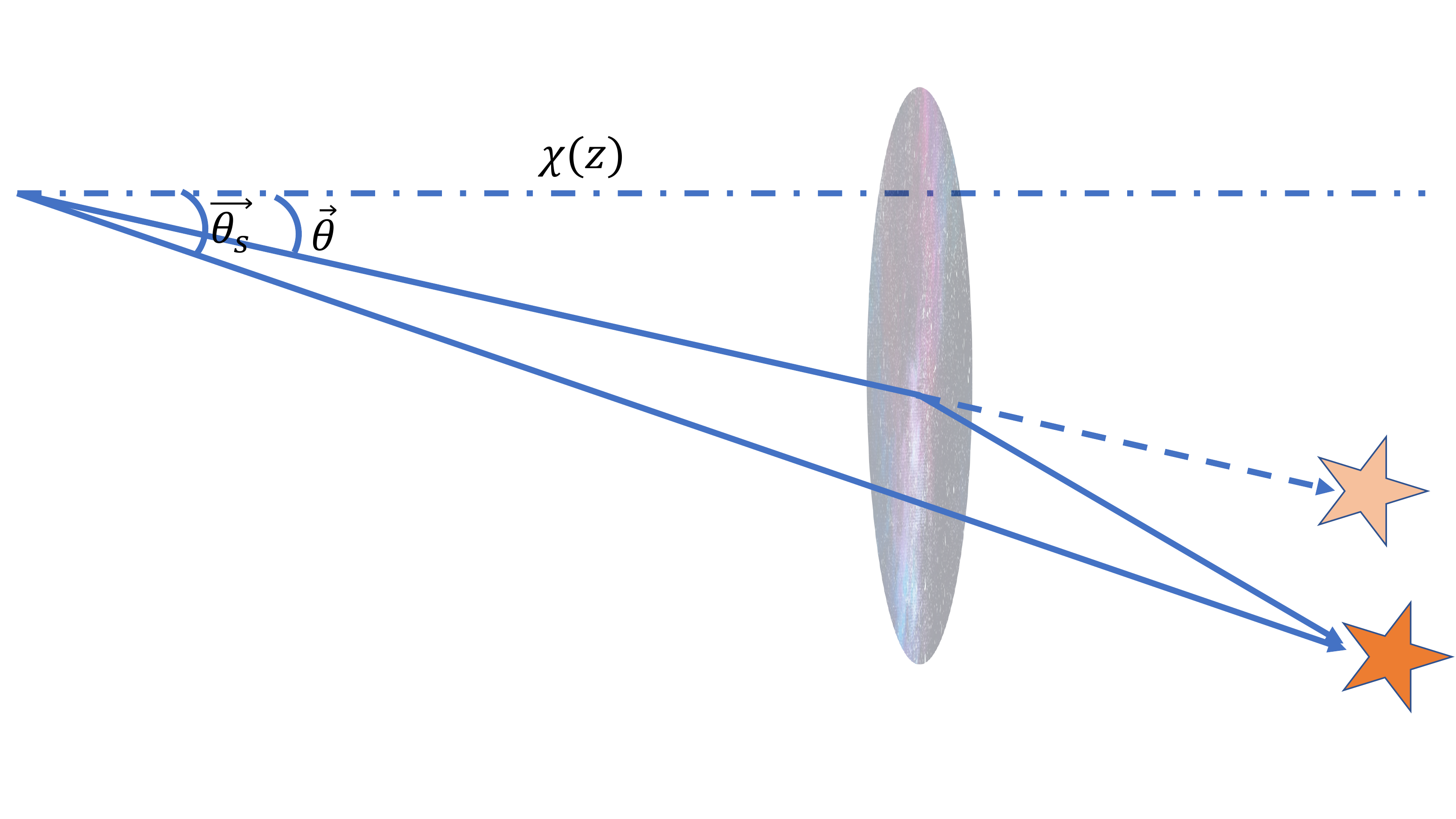}
\caption{Schematic effect of lensing of photons.}
\label{galaxy_lensing}
\end{centering}
\end{figure}

In addition to the density field, we can define the \emph{shear} field which describes the effect of lensing of photons due to matter between a galaxy and an observer. To understand the concept of shear we can start with the sketch shown in figure \ref{galaxy_lensing}.

A photon arriving at earth from a galaxy with intensity $\mathcal{I}_{true}$ from direction $\vec{\theta}_{S}$ is observed on earth coming from direction $\vec{\theta}$ and intensity $\mathcal{I}_{obs}$ such that 
\begin{equation}
	\mathcal{I}_{obs}(\vec{\theta}) = \mathcal{I}_{true}(\vec{\theta}_S).
\end{equation}
The cosmological information is contained in $\mathcal{I}_{true}$ and $\vec{\theta}_s$ and thus we need to relate these to the observed values. 
The observed angle can be related to the true angle by using the spatial component of the geodesic equation
\begin{equation} 
	\partial_\lambda^2x^i = - \Gamma^i_{\alpha \beta} \partial_\lambda x^\alpha \partial_\lambda x^\beta.
\end{equation}
Here $\lambda$ is the affine parameter for the photons. If we use the Newtonian metric given in Eq \ref{metric_newtonian} and using $\frac{d \chi}{d\lambda} = \frac{d \chi}{dt} \frac{dt}{d\lambda}$ we get 
\begin{equation}
	\vec{\theta}_S = \vec{\theta} + 2 \int^\chi_0 d\chi' \ \frac{\partial \Phi}{\partial \vec{x}} (\vec{x}(\chi')) \left( 1 - \frac{\chi'}{\chi} \right).
\end{equation}
The relation between the observed and true angles is canonically parametrised by 
\begin{equation}
	A_{ij} \equiv \frac{\partial \vec{\theta}_S}{\partial \vec{\theta}} \equiv \begin{pmatrix} 1 - \kappa - \gamma_1 & -\gamma_2 \\ -\gamma_2 & 1 - \kappa + \gamma_1 \end{pmatrix}.
\end{equation}
$\kappa$ describes how much the source image is magnified due to lensing and is called the \emph{convergence} field. The $\gamma_{1/2}$ describe the change in shape of the galaxies. Collectively the ${\gamma_{1}, \gamma_2}$ are components of the \emph{shear} field and are defined as 
\begin{eqnarray}
	& & \gamma_1 = - \frac{A_{11}  - A_{22}}{2},	\nonumber \\
	& & \gamma_2 = - A_{12}.
\end{eqnarray}
As with other observables, we are interested in the statistical properties of the shear and convergence fields and thus we need their correlation function/power spectra. 
For the shear field, since we are interested in deflection of photons, it is natural to define a new tensor, called the \emph{distortional tensor}
\begin{equation}
	\psi_{ij} \equiv A_{ij} - \delta_{ij}
\end{equation}
where $\delta_{ij}$ is the identity matrix.
The distortion tensor can be written in terms of the distribution function $W(\chi)$, normalised as $\int d \chi W(\chi) = 1$,  which parametrises the distribution of redshifts in a given survey, 
\begin{eqnarray}
	\psi_{ij}(\vec{\theta}) & = & \int^{\chi_\infty}_0 \ d \chi \ \frac{\partial \Phi}{\partial x^i \partial x^j} (\vec{x}(\chi)) g(\chi),  \nonumber \\
	g(\chi) & \equiv & 2 \chi \int^{\chi_\infty}_\chi d \chi' \ \left( 1 - \frac{\chi}{\chi'} \right) W(\chi').
\end{eqnarray}
It is assumed that the distortion is zero on average, $\langle \psi_{ij} \rangle = 0$. 
The power spectra for the distortion field and the convergence field are then given by 
\begin{eqnarray}
	P^{ijlm}_\psi(\vec{l}) & = & \int^{\chi_\infty}_0 d \chi \ \frac{g(\chi)^2}{\chi^2} \frac{l^i l^j l^l l^m}{\chi^4} P_\Phi(l/\chi), \nonumber \\
	P_\kappa(l) & = & \frac{l^4}{4} \int^{\chi_\infty}_0 d \chi \ \frac{g(\chi)^2}{\chi^6} P_\Phi(l/\chi).
\end{eqnarray}
These quantities can be computed from observed galaxy data by using ellipticity as a tracer of the shear field. 
The fundamental estimator of shear is the quadrupole moment of an image\footnote{The dipole can always be set to zero by using translation symmetry and centering the image at the origin of the $\theta_x$ - $\theta_y$ axis.}
\begin{equation}
	q_{ij} \equiv \int d^2 \theta \mathcal{I}_{obs}(\theta) \theta_i \theta_j.
\end{equation}
In a circular image $q_{xx} = q_{yy}, q_{xy} = 0$. The two canonical estimators for shear are therefore given by 
\begin{eqnarray}
	\epsilon_1 & \equiv & \frac{q_{xx} - q_{yy}}{q_{xx} + q_{yy}} \approx 2 \gamma_1, \nonumber \\
	\epsilon_2 & \equiv & \frac{2q_{xy}}{q_{xx} + q_{xy}} \approx 2 \gamma_2.
\end{eqnarray}
By calculating $\epsilon_1, \epsilon_2$ from images of galaxies, we can get a statistical measure of the shear field\footnote{It is also common to decompose the shear into an E and B mode as it is also a spin 2 field.}.

There are several other probes of cosmology such as measurements of supernovae, intensity mapping, gravitational waves etc. However, as we do not use these probes directly in the work presented here, we do not elaborate further upon them.

\section{Statistical methods for cosmology}

\subsection{The two schools}

In simple terms, the goal of any scientific experiment is to test a given hypothesis. 
There is some general model that we think describes the Universe and we want to test if this model is correct by comparing it to observational data. This can be broken down into two questions; the first is to constrain the free parameters of the model (if indeed it has any) and second is to check how well the model fits the data. 

There are two classical frameworks in statistics that can be used to test hypothesis and fit parameters with observed data: the frequentist approach and the Bayesian approach. 
The fundamental difference between the two approaches is the interpretation of probability. 
In the frequentist framework probability is interpreted as relative frequencies in experiments with many trials. 
On the other hand, the Bayesian framework defines probability as the a figure of belief in a given event or situation happening. 
This belief can be updated based on new events or data and this update is defined by Bayes rule. 

Interestingly, within physics there are some subjects that naturally lend themselves to a frequentist or Bayesian approach. 
In particle physics there is a vast amount of data from particle accelerators that can be repeated many times under controlled conditions and thus most of the statistical analysis is done in frequentist framework. 
The data sets in cosmology are very different. 
For example, the CMB is unique thus inferences drawn from the CMB photons don't have a natural interpretation in the frequentist framework and the analysis is done in a Bayesian framework. 
This is also true for other cosmological data sets and therefore from now on we will focus on Bayesian analysis. 

\subsection{Bayesian analysis}

Bayes theorem follows trivially from the product rule of probabilities which states
\begin{equation}
	P(A \cap B) = P(A|B) P(B), \label{prob_prod}
\end{equation}
or in words: the probability of A and B is given by the probability of B happening times the probability of A happening given than B has happened. 
It is obvious that reversing the order of A and B should have no effect, i.e $P(A\cap B) = P(B\cap A)$. 
Rewriting this equality using Eq \ref{prob_prod} gives Bayes theorem
\begin{equation}
	P(A|B) = \frac{P(B|A) P(A)}{P(B)}.
\end{equation}
This simple equation has been at the heart of all cosmological analysis and is responsible for what we understand about our Universe today from an empirical point of view.
The terms in this equation are usually defined as follows: 
\begin{itemize}
\item $P(A|B)$ is called the posterior probability distribution function. Obtaining this distribution is typically the end goal of a Bayesian analysis as it allows one to measure the statistical properties of the quantity of interest, in this case $A$ (which represents cosmological parameters in our case). 
\item $P(B|A)$ is called the likelihood function. In cosmological analysis this requires a forward model that can computes the theoretical predictions (such as the temperature anisotropies in the CMB) from cosmological parameters.
\item $P(A)$ is called the prior probability distribution function. This encompasses previous beliefs about $A$. For instance it could account for certain physical properties of the cosmological parameters (such as some of them having to be positive). 
\item  $P(B)$ is called the Bayesian evidence. This is the probability of getting an observable $B$ given all possible values of the parameters $A$.\footnote{Mathematically this is written as follows
\begin{equation}
	P(B) = \sum_i P(B \cap A_i) = \sum_i P(B|A_i) P(A_i). 
\end{equation}
}
\end{itemize}
The typical situation in cosmology is that we have a theoretical model $\mathbb{T}$ that describes the Universe. 
In our case it is the $\Lambda$CDM model. 
The model contains a set of free parameters, $\vec{\theta}$ (such as the densities of different particle species, the Hubble parameter etc), that have to be fit to the observed data $\mathbb{D}$. 
There is a natural way to extract information about cosmological parameters from obervational data sets using Bayes theorem as follows
\begin{equation}
	P(\vec{\theta}|\mathbb{D}) = \frac{P(\mathbb{D}|\vec{\theta}) P(\vec{\theta})}{P(\mathbb{D})}. \label{Bayes_theorem}
\end{equation}
The goal is to find the values of the cosmological parameters that best fit observed data. 
A crucial component of cosmological inference in a Bayesian framework is the likelihood function, often just called the likelihood. 
We describe how some of the likelihoods used in cosmology are derived and then briefly describe the role of priors and Bayesian evidence. 

\subsection{Likelihoods in cosmology}

The likelihood function is often written as 
\begin{equation}
	\mathcal{L}(\vec{\theta}) = P(\mathbb{D} | \vec{\theta}).
\end{equation} 
The r.h.s is the probability of getting a data set $\mathbb{D}$ given a value of $\vec{\theta}$. 
Therefore the likelihood is a function of $\vec{\theta}$ however it is not a probability distribution function of $\vec{\theta}$ as it is not normalised over $\vec{\theta}$, rather it is normalised over $\mathbb{D}$. 

In order to get the value of the parameters $\vec{\theta}$ from a given sample of data, we need to write down an estimator of $\theta$. 
A common choice for estimators is the maximum likelihood estimator (MLE)
\begin{equation}
	\vec{\theta}_{MLE} \equiv \max_{\vec{\theta}} [\mathcal{L}(\vec{\theta}) ].
\end{equation}
This can be obtained by setting the first derivative of the likelihood to zero and the ensuring the second derivative is negative.
For optimisation tasks it is often easier to work with the logarithm of the likelihoods and that is what we will focus on from now on. 
The form of the likelihood function is determined by the statistical properties of the data being used.
In cosmology we assume the data sets, for instance the temperature/polarisation anisotropies in the CMB, the distribution of galaxies etc., can be modelled as multivariate Gaussian random fields. 
Therefore we only need to estimate the mean and covariance of the data to get an accurate measure of the likelihood. 
The likelihood can then be written as follows
\begin{eqnarray}
	\mathcal{L} & = & \frac{\exp{\left[ - \frac{1}{2} \mathbb{X}^T \textbf{C}^{-1} \mathbb{X} \right] }}{\sqrt{2 \pi \textbf{C}}}, \nonumber \\
	\mathbb{X} & \equiv & \mathbb{D} - \mathbb{T}(\vec{\theta}), \nonumber \\
	\textbf{C} & \equiv & \langle \mathbb{X}\mathbb{X}^T \rangle, \label{likelihood_general}
\end{eqnarray}
here we have defined $\mathbb{T}(\vec{\theta})$ as the theory vector that is computed given the cosmological parameters and a forward model, which for us is the $\Lambda$CDM model. 
This form of the likelihood holds for any field that follows a Gaussian distribution. 
The canonical argument for assuming a Gaussian form for likelihoods is the \emph{central limit theorem} which states that the distribution of a large sample of independent events will tend to a Gaussian distribution. The covariance matrix needed to evaluate the likelihood is potentially computationally very expensive to compute and invert. 
Therefore, it is often assumed that the covariance matrix can be evaluated at a fixed point in parameter space while the theoretical prediction will be different for different parameters. 
We explore how good this approximation is for various likelihoods used in cosmology in chapter \ref{PDCM}. 

\subsection{Inference}
Having discussed the likelihood, we must remember that we are actually interested in the posterior probability distribution of the cosmological parameters which is related to the likelihood via Bayes theorem shown in Eq \ref{Bayes_theorem}. 
In addition to the likelihood we also need to address the priors and the evidence terms. 
The first thing to note is that the evidence term is only there to normalise the posterior to one and thus it can be ignored when it comes to optimisation problems related to the cosmological parameters. 
It is typically used for model selection purposes; a model with a higher Bayesian evidence is a preferred model.
For the prior one can use physical knowledge or lack of knowledge to give the form it takes.
For instance, suppose any value a parameter can take is equally likely. 
In this case the prior is assumed to be a constant value, sometimes called an uninformative or flat prior. 
Under this assumption the likelihood is proportional to the posterior and optimising the likelihood is equivalent to optimising the posterior. 
In general, when the priors have a functional form and are not just constant, the peak of the likelihood is not guaranteed to correspond to the peak of the posterior.
There is a well established method for obtaining the posterior distribution for cosmological parameter which is the Markov Chain Monte Carlo (MCMC) method.
\subsubsection{MCMC}
In order to obtain a form of the posterior we need to able to sample a distribution of points from the parameter space. 
An MCMC algorithm gives a procedure to obtain this distribution of samples. 
In order to get a reasonable estimate for the full posterior distribution we must have enough samples from the underlying distribution function. 
This can become very large in high dimensional parameter spaces. Typically the number of points scales exponentially with the number of parameters and the standard $\Lambda$CDM model has order 10 parameters. 
If we choose 1000 points for each parameter, for 10 parameters we will need to compute $10^{30}$ points to effectively sample the posterior. 
If each computation takes $10^{-3}$ seconds, it will take $10^{27}$ seconds to run such a computation which is roughly $10^{10}$ larger than the age of the Universe.
Therefore MCMC requires smarter ways to sample the underlying the distribution. 

One of the crucial properties of a Markov chain is that it converges to a stationary state where after some number of iterations, called burn-in phase, the chain contains samples from the posterior distribution. 
This guarantees that we will eventually converge to the distribution we are interested in.
The work presented in this thesis does not use MCMC explicitly at any point (although it is mentioned a few times) and thus we do not discuss the sampling methods in detail. 
Comprehensive reviews of MCMC can be found in \cite{2010LNP...800..147V, Heavens:2009nx, Cosmomc}. 

\subsubsection{Fisher analysis}

An MCMC analysis is done when we have experimental data and theoretical predictions of a model. 
However, it is often useful to get an estimate of how well an experiment can constrain parameters before building it (to get an idea of how impactful it will be). 
A method to do this is called Fisher analysis and has been used in cosmology for over 20 years \cite{Tegmark:1996bz, Bond:1997wr}. 

For constant priors, maximising the posterior is equivalent to maximising the likelihood for the parameters we are interested. 
The value of the parameters that maximise the likelihood, $\vec{\theta}_{MLE}$, are defined by
\begin{equation}
	\left[ \frac{ \partial \mathcal{L}}{\partial {\vec{\theta}}} \right]_{\vec{\theta}_{MLE}} = 0	.
\end{equation}
To find the maximum likelihood point we can Taylor expand the derivative of the likelihood around some initial guess $\vec{\theta}_0$ and then use a root finding algorithm, such as the Newton-Raphson method to approach the maximum likelihood point. 
As discussed before, it is often easier to work with the log-likelihood and so we can work with that from now. 
Expanding the log-likelihood gives
\begin{equation}
	\left[ \frac{\partial \ln (\mathcal{L})}{\partial \vec{\theta}} \right]_{\vec{\theta}_{MLE}} = \left[ \frac{\partial \ln(\mathcal{L}) }{\partial \vec{\theta}} \right]_{\vec{\theta}_{0}} + \left[ \frac{\partial^2 \ln (\mathcal{L})}{\partial \vec{\theta}^2} \right]_{\vec{\theta}_{0}} \left( \vec{\theta}_0 - \vec{\theta}_{MLE}\right)+ \mathcal{O}\left( \left( \vec{\theta}_0 - \vec{\theta}_{MLE} \right)^2 \right) \label{likli_taylor}
\end{equation}
A quantity of particular interest here is the second derivative of the log-likelihood. 
If the likelihood is being expanded around the maximum likelihood parameters, then the first derivative of the likelihood is zero and the second derivative gives a measure of how quickly the parameters change affects the likelihood value. Thus, it is actually a measure of the errors of the parameters: if the curvature is large then the data is very constraining, i.e a small change in the parameters will lead to a large change in likelihood value. 
The second derivative of the likelihood is defined as the curvature matrix,
\begin{equation}
	\mathcal{F} \equiv - \frac{\partial^2 \ln (\mathcal{L})}{\partial \vec{\theta}^2}.
\end{equation}
If we sample the likelihood many times then the mean curvature gives an estimate of the errors. 
The Fisher matrix is defined as the ensemble average (over signal and noise) of the curvature matrix
\begin{equation}
	\textbf{F} \equiv \biggl<- \frac{ \partial^2 \ln (\mathcal{L})}{\partial \vec{\theta}^2} \biggr>.
\end{equation}
The covariance matrix is given by the inverse of the Fisher matrix and gives a measure of the uncertainty in the parameters. 
In general the Fisher matrix can be computed at any point in parameter space and it will then inform us how well that set of parameters will be constrained. 
If the likelihood is Gaussian, as shown in Eq \ref{likelihood_general}, then $\sqrt{\mathbf{F}^{-1}}$ will give the lowest possible errors on the parameters\footnote{This is known as the Cramer-Rao inequality \cite{Heavens:2009nx}.}. 
It is important to realise that the Fisher information does not provide any information about how well the parameters fit the data, it merely gives the errors with which the parameters can be measured with. 
Thus, it always relies on independent information about the best fit parameter values. 

Even with this limitation the Fisher matrix can be a very useful object as it allows us to know how well a set of parameters can be measured in an experiment before doing the experiment itself. 
This is often the case in cosmology as we want to know how much constraining power a particular cosmological survey can have.
 An even more interesting question to ask is how well a given set of parameters can be measured \emph{in principle}. 
 We will return to this question several times when analysing the initial conditions of the Universe. 

\section{Overview of thesis}

The thesis has four main chapters based on the work in four papers \cite{Kodwani1, Kodwani2, Kodwani3, Kodwani4} and a concluding chapter that gives the summary of all the work presented in this thesis and potential future works. 

\begin{itemize}

\item Chapter 2 is based on \cite{Kodwani1}: Describes the most general initial conditions for the scalar perturbations in our Universe. The novelty in this work is the model independent, non-parametric, {Fisher} analysis of the decaying modes in adiabatic perturbations. 

\item Chapter 3 is based on \cite{Kodwani2}: Describes the most general initial conditions for the tensor perturbations in our Universe. We examine the decaying mode solution for tensors and see how it effects the CMB anisotropies. This has never been considered before and then we also perform a Fisher analysis similar to Chapter 3. 

\item Chapter 4 is based on \cite{Kodwani3}: Proposes a novel observable in the clustering of galaxies that arises due to screened modified gravity theories. 
The two point correlation function of different types of galaxies (i.e bright or faint galaxies) is calculated in the presence of screened modified gravity theories and a novel parity breaking component is shown to be present.

\item Chapter 5 is based on \cite{Kodwani4}: This work attempts to quantify the effect of a long standing assumption of cosmological analysis, which is that the covariance matrix used in a likelihood is fixed. This is not true in general and we quantify this effect using analytic and numerical analysis of covariance matrices used in large-scale structure analysis. 

\item Chapter 6 contains a summary of all the work and the key results from the other chapters.
It also contains a brief description of ongoing and future projects related to work presented in this thesis.

\end{itemize}

\chapter{Initial conditions of the universe: A sign of the sine mode}\label{ICOU}

\section{Introduction and historical context}

Our current understanding of the Universe builds upon a widely accepted standard big bang model, in which the Universe starts out in a hot and dense radiation dominated phase. Precise initial conditions and an explanation of the homogeneity and isotropy of the large scale Universe are required to match current observations.
An epoch of cosmological inflation has been the most widely accepted extension to the standard big bang model that could potentially resolve these issues. Most importantly, it provides a natural way to generate small perturbations in the metric and densities of particles that manifest themselves as the temperature and polarisation anisotropies in the cosmic microwave background (CMB) and density fluctuations that eventually grow into the large scale structure, which have been studied extensively over the last few decades \cite{wmap_final, refId0}.
The simplest models of inflation predict Gaussian adiabatic initial conditions for the radiation dominated era. 
However these are not the only possible initial conditions. .  
 
After neutrino decoupling at around $z\sim 10^9$, the universe contains baryons, photons, dark matter and neutrinos.
Each of these species has an equation governing its perturbations which are described by second order partial differential equations. 
In total there are 8 possible solutions for the densities of the particles that can exist in the early universe; two corresponding to each species \cite{Ma:1995ey, Bucher:1999re, Carrilho:2018mqy, Bucher:2000cd, Bucher:2000kb, Bucher:2000hy}. 
These solutions fall into two general classes, {\it adiabatic} or curvature and {\it entropy} or isocurvature fluctuations. 
The adiabatic solutions are defined as the solutions of the differential equations in which the densities and velocities of all the particle species are the same. 
These are known as curvature perturbations as they correspond to an overall shift in the curvature of spacelike surfaces. 
On the contrary, the isocurvature perturbations correspond to solutions where the fractional density and/or velocities of the particle species is not the same on spacelike surfaces.
Thus isocurvature perturbations are defined between any two species. For example, there can be a relative difference in the densities or velocities of the baryons and cold dark matter.
Canonically the isocurvature is defined as the fractional difference in particle species to the photon density.
In general there can be isocurvature between any of the particle species and therefore the most general initial conditions are given by a set of five possible linear combinations of modes: Adiabatic modes, CDM isocurvature, Baryon isocurvature, Neutrino density isocurvature and Neutrino velocity isocurvature \cite{Ma:1995ey, Bucher:2000cd, Bucher:2000kb, Bucher:2000hy}. 
There have been many attempts to constrain the amplitude of these general set of initial conditions and most studies show that the amplitude of isocurvature fluctuations must be much smaller than the amplitude of adiabatic fluctuations \cite{Moodley:2004nz, Beltran:2004uv, Lazarides:2004we, Beltran:2005xd}.
There is a further class of isocurvature known as {\it compensated isocurvature} in which there are isocurvature fluctuations due to both baryons and dark matter. 
This type of isocurvature has been shown to be more compatible with current observations \cite{Grin:2011tf, Grin:2013uya, Munoz:2015fdv}. 

In this study we do not consider isocurvature modes, instead we analyse the {\it structure} of adiabatic modes. 
Since the differential equations that govern all perturbations are second order differential equations, even for the adiabatic solution, there are two possible modes. 
One is called the {\it decaying} mode and the other is the more familiar {\it growing} mode. 
These names are motivated by the early time, super-horizon behaviour of these modes, as the decaying mode has a decaying behaviour whereas the growing mode remains constant.
The amplitude of these modes is usually set initially during a pre-radiation dominated era. Since the perturbation solution is a linear combination of each of these modes, \emph{both of these modes will be sourced by any pre-radiation phase that gives rise to adiabatic initial conditions}.

\begin{figure}
\includegraphics[scale=0.4]{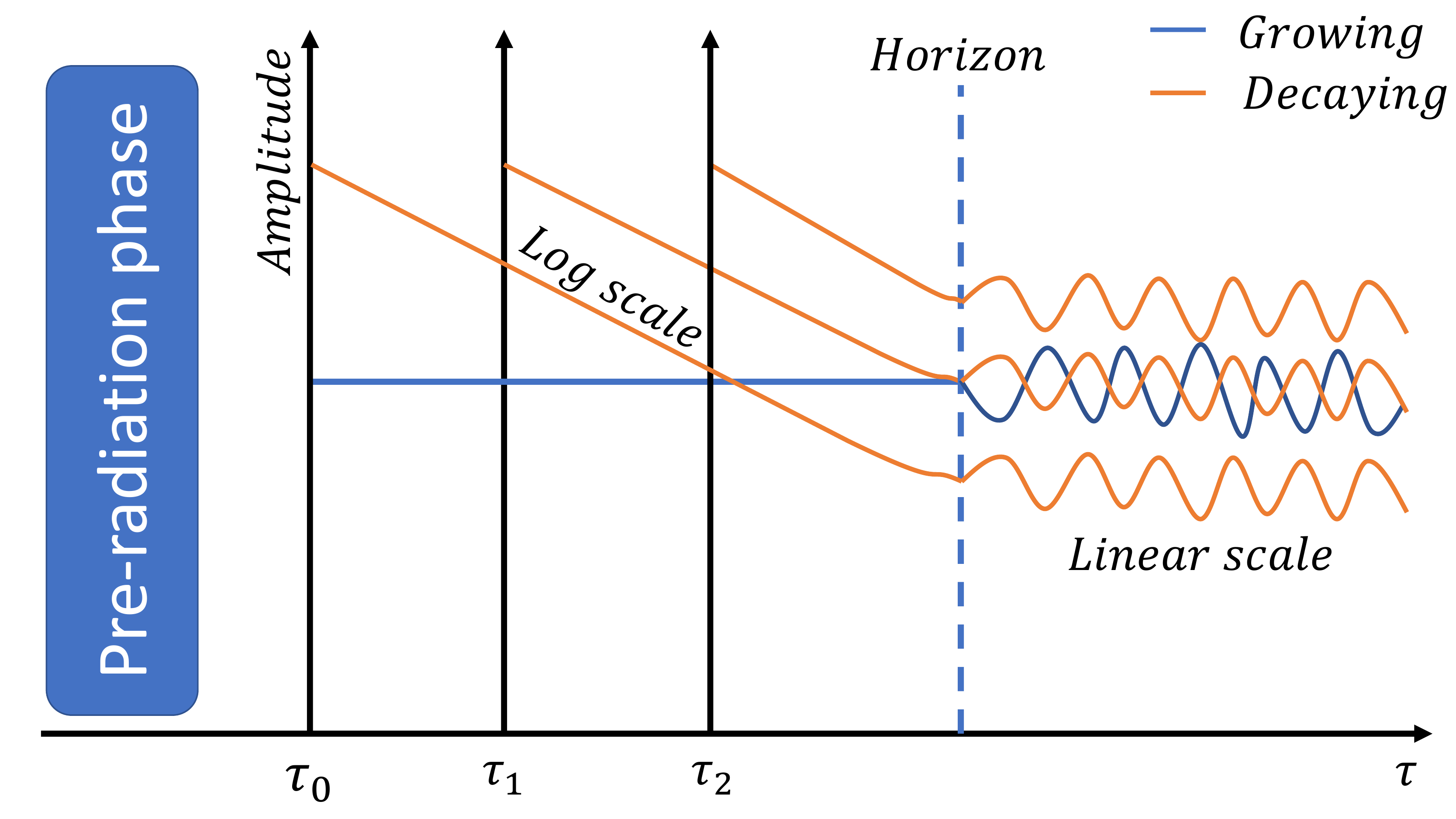}
\caption{Schematic diagram showing that in general both the decaying and growing modes should be sourced by whatever pre-radiation dominated era sets the initial conditions of the universe.
The amplitude of the decaying mode is time-dependent and therefore the amplitude of the decaying mode at recombination is very sensitive to the initial time the amplitudes are set. The amplitude is in log(linear) scale for the super(sub) horizon modes. While the numerical value of the amplitude appears to diverge super-horizon, it does not lead to divergent observable constraints.}
\label{Schematic}
\end{figure}

The decaying mode is qualitatively different to the growing mode as its amplitude is time dependent even on super-horizon scales as shown in Fig.~\ref{Schematic}.
Furthermore, since we are not directly able to measure super-horizon modes it may also be sensible to define these modes by their sub-horizon behaviour. 
On sub-horizon scales, both of these modes are described by oscillatory functions. 
In a pure radiation Universe, the decaying solution is a sine wave and the growing solution is a cosine wave. 
We will use the names {\it sine(cosine)} modes or {\it decaying(growing)} modes interchangeably throughout this {chapter}.
While it is difficult to source decaying modes from inflation, there are scenarios in which they might be generated. 
Specifically, there have been many studies of bouncing and cyclic universes in which decaying modes can be sourced. 
In particular growing modes in a pre-bounce contracting phase can become decaying modes in the post-bounce expanding phase \cite{ Bozza:2005xs, Battefeld:2005cj, Chu:2006wc, Brandenberger:2007by, Alexander:2007zm}.
There is currently no consensus on how the modes are matched across a bounce as this involves understanding the quantum behaviour of the fields causing the bounce in the large curvature regime. 
There have been some recent attempts at computing the propagation of perturbations across a bounce both classically and quantum mechanically in \cite{Gielen:2015uaa, Gielen:2016fdb} which suggest decaying modes could be present. 
More recent studies of the perturbations have gone beyond the leading order expansions and have shown that the decaying modes will also be sourced at second order in perturbation theory (for example from the neutrino velocity mode as it sources anisotropic stress) even if at leading order one only keeps growing modes \cite{Carrilho:2018mqy}. 
Instead of studying a particular scenario in detail we instead use the studies above as motivation to study decaying modes in general.

There has only been one study \cite{Amendola:2004rt} which has attempted to analyse the effect of decaying modes and our aim is to further elaborate and build on this analysis.
In this study we quantify how large the amplitude of these decaying modes can be irrespective of how they are sourced.
We do this by finding the Fisher information in each bin of $k$ in the decaying mode power spectrum, similar to what is done in studies that attempt to reconstruct the power spectrum for the growing mode \cite{Gauthier-Bucher, Bridle:2003sa, Hazra:2017joc, Kogo:2003yb, Nicholson:2009zj}. 
This gives a direct handle on the fraction of decaying modes present on all scales in the universe at the time of recombination.
We will show the constraints on the decaying mode power spectrum that come from using both the temperature and polarization angular power spectrum of the CMB.

The {chapter} is organised as follows. 
In section \ref{theory_decay} we present an intuitive explanation for the growing and decaying modes in a pure radiation universe. 
We then extend this analysis to the describe the initial conditions in general in both the Synchronous and Newtonian gauge to analyse the gauge dependence of the gravitational potentials and confirm the time dependent behaviour of decaying modes on super-horizon modes.
With the time dependence established we provide a normalisation procedure of decaying modes on subhorizon modes.
In section \ref{Analysis} we describe our formalism to constrain the power in the decaying modes using a Fisher matrix formalism and present the results.
We conclude and address possible future directions in \ref{conclusion}

\section{Theory of the decaying mode}\label{theory_decay}

\subsection{Review of radiation domination}

The equations that govern the evolution of the perturbations in standard cosmology are the perturbed Einstein equations. 
In homogenous and isotropic models of the universe, the solution to the Einstein equations is given by the Friedmann-Robertson-Lemaitre-Walker (FRLW) metric.
In the Newtonian (N) gauge, the perturbed FRLW metric for scalars is parametrised by
\begin{equation}
	ds^2_{\rm N} = a(\tau)^2\left( - d \tau_{\rm N}^2(1 + 2 \Phi) + dx^i_{\rm N} dx^j_{\rm N} \gamma_{ij}(1 - 2\Phi) \right). \label{Newtonian}
\end{equation}  
Here $a(\tau)$ is the conformal scale factor and $\gamma_{ij}$ is the flat three dimensional metric on spatial hyper-surfaces. 
This parametrisation of the metric is particularly useful to analyse the physical behaviour of perturbations as it is directly related to the gauge invariant Bardeen potentials, $\Psi = \Psi_{\rm B}, \Phi = - \Phi_{\rm B}$ \cite{Bardeen:1980kt}. 
The equation of motion for the gravitational perturbations in the presence of a pure radiation fluid in the Newtonian gauge, in the absence anisotropic stress, is given by \cite{MUKHANOV1992203}
\begin{equation}
	\Phi'' + 3\mathcal{H} (1 + c_s^2) \Phi' - c^2_s \nabla^2 \phi + (2 \mathcal{H}' + (1 + 3c_s^2) \mathcal{H}^2) \Phi = 4 \pi G a^2 \tau \delta S. \label{phi_EOM}
\end{equation}
Here $\mathcal{H} \equiv a' / a$ is the conformal Hubble parameter and $\delta S$ is a source term (See Eq.~(5.22) in \cite{MUKHANOV1992203} for full definitions). 
The source term is generated by isocurvature fluctuations and thus is zero for a pure adiabatic solution.
If we restrict ourselves to the radiation dominated era of the universe and without isocurvature, Eq (\ref{phi_EOM}) simplifies to 
\begin{equation}
	\phi_k'' + \frac{4 \phi'_k}{\tau} + \frac{k^2 \phi_k}{3} = 0,
\end{equation}
which has a simple solution
\begin{equation}
	\phi_k = A_k \frac{j_1(x)}{x} + B_k \frac{n_1(x)}{x}.
\end{equation}
The amplitudes $A_k$ and $B_k$ are set by the initial conditions for the differential equation, which are the initial conditions for our universe. 
The $k$ index shows that the amplitude can be different for different $k$'s.
Here we have defined $x \equiv \frac{k \tau}{\sqrt{3}}$. The $j_1(x)$ and $n_1(x)$ are the Bessel and Neumann functions of order 1 respectively. 
The term with the Bessel (Neumann) function is the {\it growing (decaying)} which have a cosinal and sinusoidal oscillation respectively.  
It is illuminating to look at the asymptotic limit of these modes. 
At early times on super-horizon scales, i.e. $x \ll 1$, the potential becomes 
\begin{equation}
	\phi_k(x\ll1) = \frac{A_k}{3} + \frac{B_k}{x^3} \label{decay_1}.
\end{equation}
Here we see that the decaying mode diverges as $x \rightarrow 0$. 
Furthermore, in most models of inflation the decaying mode will be suppressed by $\mathcal{O}(e^{3N})$, where $N$ is the number of e-folds, as the curvature perturbations in inflation will have their amplitudes set at a much earlier time. 
These are the main reasons behind most cosmological analysis assuming $B_k =0$. 
We also see that the growing mode is a constant on super-horizon scales. 
The usual procedure is to match the primordial curvature perturbation $\mathcal{R}_k$ to the amplitude of $A_k$, i.e $\mathcal{R}_k(\tau = 0 ) = - \frac{3}{2} \phi_k(\tau=0)$. 
Now lets analyse the large $x$ limit (sub-horizon limit)
\begin{equation}
	\phi_k(x) = - \left( A_k \frac{\sin{x}}{x^2} + B_k \frac{\cos{x}}{x^2}\right). \label{pure_radiation}
\end{equation}
Here we see that both modes simply oscillate at late times on sub-horizon scales. 
Thus if there was any remaining non-negligible amount of decaying mode amplitude on sub-horizon scales, {\it it would not decay away}. 
It is therefore sensible to ask how large the amplitude of such a decaying mode has to be to lead to observable effects (or similarly, constrained by the data). 
That is the main question we set out to answer in this {chapter}. 

\subsection{CMB anisotropies}

The angular power spectrum of the CMB anisotropies is given by \cite{durrer_cmb}
\begin{equation}
	C^{XY}_{\ell} = \int^\infty_{0} d \ln k \ P^{XY}(k) |\Delta^X_{\ell}(k) \Delta^Y_{\ell}(k)| 
\end{equation}
Here $P(k)$ is the primordial power spectrum of curvature perturbations.  
$X,Y \in \{T, E\}$ where $T,E$ stand for temperature and polarization respectively. 
$\Delta^X_{\ell}(k)$ is either temperature or polarization transfer function for adiabatic modes. 
In general, the transfer functions are computed using a line of sight approach by separating out the geometric projection effects (that depend on $\ell$) and the physical effects coming from gravitational potentials and Doppler effects \cite{Seljak:1996is}. 
On large scales the source function for temperature anisotropies is given by the gravitational potential, $\frac{\Delta T}{T} \approx \frac{1}{3} \Phi$.
This effect is caused by photons from the CMB having to climb out of a gravitational well and is called the Sachs-Wolfe (SW) effect. 
Thus, on large scales the CMB power spectrum should directly see a change in the gravitational potential, such as the change due to decaying modes in Eq. \ref{decay_1}.

We can check this explicitly by implementing the initial conditions for the decaying mode into the Boltzmann-solver CLASS \cite{CLASSII} and in the synchronous (S) gauge these are parametrised by
\begin{equation}
	ds^2_{\rm S} = a^2(\tau) \left( - d\tau_{\rm S}^2 + dx_{\rm S}^i dx_{\rm S}^j \left(\gamma^{\rm S}_{ij} + h_{ij} \right) \right). \label{Synchronous}
\end{equation}
We will focus on scalar perturbations in this {chapter} and it is canonical to separate $h_{ij}$ into two scalars: its trace $h$ and traceless $6 \eta$ parts.
The initial conditions in this gauge are given by \cite{Amendola:2004rt}

\begin{eqnarray}
h(x, \phi) &=& x^2 + f_{\rm GD}x^\frac{3}{2} \sin{\xi}, 	\nonumber	\\
\eta(x, \phi) &=&  2 - \frac{5 + 4R_\nu}{6(15+4R_\nu)} x^2  +  \frac{f_{\rm GD}}{x^\frac{1}{2}} \left[ \frac{11 - \frac{16R_\nu}{5}}{8} \sin{\xi} + \frac{5 \gamma}{8} \cos{\xi} \right],	\nonumber	\\
\delta_\nu(x, \phi)  &=& - \frac{2x^2}{3} + f_{\rm GD}x^\frac{3}{2} \left[ \left( \frac{1}{4R_\nu} - \frac{2}{5} \right) \sin{\xi} - \frac{\gamma}{4R_\nu} \cos{\xi} \right],	\nonumber	\\
\Theta_\nu(x, \phi)  &=& - \frac{23 + 4R_\nu}{18(15+4R_\nu)} kx^3 + \frac{f_{\rm GD}}{16R_\nu}kx^\frac{1}{2} \left[ \left( -3 - \frac{72}{5} R_\nu \right) \sin{\xi} + \gamma \left(3 - \frac{8R_\nu}{5} \right) \cos{\xi} \right],	\nonumber	\\
\Theta_r(x, \phi)  &=& \Theta_b = - \frac{kx^3}{18} + \frac{f_{\rm GD}kx^\frac{5}{2}}{3(25+\gamma^2)} \left(\gamma \cos{\xi} - 5 \sin{\xi} \right), \nonumber	\\
\sigma_\nu(x, \phi)  &=& \frac{4}{3(15+4R_\nu)}x^2 + \frac{f_{\rm GD}}{x^\frac{1}{2}} \left[ \frac{\gamma}{2} \cos{\xi} + \frac{11 - 16R_\nu/5}{10} \sin{\xi} \right],	\nonumber	\\
\delta_r(x, \phi)  &=& - \frac{2}{3} x^2 - \frac{2f_{\rm GD}}{3} x^\frac{3}{2} \sin{\xi},	\nonumber	\\
\delta_c(x,\phi)  &=& \delta_b = - \frac{x^2}{2} - \frac{f_{\rm GD}x^\frac{3}{2}}{2} \sin{\xi},
\label{decay_ic}
\end{eqnarray}

with the following definitions
\begin{eqnarray}
	\xi & \equiv & \frac{\gamma}{2} \log{x} + \phi; \,\,\,\gamma  \equiv  \sqrt{\frac{32}{5} R_\nu -1},	\nonumber	\\
	x & \equiv & k\tau;\,\,\,\,\,\,\,\,\,\,\,\,\,\,\,\,\,\,\,\,\,\,\,R_\nu  \equiv  \frac{\rho_\nu}{\rho_\nu + \rho_\gamma}.
\end{eqnarray} 
The amplitude $f_{\rm GD}$ is the ratio of the decaying mode to the growing mode.
We have defined the densities $\delta_i$, velocities $\Theta_i$ for each of the species $i \in \{$radiation (r), CDM (c), Baryons (b), neutrinos $(\nu)\}$. 
$\sigma_\nu$ is the quadrupole moment of the neutrino phase space density and $R_\nu$ is the relative energy density fraction of neutrinos.
The physical reason for the neutrinos having a quadrupole is that they will have anisotropic stress after they decouple. 
However this is also the case for the growing adiabatic mode \cite{Bucher:1999re, Ma:1995ey}, which can be obtained by setting $f_{\rm GD}$ equal to zero in the Eq. \ref{decay_ic}. 
We also note that the decaying mode has two independent variables $f_{\rm GD}$ and $\phi$. 
This is because for decaying modes there is an additional equation of motion for the neutrino distribution. 
This can easily be seen if one considers a pure radiation fluid coupled to neutrinos, as was pointed out in \cite{Amendola:2004rt}. 
It is known that the growing mode remains constant on super-horizon scales. 
However this is not the case for decaying modes. 
We have already seen this for a pure radiation universe in Eq.~\ref{pure_radiation}.
The initial conditions in the synchronous gauge do not make the time dependence(or independence) apparent as it appears both growing and decaying modes are time dependent. 
However the metric potentials $\eta$ and $h$ are not gauge invariant quantities. 
It is, therefore, better to analyse the time dependence in the Newtonian gauge as the metric potentials are directly related to the gauge invariant Bardeen potentials. 
We can switch to Newtonian gauge by either solving the Boltzmann equations in the Newtonian gauge or, as we are only interested in the behaviour of the gravitational perturbations, we can relate the two metrics via, 
\begin{equation}
	g_{\mu \nu(\rm N)} = g_{\alpha \beta(\rm S)} \frac{\partial x^\alpha_{(\rm S)}}{\partial x^\mu_{(\rm N)}} \frac{\partial x^\beta_{(\rm S)}}{\partial x^\nu_{(\rm N)}},
\end{equation}
where the variables with $(\rm N) / (\rm S)$ are in the Newtonian / Synchronous gauge which are defined in Eq.~\ref{Newtonian}, \ref{Synchronous} respectively. 
The relations between the metric potentials can then be calculated to be
\begin{eqnarray}
	\Psi(x, \phi) & = & \frac{1}{2k^2} \left[ \ddot{h}(x, \phi) + 6 \ddot{\eta}(x, \phi) + \frac{\dot{a}(\tau)}{a(\tau)} \left( \dot{h}(x, \phi) + 6 \dot{\eta}(x, \phi) \right) \right], \nonumber \\
	\Phi(x, \phi) & = & \eta(x, \phi) - \frac{1}{2k^2} \frac{\dot{a}(\tau)}{a(\tau)} \left[] \dot{h}(x, \phi) + 6 \dot{\eta}(x, \phi) \right].
\end{eqnarray}
Using these to evaluate the Newtonian potentials we get

\begin{eqnarray}
	\Psi(x, \phi) &=& \frac{20}{15 + 4R_\nu} + \frac{f_{GD}}{8 x^{\frac{1}{2}}} \left( 6 \gamma \cos \xi - (9-\gamma^2) \sin \xi \right)  + \mathcal{O}(x^{-\frac{5}{2}}) \nonumber \\
	\Phi(x, \phi) &=& \frac{4(5+2R_{\nu})}{15 + 4 R_\nu} + \frac{f_{GD}}{40 x^\frac{1}{2}} \left[(15 \gamma \cos \xi + (25 - 16R_\nu) \sin \xi  \right) + \mathcal{O}(x^{-\frac{5}{2}}) \nonumber \\
    \label{Newtonian_pot}
\end{eqnarray}

\begin{figure}
\begin{center}
\includegraphics[scale=0.4]{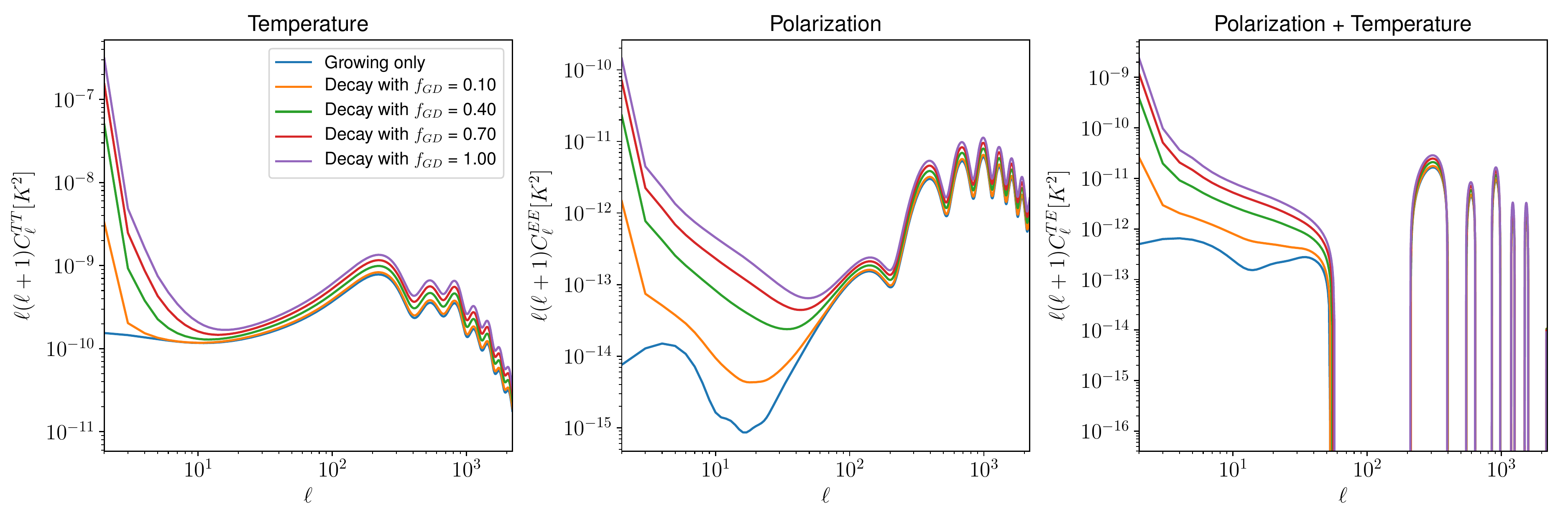}
\caption{Angular power spectrum of the CMB temperature and polarization anisotropies for the decaying modes. {The different lines correspond to different values of the decaying mode amplitude set according to the default values of initial conditions in CLASS. The blue line represents only the growing modes, i.e the fiducial value.}}
\label{decay_cls}
\end{center}
\end{figure}

We see that for the growing mode, i.e when $f_{\rm GD} = 0$, the metric potentials are constant. 
Whereas for the decaying mode, the potentials are clearly time dependent. 
Thus, we need to specify the time at which the decaying modes start evolving as the constraints we get on the amplitude will be depend on this time, as is shown in Fig.~\ref{Schematic}. To get an idea of what the power spectrum of the decaying mode looks like we have implemented these initial conditions in the CLASS Boltzmann code and the resulting power spectra for the temperature, polarization and their cross spectrum are shown in Fig.~\ref{decay_cls}.
The initial conditions in these plots are set by the default CLASS settings that can be found in \cite{CLASSII}\footnote{A summary of schemes used for setting the initial conditions is given in Fig.~10.}. 
In Fig.~\ref{decay_cls} we have assumed a power spectrum of the decaying mode to be analogous to the growing mode and set the spectral index $n_s^{\rm D} = n_s^{\rm G} = 0.96$ while the amplitude is defined by the scalar amplitude $A_s^{\rm G}$ and the fraction of decaying mode amplitude $A_s^{\rm D} = f_{\rm GD} A_s^{\rm G}$. 
We set $\phi = 0$ and the rest of the cosmological parameters are set to the fiducial values given in table \ref{fid_params}.

\begin{table}
\begin{centering}
\begin{tabular}{ p{3cm}  p{3cm}}
\hline
\hline
$A_s$ & 2.3 $\times 10^{-9}$  \\
$h$ & 0.6711  \\
$\Omega_bh^2$ & 0.022068  \\
$\Omega_{cdm}h^2$ & 0.12029 \\
$k_{*}$ & 0.05 $Mpc^{-1}$ \\
$n_s $ & 0.9619 \\
$N_{eff}$ & 3.046 \\
\hline
$\ell_{max}$ & 2500 \\
$f_{sky}$ & 1 \\
\hline
\hline
\end{tabular}
\caption{Fiducial cosmological parameters and systematic parameters - {these are the values used throughout this chapter unless stated otherwise. These are chosen to best fit the current CMB data - however as we are only interested in the difference between the decaying and growing mode amplitudes, for the purposes of this stud, the precise values of the rest of the cosmological parameters are not important.}}\label{fid_params}
\end{centering}
\end{table}

\begin{figure}
	\centering 
	\includegraphics[scale = 0.35]{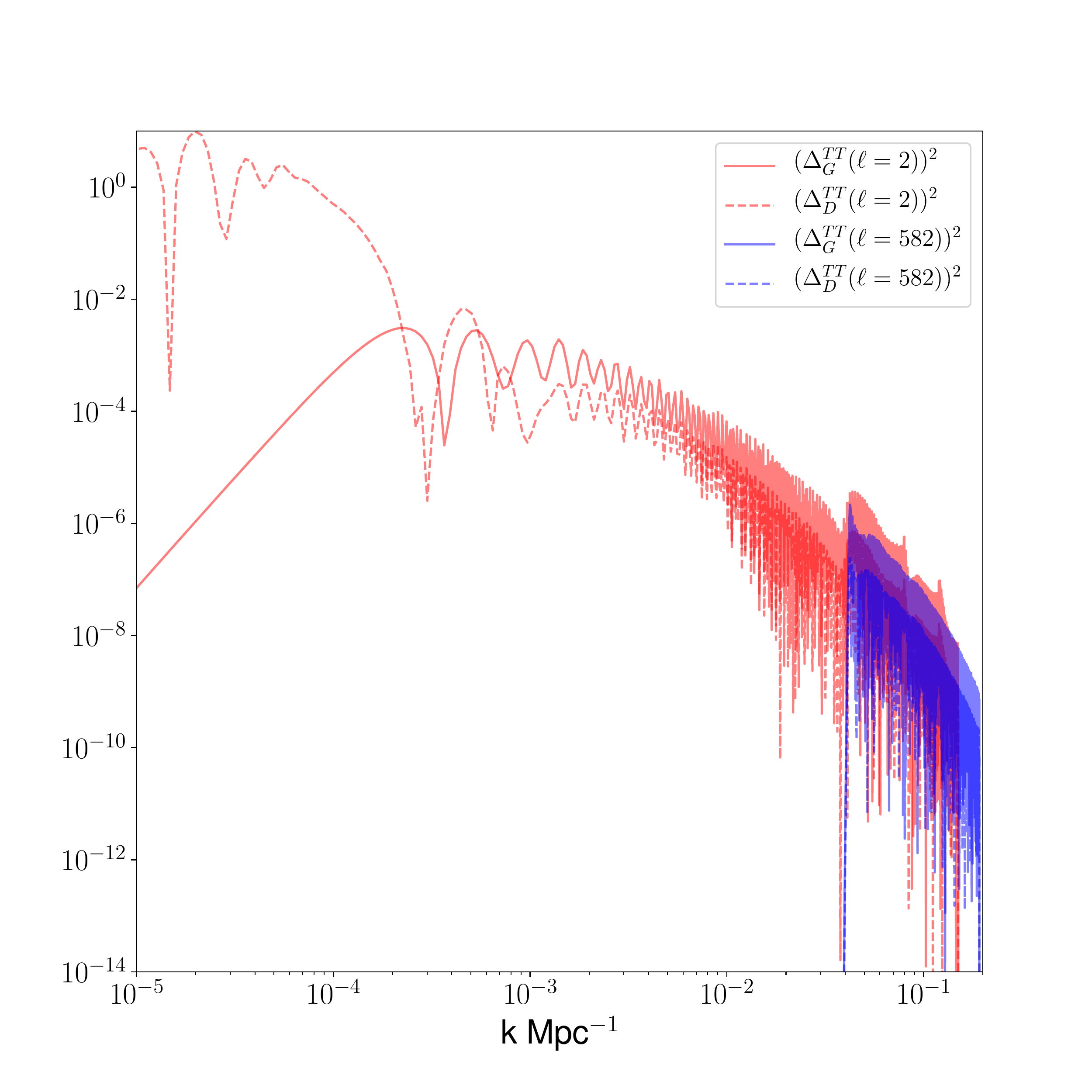}
	\caption{Transfer functions for growing and decaying modes {for $\ell$ = 2, 582. These scales are chosen to show that the decaying mode has most of its effect on large scales. Furthermore, we also notice that how different $k$ modes contribute to a given $\ell$, thus even large $\ell$'s will be effect by the large scale decaying mode perturbations.}}
	\label{transfer_plots}
\end{figure}

There is a clear divergence on large scales which comes from the divergence of the gravitational potential on super-horizon scales. 
The gravitational potential enters the $C_{\ell}$'s through the transfer function's $\Delta_{\ell}(k)$.
These are (numerically) computed using a line of sight integral \cite{Seljak:1996is} over the source function (which contains the Sachs Wolfe, Doppler and Integrated Sachs Wolfe effect terms) convolved with a projection function which is a Bessel function. 
\begin{equation}
	\Delta_{\ell}(k) \equiv \int^{\tau_0}_{\tau_i}  \ d \tau \ S_T(\tau, k) j_{\ell}(k(\tau_0 - \tau))
\end{equation} 
Here $\tau_0$ is the time at recombination and $\tau_i$ is the time at which the initial conditions are sourced. We show the transfer functions for $\ell =2, 582$ in Fig.~\ref{transfer_plots}. 

The low $\ell$'s show the divergent behaviour for the decaying mode, whereas at $\ell = 582$ we see that both modes are similar with the decaying mode having a lower amplitude. 
{The precise shape of the large scale transfer functions depends on the smoothing scales used in the numerical integrator in the Boltzmann code, however the the amplitude of the transfer function is the same up to $\mathcal{O}(\text{few}\%)$ irrespective of the numerical scheme. Furthermore, the observable effect and hence the constraining power of each of these modes comes from the effect they have on the $C_\ell$'s, which are obtained by integrating over the transfer functions. This the reason we choose the normalisation scheme shown in Eq \ref{renorm_func}, as that smooths over the oscillatory transfer functions and is most directly related to the $C_\ell$'s.}
This means a non-negligible amplitude of the adiabatic perturbations could be in decaying modes if they are generated at late times or on large scales. 
Furthermore a primordial power spectrum with a large spectral index could also allow for a non-negligible contribution of the decaying mode amplitude to the overall adiabatic perturbations. 

Instead of focusing on setting the amplitude at early times, we use a renormalising procedure to set the amplitude of the decaying modes.
There are two reasons to use this normalisation procedure. 
First, it provides a unique way to set the initial conditions as the decaying modes are time dependent and the time dependence is \emph{different} in different gauges. 
For example the time dependence of decaying mode metric potentials in the Synchronous gauge in Eq.~\ref{decay_ic} is clearly different from the metric potentials in the Newtonian gauge in Eq.~\ref{Newtonian_pot}.
Second, since both the growing and decaying solutions are described by regular (non-diverging) functions on sub-horizon scales we can set the amplitudes of the growing mode equal to that of the decaying mode deep inside the horizon.
This makes it easier to see the effect of decaying modes that are set at late times as they would naturally be normalised on sub-horizon scales. 

\begin{figure}
	\centering 
	\includegraphics[scale = 0.3]{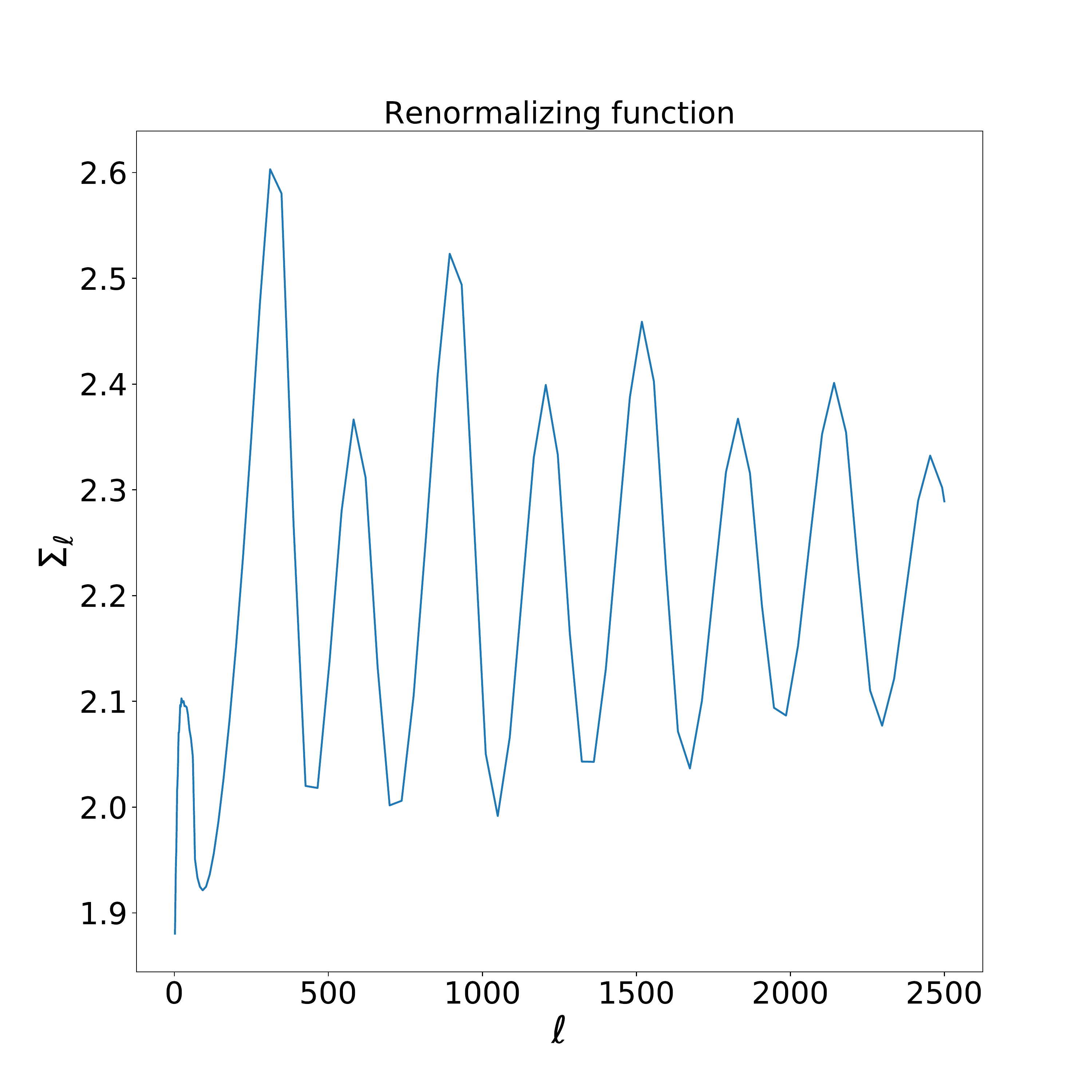}
	\caption{Renormalisation function defined in Eq (\ref{renorm_func})}
	\label{renorm}
\end{figure}

The normalisation of the two modes is done in terms of the transfer functions in $k$ space as opposed to the transfer functions in $\ell$ space as we wish to isolate the physical effects of the gravitational potentials (which show the behaviour of the growing and decaying modes) from the projection effects.
We equate the amplitudes of the decaying and growing modes on all scales below the fiducial horizon scale $k_{\rm horizon} = 3 \times 10^{-3} \mbox{ Mpc}^{-1}$. 
In practice it is not easy to do this since the transfer functions are highly oscillating functions.
Our approach is to integrate the transfer function for each $\ell$ for all $k$'s that are inside the horizon for both the growing mode and decaying mode. 
{This gives a renormalisation function which is stable up to $\mathcal{O}(10\%)$. In practice this is not very important anyway, as the constraints on the decaying modes on large scales are several orders of magnitude greater than growing modes, thus a few percent difference in the absolute amplitude of the decaying mode will not change that constraint significantly.}
The ratio of these integrals will tell us the normalisation for the decaying mode transfer function for a given $\ell$ that will ensure the decaying mode will have the same amplitude as the growing mode on sub-horizon scales. 
This would correspond to the case where the universe starts at $\tau_1$ in Fig.~\ref{Schematic}.
Thus the renormalised decaying mode transfer function can be written as 
\begin{eqnarray}
	\hat{\Delta}^D_{\ell}(k) & = & \Delta^D_{\ell}(k) \Sigma_{\ell}, \nonumber \\
	\Sigma_l & \equiv &  \frac{ \int_{k_{\rm horizon}}^{k_{\rm max}} dk \ \Delta^G_{\ell}(k)}{\int_{k_{\rm horizon}}^{k_{\rm max}} dk \ \Delta^D_{\ell}(k)}. \label{renorm_func}
\end{eqnarray}
We know \emph{a-priori} that this is the most conservative one can be as for decaying modes to have the same amplitude as growing modes on sub-horizon scales they must have a very large amplitude on super-horizon scales (at least for modes that entered that horizon at early times) and thus they will be highly constrained. 
Any early universe model that is responsible for generating the initial conditions can be renormalised in this way, thus allowing a direct comparison of the amplitudes of a model to our results by simply applying the renormalization function in Fig.~\ref{renorm}. 

\section{Analysis}\label{Analysis}
 
\begin{figure}
	\centering 
	\includegraphics[scale = 0.25]{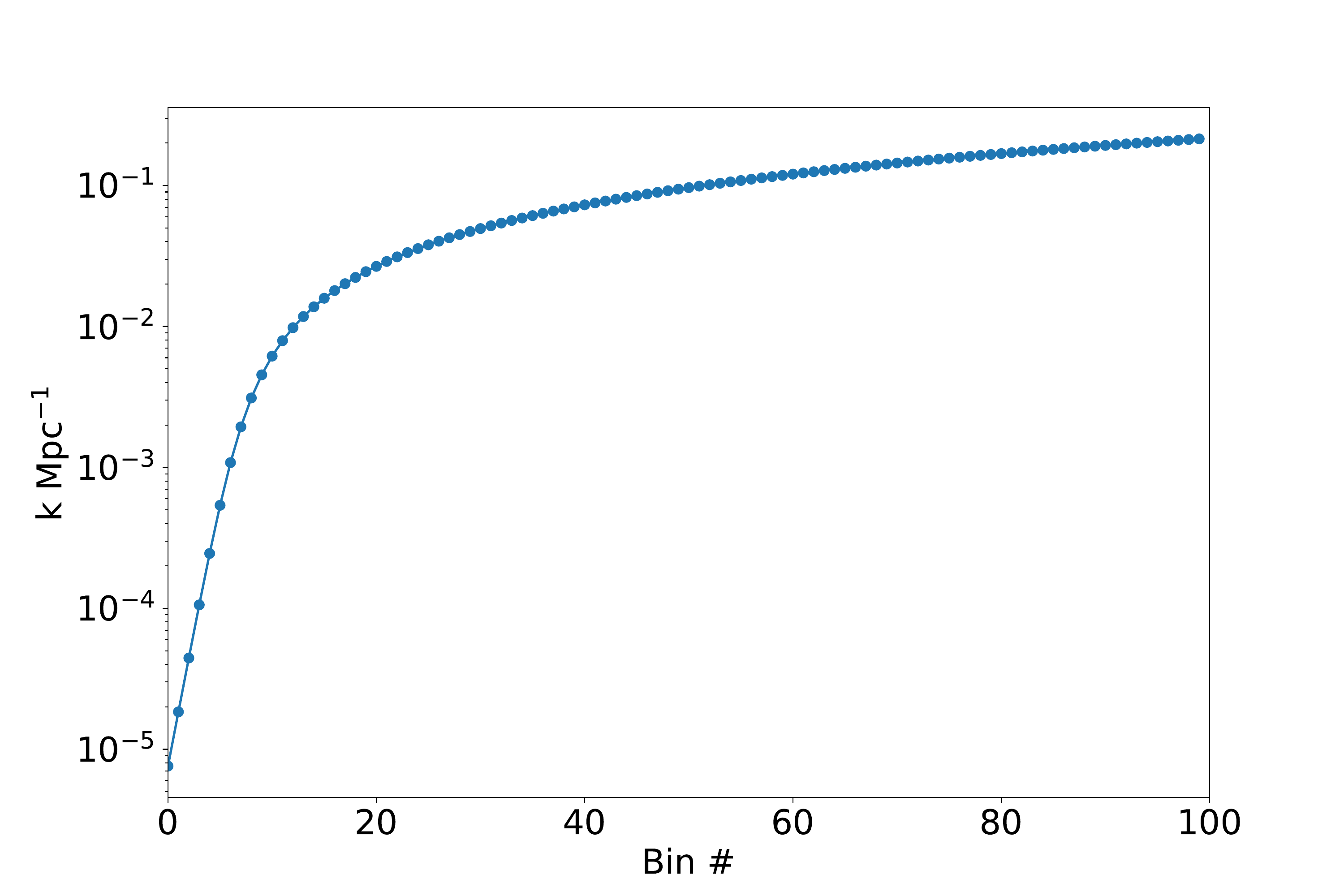}
	\caption{$k$ values at which we add power to the primordial power spectrum}
	\label{kbins}
\end{figure}

There are a variety of ways to model the primordial power spectrum. The most popular one, and the one which is normally constrained with data, is a power law with an amplitude and spectral index. There are a variety of ways to look for deviations from this. Here we take an unparameterised approach to constraining the decaying mode to keep our findings as general as possible.   
For that purpose, we model the power spectrum as a set of bins in $k$ with an independent amplitude and constrain the amplitude in each of those bins. 
The power spectrum is then given by 



\begin{equation}
	P (k, k_0, \epsilon) = 	\begin{cases} P(k)^{(G)} + \epsilon^{(G) \mbox{ or } (D) }_{k_0} & \text{if $k_0 \equiv k $} \\
	P(k)^{(G)} & \text{otherwise}
	\end{cases} 
\end{equation}

where $P(k)^{(G)} = A^{(G)}_s \left( \frac{k}{k_*} \right)^{n_s^{(G)} - 1}$. 
We choose 100 values for $k_0$ from an infrared cutoff of $3 \times 10^{-5} \ \mbox{Mpc}^{-1}$ to $3 \times10^{-1} \ \mbox{Mpc}^{-1}$ with the precise values for each bin shown in Fig.~\ref{kbins}. 

To account for the information on smaller scales we would also need to account for CMB lensing due to large scale structure which we know can change the temperature power spectrum by $\mathcal{O}(20 \%)$ on scales below $\ell \sim 3000$, thus we do not look at {larger} $\ell$'s.
This parametrisation allows us to look for features in the primordial power spectrum that can arise by either the growing mode or the decaying mode. 
In the case where the feature is due to the decaying mode, i.e $\epsilon^{(D)}$ is added to the power spectrum, we also use the decaying mode transfer functions to evaluate the $C_{\ell}$'s.
Since the $C_{\ell}$'s are a linear function of the power spectrum, the total $C_{\ell}$'s will just be the sum of the growing mode fiducial power spectrum $C_{\ell}$'s and a response due to the decaying mode being added. 
We will also consider the effect adding polarization information has on the constraints.
Since the transfer function for the decaying mode is different for polarization and temperature, the same primordial power spectrum may not be able to account for the change in temperature and polarization. 
A similar analysis has been done for parametrised isocurvature modes \cite{Bucher:2000cd} and it was shown that adding polarization significantly increases the constraining power of the CMB for the amplitude of isocurvature modes. 
In principle one could apply this un-parametrised approach to primordial isocurvature perturbations as well and we will leave this to future works.

\begin{figure}
\begin{center}
\includegraphics[scale=0.4]{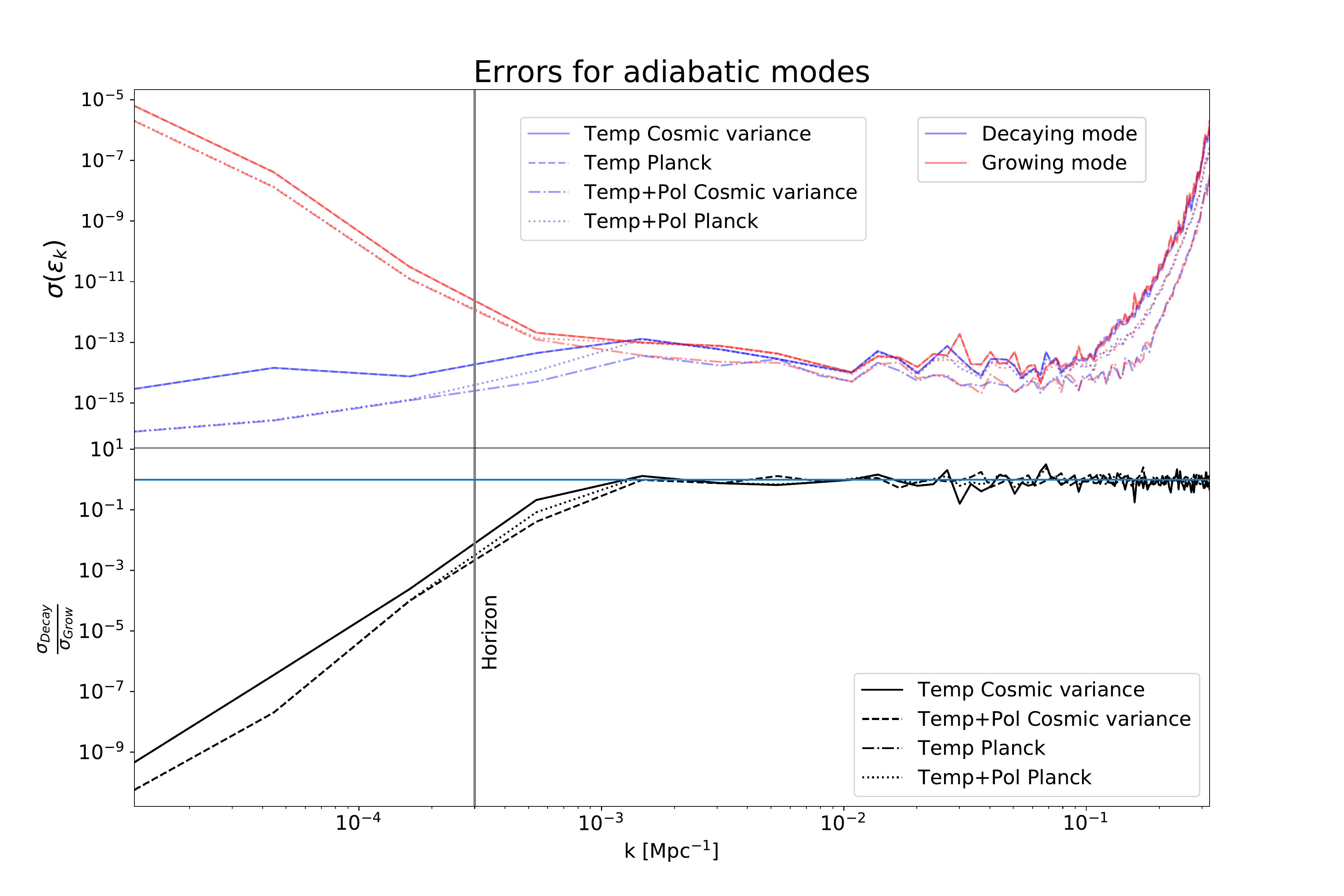}
\caption{This plot shows the errors for the decaying and growing modes in each of the 100 $k$ bins. The analysis is done for four specifications:  temperature anisotropies in a Planck like experiment and a cosmic variance limited experiment and the same analysis for temperature and polarisation data. The top plot shows the errors and the bottom plot shows the ratio of the errors of the decaying to growing modes. The vertical line is drawn at roughly the size of the horizon as inferred from the maximum scale observable by an observer at recombination. Since $\ell = 2$ is the largest mode observable in the CMB we compute the corresponding $k$ using $\ell = k \chi$. This expression is true in a flat sky, where $\chi$ is the comoving distance to recombination $\sim 10$ Gpc/h, furthermore $\ell =2$ corresponds to a mode wave with two wavelengths in a unit circle giving, thus giving a further factor of $\pi/2$ to the wavenumber giving $k\sim 3\times10^{-4} \mbox{ Mpc}^{-1}$. The horizontal line on the bottom plot is at one and we note that the ratio of the errors asymptotes to 1. This is just a manifestation of the fact that we have normalised the amplitudes (but not the phase) of both modes to be equal on sub-horizon scales.}
\label{fish_sigma}
\end{center}
\end{figure}

To answer these questions we use the Fisher information as a metric to quantify the information in the decaying modes. 
The expression for the Fisher matrix for a Gaussian likelihood with a parameter independent covariance matrix can be written as 
\begin{equation}
	F_{\alpha \beta} = \sum_{l=2}^{l_{max}} \frac{f_{sky}(2l+1)}{2} Tr \left( \mathbb{C}_l^{-1} \partial_\alpha \mathbb{C}_l \mathbb{C}_l^{-1} \partial_\beta \mathbb{C}_l \right).
\end{equation}
The matrix $\mathbb{C}$ depends on the observables being used. 
When the temperature and polarization of the CMB are being used the matrix becomes
\begin{equation}
	\mathbb{C}_l \equiv \begin{pmatrix} C_l^{TT} + N_l^{TT} & C_l^{TE} \\ C_l^{ET} & C_l^{EE} + N^{EE}_l \end{pmatrix}.
\end{equation}
The fiducial $\mathbb{C}_l$ is assumed to be that of the growing mode only as we know it fits the data with the fiducial cosmology. 
The derivatives of the $\mathbb{C}_l$ matrix will have either the growing or decaying transfer functions, depending on which mode is being constrained.
Where $N_l^{TT}, N_l^{EE}$ represent the noise covariance for temperature, polarization respectively. 
We also assume the polarization and temperature noise are uncorrelated thus the covariance between them is zero.
We model the noise for the CMB polarization and temperature as Gaussian random noise per frequency channel as given in the Planck blue book \cite{Planck:2006aa}

\begin{equation}
	N^{TT(EE)}_l =  \left( (\sigma^2_{T(E)}B^2_l)_{100} + (\sigma^2_{T(E)}B^2_l)_{143} + (\sigma^2_{T(E)}B^2_l)_{217} + (\sigma^2_{T(E)}B^2_l)_{353} \right)^{-1} \label{Planck_noise}
\end{equation} 

where $\sigma_{T(E)}$ represent the variance for temperature (polarization) and $100,143,217,353$ are the Planck frequency channels in $GHz$. 
The window function is given by $B^2_l = \mbox{exp} \left({-\frac{l(l+1) \theta^2_{beam}}{8 \ln 2}} \right) $. 
The values of the beam size and variance are given in Tab.~\ref{Planck_noise}.

\begin{table}[h!]
\begin{centering}
\begin{tabular}{ |p{1.5cm}p{2cm}p{2 cm}p{2 cm}|}
\hline
Frequency ($GHz$) & $\theta_{beam} (rad)$ &  $\sigma_T$($\mu K$ - rad) & $\sigma_E$($\mu K$ - rad)   \\
\hline
100 & 0.002763 & 0.001984 & 0.003174 \\
143 & 0.002065 & 0.001746 & 0.003333 \\
217 & 0.001454 & 0.003809 & 0.007785 \\
353 & 0.001454 & 0.011665 & 0.023647 \\
\hline
\end{tabular}
\caption{{Noise parameters used for the fisher analysis as taken from the Planck blue book \cite{Planck:2006aa}.}}\label{Planck_noise}
\end{centering}
\end{table}

If we only use the temperature spectrum from the CMB the expression for the Fisher matrix simplifies to 
\begin{equation}
	F_{\alpha \beta} = \sum_{l=2}^{l_{max}} f_{sky} \frac{2l + 1}{2} \frac{\partial_\alpha C^{TT}_l \partial_\beta C^{TT}_l}{(C_l^{TT} +N_l^{TT})^2}.
\end{equation}
It is worth noting that the derivatives of the $C_l$'s wrt the parameters $\epsilon^{(D)}_{k_0}/\epsilon^{(G)}_{k_0}$ will simply return the transfer function squared of the decaying/growing mode at $k_0$. 
The errors on the parameters $i$, $\sigma_i$ (which in our case will be the amplitudes in each $k$ bin) can be obtained by $\sigma_i = \sqrt{(F^{-1})_{ii}}$. 
We plot these variances in Fig.~\ref{fish_sigma} along with the ratio of the errors of the decaying and growing modes.
We see that most of the information is in the range $ k \sim 10^{-3}- 10^{-1} \mbox{Mpc}^{-1}$ and 
adding the polarization data increases the information content by up to 2 orders of magnitude in this range. 
Similar results for the growing mode have been found in previous studies, see for example \cite{Gauthier-Bucher, Mukherjee:2005dc, Hu:2003vp}. We note that most of the analysis done so far {focuses} on providing detailed precision on growing modes and thus have more $k$ bins in a narrower range of wavenumber. 
Our aim is to probe the errors on a much broader range of $k$'s which has not been done before, yet we still note that in regions of overlapping $k$ space we recover similar results albeit without the same level of resolution. 
{The highly oscillatory features on small scales can be further resolved by increasing the resolution of the k modes in those regions, however that requires significantly larger computation time. As the aim of this study is to get an idea of how the decaying modes effects the observable power spectra, as opposed to getting precision constraints on them, along of with cosmological parameters, we do attempt this here.\footnote{Indeed, to do this systematically one would need to do a full MCMC analysis of all the decaying mode parameters in addition to the standard $\Lambda$CDM parameters and we take this up in a future study.}}

We see that on larger scales cosmic variance dominates and most of the information is lost.
The first thing to note about the decaying mode is that the overall difference in the Fisher information from the largest to the smallest scales is much lower than the growing mode.
This is because on large scales we can see from the $C_l$'s there is a large rise in power for the decaying mode transfer functions. 
Therefore even with the large errors due to cosmic variance, the excessive power in decaying modes on large scales can be constrained.  
On subhorizon scales the errors on both modes are approximately the same as we have normalised both modes to have the same amplitudes on subhorizon scales.
The second feature of the decaying mode is that there is a large increase in Fisher information, relative to the growing mode, when polarization information is included. 
This is to be expected because, as was mentioned before, the polarization transfer functions and temperature transfer functions are different. 
The fiducial cosmology we have assumed has been fitted to the temperature and polarization data with growing mode transfer functions, thus even if we allow a lot of freedom in the primordial power spectrum, the $C_l$'s, which are a convolution between the transfer functions and the primordial power spectrum, will struggle to accommodate the decaying mode power spectrum with the temperature and polarization transfer functions at the same time.

Finally it is interesting to note that modes that are smaller than $ \sim 10^{-4}$ Mpc$^{-1}$ will be larger than the universe's horizon today and some modes that are even larger may never enter the horizon of our universe. 
Thus one has to ask the philosophical question of how modes that are beyond our observable universe can be observed, even indirectly. 
The physical mechanism for super-horizon modes effecting sub-horizon observables is through the gravitational effect of super-horizon modes on small scale structure. 
This has been at the heart of \emph{separate universe approach} of describing super-horizon perturbations in which the local, sub-horizon, modes evolve in a different \emph{universe} with different cosmological parameters such as curvature, Hubble rate etc.  
Such claims have to be backed up with careful analysis of the underlying physics, in particular the curvature of spacelike surfaces, as one has to understand how the equivalence principle, which would suggest large scale modes should not effect the curvature of spacelike surfaces, can allow for such super-horizon modes to effect the sub-horizon modes. 
There have been many attempts to address this issue and a long yet in-exhaustive list is given here \cite{Rigopoulos:2003ak, Lyth:2003im, Martineau:2005aa, Sasaki:2005ju, Tanaka:2006zp, Takamizu:2008ra}. 
Most of these attempts have focused on calculating the back-reaction of the \emph{growing super-horizon modes} through the non-linear evolution of the modes due to Einstein's equation.
It would be interesting to see whether similar calculations can be used to evolve decaying modes and understand the physical origin on their effect on sub-horizon scales.
We do not attempt to address this here and note that our current study will provide a direct way to test whether the methods used to understand super-horizon evolution of modes lead to testable predictions. 

\section{Summary and future outlook}\label{conclusion}

In this {chapter} we have analysed the constraints on the amplitudes of the primordial power spectrum across a broad range of scales for adiabatic initial conditions.
Adiabatic initial conditions have two orthogonal set of modes that can be excited when the universe starts (during radiation domination) or at later times. 
These are the sine (decaying) mode or the cosine (growing) mode.
In general both modes can be excited however most cosmological analysis assume only the cosine mode is excited and thus the constraints on the amplitudes of the primordial power spectrum is directly matched to the amplitude of the cosine mode. 

The sine mode numerically appears to diverge at early times on super-horizon scales. Special care is needed to interpret super-horizon physics, and a mapping onto physical quantities is essential.
Past work attempted to normalize the decaying mode at a super-horizon initial condition, making the allowed amplitudes for the sine mode sensitive to the numerical start time {of the} universe.
Instead of taking a parametrised approach, in this analysis we have mapped the amplitude of the primordial power spectrum to the amplitude of both modes by looking for additional power spectrum features for discrete scales.

We have calculated the Fisher information for both the sine and cosine modes using a fiducial cosmology.
The initial conditions for this cosmology are normalised to be equal for both modes on sub-horizon scales. 
We have computed the Fisher information for these modes for a cosmic variance limited experiment as well as a full sky Planck like experiment with temperature and polarization anisotropies.
Both of the modes are best constrained on scales $k \sim 10^{-3} - 10^{-1}$ Mpc$^{-1}$. 
The sine mode is almost equally well constrained on larger scales, $\sim 10^{-4} \mbox{Mpc}^{-1}$ due to the divergent growth of its amplitude, whereas the cosine mode is less well constrained on these scales as they are cosmic variance limited.
The angular power spectrum for the anisotropies of the CMB are a convolution between the primordial power spectrum and the transfer function. Therefore allowing the primordial power spectrum to be a freely varying function may allow the decaying mode to fit the observed temperature anisotropies, it is unable to fit the polarization anisotropies at the same time as they have different transfer functions. It is worth emphasising that this argument only holds when we keep the cosmological parameters fixed. If we let the cosmological parameter vary \emph{at the same time} as varying the primordial parameters one may be able to find new points in parameter space that fit the observed data that allow for non-negligible amounts of power in the sine mode.

This approach of constraining the initial conditions of the universe can be very useful in understanding the early universe models that set the initial conditions in radiation domination. 
While the simplest models of single field inflation give rise to nearly scale invariant adiabatic perturbations, alternative early universe models can give rise to localised features. 
In the context of inflation, these localised features will temporarily break the slow roll behaviour as the features usually come from (but not limited to) sharp features in the inflationary potential \cite{Starobinsky:1992ts, Chen:2016zuu, Romano:2014kla}.
Perhaps the more interesting set of models to test using our approach are those of bouncing or cyclic universes.
It is possible that cosine modes in a pre-bounce era source sine modes in the post-bounce era. 
Thus any signs of the sine mode in our current universe might also be a sign of a previous cycle of our universe. 
This intriguing possibility depends on how the perturbations are matched across the bounce. 
There are various approaches to how this matching is done however most approaches depend on the underlying model that causes the bounce \cite{Bozza:2005xs, Battefeld:2005cj, Chu:2006wc, Brandenberger:2007by, Alexander:2007zm}. 

There are various natural extensions to this paper.
We have not looked at specific models in this paper however one could try to understand what is the best way to match perturbations across a bounce and what features they give rise to in the primordial power spectrum. 
Throughout this work we have assumed the cosmological parameters for the sine and cosine mode are the same. 
This does not have to be the case, as described above, and the best way to constrain the primordial and cosmological parameters together would be to do an MCMC analysis. 
We will leave a complete MCMC analysis of the adiabatic sine and cosine modes as well as the different types of isocurvature modes in addition to the cosmological parameters to future works.
In addition to scalar perturbations, one can also ask whether the most general tensor perturbations have been understood. 
Since tensor perturbations also have a second order differential equation that is the equation of motion they also must have two independent solutions and these topics are currently being explored. 

\chapter{Initial conditions of the universe: Decaying tensor modes}\label{ICOU_tensors}

\section{Introduction}

The CMB is the dominant observational probe when it comes to constraining models of the early universe.
The temperature and polarisation anisotropies in the CMB have been observed by several experiments over the last few decades \cite{wmap_final, refId0}. 
One of the main goals of future CMB experiments is to measure the polarisation anisotropies to greater precision, especially on large scales. 
In particular, the detection of B mode polarisation in the CMB is a primary target \cite{Core1, LiteBird} as they could be a signature of primordial tensor perturbations, i.e gravitational waves, which are predicted by a large number of inflationary theories (see \cite{Finelli:2016cyd, Baumann:2008aq} and references therein).
To calculate the effect of primordial tensor perturbations on the CMB we parametrise the primordial amplitude of the perturbations using a power law power spectrum and convolve that with the transfer functions for $B$-mode polarisation using a Boltzmann solver such as CAMB\footnote{https://camb.info} or CLASS\footnote{http://class-code.net}.

One implicit assumption in current cosmological analysis that search for primordial gravitational waves is that the primordial perturbations only have a single solution, the so-called \emph{growing} solution/mode. However, since the perturbations in the early universe are described by a second order differential equation, another solution exists known as the \emph{decaying} solution/mode. 
In a recent study \cite{Kodwani2} the effect of this second, orthogonal, mode, was considered for scalar perturbations in radiation domination. By explicitly keeping the decaying scalar mode and describing the primordial power as a set of independent bins in $k$, the effects of the decaying scalar mode on the CMB anisotropies was studied.
By constraining the amplitude of the primordial power spectrum for a broad range of band powers it was found that the decaying mode is equally well constrained as the growing mode on subhorizon scales, whereas on superhorizon scales there is a divergence in the decaying mode anisotropy spectrum which means they are more constrained then growing modes by several orders of magnitude.

The aim of this {chapter} is to extend this analysis by calculating the effect of the decaying \emph{tensor} mode in radiation domination on the CMB. To our knowledge, this has not been considered before, but given the broad theoretical interest in inflationary tensor modes and potentially far-reaching theoretical implications in case of a detection, it is timely to explore the effect such a mode could have on the CMB in case it was produced in the early Universe.  
Specifically, while inflation predicts negligible decaying modes, such modes can be generated in bouncing universe scenarios and a detection could open a new window onto the novel physics that describes the beginning of our universe.
As was described in \cite{Kodwani2, Amendola:2004rt}, the decaying modes are not constant, but evolve with time outside the horizon. Therefore we must specify the time at which we start the evolution of these modes. In the case of scalar perturbations one has to be careful about which gauge we use to define the time at which we evolve the modes as they will have different dependence on time in different gauges (i.e Newtonian and Synchronous gauges). 
Tensor perturbations have a unique description in any gauge at linear order and therefore do not suffer from these ambiguities. 

Canonically the form of the primordial power spectrum (PPS), $P_T(k)$, of tensor perturbations is assumed to come from the growing mode only and can be parametrised as a power law with an amplitude, $A_T$, and spectral index $n_T$
\begin{equation}
	P_T(k) = A_T \left( \frac{k}{k_*} \right)^{n_T} \label{tensor_pps}
\end{equation}
Here $k_*$ is the pivot scale for tensor perturbations.
We relax both of these assumptions, allowing the PPS to have a decaying mode solution while its power is described by a non-parametric binned form of the power spectrum. Similar approaches have recently been used to analyse the growing mode PPS for tensor perturbations \cite{Hiramatsu:2018nfa, Farhang:2018wgm, Campeti:2019ylm}. 

The {chapter} is structured as follows: In section \ref{decay_tensor} we briefly describe the theoretical framework of primordial tensor perturbations and the form of decaying initial conditions. Section \ref{anal} describes the formalism we use to constrain the decaying tensor initial conditions using a Fisher matrix analysis and presents the results. 
Finally, we summarise in section \ref{summary}.

\section{Decaying tensor modes}\label{decay_tensor}

\subsection{Primordial perturbations}

In this section we briefly review the formalism of computing tensor perturbations in the early universe and work out the form of the decaying modes (more detailed introductions can be found in \cite{Kodama:1985bj}, \cite{MUKHANOV1992203}).
The tensor perturbations are uniquely defined by perturbing the flat Minkowski metric 
\begin{equation}
	ds^2 = a^2(\tau) \left( -d \tau^2 + (\delta_{ij} + h_{ij}) dx^i dx^j \right), \label{TT}
\end{equation}
where $h_{00} = h_{0i} = 0, |h_{in}|\ll 1$ and the perturbations are transverse and traceless $h_{ij,i} = h^i_i = 0$.
To linear order, the transverse traceless perturbations in  Eq.~\eqref{TT} are gauge invariant. 
The tensor perturbations have two polarisation states denoted by ($+, \times$). 
The equation of motion for the perturbations is given by solving the Einstein equation. 
It is easiest to solve it in Fourier space and therefore we decompose the tensor perturbations into plane waves of each polarisation mode
\begin{equation}
	h_{ij}(\tau, \textbf{x} ) = \sum_{\lambda} \int \frac{d^3k}{(2 \pi)^3} h_\lambda(\tau, \textbf{k}) e^{i\textbf{k} \cdot \textbf{x}} \epsilon^\lambda_{ij}(k).
\end{equation}
Here we have defined $\epsilon^\lambda_{in}$ to be the polarisation tensor and $\lambda \in \{ +, \times \}$.
To linear order the Einstein equations for metric perturbations is given by (in units of $c = \hbar = M_{\rm pl} = 1$) a Klein-Gordon equation for a massless scalar field, for each polarisation mode, with a source term given by the anisotropic stress term
\begin{equation}
	\ddot{h}_{k,\lambda} + 2\frac{\dot{a}}{a} \dot{h}_{k.\lambda} + k^2 h_{k,\lambda} = 2 a^2 \Pi_{ij}, \label{EOM}
\end{equation}
where the dot denotes derivatives w.r.t conformal time. 
$\Pi_{in}$ is the anisotropic stress of the fluid with stress energy tensor $T_{in} = p g_{in} + a^2 \Pi_{in}$. 
The anisotropic stress is typically generated by neutrinos free-streaming in the early universe after they decouple at $z \sim 10^9$. 
It has been shown in \cite{Weinberg:2003ur} that the effect of this anisotropic stress is to damp the effects of primordial tensor perturbations in the $B$-mode power spectrum. 
As anisotropic stress is generated by causal mechanisms it will not have an effect on superhorizon scales. 
If we look at the solutions to Eq.~\eqref{EOM}, during radiation domination, and in the absence of anisotropic stress we find
\begin{equation}
	h^{\text{rad}}_k(x) = A^{\text{rad}}_k j_0(x) + B^{\text{rad}}_k y_0(x).  \label{h_modes}
\end{equation}
Here $x = k\tau$, where $j_0(x)$ and $y_0(x)$ represent the spherical Bessel functions of the first and second kind of zero order respectively
\begin{equation}
	j_0(x) = \frac{\sin{x}}{x}, \hspace{3mm} y_0(x) = -\frac{\cos{x}}{x}.
\end{equation}
The $k$ index represents the fact the amplitude can be different for different $k$ modes.
The mode proportional to $A^{\text{rad}}_k/B^{\text{rad}}_k$ is the \emph{growing/decaying} mode.
The initial conditions are usually set when $x\ll1$ (i.e early times superhorizon scales) and in this limit the behaviour of these modes is 
\begin{equation}
	h_k^{\text{rad}}(x\ll1) = A^{\text{rad}}_k - \frac{B^{\text{rad}}_k}{x}. \label{h_ini}
\end{equation}
Moreover, if we look at solutions to Eq.~\eqref{EOM} in a matter domination phase, which starts at $z \approx 3000$, then we find
\begin{eqnarray}
	& & h^{\text{mat}}_k(x) = 3\left[ A^{\text{mat}}_k \frac{j_1(x)}{x} + B^{\text{mat}}_k\frac{y_1(x)}{x} \right], \nonumber \\
	& & j_1(x) = \frac{\sin{x}}{x^2} - \frac{\cos{x}}{x}, \hspace{3mm} y_1(x) = - \frac{\cos{x}}{x^2} - \frac{\sin{x}}{x}.
\end{eqnarray}
At late times, $j_1(x\gg 1) \rightarrow - \frac{\cos{x}}{x}$ and $y_1(x \gg 1) \rightarrow - \frac{\sin{x}}{x}$.  
So even if the decaying mode solution is ignored during radiation domination, it can still source two possible modes during matter domination. 
The key difference between these modes is that their phases during each era will be opposite, i.e the modes are orthogonal to each other. 
We have shown this schematically in Figure \ref{cmbpol}.\footnote{
Further discussion on this can be found in \cite{Turner:1993vb, Wang:1995kb, Pritchard:2004qp}.}
Our non-parametric approach allows us to analyse the effect of $k$ modes with different amplitudes and phases precisely by isolating the effect they have on the CMB anisotropies.
The two point correlation function of these modes is canonically defined as 
\begin{eqnarray}
    \sum_\lambda \langle |h_{\lambda, \textbf{k}}(\tau)|^2 \rangle & \equiv &  \frac{2 \pi^2}{2k^3} P_T(k) |T(\tau, k)|^2. \label{tensor_evo}
\end{eqnarray}
Note that we have assumed the expectation value of the $h_{\lambda, \textbf{k}}$ does not have any directional dependence (this is because we have assumed spatial isotropy, i.e. $SO(3)$ symmetry).
The correlation function is then separated into a time dependent and independent term. 
The time dependent term is often called the transfer function $T(\tau, k)$ and only tracks the \emph{time evolution of a particular tensor mode}. This is not the same as the transfer function for the CMB anisotropies which track the impact the perturbations have on the CMB photons. The growing mode transfer function is only time dependent on sub-horizon scales, while for the decaying mode it is time dependent on all scales. The time independent part is given by the PPS. 

\subsection{Review of CMB anisotropies}

The tensor perturbations will leave an imprint on temperature and polarisation anisotropies in the CMB. 
We can gain some insight into their structure by exploring the computation of the anisotropies analytically. 
The structure of the anisotropies due to primordial tensors has been studied before in \cite{Turner:1993vb, Wang:1995kb, Zaldarriaga:1996xe, Pritchard:2004qp}. 
Here we briefly review it in the presence of decaying modes. 
The Gaussian anisotropies of the CMB can be completely described by the angular correlation function, $C_\ell$, which can be written as
\begin{equation}
	C_\ell^{XY} = 4 \pi \int^\infty_0  \frac{dk}{k} P(k) |\Delta^X_\ell(k) \Delta^Y_\ell(k) |.
\end{equation}
$X,Y \in \{T, E, B\}$ are the observables (temperature and two polarisation modes) that are computed from the CMB photons. $\Delta^X_\ell(k)$ is the transfer function corresponding to the observable one is interested in. 
The temperature and polarisation transfer functions for tensor perturbations are given by \cite{Zaldarriaga:1996xe, Pritchard:2004qp}
\begin{eqnarray}
	& & \Delta^T_\ell(k) = \sqrt{\frac{(\ell + 2)!}{(\ell - 2)!}} \int^{\tau_0}_{0} d \tau S_T(k, \tau) \mathcal{P}^{T}_\ell(x), \nonumber \\ 
	& & \Delta^E_\ell(k) = \int^{\tau_0}_0 \ d\tau S_P(k, \tau) \mathcal{P}^{E}_\ell(x) \nonumber \\
	& & \Delta^B_\ell(k) = \int^{\tau_0}_0 \ d \tau \ S_P(k, \tau) \mathcal{P}^{B}_\ell(x). \label{transfers}
\end{eqnarray}
$\tau_0$ is the conformal time today. 
The leading order source functions and projection factors for temperature and polarisation are given by 
\begin{eqnarray}
	& & S_T(k, \tau) = - \dot{h}(k, \tau) e^{- \kappa} + g(\tau) \Psi(k, \tau), \nonumber \\
	& & S_P(k, \tau) = -g(\tau) \Psi(k, \tau), \label{sources} \nonumber \\
	& & \mathcal{P}^{T}_\ell(x) = \frac{j_\ell(x)}{x^2}, \nonumber \\
	& & \mathcal{P}^{E}_\ell(x) = - j_\ell(x) + j_\ell''(x) + 2\frac{j_\ell(x)}{x^2} + 4\frac{j'_\ell(x)}{x}, \nonumber \\
	& & \mathcal{P}^{B}_\ell(x) = 2 j'_\ell(x) + 4 \frac{j_\ell(x)}{x}.
\end{eqnarray}
$\Psi(k, \tau)$ is the Newtonian gravitational potential. 
$\kappa$ is the integrated Thomson cross section between $\tau$ and $\tau_0$
\begin{equation}
	\kappa = \int^{\tau_0}_\tau d \tau \ a n_e x_e \sigma_T,
\end{equation}
where we have defined $x_e$ as the ionisation fraction, $n_e$ as the electron number density and $\sigma_T$ is the Thomson cross section. 
We have also defined the visibility function $g(\tau) \equiv \dot{\kappa} e^{-\kappa}$. 

\begin{figure}[h]
  \centering
  \includegraphics[width=\linewidth]{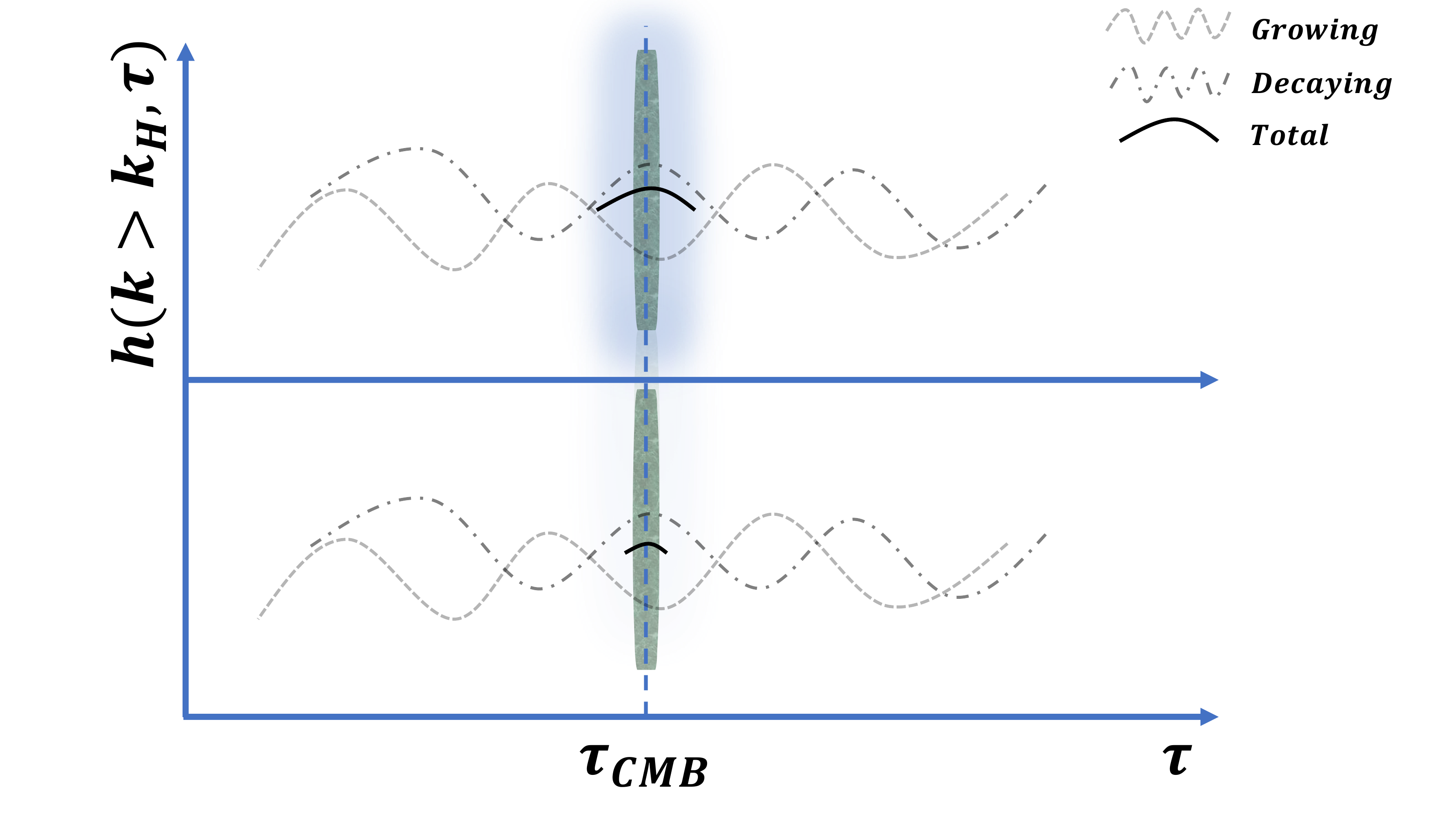}
  \caption{Schematic diagram showing the phase damping effect, as described in \cite{Pritchard:2004qp}, of primordial gravitational waves on the CMB photons. The top panel shows a situation where the last scattering screen is more diffuse then it is in the lower panel. A more diffuse screen will allow more of the gravitational wave amplitude to contribute to the production of polarisation anisotropy. }
  \label{cmbpol}
\end{figure}

From the expressions in Eq.~\eqref{transfers} and \eqref{sources} we can see that the temperature source function has two distinct features. 
The first term, proportional to $\dot{h}$, is a type of Integrated Sachs-Wolfe (ISW) effect. It describes the generation of anisotropies from the motion of photon geodesics in the presence of a time varying gravitational potential of the gravitational wave. 
The term proportional to the visibility function, $g$, will be localised to the screen of recombination, which is assumed to be almost instantaneous. 
This term is small and is almost always subdominant to the ISW term\footnote{See Figure 1 in \cite{Pritchard:2004qp} for instance.}. 
For the polarisation anisotropies, there is no ISW term. 
This is because the ISW effect changes the energy of the photon which is directly related to its temperature, not the polarisation.
The source of polarisation anisotropies will be strongly located at the surface of last scattering due to the scattering of free electrons from the local tensor quadrupole. 
Therefore, modes with $k \approx \frac{\ell}{\tau_0 - \tau_{\text{CMB}}}$ dominate the contribution to the polarisation $C_\ell$'s. 
Intuitively one would expect the anisotropies to be proportional to the width of the surface of last scattering as a larger width will lead to more polarisation being generated. This is because the finite width adds time for the generation of quadrupolar scattering, which is what fundamentally gives rise to the polarisation  anisotropy. 
In addition there will be more modes that can contribute to the anisotropy as the width increases. 
We show this schematically in the top panel of Figure \ref{cmbpol} which has a blurry CMB screen, whereas the bottom one has an almost instantaneous CMB.
Recombination is not the only screen present for the local tensor quadrupole to generate polarisation: 
Reionisation also provides another screen at which polarisation is generated and this happens at larger angular scales \cite{Zaldarriaga:1996xe}. 

If we assume the Bessel functions and their derivatives, which are usually referred to as projection factors, in Eqs.~\eqref{transfers} are approximately constant over the width of the CMB screen the $C_\ell$'s for polarisation can be calculated analytically by integrating over the source function and projection factors to get \cite{Pritchard:2004qp}
\begin{eqnarray}
	& & C^{EE/BB}_\ell \propto \int \frac{dk}{k} P_T(k) \mathcal{P}^{E/B}_\ell[k(\tau_0 - \tau_R)]^2 \dot{h}_k(\tau_{\text{CMB}})^2 \nonumber \\
	& & \times  \Delta \tau_{CMB}^2e^{- (\kappa \Delta \tau_{CMB})^2}.
\end{eqnarray}
Here we see that indeed the $C_\ell$'s are proportional to the finite time scale for recombination, $\Delta \tau_{CMB}$, and we also notice the $\dot{h}^2$ factor which will be sensitive to the initial conditions we choose. 
In particular, we see from Eq.~\eqref{h_ini} that the $C_\ell$'s will become large when $k\tau$ is small for the decaying modes (this is similar to what was seen in the case of decaying scalar perturbations \cite{Kodwani2}) and the difference in phase of $h(\tau)$, or equivalently $\dot{h}(\tau)$, will only effect the amplitude of the $C_{\ell}$'s as the tensor perturbation is evaluated locally at $\tau_{CMB}$. 
This was described in \cite{Pritchard:2004qp} as phase damping because the overall effect of multiple phases is to damp the observed perturbations (also shown schematically in Figure \ref{cmbpol}). 

\section{Analysis}\label{anal}
\subsection{$C_\ell$ results}

\begin{figure*}
\centering
\begin{subfigure}{.5\textwidth}
  \centering
  \includegraphics[width=\linewidth]{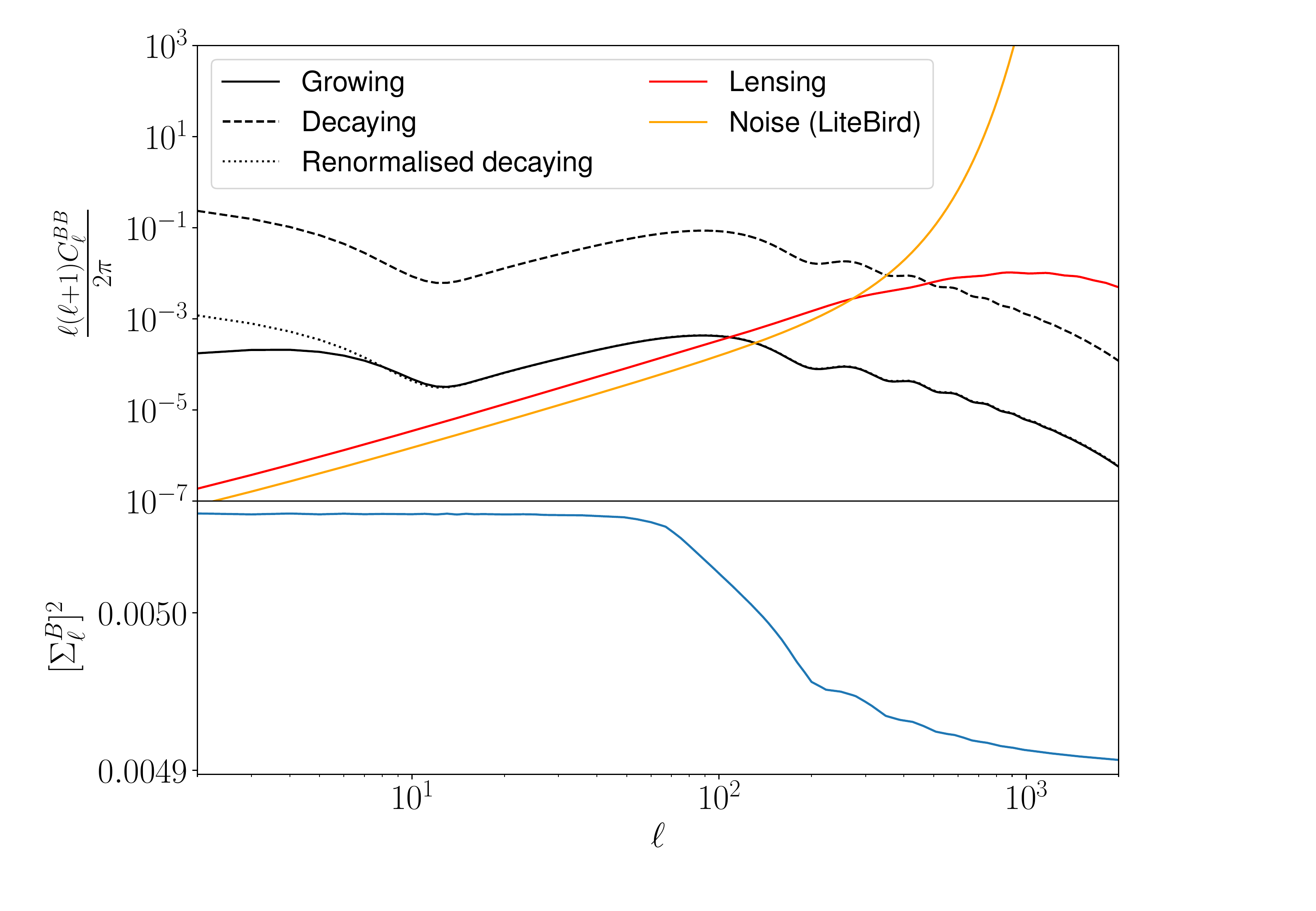}
  \caption{B mode}
  \label{clBB}
\end{subfigure}%
\begin{subfigure}{.5\textwidth}
  \centering
  \includegraphics[width=\linewidth]{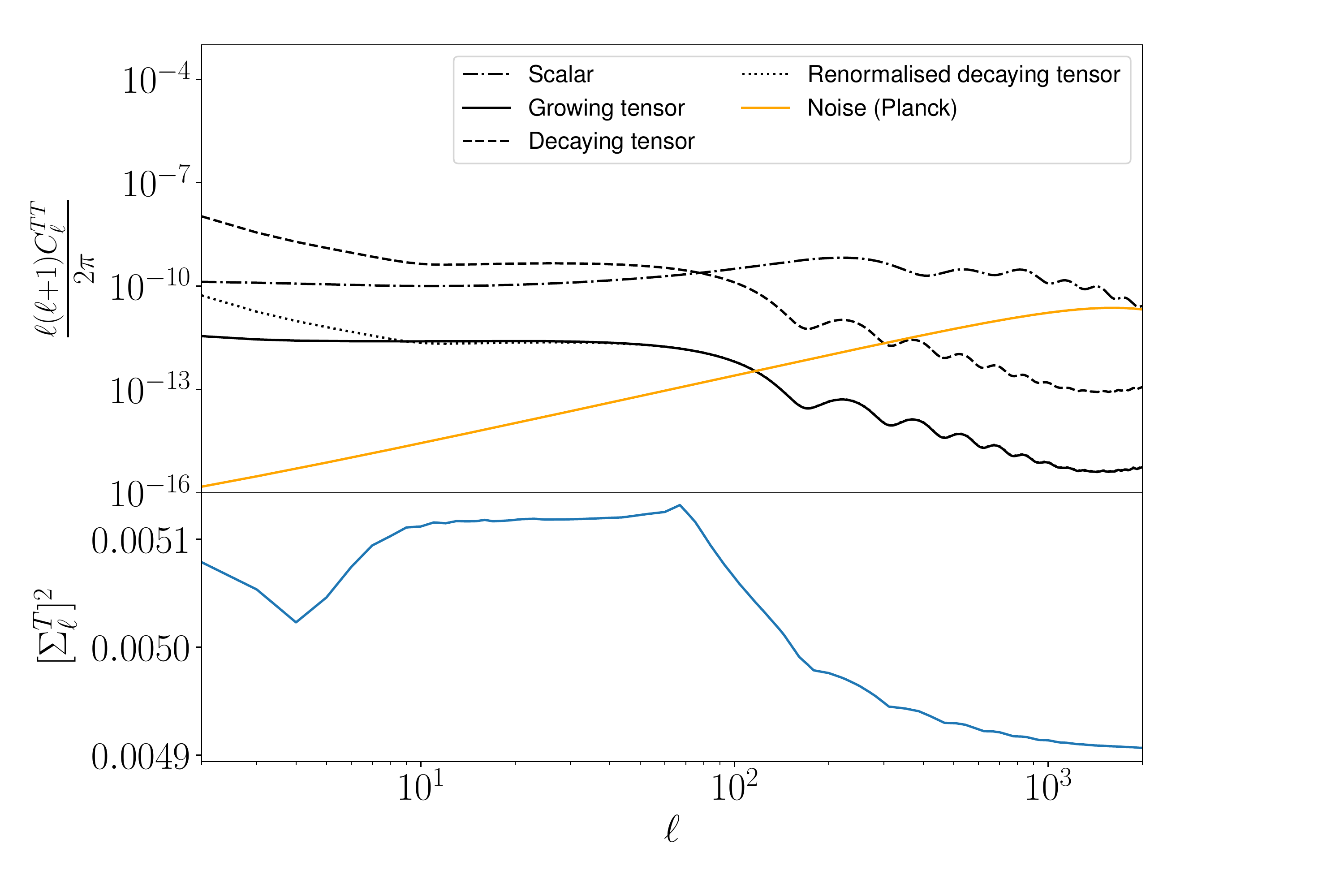}
  \caption{Temperature}
  \label{clTT}
\end{subfigure}

\begin{subfigure}{.5\textwidth}
  \centering
  \includegraphics[width=\linewidth]{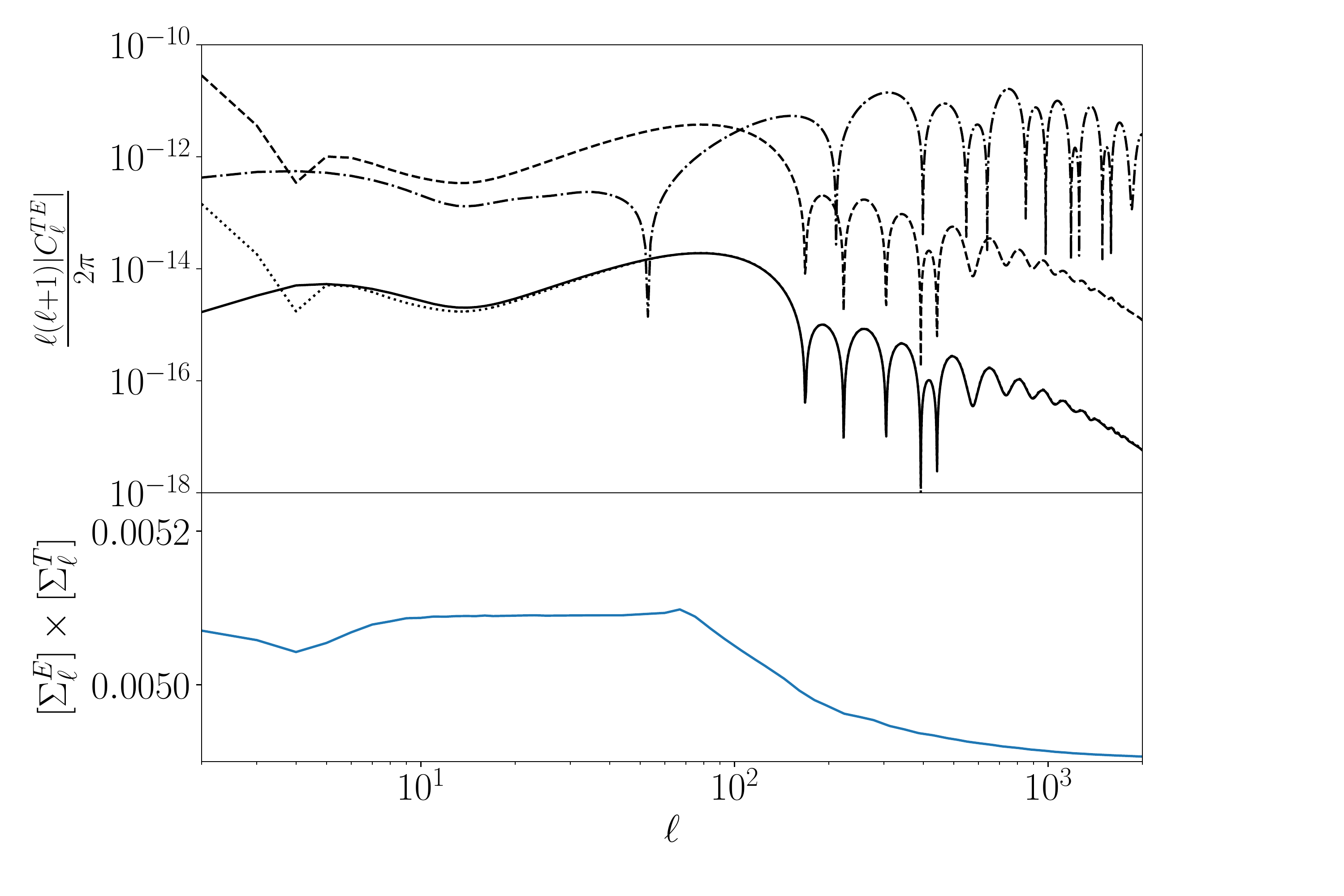}
  \caption{E mode + Temperature}
  \label{clTE}
\end{subfigure}%
\begin{subfigure}{.5\textwidth}
  \centering
  \includegraphics[width=\linewidth]{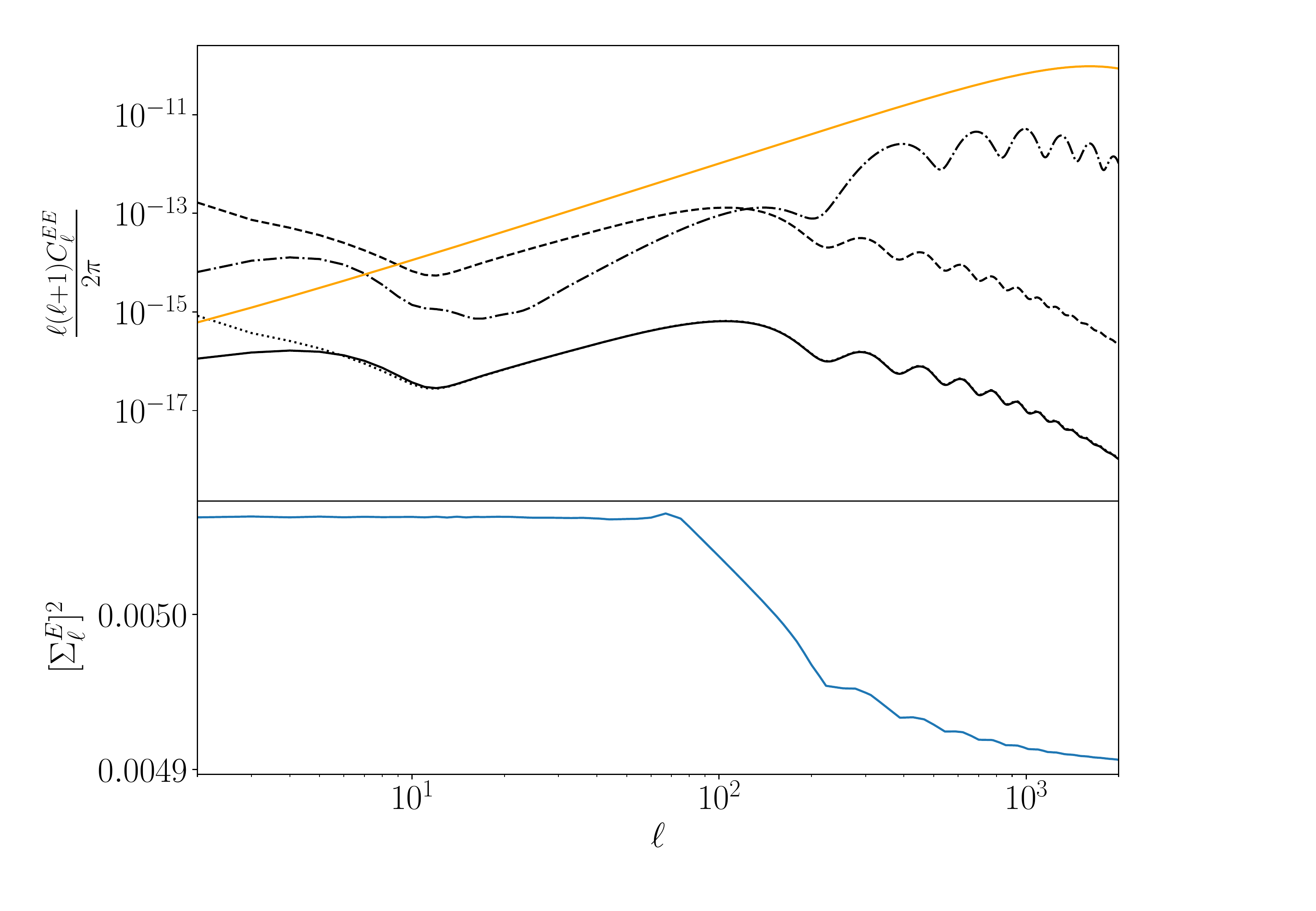}
  \caption{E mode}
  \label{clEE}
\end{subfigure}
\caption{Anisotropies for the growing and decaying tensor modes. 
The decaying mode is shown when it is normalised on superhorizon scales and when it is renormalised on subhorizon scales.
The renormalisation function for each observable is shown in the bottom panel of the plots.
The temperature and E mode polarisation spectra also show the contribution from the fiducial scalar perturbations in the standard $\Lambda$CDM cosmology with cosmological parameters given in Table \ref{cosmo_params}. Figures (\ref{clTE}, \ref{clEE}) have the same legend as Figure \ref{clTT}. The noise curves shown are for the LiteBird/Planck experiment for B/(T,E) modes which are defined in Eq.~\eqref{LiteBird_noise}/\eqref{Planck_noise}.}
\label{decaying_tensor_cls}
\end{figure*}

To constrain tensor initial conditions we will follow the standard convention of using the scalar-to-tensor ratio $r = \frac{A_T}{A_S}$ as the variable to quantify the amplitude of the tensor perturbations. 
We implement the general initial condition in Eq \eqref{h_ini} in the CLASS Boltzmann code and plot the B mode power spectra in Figure \ref{clBB}. We show the growing mode with $r = 0.05$ and the growing mode spectral index defined by the single field slow roll inflation consistency relation $n_T^{(G)} = -\frac{1}{8} r^{(G)}$.
For the decaying mode we do not assume this relation (as we do not expect decaying modes from inflation to be detectable \cite{dePutter:2019xxv}) and set both the amplitude and index of the decaying mode independently.
In Figure \ref{clBB} we show the decaying mode B mode power spectra with the same amplitude as growing modes, $r^{(D)} = r^{(G)}$ and a scale invariant power spectrum, $n_T^{(D)} = 0$.
Furthermore, as the amplitude of the decaying mode is time dependent, one needs to define a normalisation time of when the $r^{(D)}$ is set.
In Figure \ref{decaying_tensor_cls} we plot the $C_\ell$'s for two different normalisations. 
First, when the amplitudes of the modes are set using the CLASS approximation schemes as described in \cite{CLASSII}. 
Under this scheme the amplitude of the modes on superhorizon scales is set at $\sim 0.01 \tau_{CMB}$.
The other normalisation procedure is the same as the one described in \cite{Kodwani2}, where the tensor modes are renormalised such that the transfer function of both decaying and growing modes is the same on sub-horizon scales.
The renormalised transfer function for the decaying mode, $\tilde{\Delta}^{(D)}_\ell$, is defined as 
\begin{eqnarray}
    \tilde{\Delta}^{(D)}_\ell & = & \Delta^{(D)}_\ell \Sigma_\ell \nonumber \\
    \Sigma_\ell & \equiv & \frac{\int^{k_{\text{max}}}_{k_{\text{horizon}}} dk \ \Delta_\ell^{(G)}(k)} {\int^{k_{\text{max}}}_{k_{\text{horizon}}} dk \ \Delta^{(D)}_\ell(k)},
\end{eqnarray}
where $k_{\text{horizon}} = 3 \times 10^{-3} \ \text{Mpc}^{-1}$, $k_{\text{max}} = 2 \times 10^{-1} \ \text{Mpc}^{-1}$.
The renormalisation functions for each of the observables are shown in the bottom panels in Figure \ref{decaying_tensor_cls}.
The rest of the cosmological parameters are given in Table \ref{cosmo_params}.

\begin{table}
\begin{centering}
\begin{tabular}{ p{1.5cm}  p{2cm}}
\hline
\hline
$A_s$ & 2.15 $\times 10^{-9}$  \\
$h$ & 0.67556  \\
$\Omega_bh^2$ & 0.022032  \\
$\Omega_{cdm}h^2$ & 0.12038 \\
$k_{*}$ & 0.002 Mpc$^{-1}$ \\
$n_s $ & 0.9619 \\
$N_{eff}$ & 3.046 \\
$r^{(G)}$ & 0.05 \\
$n_T^{(G)}$ & - $\frac{1}{8}r^{(G)}$ \\
\hline
$\ell_{max}$ & 2500 \\
$f_{sky}$ & 1 \\
\hline
\hline
\end{tabular}
\caption{Fiducial cosmological and systematic parameters}\label{cosmo_params}
\end{centering}
\end{table}

In Figure \ref{decaying_tensor_cls} we see that when the decaying mode is normalised on superhorizon scales the anisotropies are larger than the growing mode ones. 
Furthermore, the decaying mode anisotropies for $TT$ and $EE$ can be even larger than the ones generated by \emph{scalar perturbations}\footnote{Here we are referring to the growing mode scalar perturbations, but the decaying scalar mode has a similar amplitude. See \cite{Kodwani2} for further discussion on this.}. 
For temperature we see in Figure \ref{clTT} that the decaying mode anisotropy is greater than scalar perturbations for $ \ell {\sim} 90$. 
For E mode polarisation, seen in Figure \ref{clEE}, the decaying tensor mode contribution is always larger then the scalar contribution on large scales.
This means that if the modes are sourced at very early times on superhorizon scales they could already be constrained by the temperature and E mode polarisation anisotropies from Planck and WMAP.

Next, if we look at the anisotropies for the decaying mode when they are renormalised on subhorizon scales, we see that the shape of the $C_\ell$'s is the same, but the amplitude is smaller than the decaying mode sourced on superhorizon scales by a factor of $\sim 10^{4}$, which is as expected due to the superhorizon modes being sourced at $\sim 0.01 \ \tau_{CMB}$.
In this case the decaying modes are \emph{indistinguishable} from the growing tensor modes, except for a rise in anisotropy on very large scales. 
This rise is seen because we only renormalise based on the sub-horizon amplitude and the superhorizon amplitude will generally be larger on the large angular scales. 
The physical reason behind this is that the local quadrupole generated by the tensor modes is responsible for generating the polarisation in the CMB. 
As the decaying mode varies with time on superhorizon scales, the amplitude of the mode (and hence the polarisation it generates) depends on which time the mode is sourced.
The amplitude and the time the mode is sourced are degenerate parameters when it comes to the generation of the CMB polarisation. 
To break this degeneracy one needs to be able to measure the decaying mode at least twice and thus it will be important to measure the signal from reionisation and recombination. 
In particular, since we normalise the modes at recombination, the decaying and growing modes leave an identical signal in the B mode spectrum at the recombination bump, as can be seen from Figure \ref{clBB}.
The decaying mode can be distinguished from the growing mode only by the reionisation bump where we see an increase in power from the decaying mode.
Indeed we can get an idea as to how well the decaying mode can be measured from the B mode power spectrum by comparing the difference between the growing and decaying mode power spectra at the reionisation scale. By assuming a cosmic variance limited experiment, we know the variance in the $C_\ell$'s is given by
\begin{equation}
    \sigma (\ell)^2 = \frac{2}{2\ell+1} (C_\ell^{\text{G}})^2,
\end{equation} 
where $C_\ell^{\text{G}}$ is the power spectra for the fiducial growing modes. 
The difference between the decaying and growing mode power spectra is
\begin{equation}
    \Delta C_\ell^2 \equiv (C_\ell^{G} - C_\ell^{D})^2.
\end{equation}
The $C_\ell^{D}$ is the decaying mode power spectrum with parameters give in Table \ref{cosmo_params}.
At $\ell = 2$, where the signal is largest from the decaying mode we see that $\frac{\Delta C^2_{\ell = 2}}{\sigma(\ell=2)^2} \approx 80$.
Of course this can never be achieved as the cosmic variance limit of the $\ell = 2$ mode is 5/2 and thus that will be the fundamental limit to how well we can distinguish between the growing and decaying tensor more. 
To get a complete result accounting for the full covariance between the polarisation and temperature anisotropies as well as the total sum over all the modes we compute the Fisher information matrix of the amplitude of the modes in the next section.

Before we move on to computing the Fisher information it is worth pointing out that the increase in power on superhorizon scales comes from the fact that the decaying mode has a $1/k\tau$ behaviour, which leads a divergent amplitude in the power spectrum.
We also see that the divergence in the polarisation spectra at the reionisation scale is smaller than the divergence in the temperature spectrum.
This is because of the fundamental difference between how temperature and polarisation anisotropies are generated by tensor perturbations: the temperature anisotropies are sourced continuously by tensor perturbations whereas the polarisation anisotropies are sourced at fixed screens as described above, thus more modes leave an imprint in the temperature power spectrum (and therefore increase the amplitude more).

We know that in addition to primordial gravitational waves sourcing B modes, lensing of the CMB photons can also generate B mode polarisation, which we call $C_\ell^{BB,(L)}$. 
This is given by \cite{Lewis:2006fu, Hiramatsu:2018nfa}
\begin{equation}
	C_\ell^{BB, (L)} = \frac{1}{2\ell+1} \sum_{\ell' \ell''} \left(\mathcal{S}^{(-)}_{\ell \ell' \ell''} \right)^2 C^{EE}_{\ell'} C^{\phi \phi}_{\ell''},
\end{equation}
where 
\begin{eqnarray}
	\mathcal{S}^{(-)}_{\ell \ell' \ell''} & \equiv & \left[ \frac{ (2\ell+1)(2\ell'+1)(2\ell''+1)}{16 \pi} \right]^\frac{1}{2} \times \begin{pmatrix} \ell & \ell' & \ell'' \\ 2 & -2 & 0 \end{pmatrix}\nonumber \\
	& \times & \left[- \ell(\ell+1) + \ell'(\ell'+1) + \ell''(\ell''+1) \right] ,
\end{eqnarray}
with the term in the circular brackets being a Wigner 3j symbol.

In addition to these two physical effects generating a B mode, an experiment will also have a noise contribution for the B modes. 
We parametrise the effect of the noise by white noise with a smoothing beam assumed to be Gaussian \cite{Katayama_2011}\footnote{In principle there can also be a $\ell$ dependence in the noise but we do not address that in this study.}
\begin{equation}
	N_\ell^{BB} = \exp{\left( \frac{\ell^2 \sigma_b^2}{2} \right)} \left( \frac{ \pi}{10800} \frac{w_p^{-\frac{1}{2}}}{\mu \text{K arc min}} \right)^2 \mu \text{K}^2 \text{str}. \label{LiteBird_noise}
\end{equation} 
We assume a LiteBird\footnote{A satellite mission that will aim to measure the polarisation of the CMB \cite{LiteBird}.} like experiment with $\sigma_b = 3.7 \times 10^{-3}$ and $w_p = 1 \ \mu$K \cite{LiteBird}.
The various components of lensing and noise contributions, along with the B modes from primordial tensors are shown in Figure \ref{clBB}.
In the next section we investigate this further in a model independent, non-parametric way, by computing the Fisher information.

\subsection{Fisher results}

To obtain a model independent parameterisation of the PPS we model it as a set of 100 bins in $k$ around a fiducial PPS for the standard growing mode
\begin{equation}
	P_T(k,k_0,\epsilon) = \begin{cases} P_T(k)^{(G)} + \epsilon^{(G) \text{ or } (D)}_{k_0} & \mbox{if $k_0 = k$} \\ P_T(k)^{(G)} & \text{otherwise} \end{cases}. \label{pps}
\end{equation}
where $P_T(k)^{(G)}$ takes the form in Eq \eqref{tensor_pps}. $\epsilon_{k_0}^{(D) \text{ or } (G)}$ is the amplitude of additional power coming from the decaying or growing mode at the  scale $k_0$.
We treat the $\epsilon$'s in each $k$ bin as free parameters and constrain them using the Fisher information matrix, $F_{\alpha \beta}$. 
The $k$ bins we use are shown in Figure \ref{kbins}. 

For our Fisher analysis we focus solely on the decaying modes that are normalised on subhorizon scales as those are the physical modes we can observe at the time of decoupling. 
For modes that are sourced at early time and on superhorizon scales, the Fisher constraints can be scaled accordingly depending on what time the mode is sourced. 
For instance, if the mode is sourced at $0.01\tau_{CMB}$, the constraint will increase by $\sim 10^4$ for those scales due to the $1/k\tau$ behaviour of the decaying mode on superhorizon scales during radiation domination.
We assume a Gaussian likelihood with a parameter independent covariance matrix for the $C_\ell$'s and the corresponding Fisher matrix is 
\begin{equation}
	F_{\alpha \beta} = \frac{f_{sky}}{2} \sum_{\ell = 2}^{\ell_{max}} (2\ell + 1) \text{Tr} \left( \mathbb{C}^{-1}_\ell  \partial_\alpha \mathbb{C}_\ell \mathbb{C}^{-1}_{\ell} \partial_\beta \mathbb{C}_\ell \right)
\end{equation}
where
\begin{equation}
	\mathbb{C}_\ell \equiv \begin{pmatrix} \hat{C}_\ell^{TT} & C_\ell^{TE} & 0  \\ C^{ET}_\ell & \hat{C}^{EE}_\ell & 0 \\ 0 & 0 & \hat{C}^{BB}_\ell  \end{pmatrix}. \label{covmat}
\end{equation}
There are no correlations between E, T and B modes as long both polarisations of the tensor mode are equally generated (i.e there is no breaking of parity).
The $\hat{C}_\ell$ represents the theoretical $C_\ell$ (computed from a modified version of the CLASS Boltzmann code) plus noise contributions.
For each of these modes, these are defined by 
\begin{eqnarray}
	& & \hat{C}_\ell^{TT(EE)} \equiv C_\ell^{TT(EE)} + N_{\ell}^{TT(EE)} \nonumber \\
	& & \hat{C}_\ell^{BB} \equiv C_\ell^{BB} + N_\ell^{BB} + \lambda_{(L)}C_\ell^{BB, (L)}
\end{eqnarray}
where the B mode noise is defined in Eq~\eqref{LiteBird_noise}. 
We have introduced a lensing parameter $\lambda_{(L)}$ which denotes how much the lensing B modes contribute to the signal. $\lambda_{(L)} = 0$ corresponds to a situation where the lensing signal has been completely accounted for and removed from the signal.
The T and E mode noise is modelled by Gaussian random noise in 4 frequency channels given in the Planck blue book \cite{Planck:2006aa}

The window function for the beam is defined by $B_\ell^2 \equiv \exp{ \left( - \frac{\ell(\ell+1) \theta^2_{beam}}{8 \ln 2} \right) }$ and the variance for each frequency channel is $\sigma_{T(E)}$ for temperature/polarisation. 
The numerical values are given in Table \ref{Planck_noise} and the plot of the noise curves is shown in Figure \ref{clTT} and \ref{clEE}.
Once the Fisher information matrix is computed, the errors on the parameters is simply given by $(F^{-1}_{\alpha \alpha})^{\frac{1}{2}}$.

\begin{figure*}
\centering
\begin{subfigure}{.5\textwidth}
  \centering
  \includegraphics[width=\linewidth]{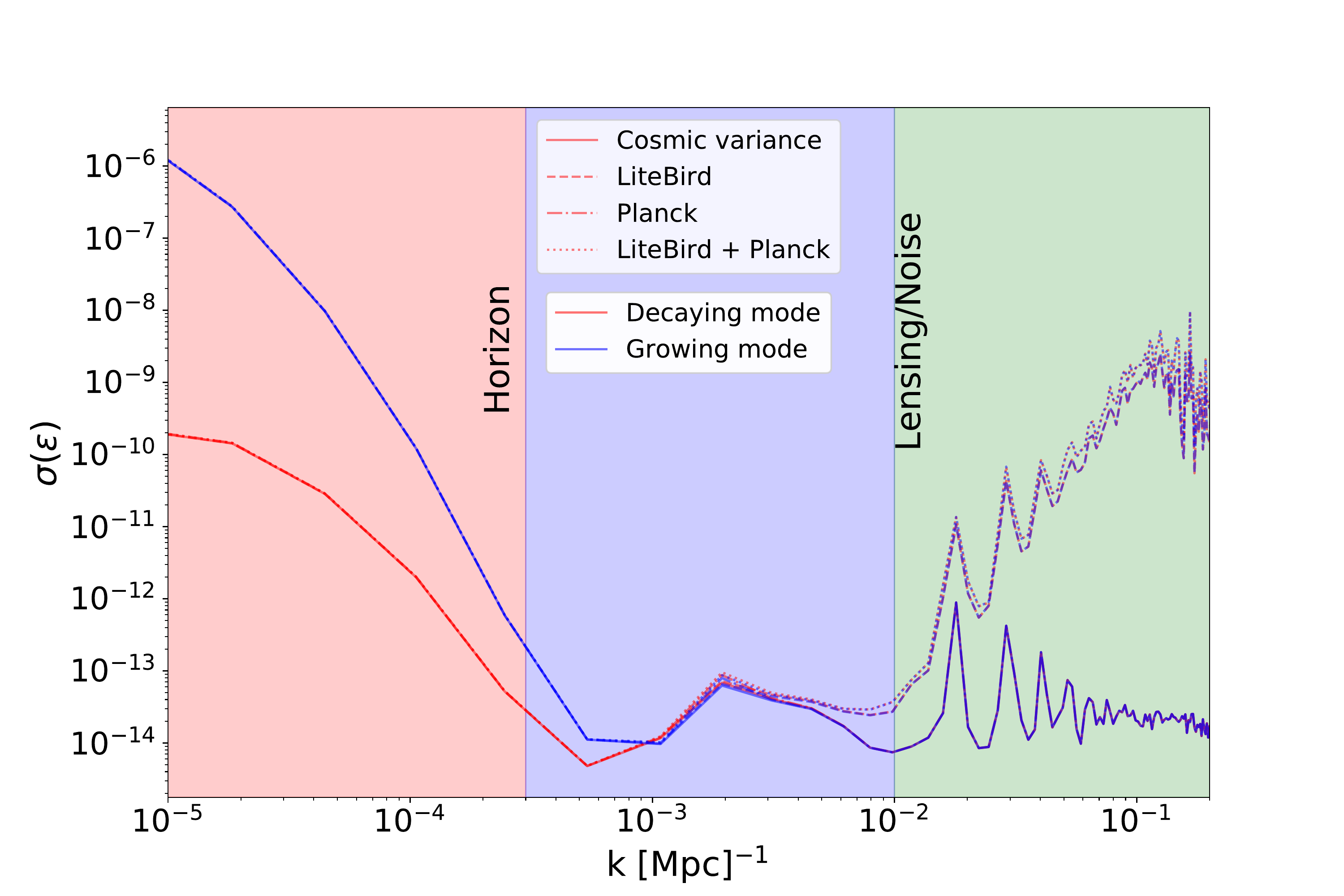}
  \caption{De-lensed B+E+T}
  \label{D_BTE}
\end{subfigure}%
\begin{subfigure}{.5\textwidth}
  \centering
  \includegraphics[width=\linewidth]{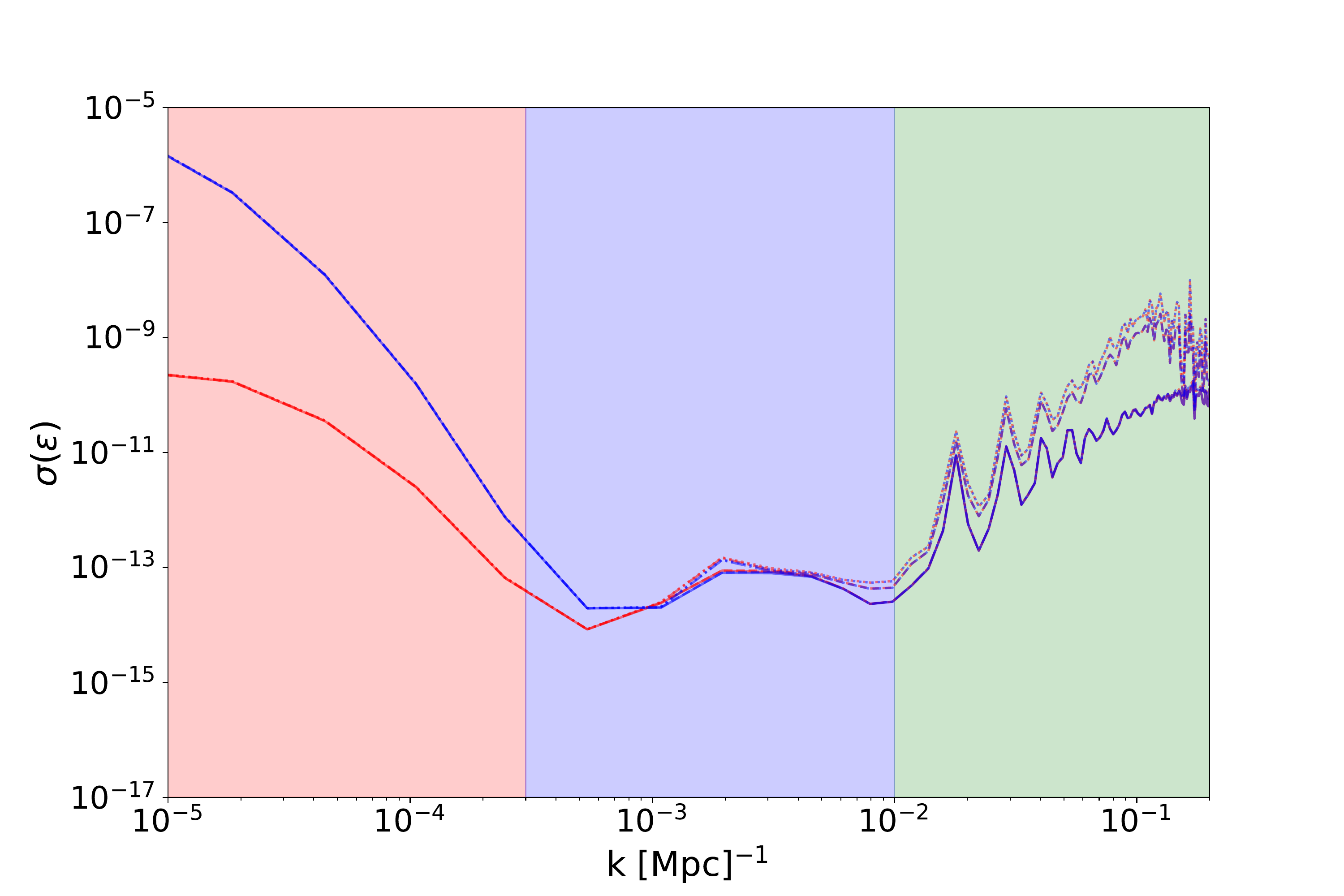}
  \caption{Lensed B+E+T}
  \label{L_BTE}
\end{subfigure}

\begin{subfigure}{.5\textwidth}
  \centering
  \includegraphics[width=\linewidth]{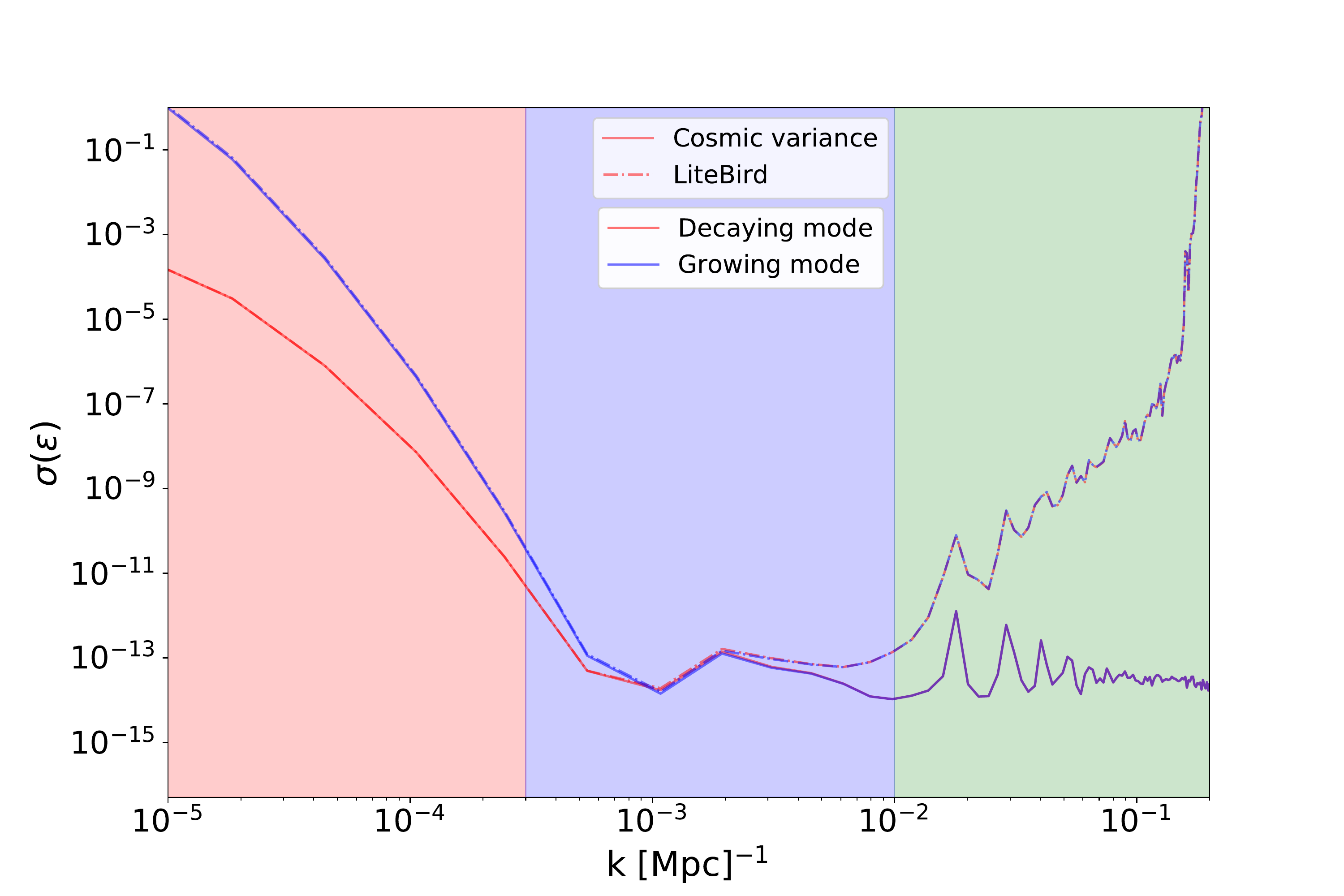}
  \caption{De-lensed B mode only}
  \label{D_B}
\end{subfigure}%
\begin{subfigure}{.5\textwidth}
  \centering
  \includegraphics[width=\linewidth]{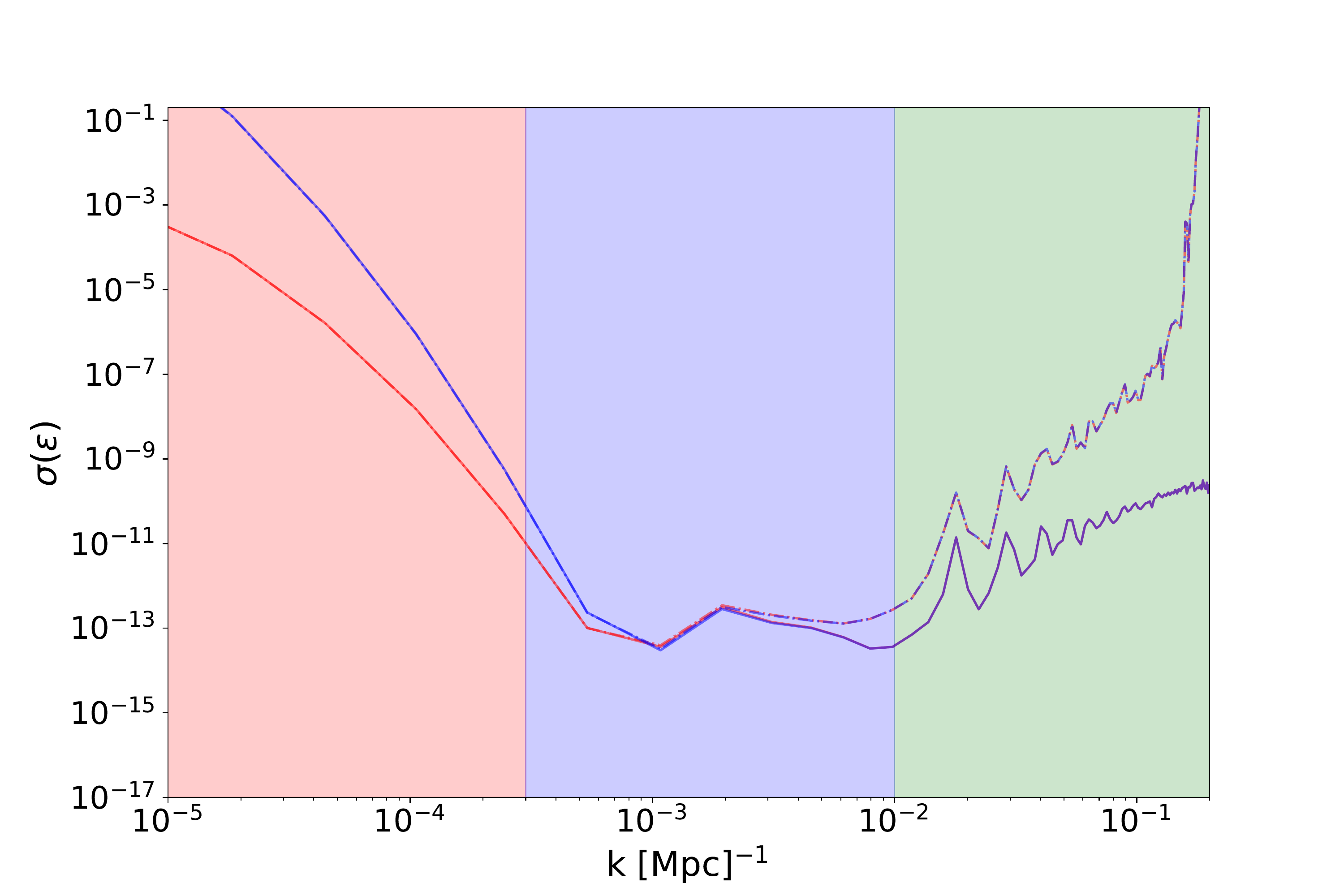}
  \caption{Lensed B mode only}
  \label{L_B}
\end{subfigure}
\caption{Errors for decaying and growing tensor modes. We have shown the errors for the four cases described in Table \ref{Resultsummary}. We have separated the noise contributions into the cosmic variance limited experiments, Planck noise for temperature and E mode polarisation and LiteBird for B mode polarisation.}
\label{fish_fig}
\end{figure*}

We show the errors on the PPS parameters $\epsilon_{k_0}$ in Eq.~\eqref{pps} for the growing and decaying tensor modes in Figure \ref{fish_fig}.
We focus on four cases which are summarised in Table \ref{Resultsummary}. 
The tracers used in the computation of the Fisher matrix are either B mode polarisation only, in which case $\mathbb{C}_\ell$  in Eq \eqref{covmat} is simply given by $\hat{C}_\ell^{BB}$, or B mode + E mode polarisation with temperature anisotropies as well. In this case we use the full $\mathbb{C}_\ell$ given in Eq \eqref{covmat}. This is denoted by T+E+B in Table \ref{Resultsummary}. For each of these cases, we consider the case when the modes are lensed/delensed with $(\lambda_{(L)} = 1) /(\lambda_{(L)} = 0) $. 

It is easiest to interpret the results in Figure \ref{fish_fig} by focusing on three different scales. 
First is the region shaded red which represents modes that are outside the horizon at the time the CMB is emitted (in fact there are scales that are larger than the observable size of the universe, thus one must be careful in how to interpret those constraints as we discuss in section \ref{summary}). 
This is also the region where cosmic variance dominates and thus the error bars increase substantially.
The region shaded in blue corresponds to scales which are subhorizon but on which the effect of lensing and noise (LiteBird experiment) for B modes is subdominant. 
Therefore the blue region is where most of the constraining power is. 
Finally, the green region is where the noise from LiteBird becomes very large and also the lensing contribution to B modes dominates over the primordial B mode signal. 

We see that the decaying mode is equally well constrained as the growing modes for all four cases we consider, except on scales below the recombination scale, $ k \lesssim 3 \times 10^{-4}$, where the decaying and growing mode amplitudes become distinguishable.
On superhorizon scales the constraint is $\sim 10^4$ larger for the decaying mode amplitude, as expected by the normalisation on subhorizon scales and the divergence of the decaying mode on superhorizon scales.
In the green region we see that the LiteBird noise dominates any signal and therefore the constraining power is reduced by 4-5 orders of magnitude. 
In the case of a cosmic variance limited experiment there is still the same amount of information in the green region as there is in the blue region when the CMB is delensed. If there is a lensing signal as well, the constraining power deteriorates by roughly 1-2 orders of magnitude.
When the temperature and E mode information is added we see that the errors on superhorizon scales, in the red region, are smaller by roughly 5 orders of magnitude. 
There are two reasons for this increase in constraining power. First, there is an increase in the $TT$ and $EE$ power spectra on superhorizon scales for the decaying mode. Second, the $TT$ and $EE$ $C_\ell$'s have different transfer functions to the $BB$, however the PPS for the decaying mode is the same for all of the observables. Therefore, the freedom in PPS is not able to compensate for the different transfer functions to the same extent when there are three observables.
 
The best constrained modes in all cases are at $k \approx 5 \times 10^{-4}$ Mpc$^{-1}$ and $\approx 7 \times 10^{-3 }$ Mpc$^{-1}$.
The physical reason behind this is that the polarisation is generated, and hence best constrained, when there is a anisotropic scattering of photons which happens at recombination and reionisation\footnote{This was also pointed out in this recent study \cite{Hiramatsu:2018nfa}.}. 
The recombination scale corresponds to a scale of $\ell \sim 80 $, which, in $k$ space corresponds to $k_{\text{recom}} \approx 6 \times 10^{-3}$ Mpc$^{-1}$.
Similarly the reionisation scale is given by $k_{\text{reion}} \approx 6 \times 10^{-4}$ Mpc$^{-1}$.

\begin{table}
\begin{centering}
\begin{tabular}{ |p{1.5cm}p{2cm}p{2 cm} p{2 cm} |}
\hline
 & Tracer used & De-lensed & Result \\
\hline
\hline
Case 1 & B+E+T &  yes & figure \ref{D_BTE} \\
Case 2 & B+E+T &  no & figure \ref{L_BTE} \\
Case 3 & B &  yes & figure \ref{D_B} \\
Case 4 & B &  no & figure \ref{L_B} \\
\hline
\end{tabular}
\caption{Summary of different cases used to compute the errors on the PPS.}
\label{Resultsummary}
\end{centering}
\end{table}

\section{Discussion \& future outlook}\label{summary}

In this {chapter} we have analysed the effect a decaying tensor mode has on the CMB temperature and polarisation anisotropies.
The decaying modes evolve on superhorizon scales and thus the amplitude of these modes is degenerate with the time at which they are sourced.
We used a Fisher matrix formalism with a non-parametric binned PPS to understand the constraints on these modes.
If the decaying modes are sourced at very early times before decoupling, then they are highly constrained. 
If they are sourced on sub-horizon scales with same power as the growing mode, then there could be an ambiguity as to which mode generates the observed $B$-mode polarisation pattern. 
The amplitudes of both modes start to become distinguishable around the reionisation bump, which suggests it could be important to measure the $B$ modes on large scales $\ell \sim 5$.
If we only look on observable scales, i.e modes that are sub-horizon at the time of decoupling, the decaying and growing modes are constrained equally well. On super-horizon scales where the decaying mode becomes distinguishable from the growing mode it is more constrained. This is because it generates more power in the anisotropies due to its $1/k\tau$ scaling.
Decaying modes generated during inflation would be highly suppressed in radiation domination. Thus, if such modes are observed, it will be a unique signature of new physics on very high energies in the early universe. In particular, bouncing models could be a source of decaying modes \cite{Kodwani2, Gielen:2016fdb}. 
There is a fundamental question that needs to be answered, however, in order to understand these modes. 
As the effect of the decaying mode is most apparent on super-horizon scales, it is worth asking how super-horizon tensor modes, specifically modes that are much larger than our current horizon, can or will effect our observable universe.
In the case of scalar perturbations it is possible the effect of these super-horizon modes will come from either a modification to overall background density, as is modelled in separate universe approached to cosmological perturbations \cite{Rigopoulos:2003ak}, or through the effects of spatial gradients \cite{Tanaka:2006zp}. 
For tensor modes, however, it is not clear what the dominant effect would be. For instance, it is possible that a large scale tensor mode modifies our patch of the universe to have an anisotropic metric, which for instance has been considered in the context of lensing in \cite{Adamek:2015mna}.
In this case the observable effect of the decaying tensor mode would actually be the presence of shear modes in the universe. More formal calculations of the shear modes can be found in \cite{Pontzen:2010eg, Ellis:1968vb, Pontzen:2009rx}. 
Recent searches for shear modes in a general class of Bianchi models can be found in \cite{Saadeh:2016sak}. 
While shear modes are highly constrained, relating the decaying modes to the constraints on shear modes will require a gauge invariant description of matching super-horizon decaying tensor modes to the shear modes.
This would be an interesting endeavour and we leave that for future works.

\chapter{Screened fifth forces in parity breaking correlation functions}\label{PBCF}

\section{Introduction}
\label{sec:intro}

The standard cosmological model, $\Lambda$ Cold Dark Matter ($\Lambda$CDM), relies on the assumption that gravity is described by a rank 2 symmetric tensor on large scales. The cosmological parameters fitted to observations such as the cosmic microwave background \cite{Adam:2015rua, Hinshaw:2012aka} and large scale structure \cite{Ivezic:2008fe, Amendola:2016saw, DESI} assume General Relativity (GR) as the theory of gravity. However, fundamental puzzles in the $\Lambda$CDM model such as the nature of dark energy and dark matter have encouraged research into the cosmological implications of modifications to GR, dubbed modified gravity. There are a plethora of modified gravity models, ranging from the addition or alteration of terms in the Einstein-Hilbert action to the explicit coupling of additional scalar, vector or tensor fields (see e.g. \cite{Clifton:2011jh} for a review). This motivates expanding the parameter space of traditional cosmological inferences, as well as designing novel probes with maximum sensitivity to new gravitational degrees of freedom.

The diversity of modified gravity models makes it inconvenient to test them individually. This has led to the development of generalised frameworks within which many theories may be tested simultaneously, for example the Effective Field Theory of Dark Energy \cite{Gubitosi:2012hu} and Parametrised Post-Friedmannian framework \cite{Ferreira:2014mja}.
Modified gravity theories may also be characterised by the \emph{screening mechanisms} they incorporate to hide the effects of new interactions at small scales. As almost all viable theories employ one of just a handful of screening mechanisms, probing such a mechanism is tantamount to probing a potentially broad class of theories. These theories may often be cast as screened scalar--tensor theories (see \cite{Khoury_rev, Jain_rev, Joyce_rev} and references therein), in which a long-range dynamical field couples universally to matter, generating a new (``fifth'') force between masses. The Lagrangian is designed so that the strength or range of the force depends on the local gravitational environment: the fifth force is suppressed in high-density regions (such as within the Solar System where the most stringent constraints exist; see \cite{Adelberger:2003zx} and references therein) but emerges at the lower densities of the Universe at large. This leads to differences in both inter-galaxy clustering and intra-galaxy morphology and dynamics between galaxies in stronger vs weaker gravitational fields. The latter class of signal has been the subject of a range of tests in recent years \cite{Hui, Jain_Vanderplas, Cepheid, Vikram, Vikram_RC, Desmond_1, Desmond_2, Desmond_3}; our purpose here is to explore the former.

This {chapter} investigates the galaxy correlation function (CF) as a probe of screened fifth forces. It is well known that standard general relativistic effects in large scale structure give rise to odd CF multiples \cite{Bonvin:2013ogt, Croft:2013taa} (or, in Fourier space, an imaginary part of the power spectrum \cite{McDonald:2009ud, Yoo:2012se}). These CFs break two distinct symmetries, at different scales and with different physical causes:

\begin{enumerate}

\item In a cluster environment, the CF is asymmetric under a swapping of spatial locations of galaxies along the line of sight. This is because galaxies behind the centre of a deep potential well appear closer in redshift space (and therefore more strongly correlated) than galaxies in front, due to the redshift induced by the gravitational potential.\footnote{See fig. 2 of \cite{Bonvin:2013ogt}.} This constitutes a breaking of the spatial isotropy symmetry group SO(3) into an SO(2) perpendicular to the line of sight, and is present for a \emph{single} population of galaxies on cluster scales $\sim1-10$ Mpc \cite{PhysRevLett.114.071103, cluster_test}.

\item The second type of symmetry breaking is present only in the cross-correlation of \emph{two different} populations of galaxies: $\langle \Delta_B(\vec{x}_1) \Delta_F(\vec{x}_2) \rangle \neq \langle \Delta_F(\vec{x}_1) \Delta_B(\vec{x}_2) \rangle $. Here $B$ and $F$ denote ``bright" and ``faint" galaxies, meaning that they trace the underlying matter field in different ways or otherwise have different properties pertinent to their clustering. For example, even if the galaxies form within the same dark matter density field and hence gravitational potential, they may form at different rates and hence end up with different final number densities. This is manifest in a difference in their bias. These differences in their spatial statistics correlate most strongly with their $z=0$ halo masses \cite{Kaiser, Sheth, Tinker}, with secondary effects deriving from other galaxy and halo properties (``assembly bias''; e.g. \cite{Gao, Wechsler, Mao}). Thus galaxy subsamples that differ in any observable that correlates with halo mass, e.g. luminosity or type, will manifest a parity-breaking CF. The formation of these different types of galaxy may be driven by the tidal field or other features of the cosmic web, making the effect a function of large-scale environment. This effect is present on larger scales than isotropy violation, $\sim$100 Mpc, and breaks the qualitatively different symmetry group $\mathbb{Z}_2$, which is parity under swapping the discrete $B$ and $F$ labels.

Although bias is the galaxy property conventionally responsible for giving the two populations different clustering, another possibility is sensitivity to a screened fifth force. Screened theories of gravity produce modifications to the gravitational force that depend on galaxies' internal properties and environments (Sec.~\ref{sec:screening}), in such a way that lower mass galaxies in lower density regions effectively feel stronger gravity. As we describe in detail in Sec.~\ref{sec:corr}, this alters the Euler equation and hence the number densities predicted by relativistic perturbation theory. We illustrate the effect schematically in Fig.~\ref{windy} where we show two galaxies at different spatial locations $x_1$ and $x_2$ in the same gravitational potential, but with different biases and screening parameters due to their different \emph{environments}. The environment is represented by the tree, which experiences wind at $x_2$ but not at $x_1$, delineating the fact that the environments are different at the two locations.

\begin{figure}
\begin{center}
\includegraphics[scale = 0.45]{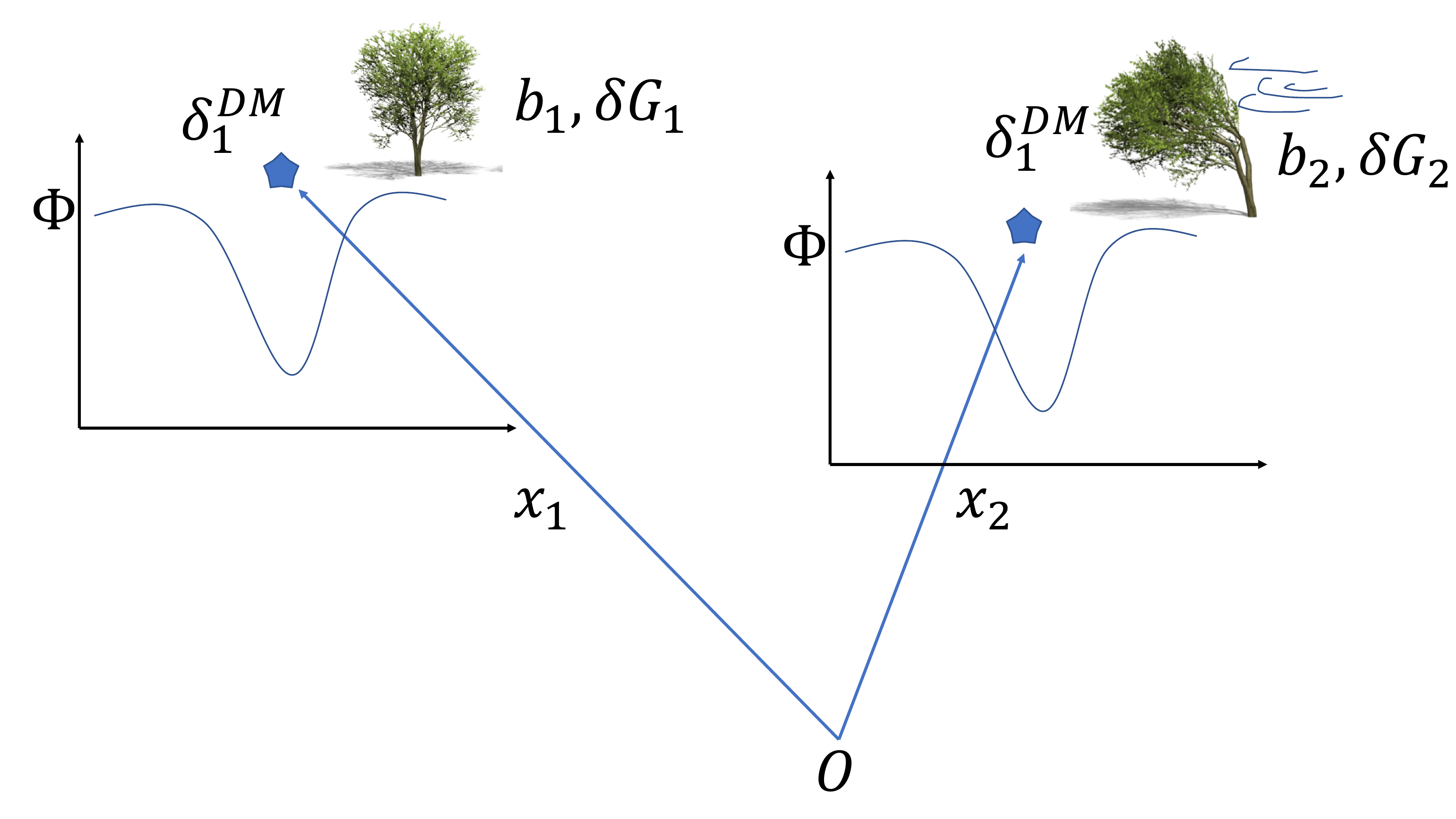}
\caption{
To tree or not to tree. 
Schematic of the effect of parity breaking in the CF. The two galaxies are located in regions of identical external dark matter density and hence gravitational potential (i.e. excluding the potential due to the galaxies' halos themselves, which are responsible for their self-screening), but the larger-scale environments are different. This is illustrated by the surroundings of the trees, which are windy at $x_2$ but not $x_1$. This difference in environment affects various properties of the galaxy that impact its clustering, including its bias, magnification bias and, of particular interest here, sensitivity to a fifth force (quantified by $\delta G \equiv \Delta G/G_N$). Swapping these parameters between the galaxy overdensities at $x_1$ and $x_2$ alters the CF, so that it is not symmetric in the galaxy labels.}
\label{windy}
\end{center}
\end{figure}

\end{enumerate}

Both of these effects are proportional to the gradient of the gravitational potential and hence receive contributions from fifth forces. In particular, as noted in \cite{Bonvin:2013ogt, Bonvin:2018ckp}, in GR the gravitational redshift term of the parity-breaking CF is precisely cancelled by the light-cone term and part of the Doppler term. This apparent coincidence is a result of the equivalence principle, whereby both light and matter feel the same potential. In contrast, theories that are conformally equivalent to GR yet include fifth forces effectively violate the equivalence principle due to the effect of the fifth force on timelike but not null geodesics. This reintroduces the redshift term, enabling relativistic effects to provide a consistency check on the validity of the Euler equation. While this effect would also be present under an unscreened fifth force, the additional effect of screening is that the two types of galaxy that enter the CF may feel \emph{different} fifth-force strengths due to their different gravitational environments endowing them with different scalar charges.\footnote{The screened fifth force that we invoke does not have to be mediated by a scalar field, but we will assume so for simplicity.} Thus timelike geodesics are affected differentially by the fifth force as a function of their trajectory, in further violation of the strong as well as weak equivalence principle. We show in Sec.~\ref{sec:corr} that this introduces interesting novel behaviour into the parity-breaking CF. We emphasise that this is not a fundamental breaking of either parity or the equivalence principle at the level of the Lagrangian. Rather, it is an effective violation stemming from differences in galaxy properties.

The structure of this {chapter} is as follows. In Sec.~\ref{sec:screening} we provide theoretical background on screened fifth forces, and in Sec.~\ref{sec:corr} we lay out the formalism for calculating the CF in their presence, paying particular attention to the relativistic parity-breaking part. Sec.~\ref{sec:results} presents our results for the dipole and octopole, specialising to a specific chameleon-screened theory, Hu-Sawicki $f(R)$, when numerical results are required. Sec.~\ref{sec:conc} provides a summary of our results and a brief discussion of future work, including prospects for testing the effects observationally. Appendix~\ref{corr_funcs} presents the full calculation of the CF, and Appendix~\ref{horndeski} shows how our work could be made more general within the chameleon paradigm by casting the chameleon action in Horndeski form. 

\section{Screened fifth forces}
\label{sec:screening}

To see the need for screening in theories with new dynamical degrees of freedom, consider the behaviour of a free light scalar $\phi$. 
The Klein-Gordon equation for the scalar field in the quasi-static limit is
\begin{equation}
	\nabla^2 \phi = 8 \pi \rho G \alpha, \label{f_Poisson}
\end{equation}
where $\rho$ is the energy density and $\alpha$ the coupling coefficient of the scalar field to matter. This is solved by $\phi = 2\alpha G M/r$, which produces the fifth force
\begin{equation}
        F_5 = -\alpha \nabla \phi = -2 \alpha^2 G M / r^2 = 2 \alpha^2 F_N. \label{eq:F5}
\end{equation}
This modifies the spatial part of the weak-field metric but not the temporal part, making it sensitive to tests of the Parametrised Post-Newtonian light-bending parameter $\gamma$. The most stringent constraint derives from the radio link to the Cassini spacecraft, which requires $\gamma < \mathcal{O}(10^{-5})$ and hence $\alpha \le \mathcal{O}(10^{-3})$ \cite{Bertotti:2003rm}. In a Friedman-Robertson-Walker (FRW) metric, the cosmological effect of the scalar field is given by

\begin{equation}
	{\phi}'' + 3 H{\phi}' + \alpha G \rho = 0,
\end{equation}

{where $H$ is the Hubble parameter and prime denotes derivative with respect to cosmic time}. For values of $\alpha$ this small, the final term is negligible and hence the fifth force is a tiny perturbation to GR dynamics.
The operation of screening may be seen from the most general equation of motion the scalar field could obey

\begin{equation}
	\Sigma^{ij}(\phi_0) \partial_i \phi \partial_j \phi + m^2_{eff}(\phi_0) \phi = 8 \pi G \alpha(\phi_0) \rho \label{gen_poisson},
\end{equation}

where $\Sigma^{ij}$ is a matrix that allows for general non-linear and non-diagonal kinetic terms. The effective mass is given by the second derivative of the potential term in the Lagrangian, and $\alpha$ is in general a function of the background field value $\phi_0$. The solution to this equation is
\begin{equation}
	\phi = 2 \alpha(\phi_0) G \frac{M}{|\Sigma(\phi_0)|r} e^{-m_{eff}(\phi_0)r},
\end{equation}
which illustrates the three qualitatively different mechanisms for removing the influence of the scalar field in high-density regions:
\begin{enumerate}
\item The field can be made short range by giving it a large effective mass at high density, i.e $m_{eff} \gg 1/R$ where $R$ is the size of the system. This is called \emph{chameleon screening} \cite{Khoury:2003rn}.
\item The amplitude of the force can be decreased by reducing the coupling to matter $\alpha(\phi_0)$. This is most commonly done via spontaneous breaking of a $\mathbb{Z}_2$ symmetry in the field configuration, in which case it is known as \emph{symmetron screening} \cite{Hinterbichler:2010es, Clampitt:2011mx}. 
\item The amplitude of the nonlinear kinetic terms can be increased, $\Sigma^{ij} \gg 1$, effectively decoupling the scalar field from matter. Depending on the precise implementation this is called \emph{kinetic} \cite{kinetic} or \emph{Vainshtein} screening \cite{Vainshtein}.
\end{enumerate}
Under any one of these mechanisms, the scalar charge of an object depends on its density and, in the case of chameleon or symmetron screening, also on its gravitational environment \cite{Brax, Falck, Zhao}. Low mass unscreened objects feel the full fifth force, which, in case the scalar field is light relative to their size, effectively causes them to feel an enhanced Newton's constant $G = G_{N} + \Delta G$. Conversely, high mass screened objects have no scalar charge and hence decouple from the fifth force and feel regular gravity. We describe this with a parameter $\delta G \equiv \Delta G/G_{N}$, which takes the value $2 \alpha^2$ for fully unscreened objects and $0$ for fully screened objects. For partial screening $\delta G$ may take any value between these limits. We show in the following section how this behaviour is manifest in cross-correlation functions when the galaxy subpopulations have different degrees of screening. In this case we will label $\delta G$ with a subscript to indicate which type of galaxy it refers to.

\section{Correlation Functions under Screened Fifth Forces}
\label{sec:corr}

The number density of galaxies traces the underlying density field in the universe on the largest scales. 
The overdensity in the number of galaxies at position $\textbf{x}$ is defined by $\Delta(\textbf{x}) \equiv \frac{N(\textbf{x}) - \bar{N}}{\bar{N}}$, where $N$ is the galaxy number density and $\bar{N}$ is the mean number density overall.
These overdensities contain a wealth of information about both the initial conditions of the universe and the distribution and properties of matter on cosmological scales \cite{Bonvin:2011bg}. 
The derivation of the main effects contributing to galaxy overdensities at linear order, including observational effects, can be found in \cite{Bonvin:2011bg, Bonvin:2013ogt,PhysRevD.84.043516}. We summarise them here. 
At a given redshift $z$, the overdensity of galaxies at an angular position $\hat{\textbf{n}}$ on the sky is given by 
\begin{eqnarray}
	& & \Delta(z, \hat{\textbf{n}}) = \Delta^{st}(z, \hat{\textbf{n}}) + \Delta^{rel}(z, \hat{\textbf{n}}) + \Delta^{lens}(z, \hat{\textbf{n}}) + \Delta^{AP}(z, \hat{\textbf{n}}) \label{start}\nonumber \\
	& & \Delta^{st}(z, \hat{\textbf{n}}) = b \delta(z, \hat{\textbf{n}}) - \mathcal{H}^{-1} \partial_r( \textbf{v} \cdot \hat{\textbf{n}}) \label{dst} \\
	& & \Delta^{rel}(z, \hat{\textbf{n}}) = \mathcal{H}^{-1} \partial_r \Psi + \mathcal{H}^{-1} \dot{\textbf{v}} \cdot \hat{\textbf{n}} \label{drel}\\
	& & - \left[ \frac{\mathcal{\dot{H}}}{\mathcal{H}^2} + \frac{2}{r\mathcal{H}} - 1 + 5 s \left( 1 - \frac{1}{r\mathcal{H}} \right) \right] \textbf{v} \cdot \hat{\textbf{n}}  \\
	& & \Delta^{lens}(z, \hat{\textbf{n}}) = (5s - 2) \int^r_0 \ dr' r' \left( \frac{r - r'}{2r} \right) \nabla^2_{\perp} (\Phi + \Psi)  \\
	& & \Delta^{AP}(z, \hat{\textbf{n}}) = \left( \partial_r - \partial_\eta \right) \left[ \Delta^{st} + \Delta^{rel} + \Delta^{lens} \right] \frac{dr (z, \hat{\textbf{n}})}{\partial \vec{\Theta}} \delta \vec{\Theta}.\nonumber  \\ \label{full_terms}
\end{eqnarray}
 $\vec{\Theta}$ is the cosmological parameter vector and $b$ is the linear bias between the galaxy density and the dark matter density $\delta$: $\Delta(\textbf{x}) = b \delta(\textbf{x})$. $\mathcal{H}$ is the conformal Hubble parameter and $\Psi$ and $\Phi$ are the weak-field metric potentials in the conformal Newtonian gauge:
\begin{equation}
	ds^2 = a^2(\eta) \left[ -\left( 1 + 2 \Psi \right) d\eta^2 + \left(1 - 2 \Phi \right) dx^i dx^j \delta_{ij} \right].
\end{equation}
$s$ describes the magnification bias that derives from the slope of the luminosity function:
 \begin{equation}
 	s \equiv \frac{2}{5} \left( \int df \ \epsilon(f) N_0(f) \right)^{-1} \int df \frac{d\epsilon}{df} \ f \ N_0(f).
\end{equation}
$N_0(f) df$ is the number density of sources with flux $f \pm \frac{df}{2}$ and $\epsilon(f)$ is the detection efficiency of those sources. 
$r$ is the comoving radial coordinate in the direction $\hat{\textbf{n}}$. 

The terms have been separated according to the physical effects that they embody. 
$\Delta^{st}$ contains the standard terms that relate the galaxy overdensity to the dark matter overdensity and the anisotropy caused by redshift space distortions.
This term is always accounted for in CF analyses.
$\Delta^{rel}$ contains the relativistic contributions such as the Doppler and integrated Sachs-Wolfe effects. This is the term which modified gravity effects alter: the acceleration terms are sourced by the Poisson equation and are therefore affected by the presence of a fifth force.
$\Delta^{lens}$ derives from the conversion of observed solid angle to physical solid angle given lensing along the line of sight. The final term describes the Alcock-Paczinski effect. We will only be interested here in the standard and relativistic terms, as it is their correlation that gives rise to the parity-breaking signal.

The expressions for the overdensities in Eqs. (\ref{start}--\ref{full_terms}) are the same in all metric theories of gravity where photons travel along null geodesics. This includes all theories conformally identical to GR, which includes most screened theories. Galaxies on the other hand are non-relativistic tracers of timelike geodesics, and are therefore directly affected by fifth forces. Their motion is governed by the Euler equation, which, in a perturbed FRW background can be written as 

\begin{equation}
	\dot{\textbf{v}} \cdot \hat{\textbf{n}} + \mathcal{H} \textbf{v} \cdot \hat{\textbf{n}} + \partial_r \Psi = 0.
\end{equation}

Here we are using the conformal Hubble parameter $\mathcal{H}$. 
This can be used in Eq. \ref{drel} to give
\begin{equation}
	\Delta^{rel}(z,\hat{\textbf{n}}) = - \left[ \frac{\dot{\mathcal{H}}}{\mathcal{H}^2} + \frac{2}{r \mathcal{H}} + 5 s \left( 1 - \frac{1}{r \mathcal{H}} \right) \right] \textbf{v} \cdot \hat{\textbf{n}}.
\end{equation}
We see that the gravitational effects in the Euler equation cancel some of the terms, which is a manifestation of the equivalence principle.
As noted in \cite{Bonvin:2018ckp}, this is no longer true in the presence of a fifth force. According to Eq.~\ref{eq:F5} a long-range fifth force behaves identically to Newtonian gravity, enabling us to capture its effect with the transformation
\begin{equation}
	\partial_r\Psi \rightarrow (1 + \delta G) \: \partial_r \Psi, \label{FF}
\end{equation}
which describes a fractional increase in the strength of gravity by an amount $\delta G$. As discussed in Sec.~\ref{sec:screening}, the logic of screening implies that this is different for different galaxies as a function of their mass distributions and environments. 
We make the simplifying assumption that $\delta G$ is a constant for each galaxy population, thereby ignoring the effect of partial screening. As we consider one population to be fully screened and the other fully unscreened, this provides an upper bound on the magnitude of their asymmetric cross-correlation. The simple modification of Eq.~\ref{FF} provides a clear intuitive picture of the physical origin of fifth force effects in the correlation function, and will also show clearly why these generate an octopole in the presence of screening, a key result of our {chapter}. Deriving the exact degree of screening of each object would require solving the equation of motion of the scalar field numerically in the presence of a given density field (e.g.~\cite{Shao}), which is beyond the scope of this work.

\begin{figure}[]
\begin{center}
\includegraphics[scale = 0.31]{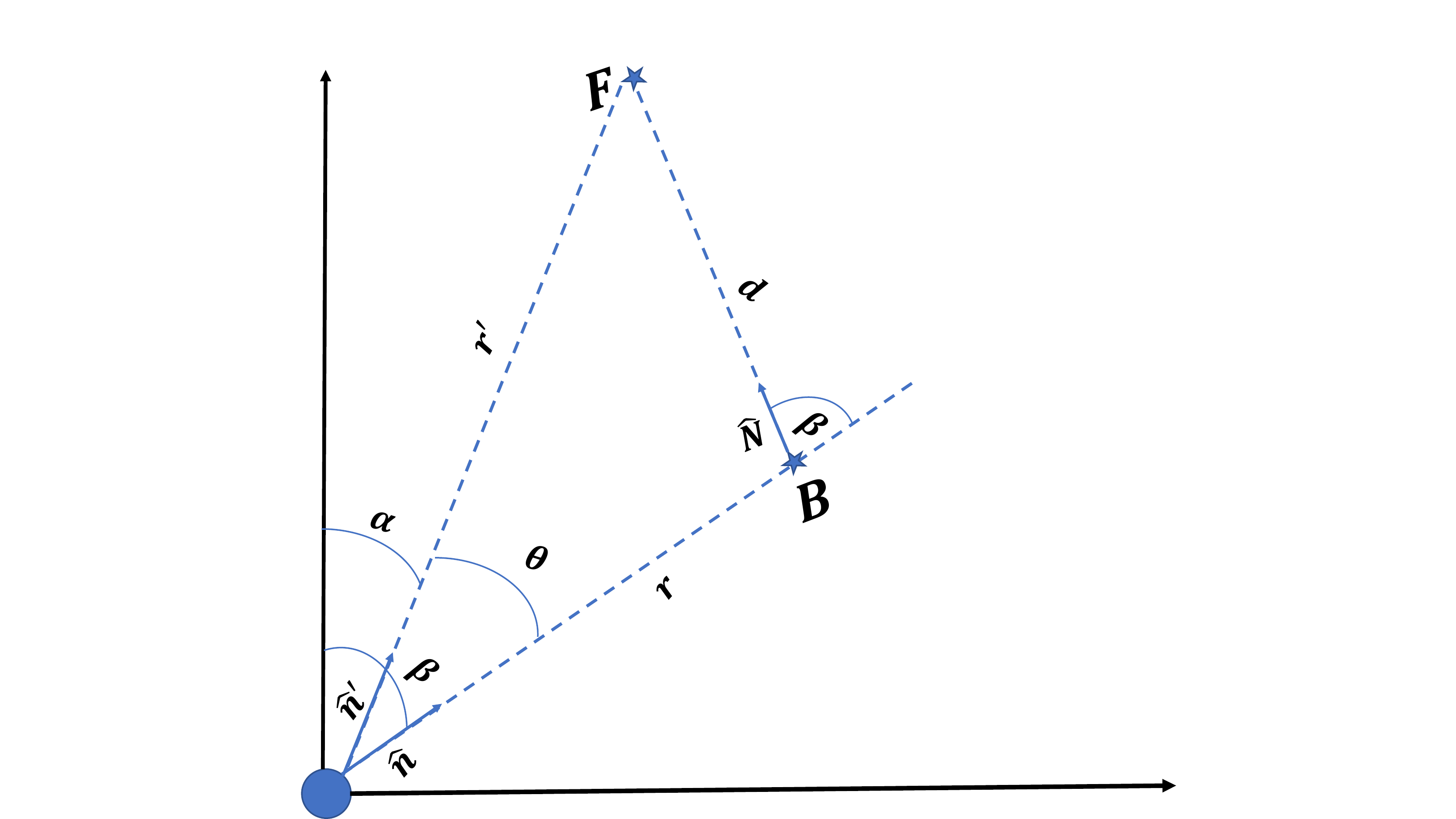}
\caption{Geometry of the CF. The observer is at the origin. $B$ and $F$ denote ``bright'' and ``faint'' galaxies; we are interested in their cross-correlation.}
\label{geometry}
\end{center}
\end{figure}

With this modification, the new expression for the relativistic part of the overdensity $\Delta^{rel(\mathcal{F})}$ is the sum of the usual relativistic term in Eq. (\ref{drel}) and an additional term $\Delta^{\mathcal{F}}$ due to the fifth force:
\begin{eqnarray}
	\Delta^{rel(\mathcal{F})}(z, \hat{\textbf{n}}) & \equiv & \Delta^{rel}(z, \hat{\textbf{n}}) + \Delta^{\mathcal{F}}(z, \hat{\textbf{n}}) \nonumber \\
	\Delta^{\mathcal{F}}(z, \hat{\textbf{n}})  & = & \zeta \left[ \frac{ \dot{\textbf{v}} \cdot \hat{\textbf{n}}}{\mathcal{H}} + \textbf{v} \cdot \hat{\textbf{n}} \right] \label{drel_screen}
\end{eqnarray}
where $\zeta \equiv \left( \frac{\delta G}{1 + \delta G} \right)$. This replaces $\Delta^{rel}$ in the total expression for $\Delta$.\footnote{Our parametrisation of modified gravity is related to that of \cite{Bonvin:2018ckp} by $\Gamma = \delta G$, $\Theta = 0$; thus, although somewhat more general, their model does not account for screening as it effectively assigns the same $\delta G$ to all galaxies.} $\zeta = 0$ in the case of complete screening (fifth force fully suppressed) and $\zeta = \zeta_{max}$ for objects that are fully unscreened. $\zeta_{max}$ is set by the coupling coefficient of the scalar field to matter; for example it is 1/4 in $f(R)$ where $\delta G = 1/3$. As $\zeta$ for a given galaxy depends on its mass distribution and gravitational environment via its degree of screening, we assign the two galaxy populations (which we denote ``bright'', $B$, and ``faint'', $F$, as in Sec.~\ref{sec:intro}) different average values $\zeta_B$ and $\zeta_F$.
We can now compute the parity-breaking CF.
The cross-correlation between the $B$ and $F$ populations is given by $\langle \Delta_B(\textbf{x}_1) \Delta_F(\textbf{x}_2) \rangle$, which we write, with the geometry of Fig.~\ref{geometry}, as
\begin{equation}
	\xi(z,z', \theta)_{BF}= \langle \Delta_B(z, \hat{\textbf{n}}) \Delta_F(z', \hat{\textbf{n}}') \rangle.
\end{equation}
Due to the assumption of statistical isotropy this depends only on the angle $\theta$ between the galaxies as they are projected on the sky. Furthermore, it is important to remember that $z, z'$ and $\theta$ are \emph{observed} redshifts and angular sizes. Converting these to \emph{physical} quantities depends on the background cosmology, although to linear order the corrections from this are already accounted for in the expressions for the overdensities.

In principle there is a CF for each term in $\Delta$, although we are only interested here in the relativistic part $\xi^{rel(\mathcal{F})}$ which is sensitive to the fifth force
\begin{eqnarray}
	\xi^{rel(\mathcal{F})}(z, z', \hat{\textbf{n}}) & = & \langle \Delta^{st}_B(z,\hat{\textbf{n}}) \Delta^{rel(\mathcal{F})}_F  (z', \hat{\textbf{n}}') \rangle \nonumber \\
	& + &  \langle \Delta^{st}_F(z', \hat{\textbf{n}}') \Delta^{rel(\mathcal{F})}_B (z, \hat{\textbf{n}}) \rangle .
\end{eqnarray}
The individual CFs are calculated in Appendix \ref{corr_funcs}; here we show the final result, expanded to leading order in $d/r \ll 1$

\begin{eqnarray}
	\xi^{rel(\mathcal{F})}(r,d, \beta) & = & \xi^{rel}(r,d, \beta) + \xi^{(\mathcal{F})}(r,d, \beta) \nonumber \\
	 \xi^{rel}(r,d, \beta) & = & \frac{2 A_s}{9 \pi^2 \Omega_m^2} \frac{\mathcal{H} D^2 f}{\mathcal{H}_0}\left\{ P_1(\cos \beta) \nu_1(d) \left[ (b_B - b_F) \left( \frac{\dot{\mathcal{H}}}{\mathcal{H}^2} + \frac{2}{r \mathcal{H}} \right) \right. \right. \nonumber \\
	& - & \left. \left. \left( 1 - \frac{1}{r \mathcal{H}} \right) \left( 5(s_Bb_F - s_F b_B) + 3f(s_B - s_F) \right) \right]  \right. \nonumber \\
	& &\left. + 2 P_3(\cos \beta) \nu_3(d) \left( 1 - \frac{1}{r \mathcal{H}} \right) f(s_B-s_F) \right\} \nonumber \\
§	 \xi^{(\mathcal{F})}(r,d, \beta) & = & \frac{2 A_s M}{9 \pi^2 \Omega^2_m} \left\{ P_1(\cos \beta) \nu_1(d) \left[ (b_F \zeta_B - b_B \zeta_F) - \frac{3}{5} f( \zeta_B - \zeta_F) \right]  \right.  \nonumber \\
	& + & \left. P_3(\cos \beta) \nu_3(d) \left[ \frac{2f}{5} (\zeta_B - \zeta_F) \right] \right\}. \label{eq:final}
\end{eqnarray}

We have defined $f \equiv \frac{d \ln D}{d \ln a}$, where $D$ is the linear growth factor. 
Throughout our calculations we have assumed these are scale-independent, following \cite{Bonvin:2018ckp}\footnote{In principle this can generalised and the growth factor made scale dependent. We plan on exploring this for more generic modified gravity theories in the modified Boltzmann code HiClass \cite{hiclass} in future works. We take the first steps in this direction by writing our screening model in Horndeski form in Appendix~\ref{horndeski}.}
$P_n$ is the Legendre polynomials of order $n$, $b_B$ and $b_F$ are the biases for the bright and faint galaxies respectively, and 
\begin{eqnarray}
	& & \nu_\ell(d) \equiv \int d \ln k \ (k \eta)^{n_s-1} \left( \frac{k}{\mathcal{H}_0} \right)^3 j_\ell(kd) T^2(k), \nonumber \\
	& & M(a) \equiv \frac{D^2}{\mathcal{H}_0 \mathcal{H}} \left( \dot{\mathcal{H}} f + \mathcal{H} \dot{f} + f^2 \mathcal{H}^2 + f \mathcal{H}^2\right). \label{nu_stuff}
\end{eqnarray}
We have dropped all terms in $\xi^{rel}$ that involve the correlation of two relativistic terms ($\langle \Delta^{rel} \Delta^{rel} \rangle$) as these are suppressed by factors of $\frac{\mathcal{H}}{k}$.

As anticipated in Sec.~\ref{sec:intro}, besides modifying the cosmological background the fifth force affects the CF in two distinct ways. The first is the reintroduction of redshift terms due to the difference induced between null and timelike geodesics (first term in $\xi^{(\mathcal{F})}$). This would also be present under a universally-coupled (i.e. non-screened) fifth force and is the type of modified gravity considered by \cite{Bonvin:2018ckp}. The second is the relative effect on the $B$ and $F$ populations due to their different sensitivities to fifth forces under a screening mechanism. This is shown by the remaining terms in $\xi^{(\mathcal{F})}$ which are proportional to $\zeta_B - \zeta_F$. We see that the fifth force introduces both a dipole (the term proportional to to $P_1\nu_1$) and an octopole (the term proportional to $P_3 \nu_3$). While the dipole is increased by both screened and universal fifth-force terms, in the absence of screening the octopole is only present if $s_B \ne s_F$. Thus, as discussed in more detail below, the octopole may provide a particularly clean probe of screening.

\section{Correlation function parameter space}
\label{sec:parameters}

The CF of Eq.~\ref{eq:final} depends on several parameters of standard $\Lambda$CDM, as well as those of modified gravity. The aim of this section is to explain and quantify the effects of these parameters. We begin by classifying them into two sets: 1) Those that affect the background cosmology and hence all CFs, and 2) Those that are specific to the parity-breaking CF and describe the environmental dependences of galaxy formation. We term the first set \emph{global} and the second set, whose members carry a $B$ or $F$ index, \emph{local}. The parameters are listed in Table \ref{fid_cosmo}, along with their fiducial values which we use throughout our analysis unless otherwise stated.

On a practical note, the global parameters take longer to evaluate as they affect background quantities such as the matter power spectrum: thus each point in parameter space corresponds to a run of a Boltzmann code. The local parameters are simply multiplicative factors in front of the functions of global parameters, making their parameter space in principle much quicker to explore.

\begin{table}[h]
\begin{centering}
\begin{tabular}{ p{3cm}  p{3cm}}
\hline
\hline
$A_s$ & 2.1 $\times 10^{-9}$  \\
$h$ & 0.7 \\
$\Omega_bh^2$ & 0.0224  \\
$\Omega_{c}h^2$ & 0.112 \\
$k_{*}$ & 0.05 Mpc$^{-1}$ \\
$n_s $ & 0.96 \\
$f_{R0}$ & $0$ \\
\hline
\hline
$(b_B, b_F)_{z = 0}$ & (1.7, 0.84) \\
$(b_B^{f(R)} b_F^{f(R)})_{z = 0}$ & (1.64, 0.8) \\
$(s_B, s_F)$ & (0.1,  0) \\
$(\delta G_B, \delta G_F)$ & (0, 1/3) \\
\hline
\hline
\end{tabular}
\caption{Global (upper) and local (lower) parameters, along with their fiducial values. The bias values used in the $f(R)$ plots are different from GR and are denoted  by the $f(R)$ superscript index.}\label{fid_cosmo}
\end{centering}
\end{table}

\subsection{Global parameters}

\begin{figure}
\begin{center}
\includegraphics[scale = 0.4]{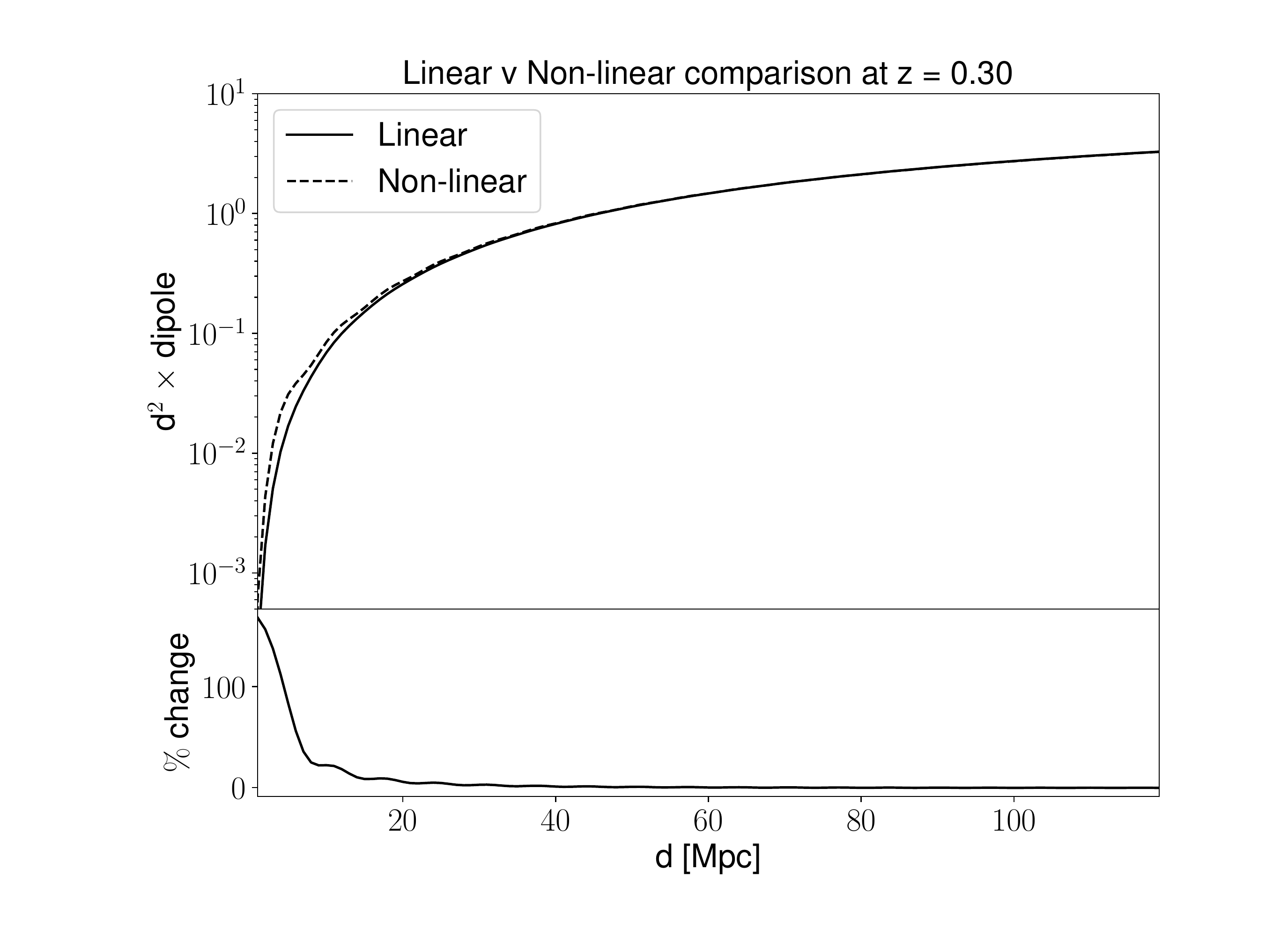}
\caption{The change to the $z=0.3$ $\Lambda$CDM dipole when a linear or non-linear power spectrum/transfer function (from Halofit) is used, relative to the linear case.}
\label{NL}
\end{center}
\end{figure}

There are three groups of global parameters. The first contains the standard $\Lambda$CDM parameters. As the effects of these on CFs have already been extensively studied \cite{Gaztanaga:2015jrs, Hall:2016bmm, Alam_mag, Giusarma:2017xmh, Alam:2017cja, Alam:2017izi}, we do not investigate them further here. The second set quantifies the effects of ``galaxy formation" physics, which generates non-linear corrections to the transfer function. We check the importance of this by using the Halofit fitting function \cite{Peacock:2000qk} to obtain the non-linear matter power spectrum, and show its effect on the dipole in Fig.~\ref{NL}. We see that on scales below $\sim$10 Mpc non-linearities lead to an $\mathcal{O}(1)$ modification, but the difference decays away rapidly on larger scales. From now on we will only use the transfer functions and power spectra with these non-linearities included.

The final set of global parameters describes the effect of modified gravity on the background perturbations, specifically the transfer function and power spectrum. To compute this effect we specialise to the case of Hu-Sawicki $f(R)$ \cite{Hu_Sawicki}, an archetypal and well-studied chameleon-screened theory known to be stable to instabilities, propagate gravitational waves at the speed of light and be capable of screening the Milky Way to pass local fifth-force tests. Although representative of the chameleon mechanism, Hu-Sawicki occupies only a small part of the full chameleon parameter space. A general chameleon model introduces three new degrees of freedom. At the level of the Lagrangian these are, for example, the $\{n, \Lambda, M\}$ of \cite{Burrage} Eq. 2.5 (see also Appendix~\ref{horndeski}). Phenomenologically they are the strength of the fifth force between unscreened objects (related to $\alpha$ in Eq.~\ref{f_Poisson}), the range of the fifth force (Compton wavelength of the scalar field) and the self-screening parameter $\chi$ (e.g. \cite{Burrage} eq. 3.2) that determines the threshold Newtonian potential at which screening kicks in. In $f(R)$ the coupling coefficient of the scalar field to matter is fixed at $\alpha = 1/\sqrt{6}$ ($\delta G = 1/3$) while the parameters $n$ and $\Lambda$ describing the field's potential are related.

In the Jordan frame, the Hu-Sawicki action is given by

\begin{equation}
S = \int d^4x \sqrt{-g} \left(\frac{M_{pl}^2}{2}(R+f(R)) + \mathcal{L}_m\right), \label{fR_HS}
\end{equation}

where

\begin{equation}\label{eq:fR}
f(R) = -m^2 \frac{c_1 (R/m^2)^k}{c_2(R/m^2)^k+1}
\end{equation}

and $\mathcal{L}_m$ is the matter Lagrangian. The mass scale is set by the average density of the Universe, $m^2 = \rho_m/(3 M_{pl}^2)$, and $c_1$, $c_2$ and $k$ are dimensionless free parameters. It is shown in \cite{Li:2011vk} that only $k$ and $\frac{c_1}{c_2^2}$ affect the matter power spectrum and thus the model only contains two relevant degrees of freedom. $k=1$ is a standard choice that we adopt here. In the Einstein frame this is a scalar--tensor theory in which the scalar field is $f_R \equiv df/dR$, the present value of which is also completely determined by $\frac{c_1}{c_2^2}$. The field at the cosmological background value of $R$, $f_{R0}$, determines the structure formation history of the Universe as well as the range and screening properties of the fifth force at a given epoch, and is effectively the theory's only degree of freedom. GR is recovered in the limit $f_{R0} \rightarrow 0$, and values in the range $\sim10^{-4}-10^{-6}$ have observable consequences in galaxy clustering, redshift space distortions, cluster abundance, intensity mapping and the matter bispectrum (see e.g. \cite{Lombriser_rev} and references therein). For the Solar System to be screened requires $f_{R0} \lesssim 10^{-6}$. Smaller values may be probed by galaxy-scale tests \cite{Vikram, Vikram_RC, Cepheid, Desmond_2, Desmond_3}, which now rule out $f_{R0} > {few} \: \times 10^{-8}$ \cite{Desmond_1}.

For one of our galaxy types to self-screen and the other not, the screening parameter $\chi$ must be between their characteristic Newtonian potentials. This can be written in terms of the background scalar field value as $\chi \simeq 3/2 \: f_{R0}$. The halo masses calculated in Sec.~\ref{sec:local} imply $|\Phi_F| \simeq 1\times10^{-6}$ and $|\Phi_B| \simeq 6\times10^{-6}$ ($c\equiv1$). The galaxies may however be partly environmentally screened, increasing the required value of $\chi$.\footnote{To determine the screening properties of both galaxy populations one would ideally solve the equation of motion for the scalar field given the mass distribution around the galaxies, or at least a proxy for the field such as the Newtonian potential \cite{Desmond_maps}. However, as we are interested here in the general effects of a screened fifth force we leave this more detailed investigation for future work.} $10^{-6} \lesssim f_{R0} \lesssim 10^{-5}$ is therefore likely to separate the galaxies by screening properties, and also causes the scalar field to mediate an astrophysical-range fifth force \cite{Hu_Sawicki}. This is therefore the range that we consider.

\begin{figure}
\begin{center}
\includegraphics[scale = 0.4]{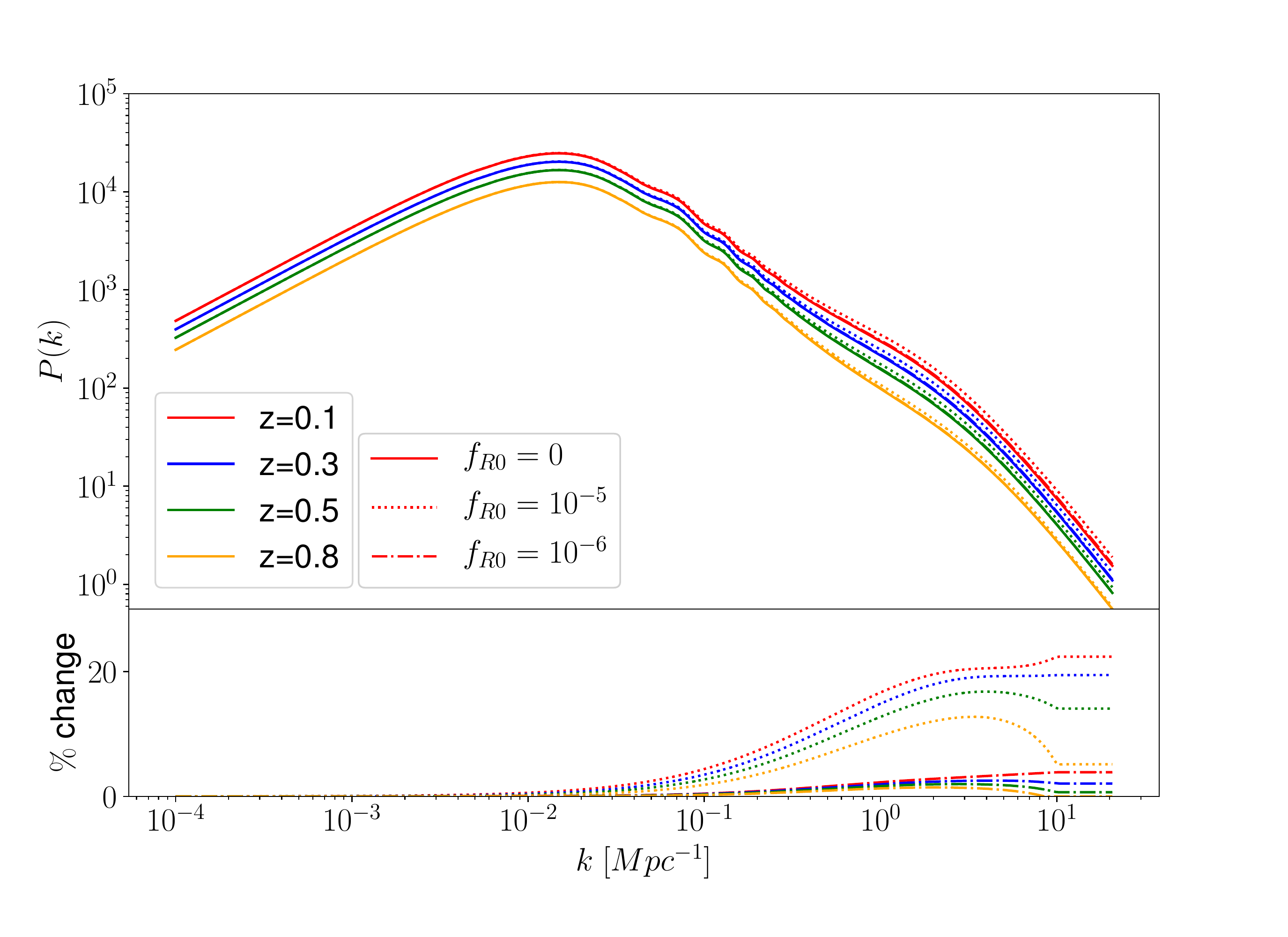}
\caption{The matter power spectrum evaluated in $\Lambda$CDM and $f(R)$ for a range of redshift and $f_{R0}$ values. The percentage difference is plotted with respect to the $\Lambda$CDM values with the fiducial parameters of Table \ref{fid_cosmo}.}
\label{mPk_fR}
\end{center}
\end{figure}

We compute the matter power spectrum using a modified version of CAMB calibrated with $N$-body simulations in the $k=1$ model \cite{Winther}, and plot this for different $f_{R0}$ values at various redshifts in Fig.~\ref{mPk_fR}. We see that the power spectrum changes by $\sim 20 \%$ on scales smaller than $\sim 1$ Mpc for $f_{R0}  = 10^{-5}$, while for $f_{R0} = 10^{-6}$ the change is only a few percent. This is propagated into the CF in Sec.~\ref{sec:results}.

\subsection{Local parameters}
\label{sec:local}

We consider three local parameters: galaxy bias $b$, magnification bias $s$ and fifth-force sensitivity $\delta G$. The first two are present in $\Lambda$CDM and give rise to the standard parity-breaking CF, while the latter is the specific focus of our study. We describe our choices for these parameters below.

\begin{itemize}

\item $(b_B, b_F)$: As our fiducial case we take the $B$ galaxies to be luminous red galaxies and the $F$ galaxies to be emission line galaxies, with biases in $\Lambda$CDM of 1.7 and 0.84 respectively at $z=0$ \cite{DESI}. These are typical populations that will be measured by the forthcoming DESI survey, which will provide the next significant improvement in measurement of the parity-breaking CF.\footnote{Forecasts for testing gravity with the parity-breaking CF for non-screened theories can be found in \cite{Bonvin:2018ckp}} Under the Sheth--Tormen model \cite{Sheth_Tormen} these biases correspond to halo masses $\sim$10$^{13} \: h^{-1} M_\odot$ and $\sim$10$^{14} \: h^{-1} M_\odot$ respectively. When showing results for $z>0$ we model the redshift dependence of the bias as \cite{Tegmark:1998wm, Fry:1996fg, Nusser:1993sx, Bonvin:2013ogt}
\begin{equation}
	b_{B / F}(z) = 1 + (b_{B / F}(z = 0) -1) \frac{D(z = 0)}{D(z)} \label{bias}
\end{equation}
with $D$ the regular growth factor.
We check how the dipole is affected by a change in the bias and as a function of redshift in Fig.~\ref{dip_local}. To account for the fact that the bias is reduced in $f(R)$ due to the action of the fifth force (e.g. \cite{Schmidt:2008tn}), we take $b_B^{f(R)}(z=0) = 1.62$, $b_F^{f(R)}(z=0) = 0.8$ when we study $f_{R0}=10^{-5}$.\footnote{Taken from fig. 14 of \cite{Arnold:2018nmv} for galaxies of mass $10^{13} M_\odot/h$ and $10^{14} M_\odot/h$ for $b_F(z=0)$ and $b_B(z=0)$. We take the mean value from both box sizes and thus change the bias by $\sim$ -4$\%$ for both types of galaxies.} We also use the growth function in $f(R)$  theory to compute the change in bias at different redshifts. This leads to a $\mathcal{O}(10\%)$ change in the bias at $z = 1$. These bias values are used for the $f(R)$ cases of Figs.~\ref{dip_oct} and \ref{pchange}.

\item $(s_B, s_F)$: If the magnification bias is the same for both galaxy populations then the octopole in $\Lambda$CDM is zero. Under this common assumption the octopole is a unique signature of \emph{screened} fifth forces: as shown in Eq.~\ref{eq:final}, even modified gravity without screening does not produce it. However, in order to gauge the relative importance of magnification bias and fifth force, we consider a plausible value of $s_B - s_F = 0.1$ \cite{Alam_mag, Durrer_1, Durrer_2}.

\item $(\delta G_B, \delta G_F)$: There are various considerations for setting the values of $\delta G_B$ and $\delta G_F$. $\delta G_B$ should be no larger than $\delta G_F$ because brighter galaxies should be more massive (and occupy denser environments), and hence more screened. Our requirement that $|\Phi_F| \lesssim \chi \lesssim |\Phi_B|$, and assumption of $f(R)$ in the background, implies $\delta G_B = 0$, $\delta G_F = 1/3$ as our fiducial choice. It is worth noting, however, that one can construct theories in which the change to the transfer function and growth rate is small but the unscreened $\delta G$ is $\mathcal{O}(1)$ or larger, in which case the predicted signal simply scales the $\Lambda$CDM result linearly with $\delta G$. The parity-breaking CF would provide maximal sensitivity to screening \textit{per se} in such a scenario, due to the insignificance of modified gravity in the background.

\end{itemize}

\section{Results}
\label{sec:results}

In this section we calculate the dipole and octopole numerically in our model. We use the Boltzmann code CAMB \cite{Lewis:1999bs} to compute the transfer functions in $\Lambda$CDM, along with the modification presented in \cite{Winther} for $f(R)$. Throughout the computations we use the fiducial parameters presented in Table \ref{fid_cosmo} and a range of redshifts, $ z \in \{0.1, 0.3, 0.5, 0.8\}$. 

\begin{figure*}
\centering
\begin{subfigure}{.5\textwidth}
  \centering
  \includegraphics[width=\linewidth]{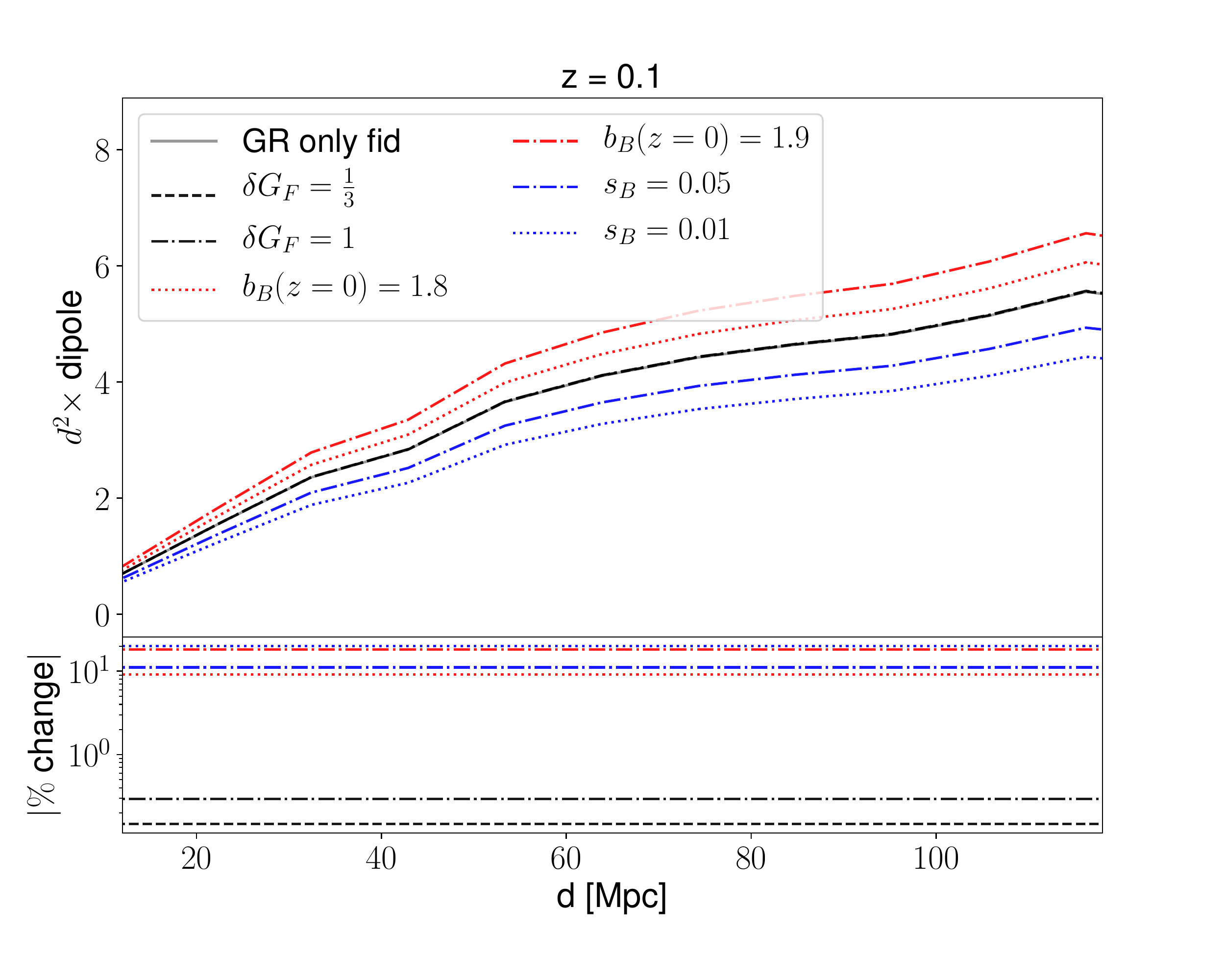}
  \caption{}
  \label{fig:sub1}
\end{subfigure}%
\begin{subfigure}{.5\textwidth}
  \centering
  \includegraphics[width=\linewidth]{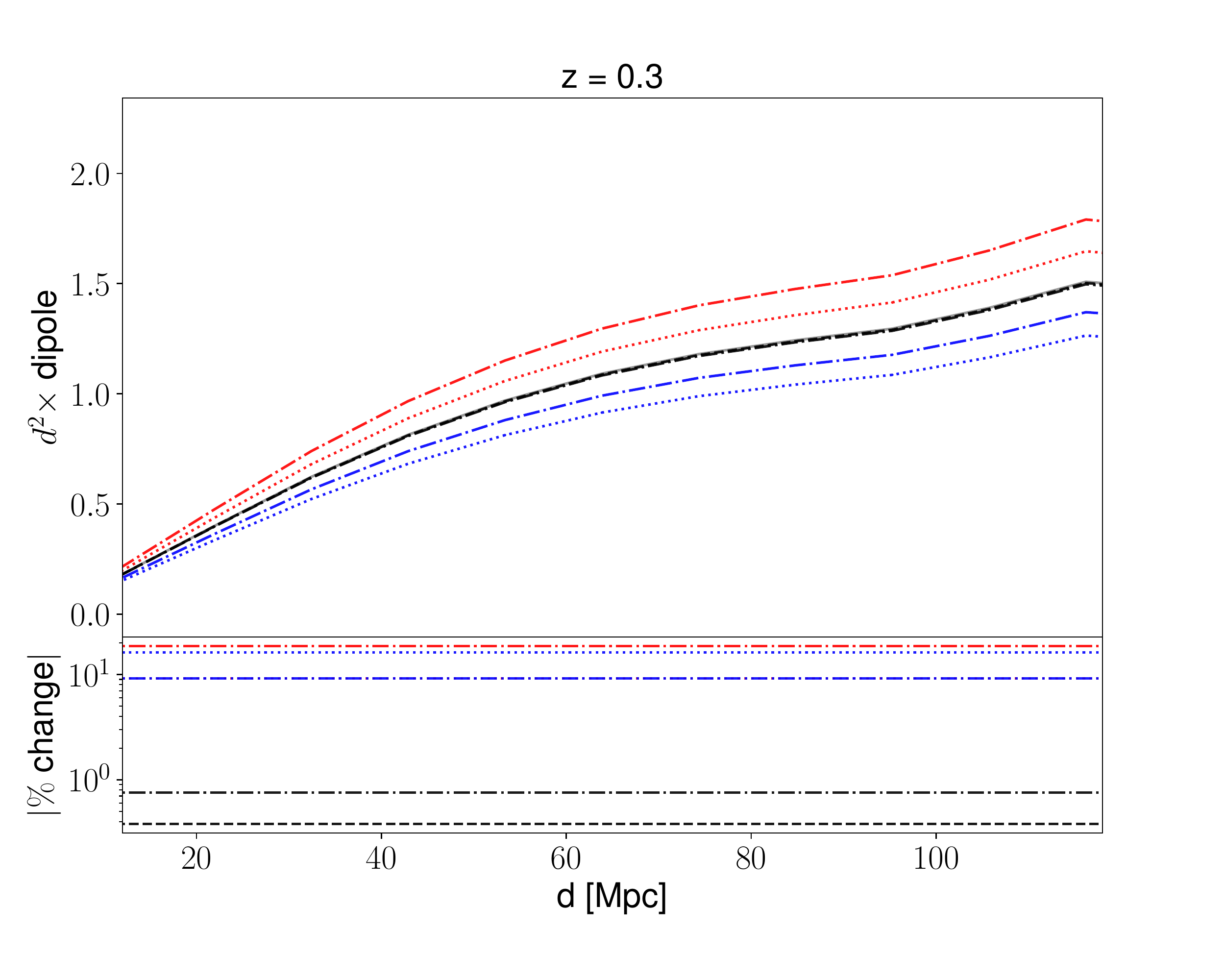}
  \caption{}
  \label{fig:sub2}
\end{subfigure}

\begin{subfigure}{.5\textwidth}
  \centering
  \includegraphics[width=\linewidth]{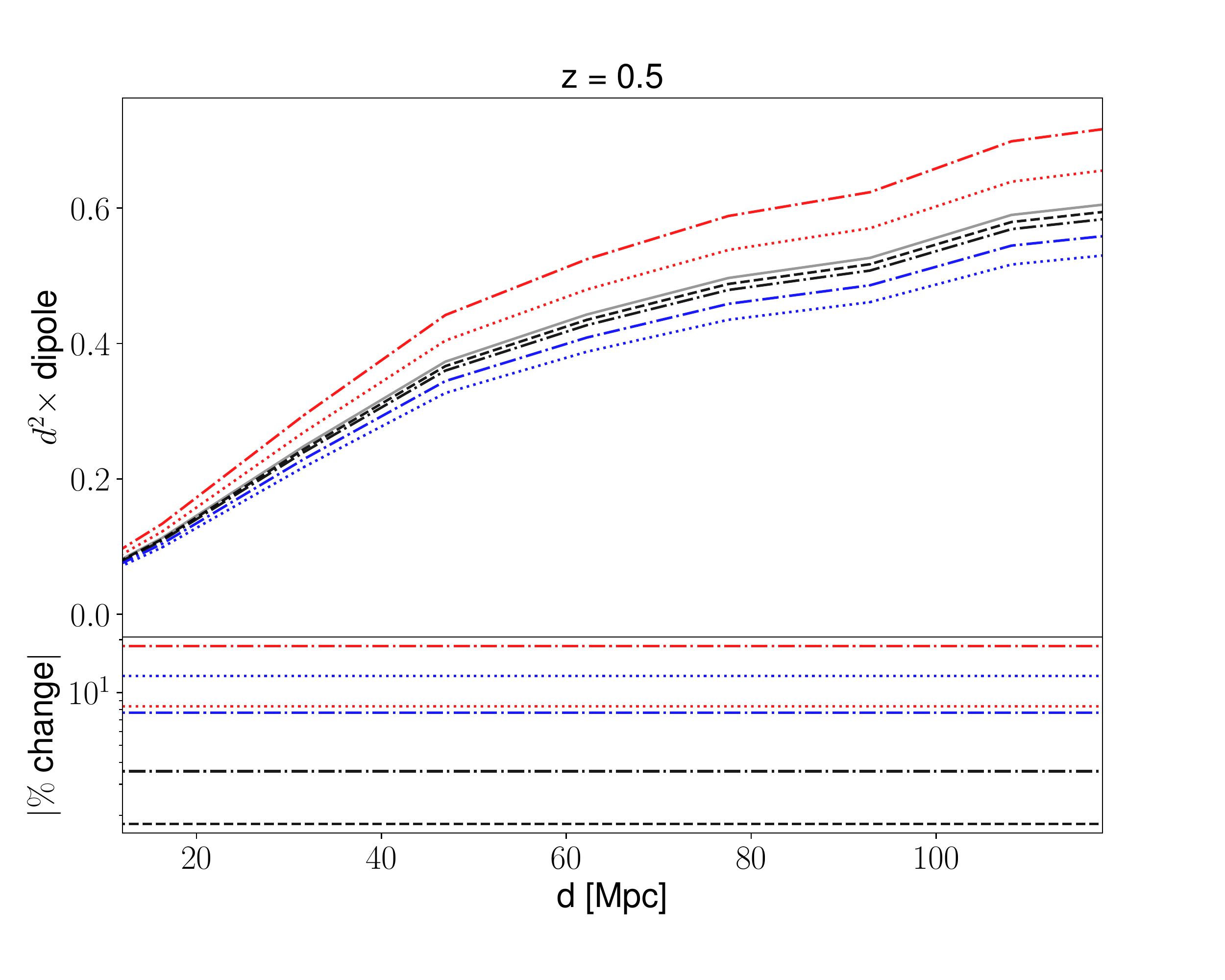}
  \caption{}
  \label{fig:sub1}
\end{subfigure}%
\begin{subfigure}{.5\textwidth}
  \centering
  \includegraphics[width=\linewidth]{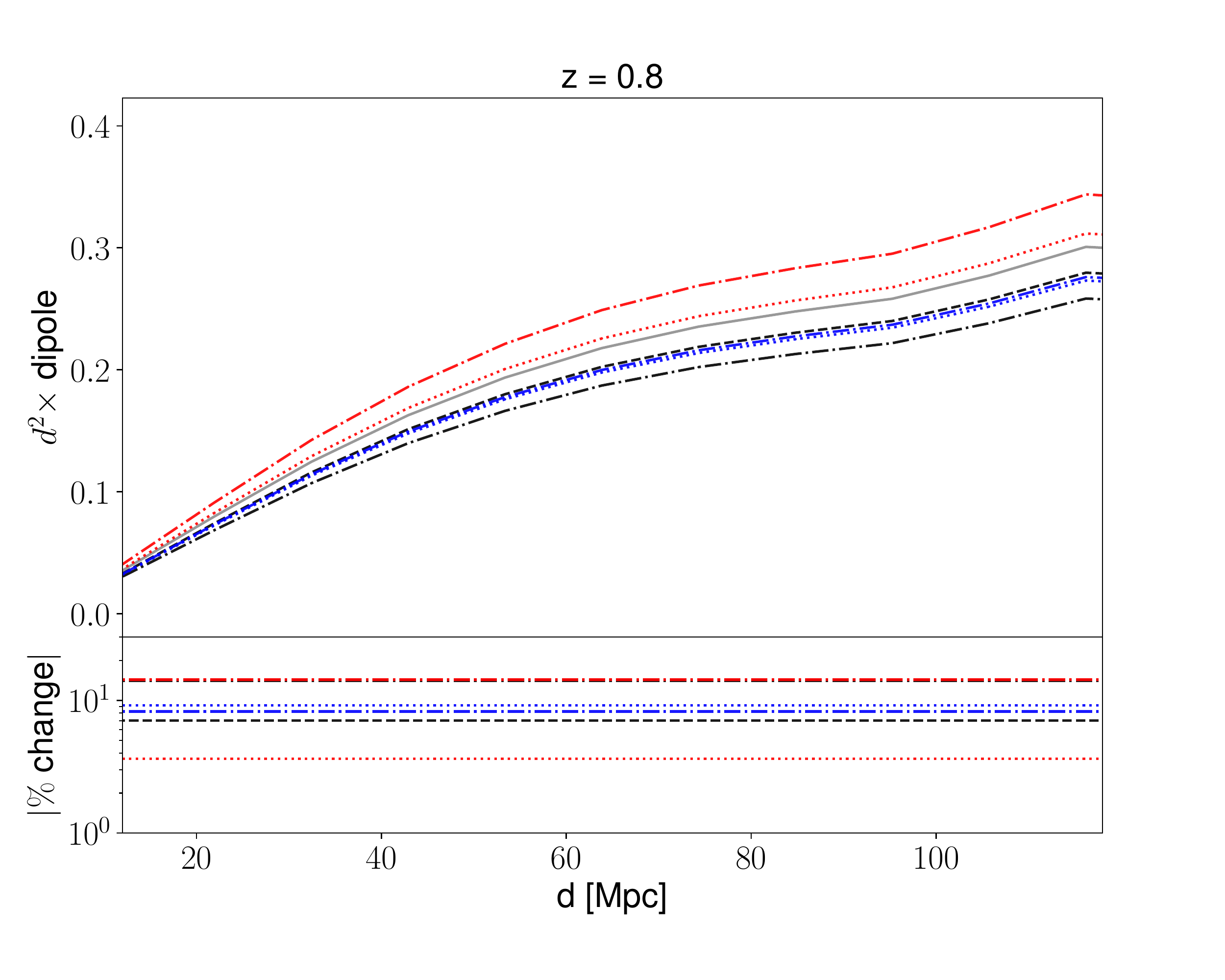}
  \caption{}
  \label{fig:sub2}
\end{subfigure}
\caption{CF dipole for various values of local parameters in a $\Lambda$CDM background, as listed in the legend of subfigure (a) and the same in all cases. Changes are defined with respect to the fiducial values listed in Table~\ref{fid_cosmo}. Each subfigure corresponds to a particular redshift, as indicated. The lower panels show the magnitudes of the percentage changes with respect to the fiducial $\Lambda$CDM dipole (grey line).}
\label{dip_local}
\end{figure*}

First we check the sensitivity of the dipole to local parameters. Fig.~\ref{dip_local} shows the dipole at various redshifts for a range of galaxy bias, magnification bias and $\delta G$ values. We assume throughout that the $B$ galaxies are completely screened while varying the screening felt by the $F$ galaxies; conversely, when investigating bias we fix $b_F$ and $s_F$. We see that $\delta G_F = 1/3$, which corresponds to complete unscreening in $f(R)$, changes the dipole by $\sim\mbox{few} \times 0.1\%$ for $z = 0.1, 0.3$, whereas at $z = 0.5, 0.8$ the change is $\sim \mbox{few} \times 1 \% - 10\%$. A $\sim 10\%$ change in the galaxy bias changes the dipole by the same amount. Decreasing the magnification bias by a factor of (2, 5) decreases the dipole by $\sim (10, 15)\%$ at all redshifts except $z = 0.8$ where the effect is slightly smaller. Interestingly, at $z = 0.8$ the percentage change due to screening is roughly the same as the effect of varying the bias in the range we consider. This shows that to able to detect the modification due to screening it will be necessary to know or model the galaxy bias to $\sim0.1-1\%$ for $z = 0.1, 0.3$, but only $\sim 5, 10 \%$ for $z = 0.5, 0.8$. It is also worth noting that the galaxy bias is typically measured in combination with $\sigma_8$ in clustering analyses, while breaking the degeneracy with $\sigma_8$ requires information from weak lensing. The magnification bias would need to be known to a within a factor $\sim$10 to constrain screening parameters at low redshift, while at higher redshift it must be known to within a factor $\sim$2. 

\begin{figure}
\centering
\begin{subfigure}{.5\textwidth}

  \includegraphics[width=\linewidth]{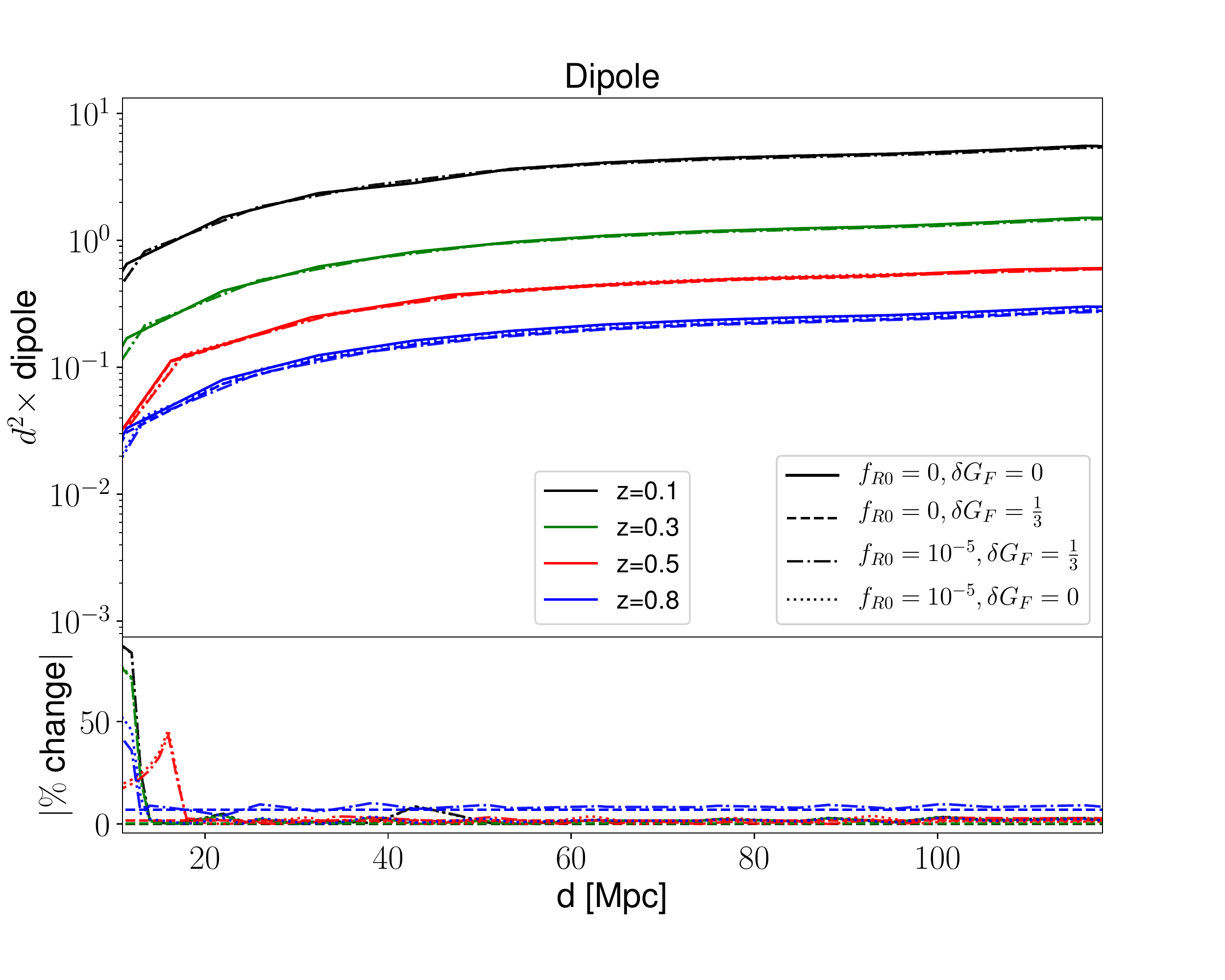}
  \caption{The dipole for a range of models involving local screening and/or modification to the cosmological background. The solid line is the fiducial $\Lambda$CDM dipole at each redshift, which the percentage differences are defined relative to.}
  \label{global_dip}
\end{subfigure}%
\begin{subfigure}{.5\textwidth}

  \vspace{3mm}\includegraphics[width=\linewidth]{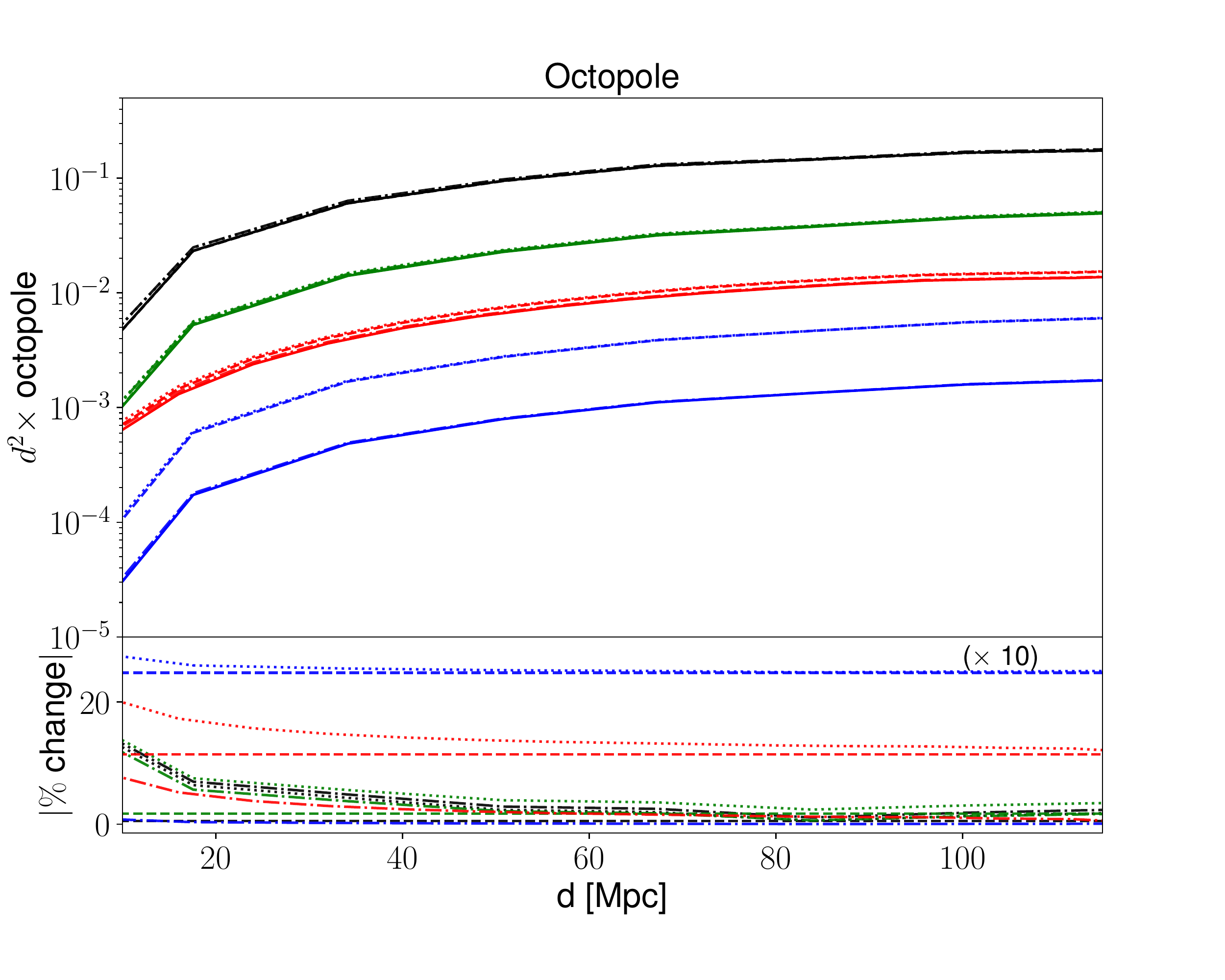}
   \caption{The absolute value of the octopole for the same models as in Fig.~\ref{global_dip}. The change in the octopole for different local parameter values can be read off from Eq.~\ref{eq:final}. We scale the blue line down by a factor of 10 in the percentage change in order for it to fit on the plot: the true value is $\sim$200\%.}
  \label{global_oct}
\end{subfigure}
\caption{Dipole and octopole components of the parity-breaking CF in $\Lambda$CDM and $f(R)$ for $0.1 < z < 0.8$.}
\label{dip_oct}
\end{figure}

Next we show in Fig.~\ref{global_dip} the dipole for $f_{R0}=10^{-5}$, which generates a cosmological-range fifth force. We see that the dipole changes by $\mathcal{O}(1)$ on scales $\lesssim 10$ Mpc, but less at higher redshift. For $z = 0.5$ and below the change is $\mathcal{O}(0.1-1\%)$ on scales $\gtrsim 20$ Mpc. Thus the effect from the change in background in $f(R)$ theories appears to dominate the effect of realistic screening values by an order of magnitude for $z \lesssim 0.5$, while at higher $z$ the screening effect can be twice as important as the change in background. Thus the most promising regime in which to search for signs of screening in the dipole is $z > 0.5$, while at lower redshift one should hope instead to detect the change due to the effect of modified gravity on the growth rate and transfer function.

In Fig.~\ref{global_oct} the octopoles for both $\Lambda$CDM (with and without screening) and $f(R)$ background with $f_{R0} = 10^{-5}$ are computed. We see that in the case of $f_{R0} = 0$ (i.e. a local screened fifth force in a $\Lambda$CDM background), the octopole changes by few $\times 1 \%$ for $z = 0.1, 0.3$, whereas for $z = (0.5, 0.8)$ it changes by $\sim (10\%, 25\%)$. This always dominates over the effect of background modification only (no local screening). It is worth bearing in mind that in $\Lambda$CDM (and non-screened modified gravity), an octopole only arises due to the difference in magnification bias between the two galaxy populations, which is typically assumed to be zero. In this case \emph{any} octopole would be a sign of screening. The results for any local parameter choices can be readily constructed as they simply multiply the curves in Fig.~\ref{global_oct}.

\begin{figure*}
\centering
\begin{subfigure}{.5\textwidth}
  \centering
  \includegraphics[width=\linewidth]{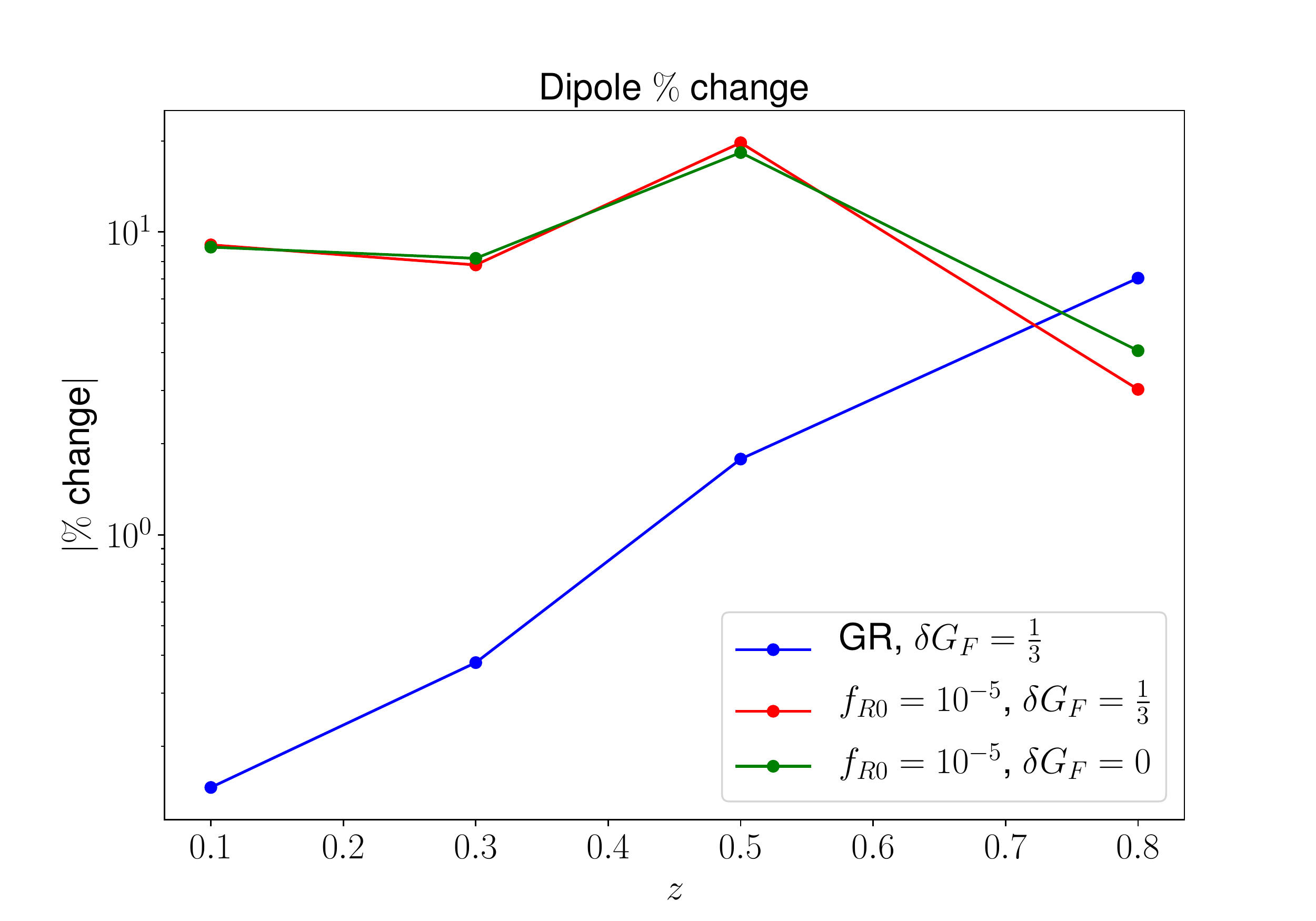}
  \caption{}
  \label{pc_dip}
\end{subfigure}%
\begin{subfigure}{.5\textwidth}
  \centering
  \includegraphics[width=\linewidth]{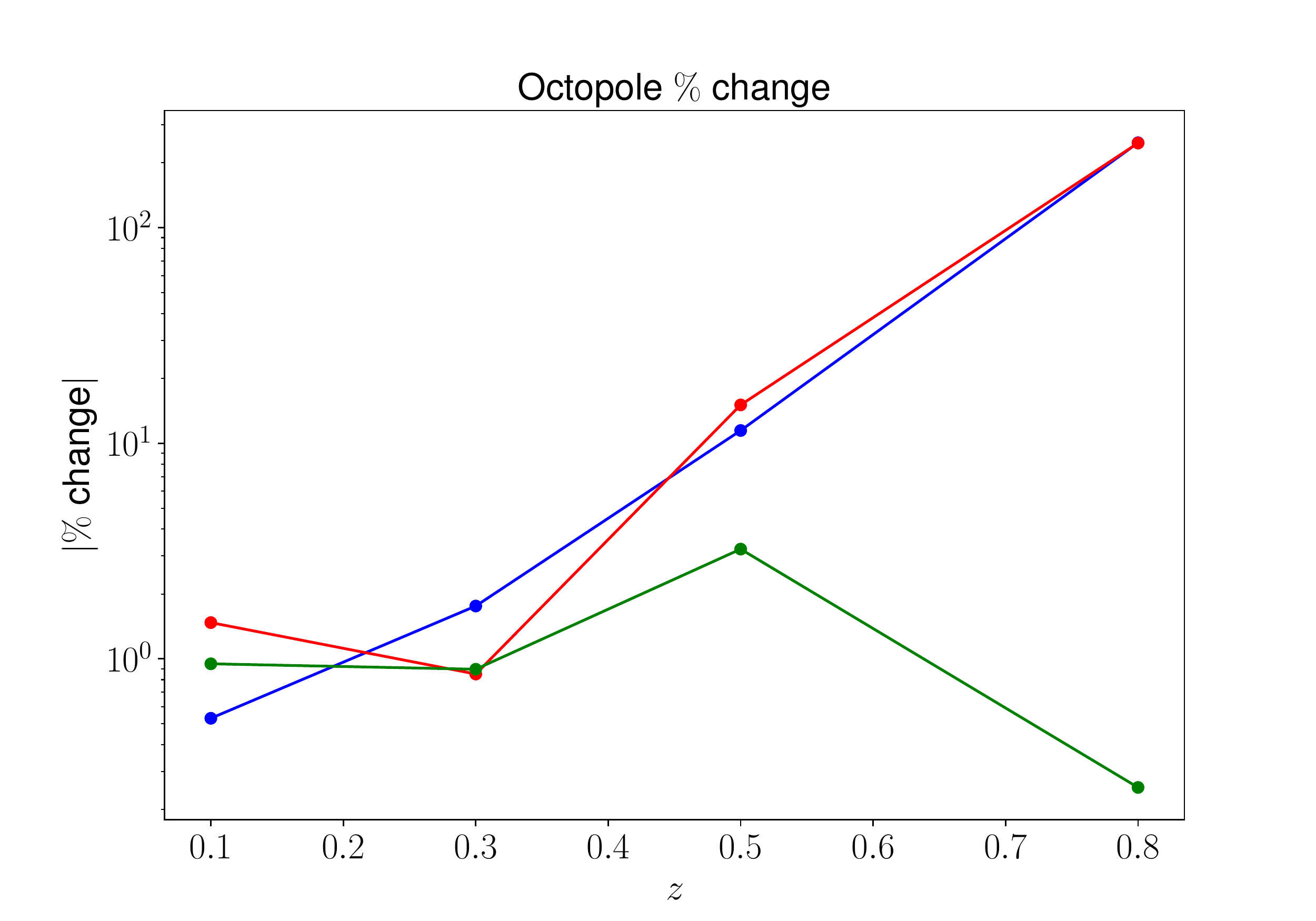}
   \caption{}
  \label{pc_oct}
\end{subfigure}
\caption{Absolute values of percentage change in the dipole and octopole under local screening and/or background modification as a function of redshift at a scale of $30$ Mpc. The percentage difference is defined relative to $\Lambda$CDM with the fiducial parameters of Table \ref{fid_cosmo}. The points, connected by straight lines, indicate the specific redshifts at which we perform the calculation. The legend is the same in the two panels.}
\label{pchange}
\end{figure*}

To summarise the effects of modified gravity, we show in Fig.~\ref{pchange} the variation of the percentage change with redshift in the range $0.1<z<0.8$, at a fiducial scale of $30$ Mpc. It is worth remembering that the octopole is always negative for our parameter choices, and the effect of screening is to reduce the dipole (hence the red line lying below the green in Fig.~\ref{pc_dip}). We see that the octopole is very sensitive to screening parameters at high redshift even with $s_B = 0.1$ (which determines the size of the octopole in our $\Lambda$CDM model). The change in dipole due to local screening is at most $\mathcal{O}(10\%)$ at high $z$, while the effect of background modification in non-screened modified gravity is roughly independent of redshift.

Current measurements of the dipole, for example from the LOWz and CMASS samples of the BOSS survey (e.g. \cite{Gaztanaga:2015jrs} fig. 7), have a signal to noise ratio (SN) of $<1$ in the dipole. In addition to the relativistic effects considered here, there are other terms that contribute to this and are in fact larger, including the wide angle and large angle effects. It is shown in \cite{Gaztanaga:2015jrs} that the current data from BOSS is only able to detect (at $\sim$2$\sigma$) the large angle effect, which is is a geometrical combination of the monopole and quadrupole and hence contains no additional physical information. Thus the detection of relativistic effects in the dipole will require data from future surveys such as DESI and SKA. As the modifications due to screening are $\mathcal{O}(10 \%)$ in the dipole at high redshift one would need SN$\gtrsim$10 to detect them, while the detection of local screening at low redshift would require an additional order of magnitude improvement (along with more precise modelling of the bias parameters). While it will be challenging for future surveys to detect the dipole at SN=100 (e.g. it is forecast in \cite{Gaztanaga:2015jrs} that DESI will reach S/N$\approx$7), it is shown in \cite{Hall:2016bmm} that a combination of SKA intensity maps and galaxy number counts can reach this sensitivity. Therefore it may be feasible for future surveys to probe both modified gravity in general, and screening in particular, through the dipole. To quantify the exact sensitivity to screening parameters, folding in uncertainties in growth rate and bias as well as cosmological variables such as $m_\nu, \Omega_m$ and $\sigma_8$, one could perform a Fisher forecast for next-generation experiments with cross-correlations of tracers. \emph{Any} measurement of the octopole may provide information on modified gravity.

Finally, it is worth noting that while we have computed the effect of the fifth force on the CF across a wide range of scales, in general we expect the force to be suppressed beyond the Compton wavelength of the field responsible for it. Given a Compton wavelength one can simply read off the change in the CF from Fig.~\ref{dip_oct} up to that scale, and assume a rapid transition back to $\Lambda$CDM beyond that. This applies only to the local fifth-force modification, however, and not the background. A completely self-consistent model would require solving the equation of motion for the field numerically given the mass distribution of the volume under consideration, while ensuring that the same fundamental theory parameters source both local screening and cosmic structure formation.

\section{Summary and future work}
\label{sec:conc}

We have calculated the effect of a screened fifth force on the parity-breaking correlation function (CF) obtained by cross-correlating two populations of galaxies that differ in properties relevant to their clustering. We show that this generates new terms in the dipole and octopole that are not present in $\Lambda$CDM, and in some cases neither in non-screened modified gravity theories. In particular, provided the magnification bias is the same between the two galaxy populations \emph{the octopole is only present under screening}. {In general this will not be true and thus the magnification bias will have to be corrected for.} Should the octopole be detected it could provide a relatively clean probe of a screened fifth force. 

The CF is also affected by cosmological modified gravity in the background, which alters the transfer function and growth rate. To model this we use a version of CAMB that has been modified \cite{Winther} to implement Hu-Sawicki $f(R)$, a canonical chameleon-screened theory. We find that Hu-Sawicki models with a fifth force on the scales in which we are interested ($10^{-6} \lesssim f_{R0} \lesssim 10^{-5}$) lead to deviations of $\mathcal{O}(10\%)$ in the dipole. To model the effect of screening we assume bright galaxies are completely screened in the Euler equation (i.e. feel GR), whereas faint galaxies feel the full fifth force. This is, of course, an approximation that is unlikely to be true in a cosmological setting, however it allows us to estimate the strength of the signal to screening.
For fifth-force strengths $\sim$10$-100\%$ of Newtonian gravity this leads to further changes in the dipole and octopole of a few percent at redshifts below $z = 0.3$, while for higher redshifts, e.g. $z = 0.8$, the dipole and octopole can change by $\sim 10-20\%$. We also show that uncertainties in the magnification and galaxy bias affect the dipole at the $\sim10\%$ level across the redshift range we consider, and therefore need to be known or modelled to this precision in order to extract information about screening.

Current state-of-the-art data from BOSS has signal to noise $<1$ in the dipole, and is not therefore able to detect these effects \cite{Gaztanaga:2015jrs}. However, upcoming DESI data will increase the signal to noise to $\sim$7, which will provide sensitivity to interesting modified gravity modifications to the cosmological background and local screening parameters at $z \approx 0.8$. The effect of screening at lower redshift may also be detectable using cross-correlations of multiple tracers, for example SKA intensity maps with galaxy number counts from Euclid or DESI \cite{Hall:2016bmm}. Another interesting prospect is the cross-correlation of galaxies with voids, which have negative bias and hence maximise the bias difference with the bright galaxy sample. This would however render the effect of screening further subdominant to the $\Lambda$CDM dipole. As the magnification and galaxy bias affect even multipoles of the CF as well, the best way to constrain the combination of bias parameters \emph{and} screening (which affects the odd multipoles) will be to do a joint inference on all multipoles simultaneously.

We have quantified modified gravity at the background level for the chameleon-screened Hu-Sawicki model of $f(R)$ only. Our analytic result in Eq.~\ref{eq:final}, however, holds for \emph{all} screened theories, including those that employ qualitatively different mechanisms such as Vainshtein. Even within the chameleon paradigm Hu-Sawicki $f(R)$ covers only a small fraction of the parameter space. In Appendix \ref{horndeski} we cast the general action for chameleon screening into Horndeski form. Therefore, a natural follow up would be to investigate the parity-breaking CF across the full chameleon (or more general) parameter space in the background, which could be achieved by implementing the general action in a modified gravity Boltzmann code such as HiClass \cite{hiclass}.

It is worth recalling here the assumptions that go into our analytic calculation of the correlation function, which may limit its scope:

\begin{itemize}

\item Our main assumption is that the effect of screening can be accounted for by a simple modification to the acceleration of the form Eq.~\ref{FF}, with constant $\delta G$. This amounts to the approximation that the two sets of galaxies we are correlating are either completely screened or completely unscreened. In reality galaxies may be partly screened, and the fifth force may be sourced by only a fraction of the matter that sources the Newtonian force. This generically reduces $\delta G$ below $1/3$ in $f(R)$, and hence reduces the magnitude of $\xi^{(\mathcal{F})}$. To account for this fully one would need to solve the scalar field equations numerically given the density field surrounding the galaxies (e.g. \cite{Shao}).

\item We have shown that the halo/galaxy bias is degenerate with the strength of screening in the dipole. It is therefore important for the bias to be known or modelled accurately in order to extract the screening signal. An order $\mathcal{O}(10\%)$ change in the bias at high redshifts\footnote{This also depends on mass.} (which is expected in $f(R)$ theory, e.g.~\cite{Schmidt:2008tn}) will lead to a change in the dipole of a similar magnitude. Thus we must be able to model the galaxy bias to percent level precision in modified gravity in order to isolate the screening signal. We have attempted to account for this by taking the bias, as a function of mass, from $f(R)$ simulations \cite{Arnold:2018nmv}, and accounting for the change in bias self-consistently as a function of redshift by using the using the $f(R)$ growth factor. The octopole however is independent of halo/galaxy bias, making it a particularly robust probe of screened fifth forces.

\item Assembly bias will generically lead to differences in the bias of the two galaxy populations at fixed halo mass, and more generally to deviations from the Sheth--Tormen prediction. It is also a function of modified gravity, as halos tend to form earlier in cosmologies with larger $f_{R0}$. A more precise understanding of this phenomenon will aid in distinguishing bias differences from effects related to screening.

\end{itemize}

The effect of the environment in which the galaxies form (represented by the trees in Fig.~\ref{windy}), could also contribute to the breaking of parity. For example, as bright galaxies are likely to have more dust around them than faint galaxies, a faint galaxy in front of a bright one will appear brighter than an identical one behind \cite{Fang:2011hc, Bonvin:2013ogt}.

The fully self-consistent way to account for all of these effects is to run numerical simulation of structure formation under screened modified gravity (e.g. \cite{Winther:2015wla, Arnold:2013nfr, Puchwein:2013lza, Arnold_1, Arnold_2}) and then compute the parity-breaking correlation function directly from the resultant galaxy density field. The advantage of our analytic approach is that it brings out the physical processes underlying such parity breaking and hence reveals novel features such as the presence of an octopole, which is not typically calculated in simulations. More detailed numerical modelling than we have performed will in any case likely be necessary to extract, validate and interpret a signal from data.

In summary, the effects of modified gravity in the parity-breaking CF could be probed in the near future with surveys such as DESI and SKA, with the best hope for constraining screening parameters coming from cross-correlation of tracers at multiple wavelengths. These analyses, augmented by numerical simulations, should be included in the fundamental physics agenda of large scale structure surveys in the coming decade.

\chapter{The effect on cosmological parameter estimation of a parameter dependent covariance matrix}\label{PDCM}

  
\section{Introduction}

 We are entering a phase in which cosmological galaxy surveys will have remarkable constraining power. This arises from the fact that they will cover large areas of the sky, to large depths and, in consequence, with a high number density \cite{2009arXiv0912.0201L, 2011arXiv1110.3193L,2013arXiv1305.5422S,Abbott:2017wau,Joudaki:2017zdt,Hikage:2018qbn}. As a result, the statistical power of these surveys will be astounding. To access their scientific potential, it will be necessary to control any systematic effects with exquisite accuracy. While there has been much focus on the instrumental and astrophysical limitations of any particular survey, care is now being given to the analysis pipeline and the approximations which are currently being used. In this {chapter} we focus on one such aspect: the covariance matrix in the likelihood analysis used for constraining cosmological parameters.

    Given a set of cosmological observations, for example in the form of a map, a galaxy catalogue, a correlation function or power spectrum, the process for estimating cosmological parameters is straightforward. One feeds the data into a likelihood function which assesses the likelihood of a given theory, given the data. With a judicious use of Bayes theorem, one explores a range of theoretical parameters (that can be cosmological but also can characterise underdetermined properties of the survey or astrophysical effects that may be present) to find the subspace of parameters which is most compatible with the data. In doing so, one also derives uncertainties on these parameters. It is therefore equally important for this procedure to produce unbiased parameter estimates and precise error bars.
    
    There are a number of effects that can plague cosmological parameter estimation. For a start there are errors due to the inaccurate modelling of simulated data, $x(p)$ and the corresponding, covariance matrix ${\bf \Sigma}$ estimated from them. These can come from not understanding the underlying model (e.g. baryonic effects in the matter power spectrum \cite{PhysRevD.87.043509, 0004-637X-672-1-19, vdb, Chisari:2018prw}, the precise nature of the galaxy-matter connection \cite{Mehta:2011xf, PhysRevD.81.083509} or the impact of intrinsic alignments \cite{Joachimi:2015mma, Troxel:2014dba}). Then there are errors due to incorrect assumptions about the statistics of the data and covariance matrix. For example there may be non-Gaussian corrections to ${\bf \Sigma}$ (i.e. corrections associated to the non-Gaussian nature of the fields being correlated). Then there are errors due to an incorrect model of the likelihood (e.g. assuming a Gaussian function of the data vector by default). Finally, even within the Gaussian approximation, there are errors that may arise from ignoring the parameter dependence of the covariance matrix. It is on this last source of errors that we will focus in this {chapter}.

    There is a growing literature on the accuracy of likelihood functions and its impact on parameter estimation. A key focus has been on how well one needs to know the covariance matrix and how Gaussian the likelihood function is. Covariance matrices are generally estimated from a large number of simulations; there is now a clear understanding of the errors associated with such a process and how to correct for them \cite{Morrison:2013tqa, Hall:2018umb, Dodelson:2013uaa, OConnell:2018oqr, Taylor:2014ota, Joachimi:2014hca, Paz:2015kwa, Pearson:2015gca, Petri:2016wlu} . A recent proposal in \cite{Heavens:2017efz} provides a novel data compression algorithm that reduces the number of simulations needed by a factor of 1000 to estimate the covariance matrix for Gaussianly distributed data. There have also been attempts at constructing better models, or approximations, for the likelihood function that include the fact that it may be non-Gaussian \cite{Sellentin:2015waz,Sellentin:2016psv}. 

    There have been attempts to analytically calculate the non-Gaussian contributions to the covariance matrix. The authors of \cite{Mohammed:2016sre} use a perturbation theory approach to compute the non-Gaussian contribution of the matter power spectrum to 1-loop and compare the results to simulations. A linear response approach to calculating the non-Gaussian contribution of the covariance matrix is presented in \cite{Barreira:2017sqa,Barreira:2017kxd}. In the case of large-scale structure observables like the cosmic shear power spectrum, it has been pointed out that that the dominant contribution to the non-Gaussian covariance is the term that comes from super-sample covariance (SSC) \cite{Barreira:2018jgd}. Separate-universe simulations have been used to evaluate the response functions for various observables such as weak lensing and the matter power spectrum \cite{Li:2015jsz,Wagner:2015gva}.

    In this {chapter}, our aim is to quantify the impact on parameter estimation of including (or not) the parameter dependence of the covariance matrix for upcoming photometric redshift surveys. Previous work on this topic has focused on the Fisher information of a parameter-dependent covariance in the two-point likelihood of Gaussian random fields \cite{Carron:2012pw}, and on the overall parameter dependence of the cosmic shear two-point covariance \cite{Eifler:2008gx}, where the authors quantified the effect on the likelihood contours of ${\bf \sigma}_8$ and $\Omega_m$, using both analytic and ray tracing simulations. Here we will focus on galaxy clustering and galaxy shear for a tomographic survey, and we will systematically analyse the effect of a parameter-dependent covariance matrix (including the effects of super-sample covariance) in terms of information content, parameter biases and final uncertainties.

    The {chapter} is structured as follows. In Section \ref{sec:theory} we present the various types of likelihoods which we will be studying and show how we can use the Fisher forecast formalism to find an estimate of the bias and errors of the model parameters; we consider a few simplified models to get an idea of what one should expect in the more general, realistic case. In Section \ref{sec:forecasts} we look at a realistic survey scenario, involving a combination of weak lensing and galaxy clustering and calculate the bias and compare the errors depending on the choice of likelihood.  In Section \ref{sec:summary} we discuss the implications of our results while in Appendix A we present a few, key, technical aspects of our calculation of the covariance matrix. 

\section{Approximating the likelihood}\label{sec:theory}

The aim of this section we quantify the impact of a parameter-dependent covariance on the bias and variance of parameter estimates. The impact of biases on the data vector has been quantified in e.g. \cite{Amara:2007as,Taburet:2008vg}, and the parameter dependence of the covariance has been explored in \cite{Eifler:2008gx, Reischke:2016ana}. In particular we calculate the bias in inferred parameters explicitly due to the parameter dependence of the covariance matrix, which is a key result that hasn't been calculated before as far as we know.

In general, for a Gaussian data vector ${\bf d}$, the likelihood is given by
    \begin{equation}\label{eq:multinorm}
      p({\bf d}|\vec{\theta})=\frac{\exp\left[-\frac{1}{2}\left({\bf d}-{\bf t}\right)^T\,{\bf \Sigma}^{-1}\left({\bf d}-{\bf t}\right)\right]}{\sqrt{{\rm det}\left(2\pi\,{\bf \Sigma}\right)}},
    \end{equation}
    where both the mean ${\bf t}$ and the covariance matrix ${\bf \Sigma}$ may in principle depend on the parameters ${\vec \theta}$.
    
    To begin with, we will start with the simple scenario, well known in the literature, in which the data is a single Gaussian sky map. It will be instructive to review some basic points about Gaussian likelihoods and the role covariance matrices play in them. In doing so, we will identify two approximations to the true likelihood function (for parameter-dependent and independent covariances) and how one might assess the difference in the parameter estimates they lead to.
 
    \subsection{Likelihoods and covariances: the case of a 2-dimensional Gaussian field}\label{ssec:theory.single_map}
      Consider the harmonic coefficients $a_{\ell m}$ of a single full-sky map. Under the assumption that they are statistically isotropic, their covariance matrix is diagonal, and given by the angular power spectrum $C_\ell$:
      \begin{equation}
        \left\langle a_{\ell m}\,a^*_{\ell' m'}\right\rangle\equiv C_\ell\,\delta_{\ell\ell'}\delta_{m m'}.
      \end{equation}
      If we further assume that the mean of the map is zero, and that it is Gaussianly distributed, its likelihood is completely determined by the power spectrum, and takes the form:
      \begin{equation}
        p(a_{\ell m}|{\vec \theta})=\prod_\ell\exp\left[-\frac{2\ell+1}{2}\frac{C^e_\ell}{C_\ell}\right]\left(2\pi C_\ell\right)^{-\frac{2\ell+1}{2}},
      \end{equation}
      where $C^e_\ell\equiv\sum_{m=-\ell}^\ell|a_{\ell m}|^2/(2\ell+1)$ is the estimated power spectrum. For simplicity, let us focus on a single multipole order $\ell$, and we will also use the goodness of fit $G=-2\log p$. Assuming flat priors on all parameters, the posterior distribution for $\vec{\theta}$ is simply given by this likelihood, and therefore
      \begin{equation}
        G(\vec{\theta}|a_{\ell m})=(2\ell+1)\left[\frac{C^e_\ell}{C_\ell}+\log(C_\ell)\right]+{\rm const.}
      \end{equation}

      It is often desirable to compress the data into a quadratic summary statistic, such as $C^e_\ell$. If we consider $C^e_\ell$ to be the data, then the likelihood is described by the Wishart distribution \cite{Hamimeche:2008ai} given by
      \begin{equation}
        G_{\rm exact}({\vec \theta}|C^e_\ell)  = (2\ell+1)\left[\frac{C^e_\ell}{C_\ell}+\log C_\ell-\frac{2\ell-1}{2\ell+1}\ln(C^e_\ell)\right] {\rm constant}, \label{eq:fulllike2D}
      \end{equation}
      where we have used the label ``exact'' to distinguish this distribution from the two approximations below. The extra term in this expression is a Jacobian/volume term that comes about due to the change of variables from $a^e_{\ell m}$ to $C^e_{\ell}$. This expression effectively tells us that $z=(2\ell+1){C}^e_\ell/C_\ell$ obeys a $\chi^2_{2\ell+1}$ distribution.
      
      For any distribution $G$, the Fisher information matrix is given by
      \begin{eqnarray}
        {\cal F}_{\mu\nu}=\frac{1}{2}\left\langle \partial_\mu\partial_\nu G\right\rangle,
      \end{eqnarray}
      where we have used the shorthand $\partial_\mu=\partial/\partial \theta^\mu \equiv \partial / \partial \vec{\theta}$. For the exact likelihood in Eq. \ref{eq:fulllike2D} we find
      \begin{eqnarray}
        {\cal F}^{\rm exact}_{\mu\nu}=\frac{2\ell+1}{2}\frac{\partial_\mu C_\ell}{C_\ell}\frac{\partial_\nu C_\ell}{C_\ell}
      \end{eqnarray}

      Until now, all our calculations have been exact in the limit of Gaussian, full-sky maps. Let us now consider a simplification that can be made if we assume that $C^e_\ell$ itself is Gaussianly distributed. This is a valid approximation in the large-$\ell$ limit, since the number of independent modes contributing to a given $C_\ell$ grows like $\ell$, and the Central Limit Theorem (CLT) eventually applies. {Of course, one has to determine what suffficiently Gaussian means - the Planck likelihood code uses $\ell < 29$ as a low $\ell$ cutoff for its likelihoods (for lower $\ell$'s it uses a pixel based likelihood), thus one can use this as a heuristic value for when Gaussianity is a good approximation.} In this approximation, and under the assumption that the covariance is parameter-independent, the likelihood takes the form
      \begin{eqnarray}
        G_{\rm PI}({\vec p}|C^e_\ell)\simeq {\tilde G}({\vec p}|C^e_\ell)=\frac{(C^e_\ell-C_\ell)^2}{{\bf \Sigma}^f_\ell},
      \end{eqnarray}
      where the label PI identifies this distribution with the case of a ``parameter-independent'' covariance matrix, and where we have defined the power spectrum covariance ${\bf \Sigma}^f_\ell$
      \begin{equation}
        \langle C^e_\ell C^e_{\ell'}\rangle-\langle C^e_\ell\rangle\langle C^e_{\ell'}\rangle\equiv{\bf \Sigma}^f_\ell \delta_{\ell\ell'}.
      \end{equation}
      Wick's theorem can be used to relate ${\bf \Sigma}^f_\ell$ to the underlying fiducial power spectrum $C^f_\ell$, finding the well-known result
      \begin{equation}
        {\bf \Sigma}^f_\ell=\frac{2C^{f\,2}_\ell}{2\ell+1}
      \end{equation}
      where $C^f_\ell$ is  a fiducial power spectrum which is {\it independent} of the cosmological parameters. One can show that this approximation can be obtained by Taylor expanding Eq \ref{eq:fulllike2D} in $\xi=(C_\ell-C^e_\ell)/C^e_\ell$ to get 
      \begin{equation}
        {\tilde G}({\vec p}|C^e_\ell) = (2\ell+1)\left[\frac{1}{2}\xi^2+\frac{2}{2\ell+1}\log C^e_\ell\right] +{\rm constant}
      \end{equation}
      
      and assuming $C^e_\ell\simeq C^f_\ell$.
      
      This approximate likelihood is similar to the original exact likelihood in two ways. First, their Fisher matrices coincide if $C^f_\ell$ is the true underlying power spectrum
      \begin{equation}
        {\cal F}^{\rm PI}_{\mu\nu}=\frac{2\ell+1}{2}\frac{\partial_\mu C_\ell}{C^f_\ell}\frac{\partial_\nu C_\ell}{C^f_\ell}.
      \end{equation}
      Secondly, in the simplest case, where our only free parameter is the amplitude of the power spectrum itself (i.e. $\theta\equiv C_\ell$), both likelihoods yield unbiased maximum likelihood estimators for this parameter:
      \begin{equation}
        \hat{C}^{\rm exact}_\ell=\hat{C}^{\rm PI}_\ell=C^e_\ell,\hspace{12pt}\langle C^e_\ell\rangle=C^f_\ell
      \end{equation}

      Consider now a different approach and take the large $\ell$ limit of Eq (\ref{eq:fulllike2D}). Applying the CLT we have that $\xi$ obeys a Normal distribution, ${\cal N}[0,\ell+1/2]$, i.e.:
      \begin{eqnarray}
        G_{\rm PD}({\vec \theta}|C_\ell)=\log({\bf \Sigma}_\ell)+\frac{(C_\ell-C^e_\ell)^2}{{\bf \Sigma}_\ell} \label{gaussparam}
      \end{eqnarray}
      where now the covariance matrix
      \begin{eqnarray}
        {\bf \Sigma}_\ell=\frac{2}{2\ell+1}C^{2}_\ell \label{2DGcov}
      \end{eqnarray}
      depends on $\theta$ (and hence the label ``PD''). Accounting for this parameter dependence, the Fisher matrix for this distribution is
      \begin{equation}
        {\cal F}^{\rm PD}_{\mu\nu}=\frac{2\ell+5}{2}\frac{\partial_\mu C_\ell}{C_\ell}\frac{\partial_\nu C_\ell}{C_\ell},
      \end{equation}
      and the maximum likelihood estimator for the power spectra is
      \begin{equation}
        \hat{C}^{\rm PD}_\ell = C^e_\ell\frac{2\ell+1}{4}\left[\sqrt{1+\frac{8}{2\ell+1}}-1\right]\\\label{eq:bias_single}
                             \simeq C^e_\ell\left[1-\frac{2}{2\ell+1}\right],
      \end{equation}
      where in the second line we have kept only the first-order term of the large-$\ell$ expansion of the first line. Therefore, $G^{\rm PD}$ reduces to $G^{\rm exact}$ in the large-$\ell$ limit.
      
      We thus see that in this particular case, while assuming a parameter {\it independent} covariance matrix may lead to unbiased parameter estimates, it necessarily leads to a mis-estimate of the parameter uncertainties (unless the chosen fiducial covariance ${\bf \Sigma}^f_\ell$ is the underlying true one). A parameter {\it dependent} covariance matrix (assuming a Gaussian approximation for the likelihood) leads to both biased parameter estimates and a misestimate of the uncertainties but there is a well defined limit in which it recovers both correctly and this limit is set by the number of modes being considered in the analysis. This point was made by \cite{Carron:2012pw}, who also show that using a parameter-dependent covariance when approximating the two-point likelihood of Gaussian random fields is formally incorrect.

      The question then arises: how important, in practice, is the parameter dependence in the process of parameter estimation from current and future data sets? In this {chapter}, we go beyond the simple analysis of 2D full-sky Gaussian fields presented here to consider the case of tomographic analyses with non-Gaussian contributions to the covariance matrix.

    \subsection{Likelihoods and covariances: the general case}\label{ssec:theory.general}
      The aim of this section is to derive approximate expressions for the parameter uncertainties and biases associated to the parameter dependence of the covariance matrix. Using the identity $\log{\rm det}{\sf M}={\rm Tr}\log{\sf M}$, let us start by writing the goodness of fit for the generic multi-variate Gaussian distribution (Eq. \ref{eq:multinorm}) as
      \begin{eqnarray}
        G_{\rm gen}({\bf d}|\vec{\theta})= \left({\bf d}-{\bf t}\right)^T {\bf \Sigma}^{-1} \left({\bf d}-{\bf t}\right)+{\rm Tr}\left(\log{\bf \Sigma}\right) \label{eq:ggen}.
      \end{eqnarray}
      Let us now define three sets of parameters
      \begin{enumerate}
        \item $\vec{\theta}_{\rm T}$: the true parameters that generate the data.
        \item $\vec{\theta}_{\rm PD}$: the maximum-likelihood parameters found by minimizing $G_{\rm PD}$, the version of $G_{\rm gen}$ in which the covariance matrix \emph{depends} on $\vec{\theta}$.
        \item $\vec{\theta}_{\rm PI}$: the maximum-likelihood parameters found by minimizing $G_{\rm PI}$, the version of $G_{\rm gen}$ in which the covariance matrix \emph{does not depend} on $\vec{\theta}$.
      \end{enumerate}
      $\vec{\theta}_{\rm T}$ generate the data in the sense that $\langle {\bf d}\rangle={\bf t}(\vec{\theta}_{\rm T})$\footnote{We note that this assumes that the theory we have, in this case $\Lambda CDM$, is the \emph{true} theory of the universe. This of course may not be true however answering that question is tangential to the goal of this {chapter} and thus we do not address this further.}  and
      \begin{equation}
        \left\langle [{\bf d}-{\bf t}(\vec{\theta}_{\rm T})]\,[{\bf d}-{\bf t}(\vec{\theta}_{\rm T})]^T\right\rangle={\bf \Sigma}(\vec{\theta}_{\rm T}).
      \end{equation}
      We will also assume that the PI likelihood uses the true covariance as the fiducial one, i.e. ${\bf \Sigma}^f={\bf \Sigma}(\vec{\theta}_{\rm T})$\footnote{Note that this choice only simplifies the calculations, but does not affect our results. Choosing any other fiducial point ($\vec{\theta}_{\rm PI}$, or $\vec{\theta}_{\rm PD}$ for example) only leads to second-order corrections.}. This will allow us to isolate the impact of the parameter dependence from the systematics effects associated with using an inaccurate covariance matrix. For brevity, we will often use the shorthand ${\bf \Sigma}_{\rm T}\equiv{\bf \Sigma}(\vec{\theta}_{\rm T})$. Note that, at this stage, we have not made any statements about the validity of either $G_{\rm PI}$ or $G_{\rm PD}$\footnote{Note however, that as pointed out in \cite{Carron:2012pw} and in Section \ref{ssec:theory.single_map} there is a clear distinction between both for Gaussian random fields, and using a parameter-dependent covariance produces fictitious information.}, and we will focus only on the comparison of their associated parameter uncertainties and on their relative bias.
    
      By definition
      \begin{equation}
        \left.\frac{\partial G_{\rm PD}}{\partial\vec{\theta}}\right|_{\vec{\theta}_{\rm PD}}=0.
      \end{equation}
      Writing $\vec{\theta}_{\rm PD}=\vec{\theta}_{\rm PI}+\Delta\vec{\theta}$ and $G_{\rm PD}=G_{\rm PI}+\Delta G$ we can expand the equation above to linear order, to find
      \begin{equation}
        \left.\frac{\partial^2 G_{\rm PI}}{\partial\vec{\theta}\partial\vec{\theta}}\right|_{\vec{\theta}_{\rm PI}}\Delta\vec{\theta}+\left.\frac{\partial\Delta G}{\partial\vec{\theta}}\right|_{\vec{\theta}_{\rm PI}}=0,
      \end{equation}
      where $\partial/\partial\vec{\theta}$ and $\partial^2/\partial{\vec{\theta}}\partial{\vec{\theta}}$ is shorthand for the parameter gradient and Hessian matrix respectively. Taking the expectation value of the equation above, the parameter bias can be estimated in this approximation as
      \begin{equation}\label{eq:bias_general}
        \Delta\vec{\theta}=-\frac{1}{2}\,\mathcal{F}^{-1}_{\rm PI}\left\langle\left.\frac{\partial \Delta G}{\partial\vec{\theta}}\right|_{\vec{\theta}_{\rm PI}}\right\rangle,
      \end{equation}
      where $\mathcal{F}_{\rm PI}$ is the Fisher matrix for the parameter-independent case, given simply by
      \begin{equation}\label{eq:fisher_PI}
        \mathcal{F}_{{\rm PI},\mu\nu}\equiv\partial_\mu{\bf t}^T\,{\bf \Sigma}^{-1}\,\partial_\nu{\bf t}.
      \end{equation}
      To evaluate $\Delta\vec{\theta}$ for the distribution in Eq. \ref{eq:ggen}, let us start by writing $\Delta G$ to first order in $\Delta{\bf \Sigma}\equiv{\bf \Sigma}-{\bf \Sigma}_{\rm T}$
      \begin{equation}\nonumber
        \Delta G= -\left({\bf d}-{\bf t}\right)^T{\bf \Sigma}_{\rm T}^{-1}\,\Delta{\bf \Sigma}\,{\bf \Sigma}_{\rm T}^{-1}\left({\bf d}-{\bf t}\right)+ {\rm Tr}\left({\bf \Sigma}_{\rm T}^{-1}\Delta{\bf \Sigma}\right).
      \end{equation}
      Differentiating with respect to $\vec{\theta}$ we obtain
      
      \begin{equation}\nonumber
        \partial_\mu\Delta G= 2\partial_\mu{\bf t}^T{\bf \Sigma}_{\rm T}^{-1}\,\Delta{\bf \Sigma}\,{\bf \Sigma}_{\rm T}^{-1}({\bf d}-{\bf t}) +{\rm Tr}\left[{\bf \Sigma}^{-1}_{\rm T}\partial_\mu{\bf \Sigma}\left({\bf \Sigma}_{\rm T}^{-1}({\bf d}-{\bf t})({\bf d}-{\bf t})^T-1\right)\right],  \label{eq:dgpre}
      \end{equation}
      where in the second line we have used the fact that $\partial_\mu\Delta{\bf \Sigma}\equiv\partial_\mu{\bf \Sigma}$. Before continuing, it is important to note that, according to Eq. \ref{eq:bias_general}, we must evaluate this expression at the parameter-independent best fit $\vec{\theta}_{\rm PI}$. This best fit satisfies
      \begin{equation}
        \left.\frac{\partial G_{\rm PI}}{\partial\theta_\mu}\right|_{\vec{\theta}_{\rm PI}}=-2\partial_\mu{\bf t}^T{\bf \Sigma}_{\rm T}^{-1}({\bf d}-{\bf t})=0.
      \end{equation}
      Now let us define $\delta{\bf d}\equiv{\bf d}-{\bf t}(\vec{\theta}_{\rm T})$ and $\delta\vec{\theta}\equiv\vec{\theta}_{\rm PI}-\vec{\theta}_{\rm T}$. To linear order in $\delta\vec{\theta}$, the equation above reads
      \begin{equation}
        \partial_\mu{\bf t}^T{\bf \Sigma}_{\rm T}^{-1}(\delta{\bf d}-\partial_\nu{\bf t}\,\delta\theta_\nu)=0,
      \end{equation}
      and we can solve for $\delta\vec{\theta}$ as
      \begin{equation}
        \delta\vec{\theta}=\mathcal{F}_{\rm PI}^{-1}\frac{\partial{\bf t}^T}{\partial\vec{\theta}}{\bf \Sigma}^{-1}_{\rm T}\delta{\bf d}
      \end{equation}
      or, equivalently
      \begin{equation}
        {\bf d}-{\bf t}(\vec{\theta}_{\rm PI})=\left[1-\frac{\partial{\bf t}}{\partial\vec{\theta}}\mathcal{F}_{\rm PI}^{-1}\frac{\partial{\bf t}^T}{\partial\vec{\theta}}{\bf \Sigma}^{-1}_{\rm T}\right]\delta{\bf d}.
      \end{equation}
      Substituting this in Eq. \ref{eq:dgpre}, making use of the fact that $\langle\delta{\bf d}\rangle=0$ and $\langle\delta{\bf d}\,\delta{\bf d}^T\rangle\equiv{\bf \Sigma}_{\rm T}$, and after a little bit of algebra, we obtain
      \begin{equation}
        \left\langle\left.\frac{\partial\Delta G}{\partial\theta_\mu}\right|_{\vec{\theta}_{\rm PI}}\right\rangle=-{\rm Tr}\left[{\bf \Sigma}^{-1}_{\rm T}\frac{\partial{\bf \Sigma}}{\partial\theta_\mu}{\bf \Sigma}_{\rm T}^{-1}\frac{\partial{\bf t}}{\partial\vec{\theta}}\mathcal{F}_{\rm PI}^{-1}\frac{\partial{\bf t}^T}{\partial\vec{\theta}}\right].
      \end{equation}
      With index notation then, the parameter bias is
      \begin{equation}\label{eq:fisher_bias}
        \Delta\theta_\mu=-\frac{1}{2}\mathcal{F}^{-1}_{{\rm PI},\mu\nu}\,\mathcal{F}^{-1}_{{\rm PI},\rho\tau}\,\partial_\rho{\bf t}^T\,{\bf \Sigma}^{-1}\,\partial_\nu{\bf \Sigma}\,{\bf \Sigma}^{-1}\,\partial_\tau{\bf t}.
      \end{equation}
      This is a key result of our {chapter}.
      
      The effect of the parameter-dependent covariance on the final parameter uncertainties can be taken into account simply by accounting for this parameter dependence when deriving the Fisher matrix. Such a calculation  yields \cite{Tegmark:1996bz}
      \begin{equation}\label{eq:fisher_PD}
        \mathcal{F}_{{\rm PD},\mu\nu}=\mathcal{F}_{{\rm PI},\mu\nu}+\frac{1}{2}{\rm Tr}\left[\partial_\mu{\bf \Sigma}\,{\bf \Sigma}^{-1}\,\partial_\nu{\bf \Sigma}\,{\bf \Sigma}^{-1}\right].
      \end{equation}
      The parameter uncertainties can then be estimated by inverting the Fisher matrix in either case.

    \subsection{Large-scale structure likelihoods}\label{ssec:theory.lss}
      Let us now specialise the discussion in the previous section to the case of a data vector made up of the collection of auto- and cross-power spectra between different sky maps, each labelled by a roman index (e.g. $a^i_{\ell m}$ labels the harmonic coefficients of the $i$-th). In general, each sky map will correspond to an arbitrary projected astrophysical field, and the discussion below covers this general case. However, here we will only consider maps from two types of tracers, the galaxy number overdensity $\delta_g$ and the cosmic shear $\gamma$, each measured in a given tomographic redshift bin. The cross-power spectrum between two fields is
      \begin{equation}
        \left\langle a^i_{\ell m}a^{j*}_{\ell'm'}\right\rangle=C^{ij}_\ell\delta_{\ell\ell'}\delta_{mm'}.
      \end{equation}
      $C^{ij}_\ell$ can be related in general to the matter power spectrum in the Limber approximation \cite{Simon:2006gm} through \cite{Bartelmann:1999yn}
      \begin{equation}\label{eq:pspec_signal}
        C^{ij}_\ell=\int_0^\infty d\chi\,\frac{q^i(\chi)q^j(\chi)}{\chi^2}P\left(z(\chi),k=\frac{\ell+1/2}{\chi}\right),
      \end{equation}
      where the window functions for $\delta_g$ and $\gamma$ are
      \begin{eqnarray}
        q^{\delta,i}(\chi) &=& b(\chi)\frac{dn^i}{dz}(\chi)\,H(\chi), \label{eq:q_delta}\\
        q^{\gamma,i}(\chi) &=& \frac{3H_0^2\Omega_M}{2\,a(\chi)}\int_\chi^{\chi_H} d\chi'\,\frac{dn^i}{dz}(\chi')\,\frac{\chi'-\chi}{\chi\chi'}. \label{eq:q_lensing}
      \end{eqnarray}
      Here $b(\chi)$ is the linear galaxy bias, $dn^i/dz$ is the redshift distribution of lens or source galaxies respectively in the $i$-th redshift bin, normalised to 1 when integrated over the full redshift range, and $\chi_H$ is the distance to the horizon.
    
      Since our data vector is an ordered list of cross-power spectra between different pairs of maps $(ij)$ at different scales $\ell$, the covariance matrix depends on 6 indices
      \begin{equation}
        {\bf \Sigma}^{ij,\ell}_{mn,\ell'}\equiv\left\langle C^{ij}_\ell\,C^{mn}_{\ell'}\right\rangle.
      \end{equation}	
      In order to account for the non-Gaussian nature of the late-times large-scale structure, we estimate the covariance matrix as a sum of both the Gaussian and super-sample covariance (SSC) contributions \cite{Takada:2013bfn,Mohammed:2016sre}
      \begin{equation}\label{eq:covar_lss}
        {\bf \Sigma}={\bf \Sigma}_{\rm G}+{\bf \Sigma}_{\rm SSC}.
      \end{equation}
      We neglect other non-Gaussian corrections from the connected trispectrum, which are known to be subdominant with respect to the SSC contribution, at least for cosmic shear studies \cite{Barreira:2018jgd}.
    
      Note that it is important to account for non-Gaussian contributions coupling different scales in order to address the relevance of the parameter dependence of the covariance matrix. As discussed in Section \ref{ssec:theory.single_map}, for a single Gaussian map, the relative bias between the PI and PD likelihoods drops like $\sim1/\ell$ for a single multipole order (see Eq. \ref{eq:bias_single}), and therefore as $\sim1/\ell_{\rm max}^2$ for a maximum multipole $\ell_{\rm max}$, becoming negligible for a sufficiently large number of modes. This will in general also be true for an arbitrary number of maps, since the same arguments hold for each of the independent eigenmaps that diagonalize $C^{ij}_\ell$. Non-Gaussian contributions to ${\bf \Sigma}$ will couple different scales, effectively reducing the number of independent modes, and therefore may enhance the impact of the parameter-dependent covariance.
    
      The Gaussian contribution to the covariance matrix can be estimated as \cite{Knox:1995dq}
      \begin{equation}\label{eq:covar_G}
        \left({\bf \Sigma}_{\rm G}\right)^{ij,\ell}_{mn,\ell'}=\delta_{\ell\ell'}\,\frac{C^{im}_\ell C^{jn}_\ell+C^{in}_\ell C^{jm}_\ell}{f_{\rm sky}(2\ell+1)}.
      \end{equation}
      Note that we account for an incomplete sky coverage by scaling the full-sky covariance by $1/f_{\rm sky}$ to account for the reduced number of available modes. This is not correct in detail, since measurements on a cut sky induce mode correlations, but it is a good enough approximation for forecasting \cite{Crocce:2010qi}. It is also important to note that the power spectra entering Eq. \ref{eq:covar_G} must contain both signal and noise contributions. The former are given by Eq. \ref{eq:pspec_signal}, while the latter are
      \begin{figure*}
        \centering
        \includegraphics[width = 0.49\textwidth]{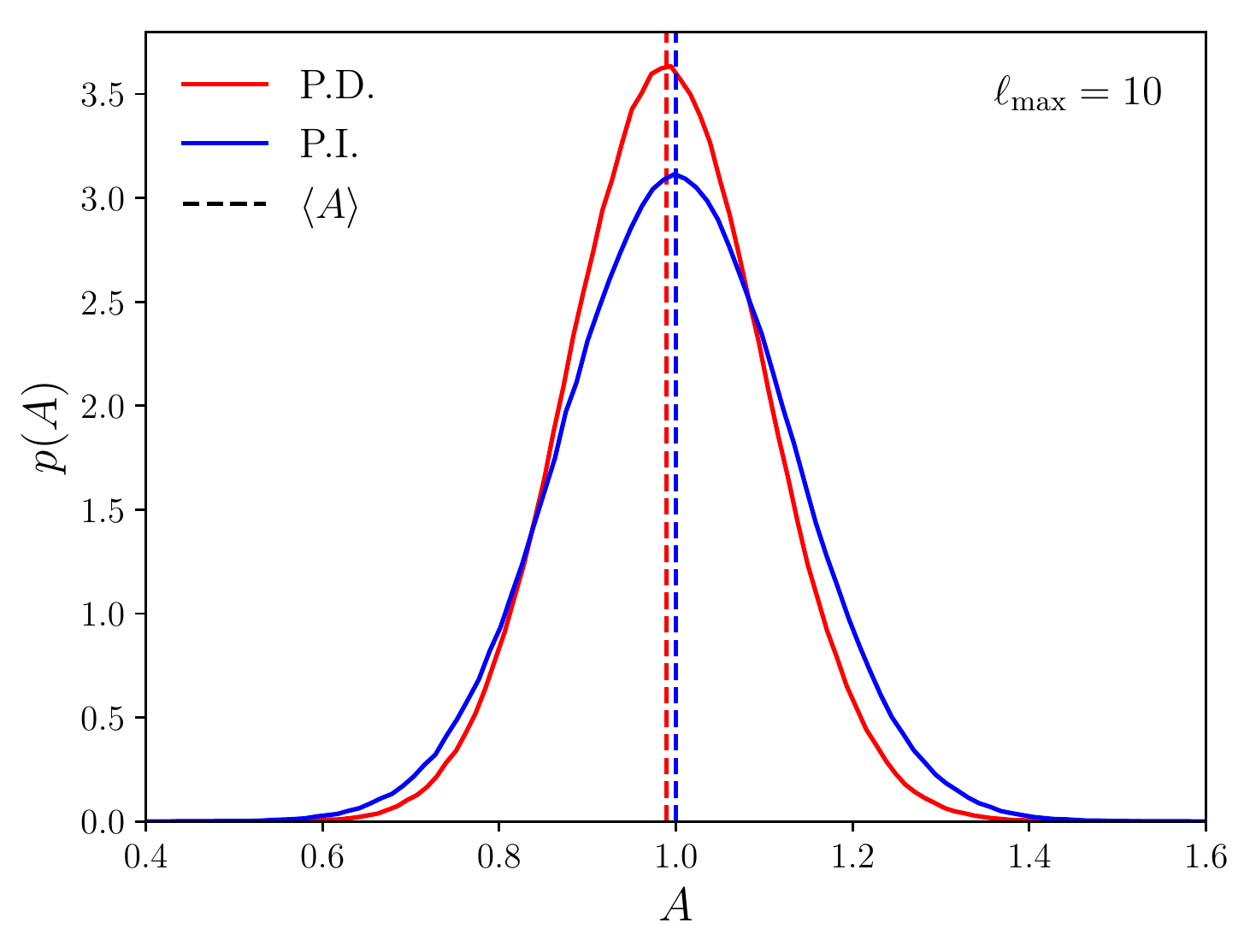}
        \includegraphics[width = 0.49\textwidth]{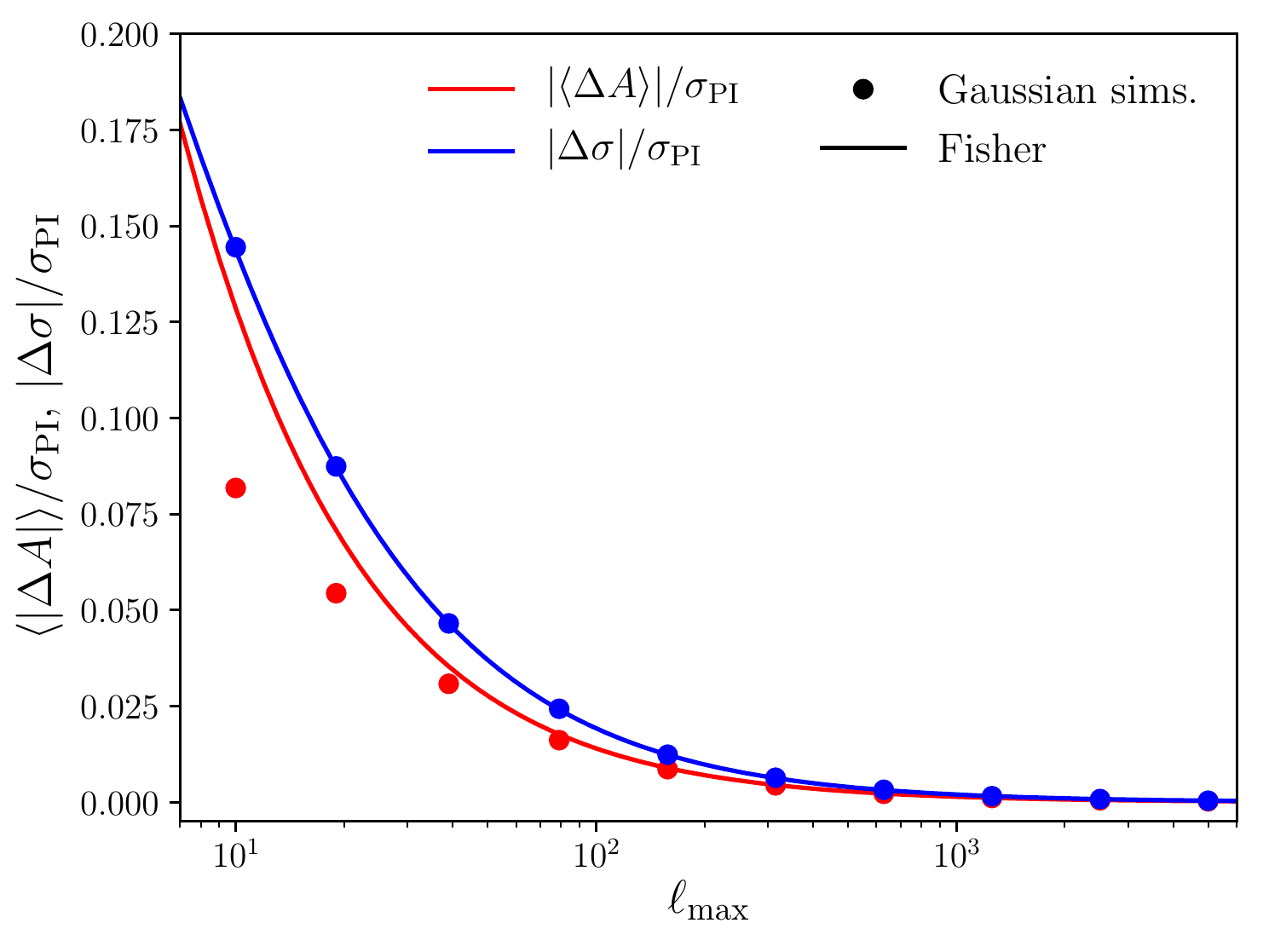}
        \caption{{\sl Left:} distribution of maximum-likelihood estimates of an overall power spectrum amplitude parameter in the case of parameter-dependent (red) and parameter-independent (blue) covariances. The parameter dependence of the covariance matrix causes a downwards relative bias with respect to the parameter-independent case, as well as a mild shrinkage of the parameter uncertainties. Results are shown for a small maximum multipole ($\ell_{\rm max}=10$ in order to highlight the effects of the parameter dependence. {\sl Right:} relative bias (red) and shift in uncertainties (blue) due to the parameter dependence of the covariance matrices as a function of the maximum multipole $\ell_{\rm max}$ (note that both quantities are normalized by the parameter-independent error bars). The red circles show the values found by directly evaluating the exact likelihood for $10^6$ simulations, while the solid lines show the values found in our Fisher matrix approximation. The Fisher approach is able to reproduce the exact result with good accuracy for both quantities, slightly over-estimating the bias for small $\ell_{\rm max}$.}\label{fig:simple}
      \end{figure*}
      \begin{equation}\label{eq:pspec_noise}
        N^{\delta,ij}_\ell=\frac{1}{n_g^i}\delta_{ij},\hspace{12pt}N^{\gamma,ij}_\ell=\frac{{\bf \Sigma}_\gamma^2}{n_g^i}\delta_{ij},
      \end{equation}
      where $n_g^i$ is the mean angular number density of objects in the $i$-th redshift bin, and ${\bf \Sigma}_\gamma=0.28$ is the intrinsic ellipticity dispersion per component.

      We compute the SSC contribution to the covariance matrix as described in appendix \ref{app:SSC}. However, ignoring this contribution for the moment, we can use the simple dependence of the Gaussian covariance with $f_{\rm sky}$ to study the importance of a parameter-dependent covariance matrix as a function survey properties. Simple inspection of Eq. \ref{eq:bias_general} shows that, since $\mathcal{F}_{\rm PI}\propto f_{\rm sky}$, the parameter bias scales like $\propto f_{\rm sky}^{-1}$. On the other hand, the effect on parameter uncertainties is
      \begin{eqnarray}
        {\bf \Sigma}_\mu & = &\sqrt{\left(\mathcal{F}^{-1}\right)_{\mu\mu}} \\
                  &=& \sqrt{\left(\mathcal{F}_{\rm PI}+\Delta\mathcal{F}\right)^{-1}_{\mu\mu}}\\
                  &\simeq &{\bf \Sigma}_\mu^{\rm PI}-\frac{1}{2}\frac{\left(\mathcal{F}^{-1}_{\rm PI}\,\Delta\mathcal{F}\,\mathcal{F}^{-1}_{\rm PI}\right)_{\mu\mu}}{{\bf \Sigma}_\mu^{\rm PI}},
      \end{eqnarray}
      where $\Delta\mathcal{F}$ is the second term in Eq. \ref{eq:fisher_PD}, and in the last line we have expanded to first order in this parameter. Since $\Delta\mathcal{F}$ does not scale with $f_{\rm sky}$, the correction to the parameter errors (given by the second term above) is $\propto f_{\rm sky}^{-3/2}$. Thus, in general \emph{the relevance of a parameter-dependent covariance matrix will decrease with the surveyed area}. This results holds also in the presence of non-Gaussian contributions to the covariance. The SSC term arises from the non-linear coupling induced between different modes by the presence of super-survey, long-wavelength modes. Its impact will therefore increase for smaller sky areas, and therefore we can expect it to induce a slightly steeper dependence on $f_{\rm sky}$ than the purely Gaussian case.
      
      The expected behaviour with the smallest scale included in the analysis $\ell_{\rm max}$ is also similar: increasing the number of modes reduces the relative impact of the parameter-dependent covariance. The non-Gaussian terms can somewhat modify this behaviour, due to mode coupling. However, the relative impact of the SSC contribution will also decrease for larger survey areas, or for more noise-dominated datasets.

    \subsection{Analytic example: the power spectrum amplitude}\label{ssec:theory.analytic}
      Before we study the importance of the parameter dependence in the covariance matrix in detail for the case of tomographic large-scale structure datasets, let us examine a particularly simpler scenario: a single sky map, and a single parameter quantifying the overall amplitude of the power spectrum (for instance, this would correspond to the case of $A_s$ or ${\bf \Sigma}_8$ for linear power spectra). In this case, we can calculate the impact of the parameter-dependent covariance exactly, and therefore it allows us to quantify the validity of the Fisher approximations derived in the previous section. Let us label the amplitude parameter $A$, such that our model for the measured angular power spectrum is related to the a fiducial one as
      \begin{equation}
        C_\ell = A C^f_\ell.
      \end{equation}
      We will assume a fiducial value of $A=1$, and consider a purely Gaussian covariance with no noise
      \begin{equation}
        {\bf \Sigma}_{\ell\ell'} = \frac{(A\,C^f_\ell)^2}{n_\ell}\,\delta_{\ell\ell'},
      \end{equation}
      where $n_\ell=\ell+1/2$.
      
      On the one hand, the parameter-dependent and independent likelihoods are (up to irrelevant constants) given by 
      \begin{eqnarray}
        G_{\rm PD}(A) &=&\sum_{\ell=0}^{\ell_{\rm max}}n_\ell\left(\frac{r_\ell}{A}-1\right)^2+2(\ell_{\rm max}+1)\ln(A)\\
        G_{\rm PI}(A)&=&\sum_{\ell=0}^{\ell_{\rm max}}n_\ell\left(r_\ell-A\right)^2,
      \end{eqnarray}
      where $r_\ell=C^d_\ell/C^f_\ell$, and $C^d_\ell$ is the measured power spectrum. The maximum-likelihood solutions for $A$ for a given realisation of the data then are
      \begin{eqnarray}
        \hat{A}_{\rm PD}&=&\frac{S_1}{2(\ell_{\rm max}+1)}\left[\sqrt{1+4\frac{(\ell_{\rm max}+1)S_2}{(S_1)^2}}-1\right]\\
        \hat{A}_{\rm PI}&=&\frac{S_1}{S_0}
      \end{eqnarray}
      where $S_n\equiv\sum_\ell n_\ell r_\ell^n$. On the other hand, our Fisher predictions for the parameter bias and the parameter-independent and parameter-dependent errors (Eqs. \ref{eq:fisher_bias} and \ref{eq:fisher_PD}) are
      \begin{eqnarray}\label{eq:bias_simple}
        \langle\Delta A\rangle&=&-(S_0)^{-1}=-\frac{2}{(\ell_{\rm max}+1)^2},\\\label{eq:sigmaPI_simple}
        {\bf \Sigma}_{\rm PI}(A)&=& (S_0)^{-1/2}=\frac{\sqrt{2}}{\ell_{\rm max}+1},\\\label{eq:sigmaPD_simple}
        {\bf \Sigma}_{\rm PD}(A)&=& {\bf \Sigma}_{\rm PI}(A)\left[1+\frac{4}{(\ell_{\rm max}+1)}\right]^{-1/2}
      \end{eqnarray}
      where $\ell_{\rm max}$ is the maximum multipole included in the analysis.
      
      To validate these results, we generate $10^6$ random Gaussian realisations of $r_\ell$ with standard deviation ${\bf \Sigma}_\ell=n_\ell^{-1}$, and compute $\hat{A}_{\rm PD}$ and $\hat{A}_{\rm PI}$ for each of them. We then evaluate the mean and standard deviation of both quantities and compare them with the approximate results in Eqs. \ref{eq:bias_simple}, \ref{eq:sigmaPI_simple} and \ref{eq:sigmaPD_simple}. The results of this validation are shown in Figure \ref{fig:simple}. The left panel shows an example of the distributions of $\hat{A}_{\rm PI}$ (blue) and $\hat{A}_{\rm PD}$ (red) for the case of $\ell_{\rm max}=10$. As discussed above, the small number of modes in this case highlights the relevance of the parameter-dependent covariance, which produces a noticeable relative bias on $A$ and a decrease in its uncertainty \cite{Hamimeche:2008ai}. The right panel then shows the relative parameter bias and shift in uncertainties normalised by ${\bf \Sigma}_{\rm PI}$ as a function of $\ell_{\rm max}$ for the simulated likelihoods (circles) and for our Fisher predictions (solid lines). The Fisher approximation is remarkably good, and remains accurate at the $10\%$ level even for low values of $\ell_{\rm max}\sim30$.

  \section{Forecasts}\label{sec:forecasts}
    We apply the results discussed in the previous section to the case of imaging surveys targeting a joint analysis of galaxy clustering and weak lensing. We start by describing the assumptions we use to quantify the expected signal and noise of these surveys and then present our results regarding the relevance of the parameter dependence of the covariance matrix.
    
    \subsection{Survey specifications}\label{ssec:forecasts.surveys}
      \begin{figure}
        \centering 
        \includegraphics[width = 0.47\textwidth]{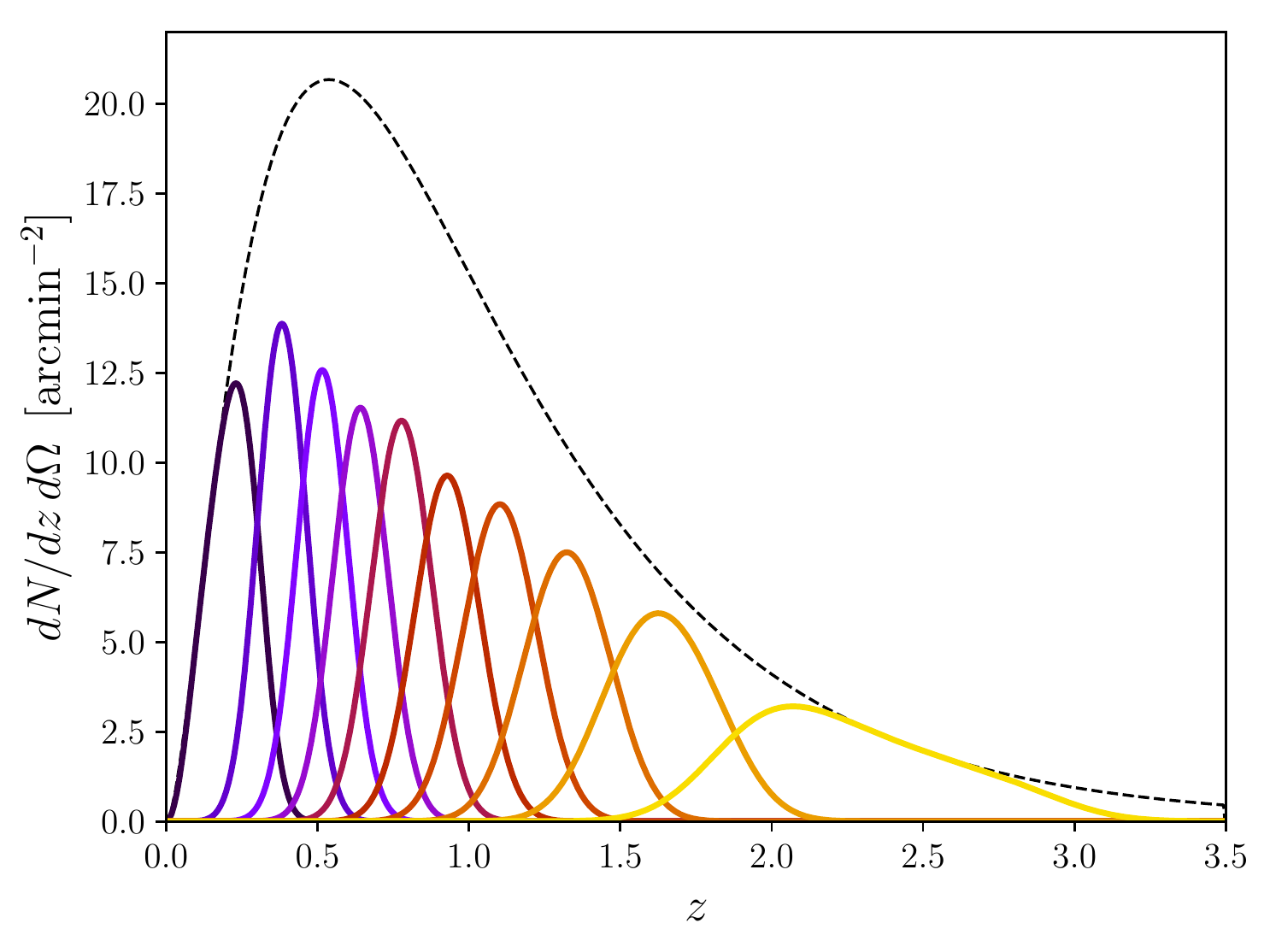}
        \caption{Redshift distribution for the overall sample assumed in this analysis (dashed black) and for each of the 10 redshift bins (solid coloured lines). For simplicity we assume the same redshift distribution for both lenses and sources.} \label{fig:dndz}
      \end{figure}
      We assume an LSST-like survey. For simplicity we assume the same redshift distribution for the clustering and lensing samples with a redshift dependence \cite{Mandelbaum:2018ouv}
      \begin{equation}
        \frac{dN}{dz}\propto z^2\exp\left[-\left(\frac{z}{z_0}\right)^\alpha\right],\hspace{12pt}
        (z_0,\alpha)=(0.12,0.7),
      \end{equation}
      and a total number density $n_g=27\,{\rm arcmin}^{-2}$. We split the total sample into 10 photometric redshift bins with approximately equal number of galaxies in each bin. The redshift distributions for each bin are calculated assuming Gaussian photometric redshift uncertainties with a standard deviation
      ${\bf \Sigma}_z=0.05\,(1+z)$. These are shown in Figure \ref{fig:dndz}.

      We produce forecasts for a set of 4 cosmological parameters: the fractional matter density $\Omega_M$, the matter power spectrum amplitude ${\bf \Sigma}_8$ and equation of state parameters $w_0$ and $w_a$. The forecasts presented below are marginalized over 6 nuisance parameters corresponding to the values of the linear galaxy bias in 6 redshift nodes $z=(0.25,0.75,1.25,1.85,2.60,3.25)$. We assume a linear dependence with redshift between these nodes, and our fiducial bias values correspond to $b(z)=1+0.84\,z$ \cite{Weinberg:2002rm}. Since the aim of this work is to explore the impact of parameter-dependent covariances, and not to produce forecast of the expected cosmological constraints, we do not consider any other sources of systematic uncertainty, such as intrinsic alignments, multiplicative shape biases or photometric redshift systematics.
      
      Finally, our fiducial forecasts assume a sky fraction $f_{\rm sky}=0.4$, as expected for LSST, and a constant scale cut $\ell_{\rm max}=3000$ for all redshift bins. A more realistic choice of scale cuts would be motivated by modelling uncertainties, by removing all scales smaller than the physical scale of non-linearities $k_{\rm NL}$, necessarily in a redshift-dependent way. By not removing these scales at low redshifts ($\ell_{\rm NL}(z=0.5)\simeq k_{\rm NL}\chi(z=0.5)\simeq400$ for $k_{\rm NL}=0.3\,h\,{\rm Mpc}^{-1}$), our fiducial choice emphasizes the role of super-sample covariance, potentially highlighting the effects of parameter dependence in the covariance. At the same time, we have seen that these effects become less relevant as we increase the number of independent modes, and therefore we will also present results as a function of $\ell_{\rm max}$ and $f_{\rm sky}$.
      
    \subsection{Results}\label{ssec:forecasts.results}
      \begin{figure*}
        \centering 
        \includegraphics[scale = 0.45]{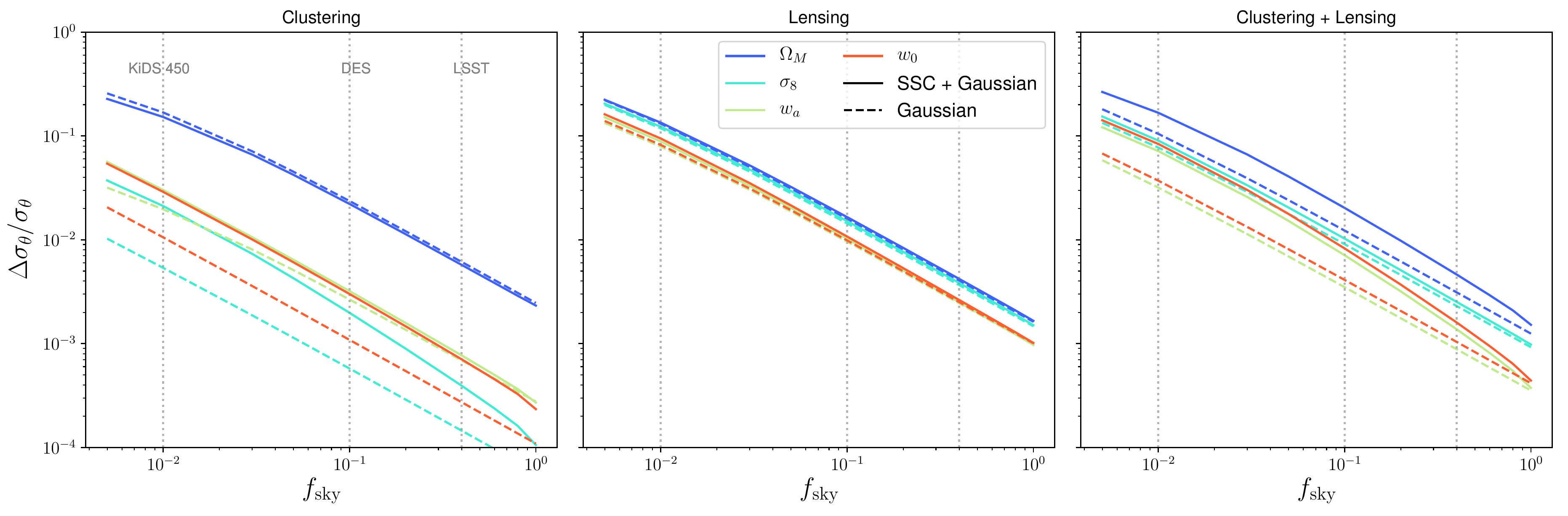}
        \caption{Relative difference in the uncertainties on late-time cosmological parameters due to the parameter dependence of the covariance matrix as a function of survey area for a fiducial $\ell_{\rm max}=3000$.  From left to right, we show results for galaxy clustering, weak lensing and for the combination of both probes. Results are shown for Gaussian covariance matrices (dashed lines) and including the super-sample covariance (solid lines). As argued in Section \ref{sec:theory}, the importance of the parameter dependence grows towards smaller $f_{\rm sky}$, as the number of modes available decreases. Nevertheless, the effect is never larger than $\sim20\%$ of the statistical errors for the smallest sky fractions ($f_{\rm sky}=0.005$), and is below $1\%$ for LSST-like areas ($f_{\rm sky}\simeq0.4$). } \label{fig:rel_error_fsky}
      \end{figure*}
      
      We compute the Fisher matrices for parameter-dependent and independent covariances as well as the bias due to the parameter dependence as described in Section \ref{ssec:theory.general}, using Eqs. \ref{eq:fisher_PI}, \ref{eq:fisher_PD} and \ref{eq:fisher_bias}. Here ${\bf t}$ is the vector of all possible cross-power spectra between two tracers ($\delta_g$ or $\gamma$) in any pair of redshift bins, calculated using Eq. \ref{eq:multinorm} (with noise power spectra given by Eq. \ref{eq:pspec_noise}), and ${\bf \Sigma}$ is the covariance matrix of this data vector, calculated as in Eq. \ref{eq:covar_lss}. To compute all angular power spectra we use the Core Cosmology Library\footnote{\url{https://github.com/LSSTDESC/CCL}} \cite{ccl}, which we also modified to provide estimates of the super-sample covariance as described in Appendix \ref{app:SSC}. Note that, we do not compute power spectra and covariance matrices for all integer values of $\ell$. Instead, we use 15 logarithmically-spaced bandpowers between $\ell=20$ and $\ell=3000$.
      
      We report our results in terms of the relative parameter biases and the relative error differences
      \begin{equation}
        \frac{\Delta\theta_\mu}{{\bf \Sigma}_{{\rm PI},\mu}},\hspace{12pt}
        \frac{\Delta{\bf \Sigma}_\mu}{{\bf \Sigma}_{{\rm PI},\mu}}\equiv\frac{{\bf \Sigma}_{{\rm PI},\mu}-{\bf \Sigma}_{{\rm PD},\mu}}{{\bf \Sigma}_{{\rm PI},\mu}},
      \end{equation}
      where ${\bf \Sigma}_{\rm PI}$ and ${\bf \Sigma}_{\rm PD}$ are computed from the inverse of the corresponding Fisher matrices. Results are reported for the 4 cosmological parameters $(\Omega_M,{\bf \Sigma}_8,w_0,w_a)$. Since the aim of this {chapter} is to study the relevance of the parameter dependence in the covariance matrix, and not to produce cosmological forecasts for future surveys, we do not report absolute errors on these parameters.
      \begin{figure*}
        \centering 
        \includegraphics[scale = 0.45]{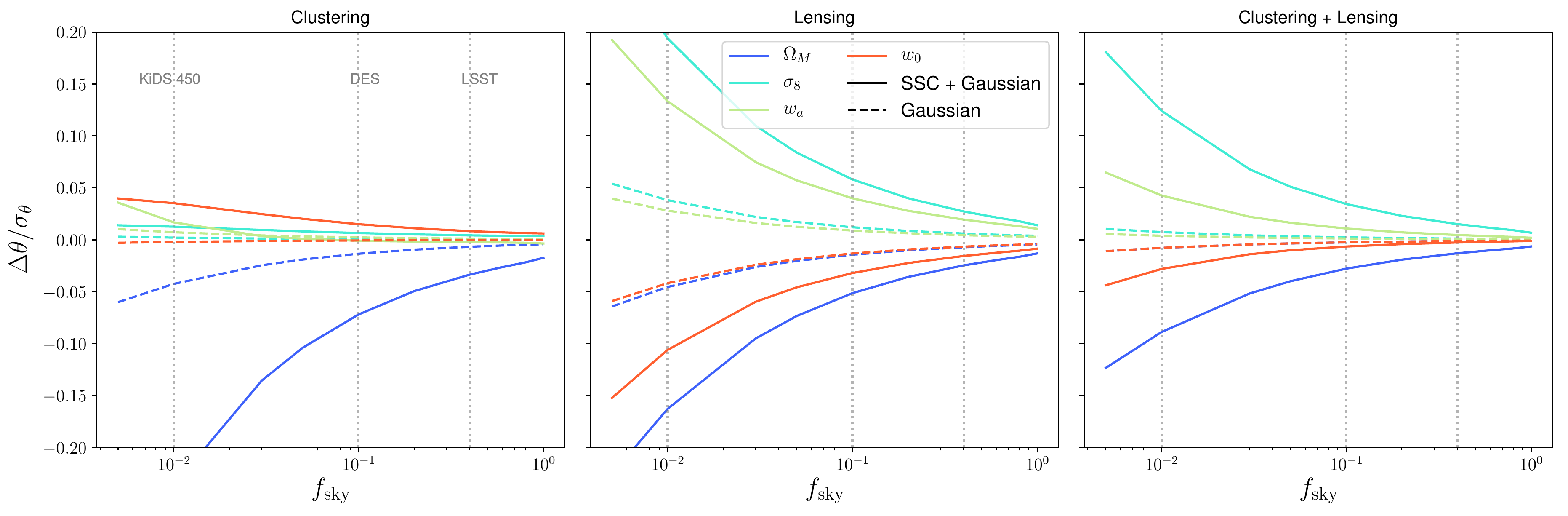}
        \caption{Same as Fig. \ref{fig:rel_error_fsky} for the relative parameter bias. The effects of a parameter-dependent covariance are always smaller than $30\%$ of the statistical uncertainties for the smallest sky areas ($\sim0.5\%$ of the sky) and become percent-level for LSST-like areas ($f_{\rm sky}\simeq0.4$).} \label{fig:rel_bias_fsky}
      \end{figure*}
      \begin{figure*}
        \centering 
        \includegraphics[width = 0.47\textwidth]{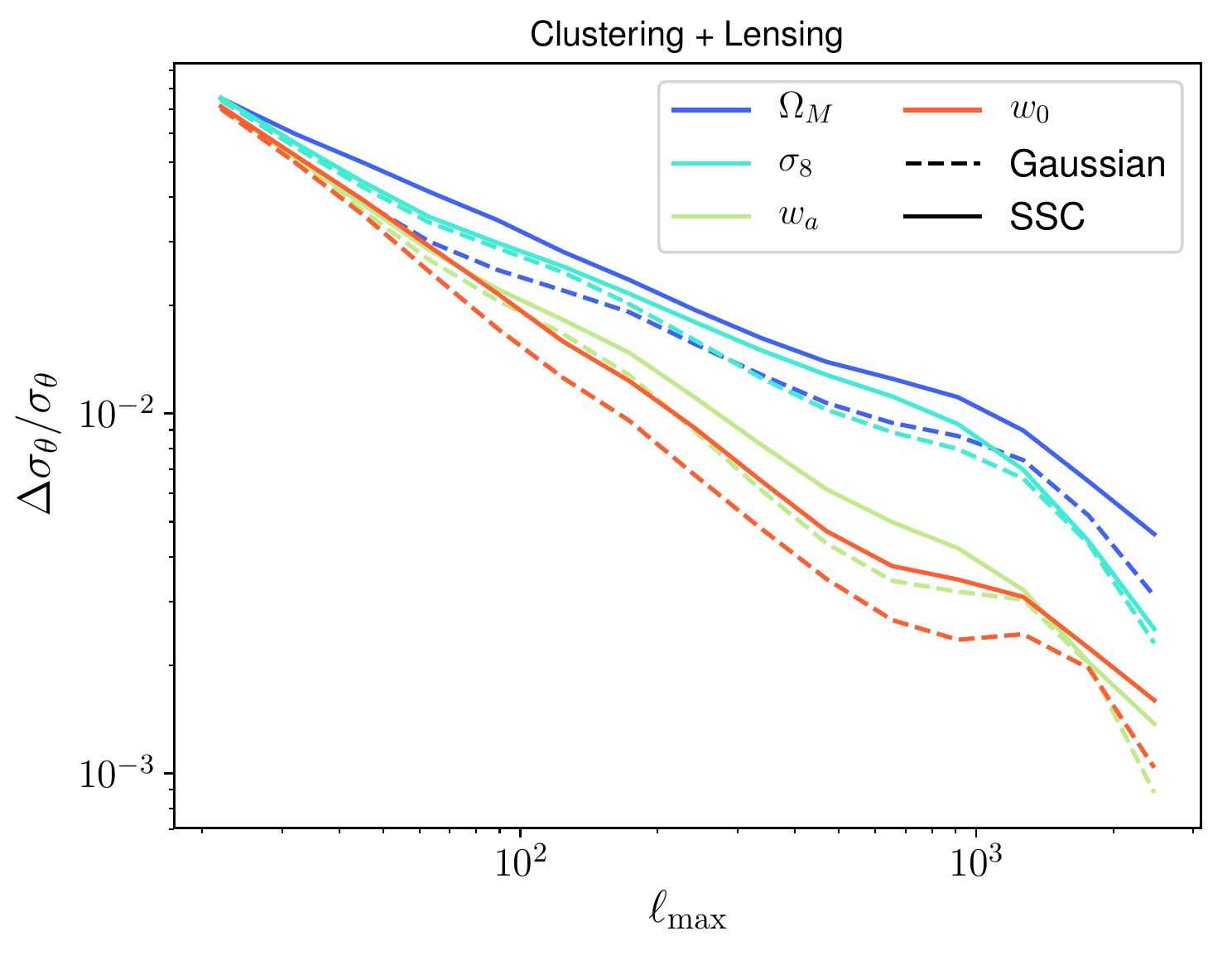}
        \includegraphics[width = 0.47\textwidth]{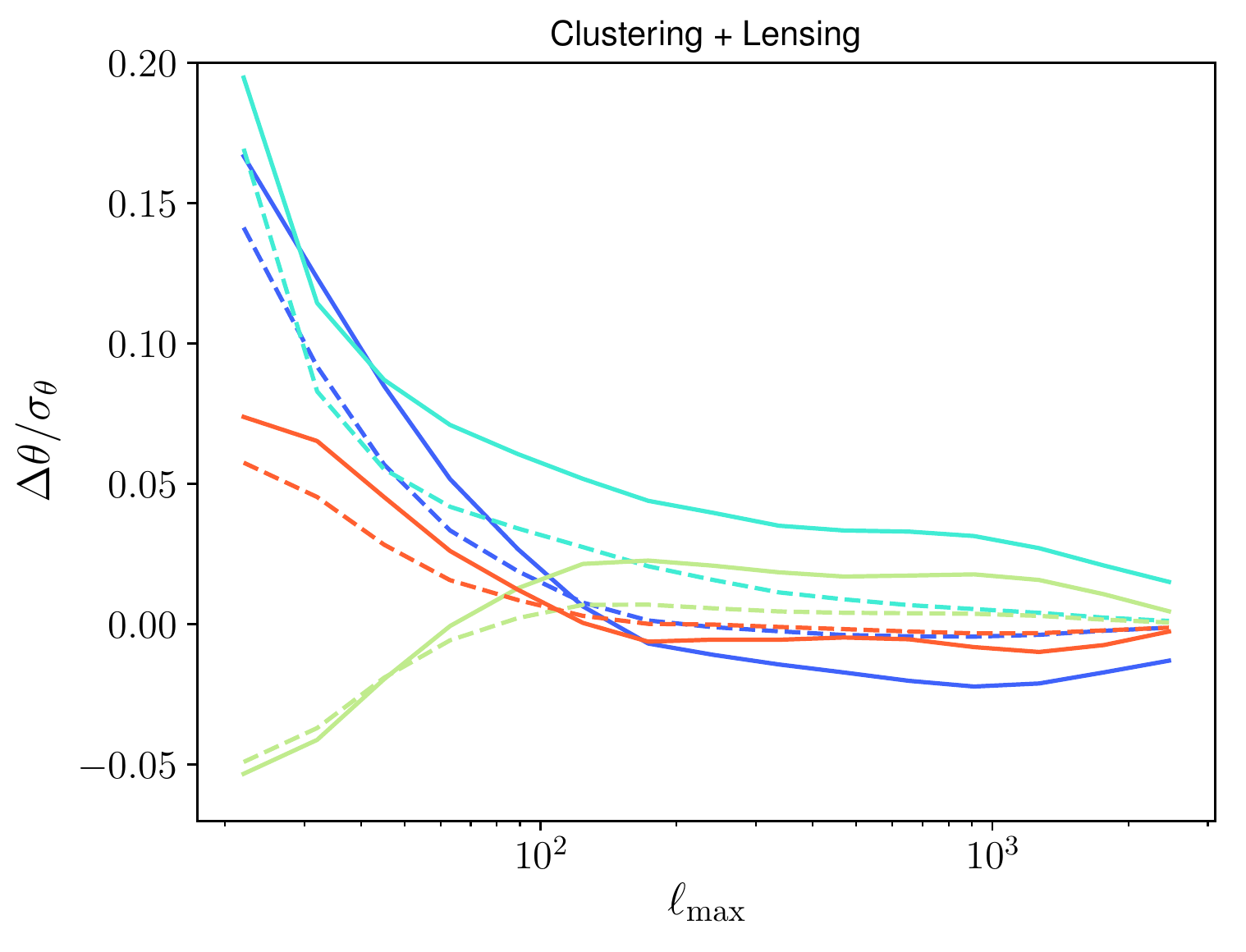}
        \caption{{\sl Left:} relative error difference due to a parameter-dependent covariance for the four cosmological parameters studied here as a function of the maximum multipole $\ell_{\rm max}$ included in the analysis for a fiducial sky fraction $f_{\rm sky}=0.4$. {\sl Right:} same as the left panel for the relative parameter bias. As expected, the impact of the parameter dependence grows as we reduce the number of independent modes used in the analysis. In all cases, the effect is smaller than $\sim20\%$ of the statistical uncertainties, even for $\ell_{\rm max}\sim20$, and becomes percent-level for realistic values ($\ell_{\rm max}\sim1000$).} \label{fig:pd_lmax}
      \end{figure*}
      
      Figure \ref{fig:rel_error_fsky} shows the relative difference in the statistical uncertainties as a function of sky area for a fiducial $\ell_{\rm max}=3000$. Results are shown for galaxy clustering, weak lensing and for the combination of both. We also show the impact of the super-sample contribution to the covariance matrix by displaying results with (solid) and without it (dashed). In all cases, the information contained in the covariance matrix is very small, and would only lead to a $\sim20\%$ suppression of the uncertainties for the smallest sky areas ($\sim0.5\%$ of the sky), corresponding to e.g. CFHTLens \cite{Erben:2012zw} or the first data release of HSC \cite{Hikage:2018qbn}. The effect becomes even less important for larger sky areas, as more independent modes become available and concentrate most of the information on the power spectrum. For LSST-like areas ($f_{\rm sky}\simeq0.4$) the effect is, at most, of the order of $1\%$ of the statistical uncertainties.
      
      The same results are shown in Figure \ref{fig:rel_bias_fsky} for the relative parameter biases, and similar conclusions hold. The effects are always smaller than $\sim0.3{\bf \Sigma}$ on small survey areas, and get suppressed to a percent fraction of the statistical uncertainties for larger areas. While these effects are small, it is interesting to note that they are enhanced by considering a more accurate model of the covariance matrix that includes the SSC term. Thus we see once again the effect of the super-sample covariance inducing a statistical coupling between modes which then reduces the effective number of independent degrees of freedom and and enhances the relative information content of the covariance matrix.

      For a fixed LSST-like sky fraction ($f_{\rm sky}=0.4$), Figure \ref{fig:pd_lmax} shows the impact of the parameter dependence of the covariance matrix as a function of the maximum multipole $\ell_{\rm max}$ included in the analysis. Results are shown for the error difference (left panel) and biases (right panel), and we also show the impact of the super-sample covariance term. Again, as expected based on our discussion in Section \ref{sec:theory}, the information content of the covariance matrix decreases as more independent modes are included in the analysis. The impact, both on the uncertainties and on the parameter biases, is smaller than $\sim20\%$ of the overall statistical error budget (and this only for $\ell_{\rm max}=20$). For more realistic scale cuts ($\ell_{\rm max}\sim1000$), the effect is suppressed to the level of ${\cal O}(1\%)$. 

  \section{Summary}\label{sec:summary}
    In the era of precision cosmology it is becoming increasingly important to understand the systematic and statistical errors that occur during parameter estimation. In this {chapter} we have analysed the effect of using a likelihood with a parameter dependent covariance matrix on the inference of cosmological parameters. The computation of covariance matrices for large-scale structure surveys is a numerically complex problem for which multiple approaches have been proposed in the literature. Therefore, assessing the need to estimate the covariance at every point of a given likelihood evaluation, rather than estimating it only once for a set of parameters sufficiently close to the maximum likelihood, is important for future surveys, where covariance estimation will become more computationally demanding.
    
    We have focused our analysis on the two main aspects of parameter inference: final statistical uncertainties and parameter biases. For multivariate Gaussian likelihoods, we have quantified the impact on the statistical errors using a standard Fisher matrix approach that accounts for the parameter dependence of both the mean and covariance (Eq. \ref{eq:fisher_PD}). We have also derived an expression to estimate the expected parameter bias by expanding the likelihood around its maximum. The resulting expression (Eq. \ref{eq:fisher_bias}) is easy to calculate and has not been presented before to our knowledge. We have evaluated the accuracy of this approximation by comparing its predictions with the analytical solutions available for a simplified case involving a single amplitude parameter and a single sky map (Section \ref{ssec:theory.analytic}). This exercise shows that our approximate estimates are indeed accurate as long as the true parameter bias is small, and that, if anything, our approximations will slightly overestimate this bias. The methods used here are therefore perfectly applicable to the case we study, and a more computationally expensive approach involving a full evaluation of the likelihood through Monte-Carlo methods is unlikely to yield different results.
    
    We have then evaluated the parameter shifts and error differences for the particular case of a large-scale structure experiment targeting the joint measurement of galaxy clustering and cosmic shear. The Fisher information for the parameter dependent covariance matrix, in contrast to the parameter independent case, does not increase with the survey area. This can be seen in Eq \ref{eq:fisher_PD} where all factors of $f_{\rm sky}$ in the covariance matrix cancel out for the second term. The associated parameter bias, on the other hand, is roughly inversely proportional to $f_{\rm sky}$ (see Section \ref{ssec:theory.lss}). This means there will always be a regime in which the parameter dependence becomes important, the question is whether it will be important for any practical value of $f_{\rm sky}$. Our study of simple single-map cases (Sections \ref{ssec:theory.single_map} and \ref{ssec:theory.analytic}) have shown that the impact of a parameter-dependent covariance decreases with multipole order (e.g. Eqs. \ref{eq:bias_simple} and \ref{eq:sigmaPD_simple}), and therefore the generic message is that \emph{the relative information content in a parameter-dependent covariance decreases as more independent modes are included in the survey}. To account for mode-coupling induced by non-linear evolution of the matter overdensities, we have included the super-sample contribution to the covariance matrix, which has been determined to be the most relevant contribution in the range of scales considered here \cite{Mohammed:2016sre}. We note that we have not considered additional nuisance parameters such as shifts in the photometric redshifts, the impact of baryons or intrinsic alignments. We have also only taken into account the super-sample covariance contribution to the Gaussian covariance matrix, neglecting all other parts of the connected trispectrum. While this has been determined to be sufficiently accurate for lensing observables \cite{Barreira:2018jgd} large-scale structure studies exploring the deeply non-linear regime will require a more careful treatment. Nevertheless, we don not expect any of these effects to change our results significantly. Furthermore it is conceivable that in situations where the parameter space is extended, to include more exotic models of dark energy for example, the likelihood may become Non-Gaussian and then the parameter dependence of the covariance may need to be taken into account. 
    
	Our findings, summarised in Figures \ref{fig:rel_error_fsky}, \ref{fig:rel_bias_fsky} and \ref{fig:pd_lmax} show that, for any current and future surveys, the parameter dependence of the covariance matrix can be safely ignored, since it only leads to changes in the statistical errors and maximum-likelihood parameters that are $\lesssim1\%$ of the statistical uncertainties. For surveys targeting very small sky fractions ($f_{\rm sky}<1\%$) or very large scales ($\ell\lesssim20$), the impact of the parameter dependence becomes more important, but the impact is at most at the $\sim0.2{\bf \Sigma}$ level, both in biases and uncertainties. Note that the these effects are sufficiently small that it strongly justifies the use of the Fisher formalism to undertake this estimate, but a more thorough analysis of these effects in cases where information content of the covariance is significant would require the use of a full likelihood exploration. Since we can expect systematic and numerical uncertainties to be at least as important, it is safe to say that \emph{the parameter dependence of the covariance matrix can be ignored in all practical cases}.

\chapter{Concluding remarks}

\section{Summary of thesis}

In this thesis we have questioned some of the basic assumptions of the $\Lambda$CDM model. 
We summarise the results and conclusions from all the chapters now. 

\begin{itemize}

\item We started by analysing the initial conditions for the perturbations that describe the formation of structure in the universe. 
In general there can be three types of perturbations; scalars, vectors and tensor perturbations. 
In chapter \ref{ICOU} we focus on analysing the scalar perturbations. 
The general scalar initial conditions can be adiabatic or isocurvature and we focused on the most general set of modes in adiabatic perturbations. 
Within adiabatic modes, we analyse the decaying modes which are usually ignored. 
This is because if inflation is assumed to happen before the radiation dominated era, then the decaying mode is suppressed by $\mathcal{O}(e^{\#_\text{efolds}})$. 
We remain agnostic to the source of perturbations and allow both modes to exist.
By parameterising the primordial power spectrum as a set of bins with independent amplitudes, we use the fisher information matrix to constrain the amplitudes in each of the modes. 
The best constrained modes, for both decaying and growing initial conditions, are in the range $k \sim 10^{-3} - 10^{-1}$ Mpc$^{-1}$. 
This is a new breakthrough as it shows the decaying modes are not highly constrained on subhorizon scales where we understand the physics well and therefore should not be neglected a priori in cosmological analysis.
The decaying mode is also highly constrained on very large scales of order $k \sim 10^{-4}$ Mpc$^{-1}$ and smaller. 
This is because the decaying mode has a $\frac{1}{k \tau}$ factor which diverges at early times and on superhorizon scales. 
Interpreting these constraints is not easy as there are modes of the order of $10^{-5}$ Mpc$^{-1}$ which will never enter our universe and thus it is not clear how they will effect the physics of inside our patch of the universe. We speculate on potential mechanisms for how this might happen, for instance in a \emph{separate universe approach}, the effect of large scale scalar modes on our local patch of the universe might be change the background cosmological parameters such as the matter density in our universe. 

\item Most models of the early universe predict that there should be primordial tensor perturbations. 
These leave an imprint into the temperature and polarisation anisotropies of the CMB. 
Like the scalar perturbations, the equation of motion for the tensor perturbations is also a second order differential equation and thus has two orthogonal solutions. 
We analyse the effect the decaying mode tensor perturbation has on the CMB anisotropies in chapter \ref{ICOU_tensors}. 
By performing the same analysis as we did for the scalar modes we find that the best constrained modes for growing and decaying initial conditions are on scales of $k \sim 5 \times 10^{-4}$ Mpc$^{-1}$ and $k \sim 7 \times 10^{-3}$ Mpc$^{-	1}$. 
The physical reason for this is that the tensor modes generate polarisation in the CMB photons during reionisation and recombination (due to the scattering of the photons from the local quadrupole of the gravitational waves) and these are the scales that are best constrained. 
This also means that the decaying and growing modes are indistinguishable if they are sourced at horizon crossing at the time of recombination. 
The only way to distinguish the decaying mode from the growing mode is by measuring the $B$ mode signal at reionisation. 
This qualitatively new understanding is crucial for our understanding of fundamental physics, for instance we will need to able to distinguish between growing and decaying modes to determine whether or not inflation happened. 
The interpretation of large scale tensor modes is also problematic. 
We speculate that the physical effect of large scale tensor modes might be to give rise to an anisotropic background metric to our universe, for instance in the form of one of the Bianchi models. 
This would mean the observables of large scale tensor modes might actually be the presence of shear modes. 
We leave the mathematical formulation of this equivalence for future works. 

\item The next assumption of the $\Lambda$CDM model we challenge is the theory of gravity itself. In chapter \ref{PBCF} we present a novel test of gravity on galactic and cosmological scales. Cross-correlating two different types of galaxy gives rise to parity breaking in the correlation function that derives from differences in the galaxies' properties and environments. This is typically associated with a difference in galaxy bias, describing the relation between galaxy number density and dark matter density, although observational effects such as magnification bias also play a role. In this chapter we show that the presence of a screened fifth force adds additional degrees of freedom to the correlation function, describing the effective coupling of the force to the two galaxy populations. These are also properties of the galaxies' environments, but with different dependence in general to galaxy bias. {We show that the parity-breaking correlation function can be calculated analytically, under simplifying approximations, as a function of fifth-force strength and the two populations' fifth-force charges, and explore the result numerically using Hu-Sawicki $f(R)$ as a toy model of chameleon screening.} We find that screening gives rise to an octopole, which, in the absence of magnification bias, is not present in any gravity theory without screening and is thus a qualitatively distinct signature. The modification to the dipole and octopole can be $\mathcal{O}(10\%)$ and $\mathcal{O}(100\%)$ respectively at redshift $z \gtrsim 0.5$ due to screening, but decreases towards lower redshift. The change in the background power spectrum in $f(R)$ theories induces a change in the dipole of roughly the same size, but dominant to the effect of screening at low $z$. While current data is insufficient to measure the parity-breaking dipole or octopole to the precision required to test these models, future surveys such as DESI, Euclid and SKA have the potential to probe screened fifth forces through the parity breaking correlation function. 

\item In chapter \ref{PDCM} we test the assumptions made in cosmological data analysis and parameter inference.     Cosmological large-scale structure analyses based on two-point correlation functions often assume a Gaussian likelihood function with a fixed covariance matrix. We study the impact on cosmological parameter estimation of ignoring the parameter dependence of this covariance matrix, focusing on the particular case of joint weak-lensing and galaxy clustering analyses. Using a Fisher matrix formalism (calibrated against exact likelihood evaluation in particular simple cases), we quantify the effect of using a parameter dependent covariance matrix on both the bias and variance of the parameters. We confirm that the approximation of a parameter-independent covariance matrix is exceptionally good in all realistic scenarios. The information content in the covariance matrix (in comparison with the two point functions themselves) does not change with the fractional sky coverage. Therefore the increase in information due to the parameter dependent covariance matrix becomes negligible as the number of modes increases. Even for surveys covering less than 1\% of the sky, this effect only causes a bias of up to ${\cal O}(10\%)$ of the statistical uncertainties, with a misestimation of the parameter uncertainties at the same level or lower. The effect will only be smaller with future large-area surveys. Thus for most analyses the effect of a parameter-dependent covariance matrix can be ignored both in terms of the accuracy and precision of the recovered cosmological constraints. 

\end{itemize}

\section{Ongoing work}

There are various natural follow up studies to the work present in this thesis, several of which are currently in progress. 

\subsection{Future of initial conditions}

The initial conditions for the scalar perturbations only focused on adiabatic perturbations. 
In principle we can also perform the same analysis for isocurvature perturbations by allowing their primordial power spectrum to also be a set of independent bins in $k$ space. 
Furthermore, to find the modes that are best constrained a principal component analysis can be performed. 
Finally, in our analysis so far we have only used the CMB as a probe to constrain the decaying modes. 
In principle one can also use other data sets such as galaxy clustering and lensing and compute the Fisher matrix for those as well. 

Since we have real data from Planck we can constrain the amplitude of the decaying modes directly from data as {opposed} to using a Fisher approach. 
In addition, we assume that the rest of the cosmological parameters are fixed when the amplitudes of the decaying modes were analysed. 
While this was a more conservative approach, as we effectively reduce the degrees of freedom to fit to the data by fixing the cosmological parameters, it is not the most accurate approach. 
In general, we should constrain the decaying mode along with the cosmological parameters and then marginalise over the cosmological parameters to get the final posteriors for the decaying mode amplitude. 

\subsection{Advanced statistical methods}

New data sets such as those from large galaxy surveys and weak lensing surveys are becoming increasingly large and complex. 
Machine learning techniques are ubiquitous in almost every field now, from applications in self driving cars to analysing speech and text. 
The fundamental use of machine learning is in making \emph{predictions}\footnote{A very nice book describing an economist view on the use of machines can be found in this book: http://predictionmachines.ai }. 
Machine learning techniques have been used in cosmology to solve a range of prediction problems, for instance to predict the formation of large scale structure as supposed to using costly N-body simulations \cite{Berger:2018aey}. 

\begin{figure}[h!]
\begin{centering}
\includegraphics[scale=0.2]{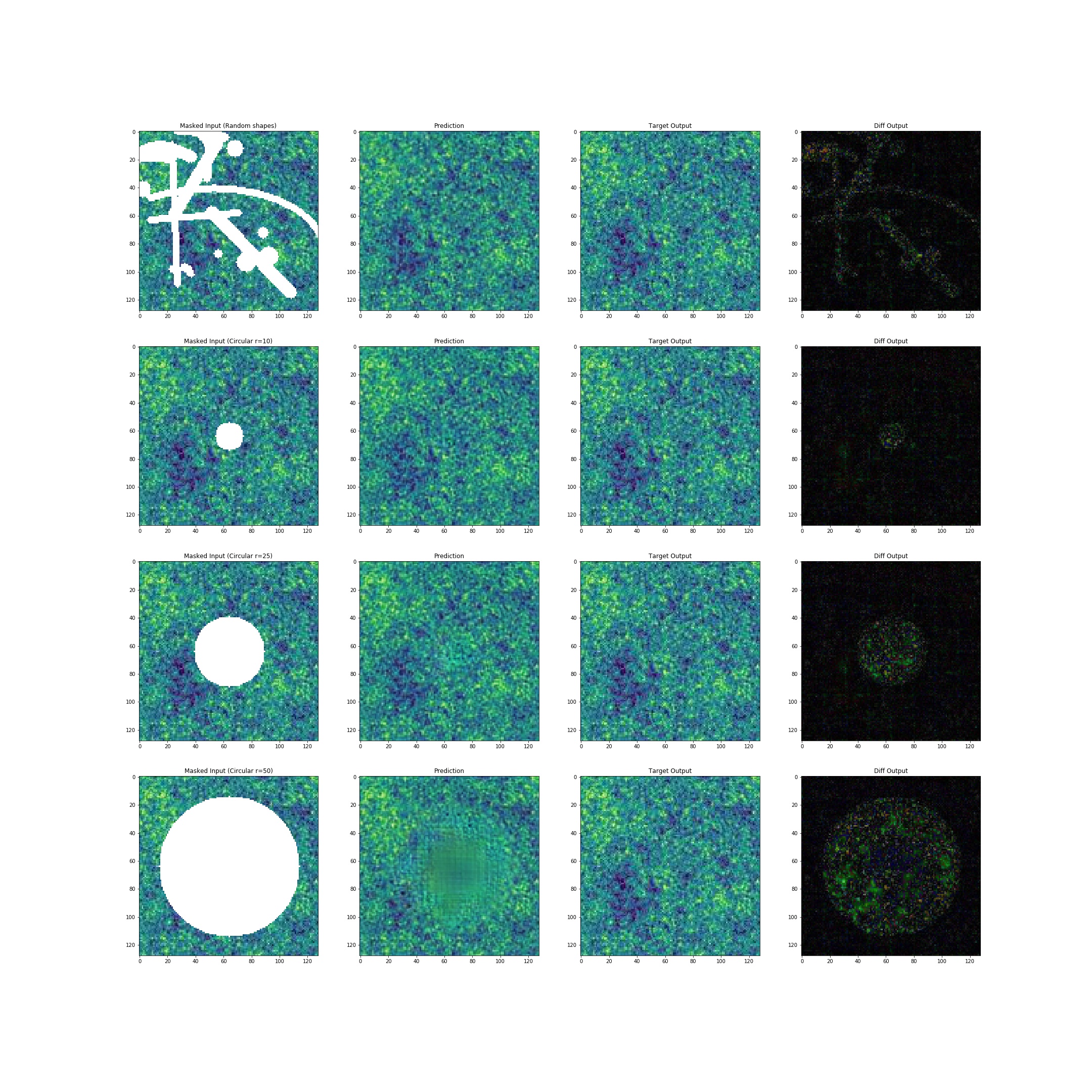}
\caption{Reconstructing CMB images}
\label{Gans}
\end{centering}
\end{figure}

We have attempted to use it to analyse images of the CMB temperature anisotropies. 
Images of the CMB can often be masked due to foreground sources or telescope noise. 
This leaves images with holes/missing values and this limits the amount of information available to construct the power spectra for temperature anisotropies. 
A method to reconstruct masked images can potentially increase the amount of information in CMB maps and thus increase constraining power. 
In an upcoming project we use a variant of Generative Adversarial Networks \cite{NIPS2014_5423} to reconstruct CMB images and temperature power spectra. 
Examples of the reconstructed images can be seen in figure \ref{Gans}. 

\section{Final thoughts}

\begin{figure}[h!]
\begin{centering}
\includegraphics[scale=0.45]{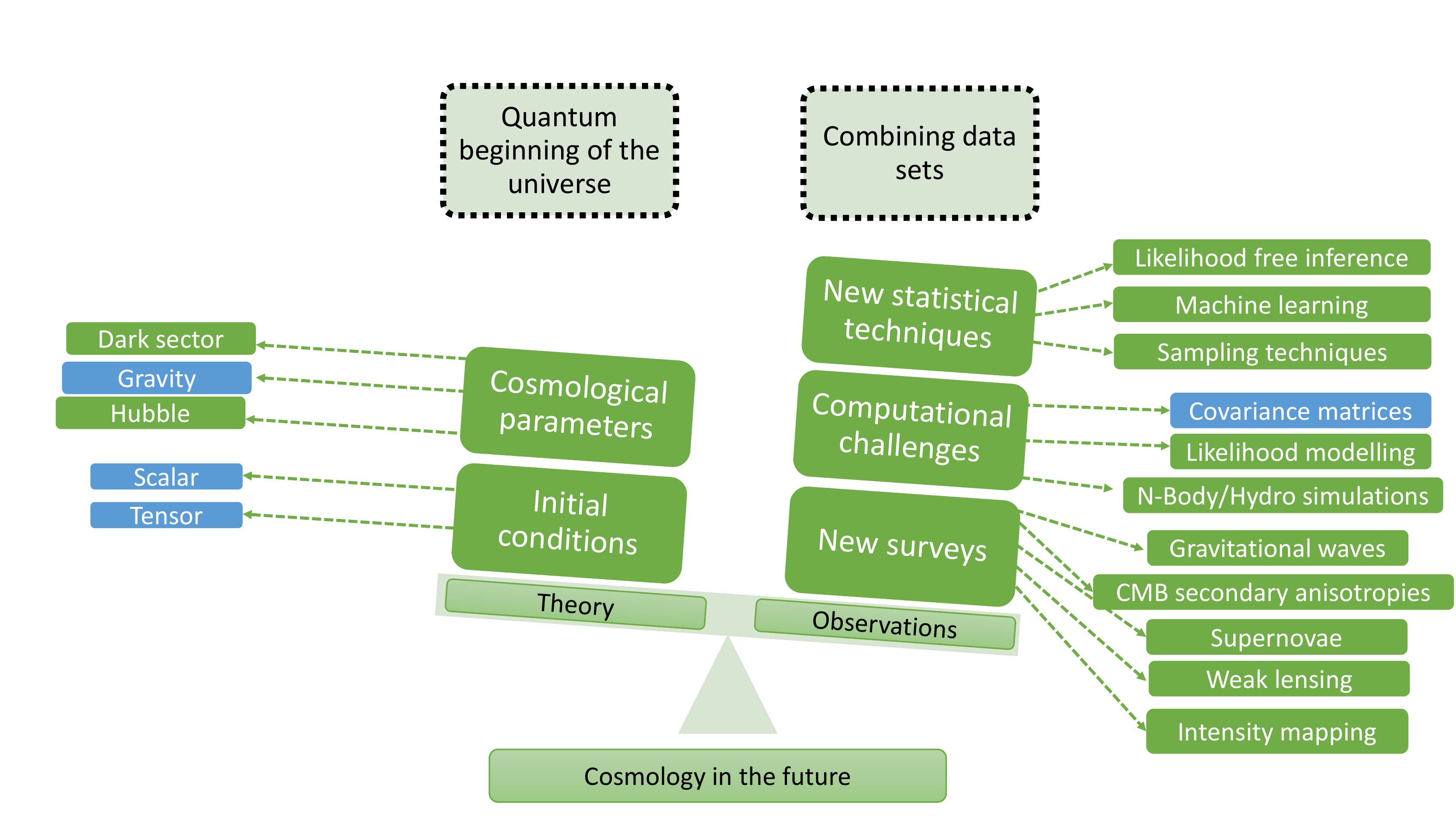}
\caption{Schematic diagram showing a non-exhaustive list of areas in cosmology that are likely to be developed in the future. The topics shaded in blue are the ones that have been partly addressed in this thesis. As the sketch shows, there are many more topics that need to be addressed. With the advent of many new observational surveys such as DESI \cite{DESI}, SKA \cite{Bull:2018lat}, CHIME \cite{chime}, LSST \cite{Mandelbaum:2018ouv} and many more that will probe various different aspects of cosmology, it is likely that the progress in the future will be driven by tackling challenges related to observational cosmology. There are many areas that need to be explored in theoretical cosmology as well. Predominantly they are related to finding a model that describes the beginning of the universe, beyond the radiation domination era, and a model that can describe the dark sector physics. Ideally we would want a model that can describe both the early universe and the late universe (dark sector) physics. While coming up with these models can be difficult, generalised frameworks exist in which the phenomenology of a class of models can be worked out, for instance the Horndeski parametrisation of scalar tensor theories \cite{Kobayashi:2019hrl} and the effective field theory of inflation \cite{Cheung:2007st}. At the top of the figure, in dotted boxes, are two areas which may be crucial to understand. The first is the theoretical framework to test for entanglement in the early universe. This can in principle be done in a model independent way and can definitively answer the question of whether the structure in the universe originates from quantum fluctuations. Understanding how to combine datasets that measure different properties of the universe on large scales to infer cosmological parameters is going to be the other crucial question that will need answering to harness the full power of future observations. }
\label{Schematic_cosmo}
\end{centering}
\end{figure}

Just over 50 years ago Penzias and Wilson made the first experimental measurement that could directly be used to test cosmological models{\footnote{One could even argue that Hubble starting testing the basic models of cosmology by studying the expansion of the universe - however this was far from the precision science we study today.}}. 
The field of cosmology has evolved tremendously since then, both with new theoretical understandings and new observational probes. 
This thesis has only focused on a small subset of both theoretical and observation problems in our current understanding of cosmology. 
A schematic diagram of the major problems in our current understanding of cosmology {is} shown in figure \ref{Schematic_cosmo}. 
In the future there will be many new cosmological probes, from gravitational waves observations to new galaxy surveys, that can help shed light on the open problems in cosmology. 
To harness the full power of different datasets it will be crucial to have a sound understanding about how to combine different data sets, particularly when they measure different quantities such as the lensing of galaxies and the polarisation anisotropies in the CMB. 
At present there is a tension between the value of the Hubble parameter inferred from large scale cosmological probes such as the CMB and local measurements of the expansion of the universe such as supernovae measurements \cite{Riess:2016jrr, Aghanim:2018eyx, Riess:2018byc, Bennett:2014tka, Desmond:2019ygn}. 
This is just one example of a case where having a thorough understanding of how data sets should be combined could be very important for understanding the true nature of this tension, i.e if it is a physical effect or a misunderstanding of the observations. 

On the theory side, understanding the very early universe has a been a topic that has been dominated by the inflationary paradigm. 
While this has been hugely successful, the observational evidence of inflation is still limited. 
As we have pointed out in this thesis, it is possible that novel early universe physics, such as bouncing/cyclic models, could also explain current CMB observations. 
As constraining individual models can be time consuming, it is important to find general frameworks/questions that can augment our understanding of the universe and its origins. 
An example of such a question is whether the universe started in a quantum state. 
Most theorists will believe this to be true at some level, regardless of what model is used to describe the early universe \cite{Brandenberger:1990bx, Kiefer:1998qe, Koksma:2011xia}. 
Testing such a general question can be difficult without focusing on a particular model. 
A novel approach to this could be to look for signs of entanglement between the quantum fields in the early universe. 
As entanglement has no classical analog, detecting it could lead to a fundamental breakthrough in our understanding about the quantum nature of our universe. 
Recent attempts at calculating the effect of entanglement on observables such as the CMB anisotropies can be found in \cite{Albrecht:2014aga, Bolis:2016vas, Bolis:2018jmo, Bolis:2019fmq}.
Many theories, from inflation, modified gravity and cyclic models, can give rise to primordial perturbations from quantum fluctuations. 
Thus, \emph{going forward it will be important to analyse the observations from various cosmological probes consistently and objectively while keeping an open mind about what theory best describes the observations in order to find out the true origin and evolution of our universe.}

\appendix
\chapter{Computing the SSC using the Halo model}\label{app:SSC}

   This appendix presents a complete description of the steps taken to estimate the super-sample contribution to the large-scale structure covariance matrix. 
    
    We follow the description in \cite{Krause:2016jvl}, and write the SSC in terms of the window functions of the tracers involved
    \begin{equation}\label{eq:ssc_full}
      \left({\bf \Sigma}_{\rm SSC}\right)^{ij,\ell_1}_{mn,\ell_2}=\int d\chi\frac{q^i(\chi)q^j(\chi)q^m(\chi)q^n(\chi)}{\chi^4} {\cal R}(k_{\ell_1},z){\cal R}(k_{\ell_2},z)\,P(k_{\ell_1},z)P(k_{\ell_2},z) \sigma_b(f_{\rm sky},z), 
    \end{equation}
    where the window functions $q^p$ are given by Eqs. \ref{eq:q_delta} and \ref{eq:q_lensing} for galaxy clustering and weak lensing respectively, $k_\ell\equiv(\ell+1/2)/\chi$, $z$ is the redshift at the comoving radial distance $\chi$ and $P(k,z)$ is the matter power spectrum. ${\cal R}(k,z)$ is the response of the matter power spectrum to a long-wavelength density perturbation $\delta_b$
    \begin{figure}
      \centering
      \includegraphics[scale = 0.7]{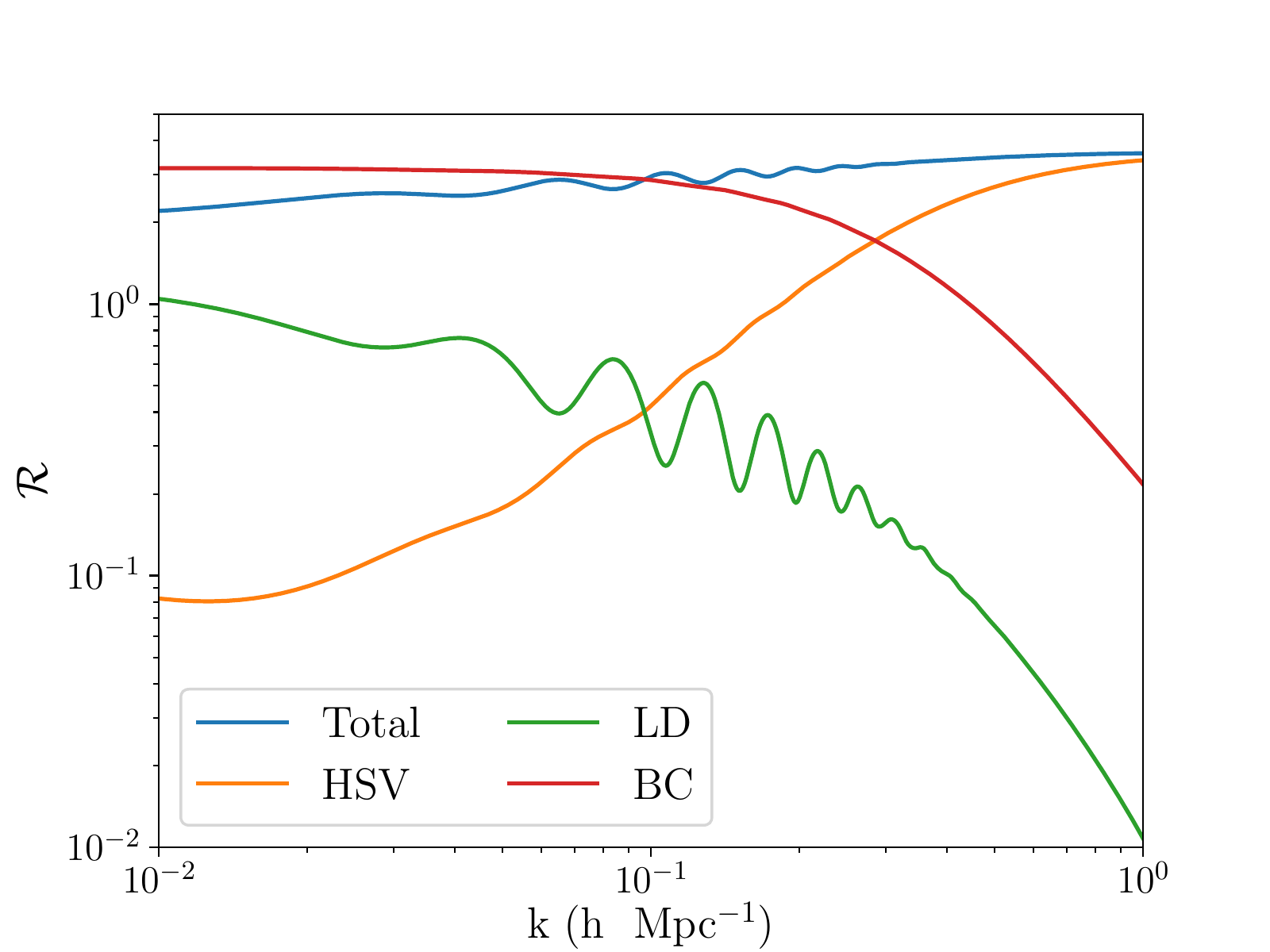}
      \caption{Different contributions to the power spectrum response at $z=0$ computed using the halo model (see Equation \ref{eq:response_halomod})}
      \label{fig:Response}
    \end{figure}
    \begin{figure}
      \begin{center}
        \includegraphics[scale = 0.7]{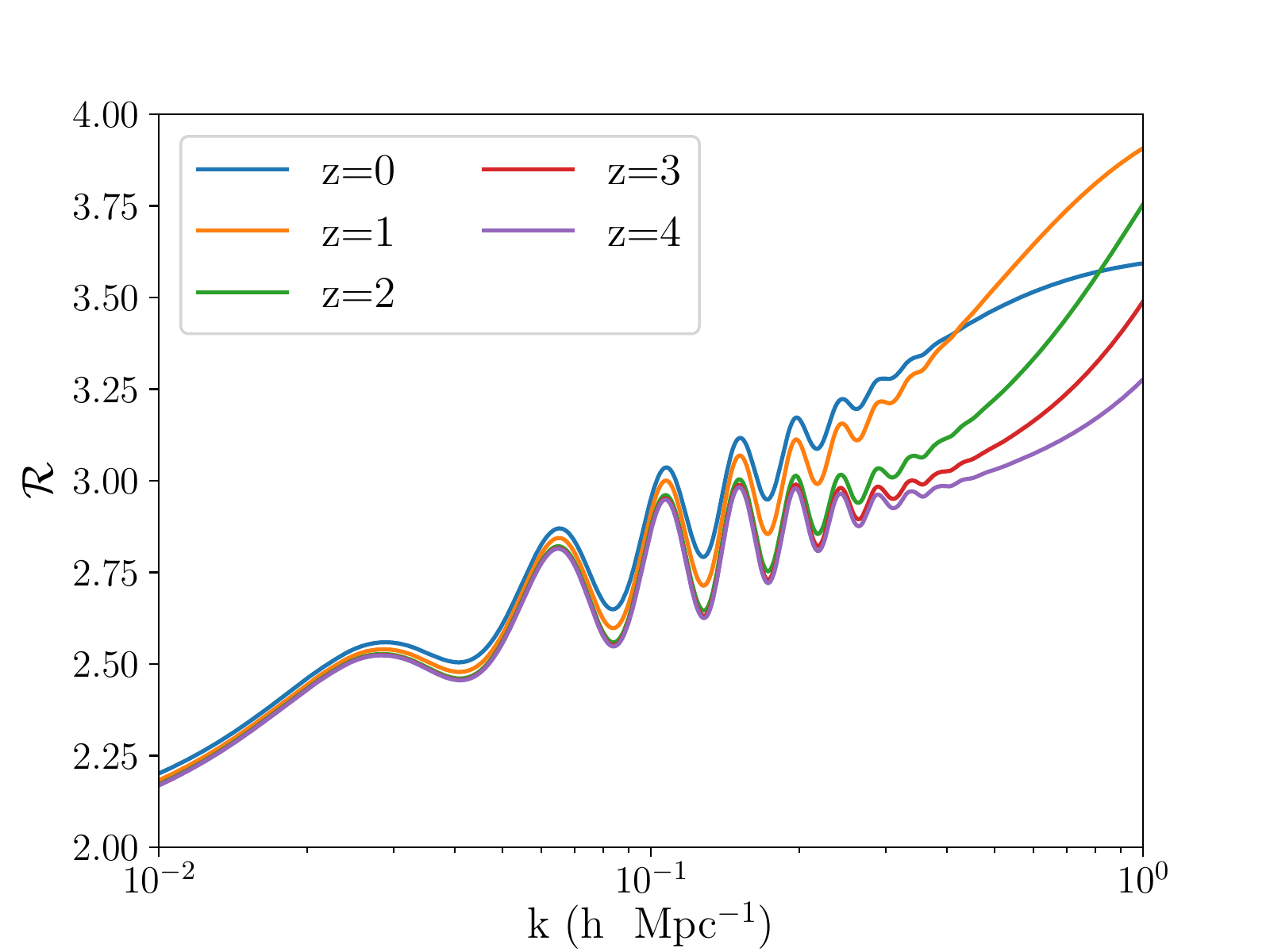}
        \caption{Power spectrum response as a function of redshift computing using the halo model.}
        \label{fig:Response_z}
      \end{center}
    \end{figure}
    \begin{equation}
      {\cal R}(k,z)\equiv\frac{d\log P(k,z)}{d\delta_b},
    \end{equation}
    and $\sigma_b$ is the projected variance of the density field within the footprint of your survey. For simplicity we assume a connected disc with an area corresponding to the sky fraction covered, in which case $\sigma_b$ can be estimated as
    \begin{equation}
     \sigma_b(f_{\rm sky},z)=\int dk_\perp^2 P_L(k_\perp,z)\left|\frac{2J_1(k_\perp R_s)}{k_\perp R_s}\right|^2,
    \end{equation}
    where $P_L$ is the linear matter power spectrum, $J_1$ is the order-1 cylindrical Bessel function and
    \begin{equation}
      R_s\equiv \chi(z)\,\theta_s,\hspace{12pt}\theta_s\equiv{\rm arccos}(1-2f_{\rm sky}).
    \end{equation}
    The expressions above reproduce the results of previous work by \cite{Krause:2016jvl}, in the simplified scenario of linear galaxy bias. We refer the reader to these works for further details.
    
    A standard approach to compute the power spectrum response ${\cal R}$ is to make use of so-called ``Separate-Universe'' simulations \cite{Li:2014sga, Hamilton:2005dx} in which the effects of a long-wavelength mode are modelled as modifying the effective background cosmology. Since our results will not depend strongly on the details of this calculation, we proceed as in \cite{Krause:2016jvl} and estimate ${\cal R}$ using a halo model approach. 
    
    \begin{figure}
      \begin{center}
        \includegraphics[scale = 0.7]{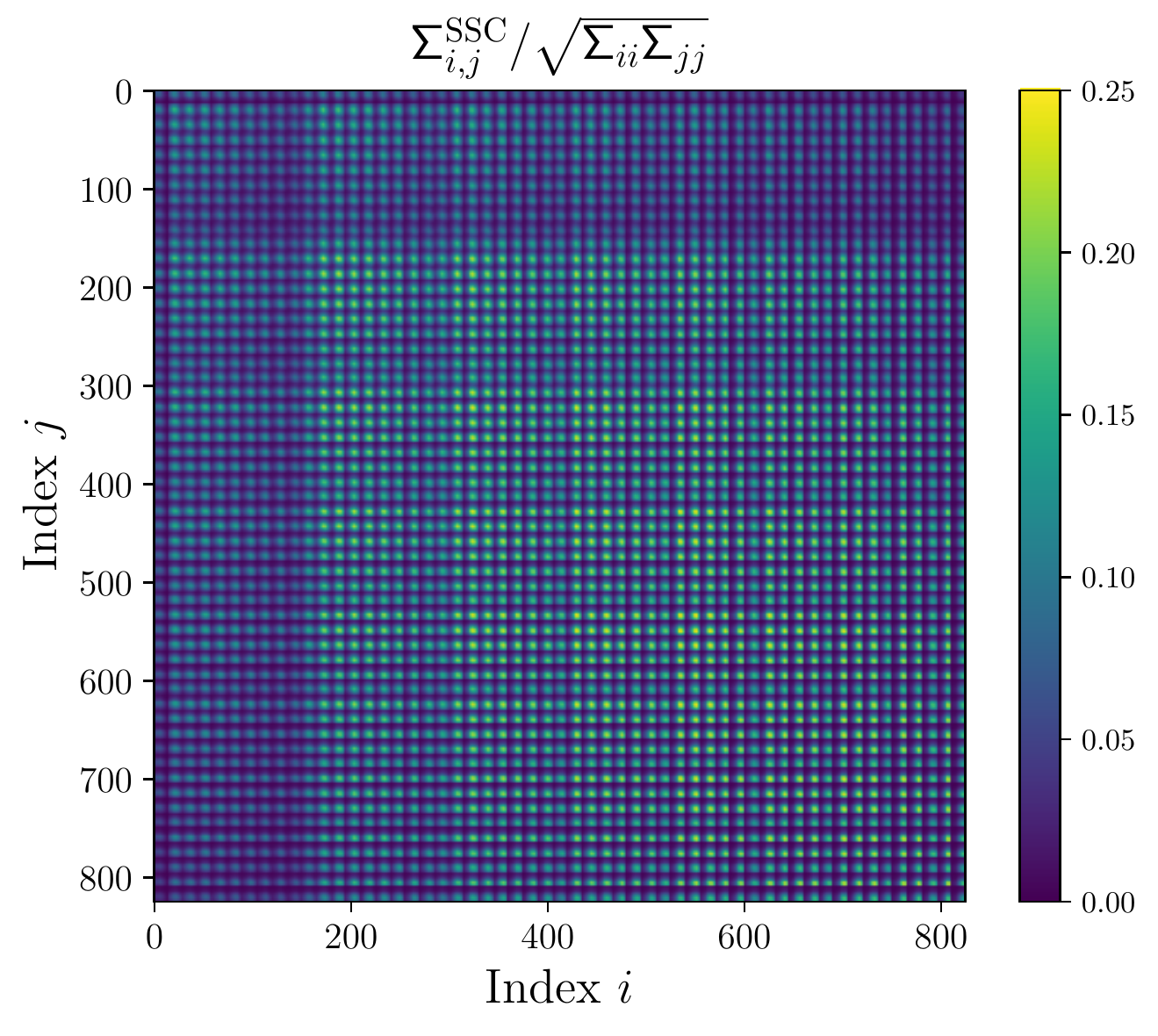}
        \caption{Contribution from the SSC term to the lensing-only part of the correlation matrix of the data vector described in Section \ref{ssec:forecasts.surveys} for a survey covering 10\% of the celestial sphere.}
        \label{fig:full_cov_lensing}
      \end{center}
    \end{figure}
    As a brief review and to introduce the notation, we provide the details of the Halo model below \cite{Li:2014sga}. In the Halo model the matter power spectrum can approximated as 
    \begin{eqnarray}
      P(k)  & = & P_{1h}(k) + P_{2h}(k)	\nonumber	\\
      & \equiv & \mathcal{I}^0_2(k) + \left[\mathcal{I}^1_1(k)\right]^2P_L(k)
    \end{eqnarray}
    where $P_{1h}(k)$ represents the contribution from correlations within a single halo and $P_{2h}(k)$ is the contribution from separate halos correlated by the linear power spectrum. Explicitly these are defined in terms of Halo model kernels $\mathcal{I}^\alpha_\beta(k)$.
    \begin{equation}
      \mathcal{I}^\alpha_\beta(k) \equiv \int dM\,\frac{dn}{dM}\,b^\alpha(M)\,\left[\frac{M}{\bar{\rho}_M}u(k|M)\right]^\beta
    \end{equation}
    where $dn/dM$ is the Halo mass function and $b(M)$ is the halo bias. Both quantities were estimated using the fits from \cite{Tinker:2008ff}. $u(k|M)$ is the Fourier transform of the halo density profile for a halo of mass $M$, which we model assuming a NFW profile with a concentration-to-mass relation given by \cite{Li:2014sga}.
    
    In terms of these quantities, the response of the power spectrum to the background density is given by \cite{Li:2014sga}
    \begin{eqnarray}\label{eq:response_halomod}
      & & {\cal R}(k) = {\cal R}_{\rm BC}+{\cal R}_{\rm HSV}+{\cal R}_{\rm LD}\\
      & & {\cal R}_{\rm BC}(k) = \frac{68}{21} \frac{P_{2h}(k)}{P(k)}\\
      & & {\cal R}_{\rm HSV}(k) = -\frac{1}{3} \frac{d \log [k^3 P_{2h}(k)]}{d \log{k}} \frac{P_{2h}(k)}{P(k)}\\
      & & {\cal R}_{\rm LD}(k) = I^1_2(k)
    \end{eqnarray}
    
    A few comments on these terms are in order.
    \begin{itemize}
      \item The beat coupling (BC) term is the one that has been studied extensively in the literature. It represents the fact that a short wavelength mode will grow more in the presence of a larger background density (i.e the large scale mode outside the window). 
      \item The halo sample variance (HSV) term shows the change in the number of Halos in the presence of a large scale density mode. 
      \item A large scale density mode will change the local scale factor and thus change the physical size of a mode that would otherwise be smaller. This is known as the linear dilation (LD) effect. 
    \end{itemize}
    These different contributions at $z=0$ are shown in Figure \ref{fig:Response}, and the mild redshift evolution of the overall response is presented in Figure \ref{fig:Response_z}. 

    Using this response we can compute the SSC matrix as shown in Figure \ref{fig:full_cov_lensing}. It is worth noting that, when evaluating the parameter dependence of the covariance matrix, we vary all terms entering Eq. \ref{eq:ssc_full} except for ${\cal R}$, which we keep fixed.

\chapter{Derivation of screened CF}\label{corr_funcs}

In this section we calculate the relativistic component of the two-point CF, $\xi^{rel(\mathcal{F})}$, in a theory with a screened fifth force:
\begin{equation}
	\xi^{rel(\mathcal{F})}(z, z', \theta) = \langle \Delta^{st}_B(z, \hat{\textbf{n}}) \Delta^{rel(\mathcal{F})}_F (z', \hat{\textbf{n}}') \rangle + \langle \Delta^{st}_F(z', \hat{\textbf{n}}') \Delta^{rel(\mathcal{F})}_B (z, \hat{\textbf{n}}) \rangle \label{corr_in}.
\end{equation}
We can substitute the expressions in Eqs (\ref{dst}, \ref{drel}) into Eq (\ref{corr_in}). We work in Fourier space and use the following convention for the Fourier transform of some function $f$:
\begin{equation}
	FT[f(x, \eta)] \equiv \frac{1}{(2\pi)^3} \int d^3k \ e^{-i \textbf{k} \cdot \textbf{x}} \ \mathbb{F} (\textbf{k}, \eta)
\end{equation}
We describe the Fourier transform of the density and velocity as
\begin{eqnarray}
	& & FT[\delta(x,\eta)] = \mathbb{D}(k, \eta) \nonumber \\
	& & FT[v(x,\eta)] = \mathbb{V}(k, \eta). \nonumber \\
\end{eqnarray}
These can be directly related to the transfer functions for the metric potential and the underlying initial metric perturbations $\Psi_i$. 
\begin{eqnarray}
	& & \mathbb{D}(\textbf{k}, \eta) = T_D(k, \eta) \Psi_i(\textbf{k}) \nonumber \\
	& & \mathbb{V}(\textbf{k}, \eta) = T_V(k, \eta) \Psi_i(\textbf{k}) \nonumber \\
	& & T_\Psi = T_\Phi = \frac{D(a)}{a} T(k) \nonumber \\
	& & T_D = - \frac{2a}{3 \Omega_m} \left( \frac{k}{\mathcal{H}_0} \right)^2 T_\Psi = - \frac{2}{3 \Omega} \left( \frac{k}{\mathcal{H}_0} \right)^2 D(a) T(k) \nonumber \\
	& & T_V = - \frac{\dot{T}_D}{k} = \frac{2a \mathcal{H}}{3 \Omega_m \mathcal{H}_0} \frac{k}{\mathcal{H}_0} \left[ T_\Psi + \mathcal{H}^{-1} \dot{T}_\Psi \right]  = \frac{2}{3 \Omega_m} \frac{\mathcal{H}}{\mathcal{H}_0} \frac{k}{\mathcal{H}_0} f(a)D(a)T(k)
\end{eqnarray}
where in the last line we have followed the standard convention, following \cite{Bonvin:2018ckp},  of splitting the time-dependent component of the transfer function into the linear growth factor $D(a)$ and the scale-dependent component into a time-dependent transfer function $T(k)$. 
In general modified gravity theories, this type of separation may not be possible as the growth can be scale dependent however we don't include that in this analysis and leave that to future works. 
We need only calculate one of the terms in Eq. (\ref{corr_in}) as the other will be related to this under $B \leftrightarrow F$, $z \leftrightarrow z'$, $\hat{\textbf{n}} \leftrightarrow \hat{\textbf{n}}'$. 
\begin{equation}
	\langle \Delta^{st}_B(z, \hat{\textbf{n}}) \Delta_F^{rel(\mathcal{F})} (z', \hat{\textbf{n}}') \rangle = \langle \Delta^{st}_B(z, \hat{\textbf{n}}) \Delta_F^{rel} (z', \hat{\textbf{n}}') + \langle \Delta^{st}_B(z, \hat{\textbf{n}}) \Delta_F^{\mathcal{F}} (z', \hat{\textbf{n}}') \rangle
\end{equation}
We compute each of these terms individually. 
\begin{eqnarray}
	& & \langle \Delta^{st}_B(z, \hat{\textbf{n}}) \Delta_F^{rel} (z', \hat{\textbf{n}}') \rangle =\mathcal{T}^{(1)} - \mathcal{T}^{(2)}  \\ 
	& & \mathcal{T}^{(1)} = \int \frac{d \ln k}{(2 \pi)^3} (k \eta_0)^{n_s -1} G(r') T_V(k,r') b_B T_D(k,r) \mathcal{I}^{(1)}  \\
	& & \mathcal{I}^{(1)} = \int d \Omega_k e^{i\textbf{k}( \textbf{x} - \textbf{x}')} (i \hat{\textbf{k}} \cdot \hat{\textbf{n}}')  \\
	& & \mathcal{T}^{(2)} = \int \frac{d \ln k }{(2 \pi)^3} (k \eta_0)^{n_s -1} G(r') T_V(k,r') \frac{k}{\mathcal{H}(r)} T_V(k,r) \mathcal{I}^{(2)}  \\
	& & \mathcal{I}^{(2)}\int d \Omega_k e^{i \textbf{k}( \textbf{x} - \textbf{x}')} (i \hat{\textbf{k}} \cdot \hat{\textbf{n}}') (\hat{\textbf{k}} \cdot \hat{\textbf{n}})^2
\end{eqnarray}
where we have defined 
\begin{equation}
	F(r) \equiv \frac{ \dot{\mathcal{H}}}{\mathcal{H}^2} + \frac{2}{r\mathcal{H}(r)} + 5s_B(r) \left( 1 - \frac{1}{r \mathcal{H}(r)} \right).
\end{equation}
To compute the angular integrals we use the following identities
\begin{eqnarray}
	e^{i \textbf{k} (\textbf{x}' - \textbf{x})} & = & e^{i d\textbf{k} \cdot \hat{\textbf{n}} } = 4 \pi \sum_{LM} j_L(kd) Y^*_{LM}(\hat{\textbf{k}}) Y_{LM}(\hat{\textbf{n}}) \nonumber \\
	\hat{\textbf{k}} \cdot \hat{\textbf{n}} & = & \frac{4 \pi}{3} \sum^1_{m = -1} Y^*_{1m} (\hat{\textbf{k}}) Y_{1m} (\hat{\textbf{n}}) \nonumber \\
	(\hat{\textbf{k}} \cdot \hat{\textbf{n}})^2 & = & \frac{8 \pi}{15} \sum_{m = -2}^2  Y^*_{2m}(\hat{\textbf{n}}) Y_{2m}(\hat{\textbf{k}}) + \frac{1}{3},
\end{eqnarray}
and we also make use of the \emph{Gaunt} integral formula
\begin{eqnarray}
	G^{l_1, l_2, l_3}_{m_1, m_2, m_3} & \equiv & \int d \Omega Y_{l_1, m_1}(\hat{n}) Y_{l_2, m_2}(\hat{n}) Y_{l_3, m_3} (\hat{n}) \nonumber \\
	& = & \sqrt{\frac{(2l_1+1)(2l_2+1)(2l_3+1)}{4 \pi}} \begin{pmatrix} l_1 & l_2 & l_3 \\ 0 & 0 & 0 \end{pmatrix} \begin{pmatrix} l_1 & l_2 & l_3 \\ m_1 & m_2 & m_3 \end{pmatrix} \nonumber \\
\end{eqnarray}
where we have defined the usual Wigner 3j symbol.
Now we compute the angular integrals as
\begin{eqnarray}
	\mathcal{I}^{(1)} & = & - 4 \pi \cos \alpha j_1(kd) \nonumber \\
	\mathcal{I}^{(2)} & = & 4 \pi \left[ - \frac{2}{5} \sin \alpha \sin \beta \cos \beta \left[ j_1(kd) + j_3(kd) \right] + \frac{3}{5} \cos \alpha \cos(2 \beta)  \left[ \frac{j_3}{2} - \frac{j_1}{3} \right] \right. \nonumber \\
	& + & \left.\frac{1}{10} \cos \alpha \left[ j_3(kd) - j_1(kd) \right] \right].
\end{eqnarray}
Using these we can now compute the $k$ integrals
\begin{eqnarray}
	\mathcal{T}^{(1)} & = & \frac{2A_s}{9 \Omega^2 \pi^2} G(r') D(r)D(r') b_B(r) \cos \alpha \nu_1(d) f(r') \nonumber \\
	\mathcal{T}^{(2)} & = & \frac{2A_s}{9 \pi^2 \Omega^2_m} G(r') D(r) D(r') f(r) f(r') \left[ \frac{2}{5} \sin \alpha \sin \beta \cos \beta \left[ \nu_1(d) + \nu_3(d) \right]  \right.\nonumber \\
	 & + &  \left.  \frac{3}{5} \cos \alpha \cos(2 \beta) \left[ \frac{\nu_3(d)}{2} - \frac{\nu_1(d)}{3} \right] \frac{1}{10} \cos \alpha \left[\nu_3(d) - 4\nu_1(d) \right] \right]. \nonumber \\
\end{eqnarray}
Thus the final answer for the two-point CF is 
\begin{eqnarray}
	& & \langle \Delta^{st}_B(z, \hat{\textbf{n}}) \Delta_F^{rel}(z', \hat{\textbf{n}}') \rangle = \frac{2 A_s G(r') D(r) D(r') f(r')}{9 \pi^2 \Omega_m^2} \left[ \frac{2}{5} \sin \alpha \sin \beta \cos \beta [\nu_1(d) + \nu_3(d)]  \right. \nonumber \\
	& & \left. + \frac{3}{5} \cos \alpha \cos(2 \beta) \left[ \frac{\nu_3(d)}{2} - \frac{\nu_1(d)}{3} \right] + \frac{1}{10} \cos \alpha \left[ \nu_3(d) - 4 \nu_1(d) \right]  + b_B \nu_1(d) \cos \alpha\right]. \nonumber \\ \label{rel_temp1}
\end{eqnarray}
$\langle \Delta^{st}_F(z', \hat{\textbf{n}}') \Delta_B^{rel}(z, \hat{\textbf{n}}) \rangle$ is simply given by the relabelling of the indices and angles (which gives an rise to a sign difference in this term),
\begin{eqnarray}
	& & \langle \Delta^{st}_F(z', \hat{\textbf{n}'}) \Delta_B^{rel}(z, \hat{\textbf{n}}) \rangle = - \frac{2 A_s G(r) D(r') D(r) f(r)}{9 \pi^2 \Omega_m^2} \left[ \frac{2}{5} \sin \beta \sin \alpha \cos \alpha [\nu_1(d) + \nu_3(d)] \right.\nonumber \\
	& & \left.  + \frac{3}{5} \cos \beta \cos(2 \alpha) \left[ \frac{\nu_3(d)}{2} - \frac{\nu_1(d)}{3} \right] + \frac{1}{10} \cos \beta \left[ \nu_3(d) - 4 \nu_1(d) \right]  + b_F \nu_1(d) \cos \beta\right]. \nonumber \\ \label{rel_temp2}
\end{eqnarray}
Next we compute the CF between the standard term for $B$ galaxies and the fifth-force term for $F$ galaxies. 
\begin{eqnarray}
	& & \langle \Delta^{st}_B(z, \hat{\textbf{n}}) \Delta_F^{\cal F}(z', \hat{\textbf{n}}') \rangle = A \int \frac{d^3k}{(2 \pi)3} e^{i \textbf{k} \cdot ( \textbf{x}' - \textbf{x})} \frac{(k \eta_0)^{n_s -1}}{k^3} \left[ b_B T_D(k,r) - \frac{k}{\mathcal{H}(r) }(\hat{\textbf{k}} \cdot \hat{\textbf{n}})^2 T_V(k,r) \right]  \times \nonumber \\
	& &  \left[ \zeta_F(r') i(\hat{\textbf{k}} \cdot{n}') \left( \mathcal{H}^{-1} (r') \dot{T}_v (k,r) + T_v(k,r') \right) \right] \nonumber \\
	& & \equiv  {\cal T}^{(3)} - \mathcal{T}^{(4)},
\end{eqnarray}
where $\zeta_F \equiv \frac{\delta G_F}{1 + \delta G_F}$ is the fifth-force sensitivity for faint galaxies. Here we define
\begin{eqnarray}
	\mathcal{T}^{(3)} & = & A \int \frac{d \ln k}{(2 \pi)^3} \zeta_F(r') \left[ \mathcal{H}^{-1}(r') \dot{T}_V(k,r') + T_V(k,r') \right] b_B T_D(k,r) \mathcal{I}^{(1)} \nonumber \\
	\mathcal{I}^{(1)} & \equiv  & \int d \Omega_k e^{i \textbf{k} (\textbf{x}' - \textbf{x})} (i \hat{\textbf{k}} \cdot \hat{\textbf{n}}') \nonumber \\
	\mathcal{T}^{(4)} & = & A \int \frac{ d \ln k}{( 2\pi)^3} \zeta_F(r') \left[ \mathcal{H}^{-1} (r') \dot{T}_V(k,r') + T_V(k,r') \right] \mathcal{H}^{-1}(r) k T_V(k,r) \mathcal{I}^{(2)} \nonumber \\
	\mathcal{I}^{(2)} & \equiv & \int d \Omega_k e^{- i \textbf{k} (\textbf{x}' - \textbf{x})} ( i \hat{\textbf{k}} \cdot \hat{\textbf{n}}') ( \hat{\textbf{k}} \cdot \hat{\textbf{n}})^2.
\end{eqnarray}
The other ingredients we need are the transfer functions
\begin{eqnarray}
	& & \dot{T}_V = \frac{2}{3 \Omega_m \mathcal{H}_0} \left( \frac{k}{\mathcal{H}_0} \right) \left[ \ddot{a} T_\Psi + \dot{T}_\Psi \left[ \ddot{a} \mathcal{H}^{-1} + \dot{a} - \dot{a} \dot{\mathcal{H}} \mathcal{H}^{-2} \right] + \dot{a} \mathcal{H}^{-1} \ddot{T}_\Psi \right] \nonumber \\
	& = & \frac{2}{3 \Omega_m} \left( \frac{k}{\mathcal{H}_0}\right) T(k) \left[ \frac{\dot{\mathcal{H}}}{\mathcal{H}_0} f(a)D(a) + \frac{\mathcal{H}(a)}{\mathcal{H}_0} \left( \dot{f}(a)D(a) + f(a)^2 D(a) \mathcal{H}\right)  \right],
\end{eqnarray}
where $ f(a) \equiv \frac{ d \ln D(a)}{d \ln a}$.  Further we can use $\dot{D} = fD\mathcal{H}$. We then put these into $\mathcal{T}^{(3)}$ and $\mathcal{T}^{(4)}$:
\begin{eqnarray}
	\mathcal{T}^{(3)} & = & \frac{2 A b_B \zeta_F \cos \alpha}{9 \pi^2 \Omega_m^2} D(r) D(r') \int \frac{d \ln k}{(2 \pi)^3} \frac{(k \eta_0)^{n_s-1}}{k^3} \left( \frac{k}{\mathcal{H}_0} \right)^3 T(k)^2 \left[ \frac{\mathcal{H}(r')}{\mathcal{H}_0} f(r') \right. \nonumber \\
	& & \left. + \frac{\dot{\mathcal{H}(r')}}{\mathcal{H}_0} f + \frac{\mathcal{H}(r')}{\mathcal{H}_0} \dot{f}(r') + \frac{\mathcal{H}^2(r')}{\mathcal{H}_0(r')} f^2 \right] j_1(kd) \nonumber \\
	& = & \frac{2 \cos \alpha A b_B \zeta_F}{9 \pi^2 \Omega_m^2} M(r') \nu_1(d),
\end{eqnarray}
where we have defined 
\begin{eqnarray}
	& & \nu_\ell(d) \equiv \int \frac{dk}{k} (k \eta)^{n_s -1} \left( \frac{k}{\mathcal{H}_0} \right)^3 j_\ell(kd) T^2(k) \nonumber \\
	& & M(r') \equiv \frac{D^2(r')}{\mathcal{H}_0 \mathcal{H}} \left( \dot{\mathcal{H}(r')} f(r') + \mathcal{H}(r') \dot{f}(r') + f(r')^2 \mathcal{H}(r')^2 + f(r') \mathcal{H}(r')^2 \right).
\end{eqnarray}
Now we compute $\mathcal{T}^{(4)}$ as 
\begin{eqnarray}
	\mathcal{T}^{(4)} & = & A \int \frac{ d \ln k}{(2\pi)^3} \zeta_F(r') \left[ \frac{2}{3 \Omega_m} \left( \frac{k}{\mathcal{H}_0} T(k) \left[ \frac{\dot{\mathcal{H}}(r')}{\mathcal{H}_0} f(r') D(r')  \right. \right. \right. \nonumber \\
	& & \left. \left. \left. + \frac{\mathcal{H}(r')}{\mathcal{H}_0} \left( \dot{f}(r') D(r') + f(r')^2 D(r') \mathcal{H}(r') \right) \right] \right) \right] \nonumber \\
	& \times & b_B\left( \frac{k}{\mathcal{H}_0} \right) \left( \frac{k}{\mathcal{H}(r)} \right) \left[ \frac{2}{3 \Omega_m} \frac{\mathcal{H}(r)}{\mathcal{H}_0} f(r) D(r) T(k) \right] \nonumber \\
	& \times & 4 \pi \left[ - \frac{2}{5} \sin \alpha \sin \beta \cos \beta [j_1(kd)  + j_3(kd) ] + \frac{3}{5} \cos \alpha \cos(2 \beta) \left[ \frac{j_3(kd)}{2} - \frac{j_1(kd)}{3} \right] \right. \nonumber \\
	& & \left. + \frac{1}{10} \cos \alpha \left[ j_3(kd) - j_1(kd) \right] \right] \nonumber \\ 
	& = & \frac{2 A \zeta_F f(r') M(r')}{9 \pi^2 \Omega_m^2 \mathcal{H}(r')} \left[ - \frac{2}{5} \sin \alpha \sin \beta \cos \beta(\nu_1(d) + \nu_3(d)) + \frac{3}{5} \cos \alpha \cos (2 \beta) \left( \frac{\nu_3(d)}{2} - \frac{\nu_1(d)}{3} \right) \right. \nonumber \\
	& & \left. + \frac{1}{10} \cos \alpha (\nu_3(d) - \nu_1(d)) \right]. \nonumber \\
\end{eqnarray}
Putting these pieces together, we find
\begin{eqnarray}
	& & \langle \Delta^{st}_B(z,\hat{\textbf{n}}) \Delta^{(F)}_F(z', \hat{\textbf{n}}') \rangle = \frac{2A_sM(r') \zeta_F}{9 \pi^2 \Omega_m^2} \left[ b_B \cos \alpha \nu_1(d) \right.  \nonumber \\
	& & \left. + \frac{f(r')}{\mathcal{H}(r')} \left[ \frac{2}{5} \sin \alpha \sin \beta \cos \beta (\nu_1(d) + \nu_3(d))  - \frac{3}{5} \cos \alpha \cos(2 \beta) \left( \frac{\nu_3(d)}{2} - \frac{\nu_1(d)}{3} \right) \right. \right. \nonumber \\
	& & \left. \left.- \frac{1}{10} \cos \alpha (\nu_3(d) - \nu_1(d)) \right] \right]. \nonumber \\ \label{FF_temp1}
\end{eqnarray}
$\langle \Delta^{st}_F(z',\hat{\textbf{n}}') \Delta^{(F)}_B(z, \hat{\textbf{n}}) \rangle$ is given by the appropriate relabelling of the indices and angles:
\begin{eqnarray} 
	& & \langle \Delta^{st}_F(z',\hat{\textbf{n}}') \Delta^{(F)}_B(z, \hat{\textbf{n}}) \rangle = -\frac{2A_sM(r) \zeta_B}{9 \pi^2 \Omega_m^2} \left[ b_F \cos \beta \nu_1(d) \right. \nonumber \\
	& & \left.+ \frac{f(r)}{\mathcal{H}(r)} \left[ \frac{2}{5} \sin \beta \sin \alpha \cos \alpha (\nu_1(d) + \nu_3(d)) - \frac{3}{5} \cos \beta \cos(2 \alpha) \left( \frac{\nu_3(d)}{2} - \frac{\nu_1(d)}{3} \right)  \right. \right. \nonumber \\
	& & \left. \left. - \frac{1}{10} \cos \beta (\nu_3(d) - \nu_1(d)) \right] \right]. \nonumber \\ \label{FF_temp2}
\end{eqnarray}
By expanding (Eq.~\eqref{FF_temp1}, Eq.~ \eqref{FF_temp2}) and (Eq.~\eqref{rel_temp1}, Eq.~\eqref{rel_temp2}) to leading order in $d/r$ we obtain the expressions in Eq.~\eqref{eq:final} for $\xi^{rel}$ and $\xi^{(\mathcal{F})}$ respectively.

\chapter{General chameleon screening in Horndeski theory}\label{horndeski}

This section casts generic chameleon-screened scalar--tensor theories to Horndeski form. We anticipate that this will be useful for implementing a more general theory than $f(R)$ in the background.

The general action for a scalar--tensor theory that is immune to instabilities and has second order equations of motion is the Horndeski action, given by \cite{Horndeski, Bellini:2014fua}:
\begin{eqnarray}
	S & = & \int d^4x  \sqrt{-g} \left[ \sum_{i=2}^5 \mathcal{L}_i + \mathcal{L}_m[g_{\mu \nu}] \right] \nonumber \\
	\mathcal{L}_2 & = & G_2(\phi, X) \nonumber \\
	\mathcal{L}_3 & = & -G_3(\phi, X) \Box \phi \nonumber \\
	\mathcal{L}_4 & = & G_4(\phi, X) R + G_{4X} (\phi, X) \left[ (\Box \phi)^2 - \phi_{; \mu \nu} \phi^{; \mu \nu} \right] \nonumber \\
	\mathcal{L}_5 & = & G_5(\phi, X) G_{\mu \nu} \phi^{; \mu \nu} - \frac{1}{6} G_{5X}(\phi, X) \left[ (\Box \phi)^3  + 2 \phi_{; \mu}^\nu \phi_{;\nu}^\alpha \phi_{; \alpha}^\mu - 3 \phi_{; \mu \nu} \phi^{; \mu \nu} \Box \phi \right]. \label{horndeski_eq}
\end{eqnarray}
This is written in the Jordan frame, in which the Lagrangian components $\mathcal{L}_i$ determine the dynamics of the metric and the scalar field $\phi$. $X$ is the canonical kinetic term of a scalar field $- \frac{1}{2} g_{\mu \nu} \partial^\mu \phi \partial^\nu \phi$. 
The $G_i$ are free functions of the scalar and its kinetic term, and we denote their derivatives by $G_{iX} \equiv \partial_X G_i$.

Any scalar field minimally coupled to gravity has the action 
\begin{equation}
	\tilde{S} = \int d^4 x \sqrt{- \tilde{g}} \left[ \frac{M_{pl}^2}{2} \tilde{R} - \tilde{g}_{\mu \nu}\frac{1}{2} \partial^\mu \phi \partial^\nu \phi - V(\phi) - \mathcal{L}_m (g_{\mu \nu}) \right], 
\end{equation}
where variables are in the Jordan frame unless denoted by a tilde, in which case they are in the Einstein frame. In the Einstein frame, the scalar is decoupled from the metric and hence the gravitational part of the action is the same as in GR. Working in Planck units, we transform this action to the Jordan frame with the conformal transformation
\begin{eqnarray}
	\tilde{g}_{\mu \nu} &  = & e^{-2 \alpha \phi} g_{\mu \nu} \nonumber \\
	\tilde{g} & = & e^{-8 \alpha \phi} g \nonumber \\
	\tilde{R} & = & e^{2 \alpha \phi} \left[R - 6 \alpha^2 g_{\mu \nu} \partial^\mu \phi \partial^\nu \phi - 6 \alpha \Box \phi \right]
\end{eqnarray}
This yields the Jordan-frame action
\begin{eqnarray}
	S & = & \int d^4x \sqrt{-g} \: \exp\left( -2 \alpha \phi \right) \: \left[ \frac{1}{2} R + 3 \alpha \Box \phi - \left( \frac{1}{2} + 3 \alpha^2\right) g_{\mu \nu} \partial^\mu \phi \partial^\nu \phi \right.. \nonumber \\
	& & - \left. \exp (-2 \alpha \phi) V(\phi) - \mathcal{L}_m (g_{\mu \nu}) \right].
\end{eqnarray}
In this frame the scalar field has the Poisson equation $\Box \phi = \partial_\phi V(\phi) + \alpha \rho$.
The effective potential is then $V_\text{eff}(\phi) = V(\phi) + \exp(\alpha \phi) \rho$. To implement the chameleon mechanism $V(\phi)$ is chosen such that $V_\text{eff}$ has a sharp minimum, corresponding to high mass, in regions of high density, and a shallow minimum, corresponding to low mass, in regions of low density. A canonical example is $V(\phi) = \Lambda^{4+n}/\phi^n$ with $\Lambda$ an energy scale and $n$ an as-yet undetermined exponent \cite{Khoury:2013yya, Burrage}.\footnote{Note however that only some choices for $n$ result in chameleon screening. The mass is an increasing function of density, as required, if $n > 0$, $-1 < n < 0$ or $n$ is an even negative integer. $n = 0$ is simply a cosmological constant, $n = -1, -2$ does not make mass a function of density, and there is no minimum of $V_\text{eff}$ when $n = -3, -5, -7, ...$.}

Comparing to the Horndeski form (Eq (\ref{horndeski_eq})) we find: 
\begin{eqnarray}
	&G_2 = & - \exp\left( -4 \alpha \phi \right) V(\phi) + \exp\left( -2 \alpha \phi \right) X \left( 1 + 6 \alpha^2 \right) \nonumber \\
	&G_3 = & -3 \alpha \exp \left( -2 \alpha \phi \right) \nonumber \\
	&G_4 = & \frac{1}{2} \exp \left( -2 \alpha \phi \right) \nonumber \\
	&G_{4X} = & G_5 = G_{5X} = 0
\end{eqnarray}
We note that the recent neutron star merger that constrains the speed of gravitational waves to be the same of the speed of light implies $G_5, G_{4X} = 0$ \cite{Noller:2018eht} and thus our action is almost as general as possible given this constraint.
The $f(R)$ Hu-Sawicki model corresponds to the range $-1 < n < -1/2$ \cite{Burrage}, with $k=1$ corresponding to $n=-1/2$. Future work could explore the full parameter space of chameleon screening by implementing this action in a Boltzmann code such as HiClass \cite{hiclass}.

\addcontentsline{toc}{chapter}{Bibliography}
\renewcommand{\bibname}{References}
\bibliography{ref_dphil}        
\bibliographystyle{unsrt}  

\end{document}